# INTERLANGUAGES AND SYNCHRONIC MODELS OF COMPUTATION.

Alex V Berka.
Isynchronise Ltd.

In memory of
Basia and Franek Berka.

**SUMMARY**

It is suggested that natural language has a major structural defect, and is inappropriate as a template for formal and programming languages, and as a basis for deriving models of computation. A novel language system has been developed, which has given rise to promising alternatives to standard formal and processor network models of computation, and to the systolic programming of reconfigurable arrays of Arithmetic Logic Units. A textual structure called the *interstring* is proposed, which is a string of strings of simple expressions of fixed length. Unlike a conventional tree language expression, an interstring linked with an abstract machine environment, can represent sharing of sub-expressions in a dataflow, and a program incorporating data transfers and spatial allocation of resources, for the parallel evaluation of dataflow output. Formal computation models called the *α-Ram* family, are introduced, comprising synchronous, reconfigurable machines with finite and infinite memories, designed to support interstring based programming languages (*interlanguages*). Distinct from dataflow, visual programming, graph rewriting, and FPGA models, α-Ram machines' instructions are bit level and execute in situ without fetch. They support high level sequential and parallel languages without the space/time overheads associated with the Turing Machine and λ-calculus, enabling massive programs to be simulated. The elemental devices of one α-Ram machine, called the *Synchronic A-Ram*, are fully connected and simpler than FPGA look up tables. With the addition of a mechanism for expressing propagation delay, the machine may be seen as a formal model for sequential digital circuits and reconfigurable computing, capable of illuminating issues in massive parallelism.

A compiler for an applicative-style, interlanguage called *Space*, has been developed for the Synchronic A-Ram. Space can express coarse to very fine grained MIMD parallelism, is modular, strictly typed, and deterministic. Barring operations associated with memory allocation and compilation, Space modules are referentially transparent. A range of massively parallel modules have been simulated on the Synchronic A-Ram, with outputs as expected. Space is more flexible than, and has advantages over existing graph, dataflow, systolic, and multi-threaded programming paradigms. At a high level of abstraction, modules exhibit a small, sequential state transition system, aiding verification. Composable data structures and parallel iteration are straightforward to implement, and allocations of parallel sub-processes and communications to machine resources are implicit. Space points towards a range of highly connected architectural models called *Synchronic Engines*, with the potential to scale in a globally asynchronous, locally synchronous fashion. Synchronic Engines are more general purpose than systolic arrays and GPUs, and bypass programmability and resource conflict issues associated with processor networks. If massive intra chip, wave-based interconnectivity with nanosecond reconfigurability becomes available, Synchronic Engines will be in favourable position to contend for the TOP500 parallel machines.

## Table of Contents.

**Chapter 6. Space Lines and Base Set.**

**Chapter 7. Space Construct Set.**

**Chapter 8. Architectures for Synchronic Computation.**

**Appendix B. Miscellaneous Earth and Space programs.**

# Chapter 1.

## THE SPATIAL CHALLENGE TO TREE AND GRAPH BASED COMPUTATION.

### 1.1 INTRODUCTION.

Consider the hypothesis that trees and graphs have not in themselves alone, revealed an optimal linguistic environment in which to represent formal structures that possess shared parts, and require some form of computation or transformation, such as dataflow. The current work may be summarised as an attempt to identify such an environment, and then to use it as a foundation for a novel computational paradigm, incorporating low level and intermediate formal models, up to and including massively parallel programming models and machine architectures. Described in this report, the implementation of a viable, general purpose parallel programming environment on top of a simple, highly connected formal model of computation, without excessive space or time overheads, provides a foundational framework for reconfigurable synchronous digital circuits, and coarse grained arrays of ALUs (CGAs). In so doing, an alternative to the systolic approach to programming and controlling CGAs is attained, which delivers a novel paradigm of general purpose, high performance computation.

As an introduction to a new branch in computer science, the report refers to future work to provide context. Taking a more mathematical perspective than the main body of the report, 9.5 outlines how the interstring structure defined in chapter 2, together with the $\alpha$-Ram models described in chapter 3, suggest an alternative to the constructive type theoretical approach for introducing time and computation into mathematical discourse. A future paper will argue that type theory's overall agenda is constrained by focusing on conventional tree based logics and calculi.

The report questions two outlooks associated with the multi-processor paradigm of parallel computing. Firstly, that the Von Neumann sequential thread and architectural model, are suitable building blocks respectively, for a general purpose parallel programming model, and a parallel computing architecture. Secondly, that the absence of faster than light communication, suggests that asynchrony and non-determinism are fundamental to parallel programming frameworks. Without originally intending to do so, the consideration of linguistic issues has led to an espousal for synchronous and deterministic approaches to parallel programming, and highly connected aggregates of ALUs as parallel architectures.

In chapter 8, a set of mostly synchronous architectural models with low area complexity high speed interconnects called *Synchronic Engines* are outlined, possessing spatially distributed, yet deterministic program control. Synchronic Engines are embryonic efforts at deriving architectures from a formal model of computation called the *Synchronic A-Ram* defined in chapter 3, inspired by the interlanguage environment presented in chapter 2.

An interstring is a set-theoretical construct, designed for describing many-to-many relationships, dataflows, and simultaneous processes. It may be represented as a string of strings of symbol strings, where the innermost strings are short and have a maximum length. Interstring syntax is confined to a strictly limited range of tree forms, where only the rightmost, and the set of rightmost but one branches are indefinitely extendable. In conjunction with a simple, denotational machine environment, an interstring can efficiently express at an intermediate syntactic/semantic level, sharing of subexpressions in a dataflow, data transfers, spatial allocation of machine resources, and program control for the parallel processing of complex programs. Languages based on interstrings are called *interlanguages*[1]. Although not incorporated in the current implementation, an interlanguage compiler may duplicate the implicit parallelism of Dataflow Models (see 2.3.3), where arithmetic operations from differing layers in a dataflow are triggered simultaneously, if outputs from operations in earlier layers become available soon enough.

In contrast with dataflow and visual programming formalisms, interlanguages are purely textual, making them directly amenable for digital representation and manipulation. The report explains how interlanguages, and more generally interlanguages based on more deeply nested string structures, where some inner strings are restricted to having a maximum length, are also useful for representing data structures intended to be processed in parallel.

The Synchronic A-Ram is a globally clocked, fine grained, simultaneous read, exclusive write machine. It incorporates a large array of registers, wherein the transmission of information between any two registers or bits occurs in constant time. Although problematic from a physical standpoint, it will be argued that this assumption facilitates a conceptual advance in organising parallel processing, and can be worked around in the derivation of feasible architectures by various means, including the use of emerging wave based interconnect technologies, and permitting differing propagation delays across variable distances within a synchronous domain. Less optimal, purely wire based platforms, and globally asynchronous, locally synchronous (GALS) strategies may also be considered.

[1] The interlanguage environment introduced here, has no relation to Selinker's linguistics concept concerned with second natural language acquisition.

In a succession of Synchronic A-Ram machine cycles, an evolving subset of registers are active. Subject to some restrictions, any register is capable of either holding data, or of executing one of four primitive instructions in a cycle: the first two involve writing either ‘0’ or ‘1’ to any bit in the register array, identified by instruction operands, the third instructs the register to inspect any bit in the register array, and select either the next or next but one register for activation in the following machine cycle, and the fourth is a jump which can activate the instruction in any register in the following machine cycle, and also those in subsequent registers specified by an offset operand. Whilst the model’s normal operation is relatively simple to explain, it’s formal definition incorporates error conditions, and is somewhat more involved than that of a Turing Machine.

In common with assembly languages, schematic representations used for VLSI design and programming FPGAs, the hardware description languages VHDL and Verilog, and configuration software for systolic dataflow [1] [2] in coarse grained reconfigurable architectures, interlanguages may be characterised as *spatially oriented*. A programming language is spatially oriented if (i) there is some associated machine environment abstract or otherwise, and (ii) a program instruction or module, is linked in some way before runtime with that part of the machine environment, in which it will be executed in.

Vahid [3] and Hartenstein [4] stress the need for educators to consider spatially oriented languages, as important as conventional, non-spatial software languages in computer science curricula, because they are fundamental for expressing digital circuits, dataflows and parallel processes generally. The attitude that software and hardware may be studied in isolation from each other, is profoundly misguided. This report contains an account of how a high level, spatial language can easily deal with communication, scheduling, and resource allocation issues in parallel computing, by resolving them explicitly in an incremental manner, module by module, whilst ascending the ladder of abstraction. In what is in my view the abscence of viable alternatives, it can be conjectured that parallel languages *have* to be spatial.

In 1.2, it is discussed how an non-spatial language and compiler system that attempts to deal with allocation and contention implicitly, is subject to a particular kind of state explosion, resulting from transforming a collection of high level non-spatial processes, into the lowest level, machine-bound actions. Lee in [26] argues non-deterministic multi-threading introduces another kind of state explosion, making the establishment of program equivalence between threads intractable.

*Space* is a programming interlanguage for the Synchronic A-Ram, and may describe algorithms at any level of abstraction, with the temporary exceptions of virtual functions and abstract data types. Moreover, it is possible to incorporate parallel iteration and typed data structures, without adding the overheads and deadlocks to programs, that are associated with conventional dataflow or graph based programming environments (see 2.3.3 and 2.3.4). An interlanguage compiler produces code that at runtime, is capable of generating massive operational parallelism at every level of abstraction.

Providing a simple programming methodology is adhered to, Space's runtime environment, perhaps surprisingly, does not need to consider resource contention, deadlocks, and Synchronic A-Ram machine errors, because these issues have been implicitly dealt with at compile time. Race and time hazards are resolved by local synchronisation mechanisms. These features are scalable, and conceptually represent significant advantages over multi-threading on processor networks.

### 1.1.1 INTERCONNECT AND SYNCHRONISATION TECHNOLOGIES, AND RELATED WORK IN RECONFIGURABLE COMPUTING.

Reference is made to David Miller's work in 1.2.2, on using light as a means of synchronising room sized systems to nanosecond/picosecond intervals, of relevance to the construction of very large, globally clocked computers. In 8.3, the prospects of implementing a highly interconnected massive array of small computational elements, using either an optically or spintronically based network architecture are discussed. In 8.4, it is also explained how global synchrony can be relaxed in Synchronic Engines, to allow greater scalability. Massively parallel programs would still be conceived as globally clocked processes, aiding programmability, but would to a large extent run asynchronously.

The apparent lack of wave-based intra-chip connections allowing reconfigurable connectivity on the order of nanoseconds, indicates that more efficient Synchronic Engines may not be fully realisable in the short to mid term. In 8.2.1, a photonic connection system is described, in which microsecond switching between large numbers of nodes without chip area explosion, seems within reach. In 8.2.2, a spin-wave technology is outlined, that may enable nanosecond data exchange times for nano architectures incorporating millions of devices. A comparison between interlanguage programming on currently buildable Synchronic Engines, and multi-threading on multi-processor networks on standard industry benchmarks, will become available further down the research path.

The consideration of using silicon alone to realise less efficient machines, revealed a close relationship between the current approach and the field of reconfigurable computing, which was only fully appreciated in the final stages of writing this report. The action of a Synchronic A-Ram register is more primitive than a logic gate or FPGA look up table, and the register array's bits are in a sense, fully connected. It will be argued in a future paper, that if propagation delay were introduced into the definition (see 3.5.2), the model is fundamental to physical reconfigurable computing. Synchronic A-Rams are finer grained and more connected, and may therefore simulate FPGAs and CGAs without the inefficiencies that conventional reconfigurable models would have simulating each other.

Further, spatial computation based on systolic processing, on grids of coarse grained functional units, that might be termed *systolic spatialism,* lacks an abstract model, beyond the coarse grained, systolic grid itself. The approach suffers from being domain restricted; the developer is obliged to cast every program as a Digital Signal Processing-like collection of pipes or streams [5]. Systolic spatialism is however, well matched to silicon's restricted, planar connectivity.[2] It is an effective approach for maximising utilization and performance in wire-based parallel architectures, for applications that can be cast as streams [1] [2] [7].

Interlanguages form the basis for developing a new class of more general purpose programming models for wire based FPGAs and Coarse Grained Arrays of ALUs. There is a concern that the interlanguage model might lead to lower efficiency of runtime resource utilisation compared with purely systolic approaches, unless compensatory mechanisms are introduced (see 8.4).

[2] When wave based technologies allowing three dimensional connectivity become available, systolic programming and hardware may scale to some extent depending on the application, by increasing the dimensionality of the systolic grid.

Alternative kinds of programming environments for FPGA and reconfigurable platforms require a significant amount of hardware expertise from the developer [6], do not port to new architectures [7], and do not adequately support general purpose parallelism [8]. Sequential language environments for reconfigurable platforms might offer the prospect of parallelizing the software base, but by their nature do not allow the expression of parallel algorithmics. Their compilers [9] [10] rely on reassembling dataflows from arithmetic operations and loop unrolling, for parallelization. They cannot transform inherently sequential algorithms, which might appear anywhere in the spectrum of abstraction, into efficient parallel programs. Languages that do offer extensions for multi-threading on reconfigurable fabrics [11] [12], inherit the limitations of multi-threading (see the next section).

The authors in [3] [4] stress the severe overheads arising out of instruction fetch in processor networks, that are bypassed in spatial computing, because instructions are executed in situ. In the next section, I examine further the case against processor networks, in that they lack a good high level programming model and theoretical basis, and discuss the impact of their ubiquity in fields of application. A reader familiar with these issues, may move directly to 1.3, for an itemised introduction to the report.

## 1.2 MULTITHREADING AND PROCESSOR NETWORKS.

Whilst there is no accepted formal definition of the expression *parallel process*,[3] it is so often employed in an asynchronous context, that to use the term to refer to globally clocked computation, might cause confusion. The original meaning of the term *parallelism,* referring to simultaneous or overlapping computations [13], has to some extent been expropriated, making it synonymous with an asynchronous, non-deterministic style of computing. This report hopes to contribute to the recovery of the original usage of the term.

In the early noughties, traditional approaches for improving the performance of processors became less productive. Increasing clock speed outpaced silicon-based global synchronisation and memory access times, generating unacceptable heat losses, whilst expanding instruction parallelism ate up a nonlinear amount of chip real estate. The parallel multi-core concept has been selected by the ICT sector, somewhat arbitrarily as it will be argued, as the new general purpose architecture for maintaining performance improvements.

[3] A formal distinction between sequential and parallel processes in the A-Ram model, is presented in 3.2.

Until recently, the architectural model embodied in modern personal computers and mobile devices, was Von Neumann's theoretical concept of a processor combined with a random access memory, in which a program's execution consists of instructions being accessed singly from memory and executed. The approach has mutated in various ways over decades, the most recent being the multi-core model, consisting of a collection of processors on a chip or multi-chip module, with a communication network, shared and non-shared memories, including special purpose SIMD accelerator cores, sometimes known as Graphic Processing Units (GPUs), for multimedia processing.

Looked at more abstractedly, multi-core is an instantiation of the main theoretical contender to succeed the Von Neumann concept, being the globally asynchronous Von Neumann network, which may be characterised as a distributed array of processors with individual clocks, connected up in a variety of topologies. A Von Neumann network is programmed by multi-threading, in which conventional Von Neumann programs are concatenated to run on the processor array, occasionally communicating with each other. There are two approaches to multi-threading, which are embodied by shared memory and message passing languages. The former allow threads to communicate using memory locations that are shared between processors, which are usually found in multi-core and embedded architectures. Message passing assumes only local memory is available to a processor, relying on messages to pass on results between threads via a connection network between processors, and has been used for programming larger processor networks. Java and C# have multi-threading extensions, and are examples of shared memory languages, whilst Occam and Erlang are explicitly parallel, message passing languages.

Most theoretical work has focused on message passing models, possibly because shared memory models are state based, and perceived difficult to formalize and work with because of the state explosion associated with multi-threading (see 1.2.1), although there have been attempts [14] [15]. The class of (mostly) stateless models known as *process algebras*, have been devised in order to theorise about message passing networks, and are critiqued in the next section. Less mathematically oriented, and more practical models have also been suggested, for programming/performance evaluation purposes, including the Single Program Multiple Data (SPMD) model the PRAM [16], the Bulk Synchronous Processor [17], and more practical, low level PVM for shared memory [18], and MPI for message passing [19].

But the Von Neumann network has until recently, struggled to enter a mainstream demanding the best available cost-performance ratios, because of what appear to be inherent problems in the model. Early doubts were expressed by proponents of Dataflow Models [20] concerning memory latency and synchronisation issues. In a survey from 1998, some of whose conclusions have not been superseded in my view, Talia and Skillicorn [21] observed that historically, the development of the Von Neumann network and associated models was ad hoc, because it preceded the development of a general purpose, theoretical model of parallel computation. It turned out to be difficult to devise programming models for Von Neumann networks, that shielded the programmer from tedious low level tasks of efficiently assigning threads and communication links to processors, across different networks.

In the field of scientific computing, a wide range of computationally intensive numerical algorithms have been usefully implemented within the multi-threaded/processor network framework, due either to the easily programmable SIMD/SPMD character or readily decomposable nature of some algorithms, or because of toleration of high programming, energy and hardware costs in order to achieve higher performance. But in general the larger the Von Neumann network, a greater difficulty is encountered in programming. Strohmaier et al [22] claim that "even today, most users consider programming tools on parallel supercomputers to be inadequate". Parallel languages for processor networks are not conducive to formalisation. Reasoning and verification are much more difficult than for sequential programming, and are subject to software bugs over and above the usual nonparallel bugs, concerning contention issues for shared resources such as memory and processor time.

A worldwide expenditure of billions of euros on research over four decades, has failed to deliver a multi-threaded programming environment that offers portability, wherein the orchestration of concurrent computations, the scaling of applications to higher processor counts, and the elimination of deadlocks, is manageable for the average programmer. The next section will present a philosophical overview of one standpoint that contributed to these difficulties remaining unresolved. Programming environments for multi-core architectures are further discussed in 1.2.2.

### 1.2.1 TRAGEDISM AND CONCURRENCY THEORY.

There is an outlook in Computer Science, that will be termed *tragedism*, which has focused interest on Von Neumann networks, and in my view held back the emergence of a viable model of parallel computation. Tragedism is an over-pessimistic view concerning constraints placed on parallel computation models, by conceptual and physical factors. The upper limit on the speed of light, and the apparent absence of a clock for the universe, entail that sufficiently large parallel machines must physically operate asynchronously. Tragedism holds that as a consequence, parallel programming environments should eschew a global clock. Furthermore, conceptual issues entail that an asynchronous environment is either necessarily, or effectively non-deterministic, and that programming languages should either implicitly or explicitly include a means for processes to make non-deterministic choices.

The next section questions the antecedent for tragedism; namely the extent to which physical factors present barriers to implementing synchronous systems, even on a scale sufficiently large for supercomputing and wide area networks. In this section I overview one aspect of Computer Science associated with tragedism, *Concurrency Theory*, which has provided the basis for the MPI paradigm of distributed, high performance computing [19]. MPI is the de facto standard environment for large scale parallel computation. Such a status has been achieved despite a perception that MPI deals poorly with applications for large data sets, in which the data an individual processor needs to call on, is larger than the space available in the processor's memory.

Message passing formalisms in Concurrency Theory are so-called *process algebras*, including CCS [23], CSP [24], and the π-calculus [25]. Although these models were originally designed to theorise about, and potentially optimise multi-threaded processes running on asynchronous Von Neumann Networks, part of their more general philosophical justification is employed, in a somewhat circular manner, in order to support the multi-threaded/Von Neumann network paradigm itself.

Proponents of process algebras take as fundamental that parts of a computing environment are disconnected, at a level of granularity that is sometimes left ambiguous. There is an additional, unacknowledged linguistic factor in my view, that disconnects environments, imposed by the use of non-spatial, tree expressions in process algebras. This report sets out the case that the interlanguage environment is superior to the non-spatial tree as a language system for representing processes. Because a tree expression alone cannot directly connect together shared subexpressions, it cannot directly express a sharing of process subcomponents in higher level components. Therefore an impetus is given to a pessimistic outlook that elements of a computing environment are unavoidably separate and disconnected from each other.

Resuming the tragedist argument, there is a quasi-political consideration, that because individuals, corporations and sovereign nation states have a reasonable desire to keep their internal workings private, and not subject to centralised control, this aspiration should be reflected in the design and modeling of computer networks. There are correspondences between programming models and political ideologies, in that they are both concerned with the question of which methodologies and levels of abstraction might be employed, to effect change in large complex systems. Privacy and freedom are absolutely legitimate concerns, but it is not obvious why politics has any relevance to the purely technical issues relating to the foundations and the implementation of high performance computing [4].

These factors make it impractical for all activities, and their potential simultaneity, to be transparent to a single observer. An environment is rather a disconnected collection of nodes, whose internal process is essentially a Von Neumann thread, opaque to other nodes. Unfortunately, multi-threading introduces state explosion [26], making it impossible in many situations, to predict either the component's or the system's behaviour. There is a need in abstract message passing formalisms, to introduce explicitly non-deterministic choice operators into the formal specification of the behaviour of a network of device nodes, in order to reflect it's effective indeterminacy. In shared memory programming, non-determinism arises implicitly, out of the use of locks on memory cells, condition variables, and monitors.

Non-determinism has the supposed additional benefit of abstracting a node's internal operations, which enables one way of treating a node's functionality as a black box. Non-determinism reinforces the notion of nodes' internal workings being in some sense fundamentally mysterious, and the inevitability of the asynchronous, compartmentalised behaviour of Von Neumann networks. State explosion and the loss of determinism, has the unfortunate effect of enabling *deadlock*, which occurs when two "simultaneous", independent processes wait for each other to release resources, which can only occur when both processes complete, resulting in each process being stuck in an infinite loop.

[4] If there is a dependency, then perhaps it should be in the opposite direction. Parallel computing is pertinent to control theory, and in the industrial field of Production Management, is also relevant to the optimisation and processing of Bills of Materials. A good theory of deterministic parallelism, has the potential to be a conceptual antidote to the political form of tragedism. Political tragedism characterises the notion of collective action for the common interest, that might be said to be formulated at a high level of abstraction, as always being in some sense incoherent, or impossible in practice.

Simultaneity is problematic for concurrency; it's existence is either denied, or the extensions to a formalism to account for it, are flawed. There are two principal means of assigning semantics to events in process algebras, that might in some sense be "simultaneous". In the *interleaving semantics* approach, the simultaneous occurrence of two events is not distinguished from their occurrence in arbitrary sequence. Proponents of *true concurrency* [27] [28], argue that the interleaving semantics approach by itself, leads to unintuitive representations of physical processes, and to state explosion, and that simultaneous events need to be characterised in terms of smaller state sets. This approach has generated true concurrency semantics for Event Structures [29], and for the synchronous process algebra SCCS [30].

Event Structure diagrams are cumbersome instruments for representing fine grained deterministic computations in synchronous environments, for example logic gate dataflows in synchronous digital circuits. Simultaneity cannot be flagged explicitly, but arises out of mathematically establishing certain conditions about diagrams. Milner attempted to introduce explicit simultaneity in SCCS. He postulated that the simultaneous occurrence of two events $a$ and $b$ should be interpreted as the result $c$ of a commutative, associative product, $a*b=c$ within an Abelian group of actions/events. But it is difficult to see how two simultaneous, primitive machine instructions can be mapped onto a single primitive instruction, or two separate operand routings be mapped onto one operand routing. It is not obvious how any "true concurrency" mapping of two or more state changes into a single state change can be achieved, without either losing information, or creating an exponential increase in the state space.

Concurrency's agenda can seem difficult to pin down. I would summarise it as the view that because large enough parallel computers must operate without a global clock, efforts should focus on formally modeling a collection of asynchronous processes interacting non-deterministically. Such an approach should lead to a straightforward and decentralized method for avoiding deadlock and resource contention between machines, and thereby attain an efficient and easily programmable form of distributed computing, including high performance computing.

In [31] however, Peyton-Jones acknowledges that “today’s dominant technology for concurrent programming – locks and condition variables – is fundamentally flawed”. The problem is that concurrency does not support modular programming; a composition of modules that have individually been shown to be deadlock free, is not guaranteed to be deadlock free itself. Moreover modules are often delayed by having to wait for locks to be released, resulting in poor performance. Peyton-Jones refers to a technique called Software Transactional Memory (STM), based on ideas by Knight in [33], which circumvents this problem. However, STM can generate starvation [34], and is limited by there being no facility for multiple simultaneous updates of a data record (see 1.2.3). In addition, it does not address concurrency’s other sources of poor programmability and performance.

In a paper entitled “What are the Fundamental Structures of Concurrency? We still don’t know.”, Abramsky in [32] notes that there are too few constraints in defining process algebras, resulting in a plethora of notations, and a lack of a “grand theory” of concurrency. Abramsky expresses the hope that ideas from physics will come to the rescue, in finding an optimal formalism, and a grand theory.

There is considerable evidence that the likelihood of concurrency achieving it’s aims is remote. Explicit asynchrony and non-determinism introduce severe difficulties into high level programming frameworks. How can one sensibly conceptualise coarse grained processes running at the same time, without some notion of global time? Simultaneous processes are relatively easy for the programmer to visualise, whereas asynchronous processes are not. Lee in [26] emphasises that non-determinism introduces state explosion, which makes the establishment of program equivalence between threads intractable. The avoidance of resource starvation, race hazards, deadlocks, and livelocks is similarly problematic. He makes the case for what is in my view a limited containment of the effects of tragedism, by using coordination languages in which non-deterministic operators are only sparingly and judiciously used.

Non-determinism may be attractive to some for philosophical reasons, and permit a high level of abstraction, but at the cost of much effort in the understanding of programs, containing multiple or nested instances of choice operations. This in turn has an impact on getting processes to co-operate, and on the avoidance of resource contention, with negative consequences for overall programmability and performance. There are a number of deterministic programming approaches implemented on asynchronous and synchronous hardware, including Dataflow Models, spatial models [35], Globally Asynchronous Locally Synchronous (GALS) approaches [36], and the Space language described in chapters 5-7, that constitute counterexamples to tragedism.

After nearly three decades, the paradigm has delivered neither a viable programming model for processor networks, nor by implication, an effective approach to high performance computing. In 1.2.3, it is argued that tragedism has also contributed to the dominance of low bandwidth asynchronous internet protocols. The mindshare the approach has enjoyed, has obscured the possibility that synchronous determinism holds the key to settling the conceptual problem of parallelism.

### 1.2.2 The synchronisation of large networks.

The physical and technical problems involved in synchronising massively parallel systems, even on a planetary scale, can be over emphasised (see 1.2.3). The following considerations might support the case that large deterministic synchronous machines, have long machine cycles, synchronisation overheads, and poor performance:

i. Deterministic centralised control and global clock pulse generation has to occur in a localised area, which is vulnerable to being a single point of failure, and requires signals to travel across the entire machine, impacting on machine cycle time.
ii. Deterministic control can impose overheads, associated with establishing the initiation and termination of multiple threads/sub-programs.
iii. As alluded to in the previous section, there is a more general appeal to Relativity Theory in physics, which denies a notion of universal time, in order to question the feasibility of enforcing simultaneity over chip, wafer and larger scales. Shrinkage of device size, entails that machine cycle time is becoming so short, and physical machine size relatively so large, that different parts of the machine or network might begin to experience relativistic effects with respect to one another.

These factors are consistent with the tragedist approach having a better potential for reducing the size and cycle time of component devices, and improving overall performance. But they may be taken issue with:

i. Large computer architectures can be clocked, if there is sufficient investment in synchronisation mechanisms. The establishment of a global clock, does not require an signals to travel across the entire radius of an environment, or through wires. Optoelectronic technology [37] enables very short cycle times. In [38] Miller states, “It is likely possible to retain absolute timing accuracy in the delivery of optical clock signals of ~10-100 picoseconds over a computer room (tens of meters) without any special technology.”
ii. Optical Micro Electrical Mechanical technology enables microsecond data exchange times, potentially for architectures incorporating tens of thousands of devices, and spin-wave technology may soon enable nanosecond data exchange times for nano architectures incorporating millions of devices (see 8.2.2).
iii. Deterministic control does not have to be spatially localised for existing dataflow models, and the same applies to the new class of machines described in this report. It is argued in 6.13, that the new machines’ need for determining multiple thread initiation and termination, does not require signals to traverse the entire width of a machine.
iv. To make an appeal to Relativity Theory is potentially valid, but there is no widely accepted physical theory explaining both relativity, and quantum phenomena such as particle entanglement. It cannot with certainty, be stated that no global clock for the universe, or method for accessing such a clock, exists. With regard to relativistic effects, regular synchronisations of parts of a machine against a global clock, can neutralise the emergence of any small temporal discrepancies.

Ultimately, the desirability of large scale determinism and synchronisation, should be judged by performance gains. Optoelectronics promises to be an enabling technology for synchronisation, at least for picosecond cycles in room sized systems. In chapter 8, the way in which requirement for a global clock for Synchronic Engines may be relaxed for scaling purposes is discussed, by using a GALS approach. GALS raises the prospect of architectural scaling, whilst retaining synchronous, deterministic program semantics.

### 1.2.3 THE INTERNET, AND GLOBAL TIME.

The discussion in this section draws out how attitudes to fundamental aspects of parallelism, influence and are influenced by, the design and control of telecommunications networks. The principal protocol of the internet is asynchronous packet switching (IP), in which programmed control of a network's communications does not exist at the highest levels of abstraction, and has no centralised location. Control of a network is rather at lower levels of abstraction, distributed amongst the network's routers and server nodes. Each communication is broken down into packets which individually may be transmitted to a destination by any route through the network, and arrive in any order, dependent on the degree of local traffic associated with, and availability of, any one of a very large number of nodes through time. Therefore an IP based network's overall routing behaviour is far more difficult to predict or monitor than say that of a circuit switched network, and is effectively non-deterministic.

Although IP is still with us today when bandwidth is the prime requirement, it was originally developed when the physical network infrastructure was minimal. The main aim was to ensure the survivability of a communications network in the event of war, where bandwidth was a secondary consideration. The asynchronous and non-deterministic nature of IP, has somewhat arbitrarily provided an application area for concurrency and process algebras, and an intermutual justification for multi-threading on processor networks.

Synchronous, circuit switched communications are spatially oriented, and are programmed at a high level of abstraction. Circuit switching is the basis of mobile voice telecommunications, reliably achieving colossal, in order transfer of data. Dynamic Circuit Networks [39] have renewed interest in circuit switching for non-voice communication, and confirmed that they are able to support superior bandwidth compared with IPs, providing the necessary network infrastructure is in place.

Battles for important standards often occur away from the public gaze. There was an attempt by Cisco in the late 1990s, apparently blocked by a corporation [40], whose technological and commercial strategy was asynchronously oriented, to displace the proposed IPv6 upgrade, with a higher bandwidth standard, capable of supporting synchronous circuit switching, called Multi Protocol Label Switching [41]. It is conceivable that the internet's current bandwidth problems ensuing from year on year exponential traffic growth [42], could have been ameliorated if Cisco had won the contest.

A major justification for packet switching, has been that it makes more effective use of limited telecommunications resources than does circuit switching, because no dedicated circuit needs to be mechanically established for a session. The extent to which this factor applies now is questionable, given that telecommunication links are on high bandwidth fibre optic cables even at local area and interboard levels, able to support multiple channels on the same hardware through the use of time and wavelength division multiplexing.

With sufficient investment, the World Clock, currently defined retrospectively by a network of clocks, could be wirelessly delivered using the Network Time Protocol [43] to every point on the planet's surface, through the GPS and Galileo satellite networks [44] [45] [46]. Every earthly event of interest, on the time scale of human activity and beyond, could be time stamped, not just retrospectively, but to microsecond periods in a good approximation of realtime. High frequency trading in stock markets does operate within shorter timeframes, using IP and transactional memory based protocols. However, there is a question as to whether this kind of market activity is parasitical, rather than socially or economically useful [47].

Globally known circuit routing is better for tracking and inhibiting criminal activity on the internet. Packet switching however has one significant benefit over circuit switching, in that the relative lack of globally known routing, facilitates the ability of individuals in totalitarian regimes to communicate with the outside world without detection. A hybrid system such as Multi Protocol Label Switching, capable of supporting both approaches, has the potential to offer a good balance.

The potential benefits of an improved World Clock are significant. There is the prospect that deadlock-prone, inefficient and unnecessarily complex, distributed programming for web services and cloud computing, could be superceded by synchronized and spatially oriented forms of parallel programming. A forthcoming paper will describe a synchronous protocol for database updates called ISYNC-UP, which will have a advantage over the transactional memory approach, because it will allow a plurality of locations to request and have processed, transaction updates for the same account without waiting or contention, in one or a small multiple of World Clock timesteps. As is the current practice, regular synchronisations with the World Clock, can neutralise the emergence of any small temporal discrepancies within planet-bound and orbital vehicles travelling at speed for prolonged periods.

A purely circuit switched internet would likely confine the need for asynchronous and non-deterministic programming frameworks, to specialist military applications where the reliability of communication links cannot be guaranteed, and to those networks incorporating interplanetary probes operating at astronomical distances. But upgrading the planet's synchronicity, and the functionality and bandwidth of it's principal communication system, might be opposed by organisations committed to asynchronous strategies.

### 1.2.4 Programming models for multi-core architectures

There is concern in the ICT sector that the multi-core effort might share the fate of the multi-billion Itanium initiative in the early noughties, that was centered around the Very Long Instruction Word architectural concept. Particularly so, given that multi-threading extensions in Java and C#, offer some prospect of parallelising object oriented applications[5], which form a major part of the existing proprietary software base. Unyielding issues with Von Neumann networks, such as decreasing speedup with increasing thread count, have not therefore impeded attempts to devise novel forms of multi-threaded programming, to save the paradigm.

Flynn in [48] notes that "the new emphasis on multi-core architectures comes about from the failure of frequency scaling, not because of breakthroughs in parallel programming or architecture". An influential overview from the EECS department at Berkeley [49] stated: "We concluded that sneaking up on parallelism via multi-core solutions was likely to fail, and that we *desperately* (their italics) need a new solution for parallel hardware and software." Asanovic et al propose that a simplified Von Neumann core, may prove to be a more suitable building block for parallel systems. They argue that the study of algorithmic methods called *dwarfs*, might yield a viable environment for the many-core architecture, which envisions hundreds or thousands of rudimentary cores on a single chip, as opposed to a few, large complex cores found in current commercially available desktop machines.

[5] A critique of object oriented programming from a spatial and interlinguistic perspective, is the subject of future research. Early investigation suggests that encapsulation/virtualisation combined with SPR (see 2.1), contributes to the phenomenon of code bloat. Further, encapsulation/virtualisation is not conducive to the sharing of methods and subexpressions in dataflow, and Service Oriented Architecture is a problematic fix.

The topology of wire based multi-core/many-core networks cannot be made fully reconfigurable in constant time, because of silicon area complexity constraints (see 8.2). In addition to the absence of a viable theory of parallelism based on the Von Neumann paradigm, lack of reconfigurability is another factor contributing to poor software portability between multi-core architectures. In the long term, reconfigurable networks based on intra-chip optical [50] [51], wireless [52] or spintronic (see 8.2) interconnects are feasible. This might to some extent improve portability. It is possible that some kind of standard programming model for multi-cores could emerge, that would shield the programmer from at least some of the low level issues.

Microsoft's current programming environment in development, for multi-cores/ many-cores is called Dryad [53]. Programs are characterised as a coarse grained, data parallel graph, without selection or program control at the highest level of abstraction. Presumably some early features have been incorporated into Windows 7 support for multi-cores. It will be interesting to observe the operating system's ability to scale an application's performance on variable core counts.

Google's programming model for it's vast processor network, Mapreduce [54], is not strictly speaking general purpose, but is a specialised data parallel model, aimed at massive database processing, and the simultaneous execution of huge numbers of searches. Java is not ideally suited to high performance, but has built-in support for multi-threading, offering a non-deterministic, shared memory style of parallelism. In an attempt at controlling resource contention and deadlocks, there is an approach for incorporating process algebra operators into Java programming methodology, called CSP for Java [55].

Programming environments for heterogeneous many-cores, incorporating a mixture of GPU and Von Neumann cores, are also in development [56] [57] [58]. These offer the advantage of separating easily programmable and efficient SIMD and SPMD parallelism, from the quagmire of multi-threaded programming, but present no fundamentally novel solutions for general purpose programming.

Thanks to Lee in [26], there is an explication of the state explosion in thread interleavings in non-deterministic programming, suggesting that the creation of bug-free code within that framework is virtually impossible. But there is another perspective on state explosion arising from multi-threading. The compilation and linking of a sequential, high level program into machine code at the lowest level of abstraction, generates an enormous increase in the size of the state transition system. The new state system can be mapped to a sequential machine environment easily, because of a relative lack of concurrency related contentions. In a multi-threaded environment, compilation results in an array of vast transition systems that must be made to overlap and work together, because threads have to communicate and share resources. The problem of mapping compiled multi-threaded code to hardware, in such a way as to preserve program semantics and avoid low level contentions, is therefore subject to a further combinatorial explosion, and is significantly harder to automate.

This phenomenon shifts the burden to the programmer, who must be highly competent in order to extract acceptable, reliable performance from a particular language and architecture combination. It is not surprising that speedup decreases with increasing thread count, or that cost-performance ratios for application development on Von Neumann network architectures are dismal. The report argues for a spatial approach, where performance scales with machine resources, and where state explosion is avoided by deterministic programming. Scheduling, allocation and contention issues are resolved, explicitly and reliably by the programmer in an incremental and holistic manner, module by module, whilst ascending the ladder of abstraction.

## 1.3 A NEW APPROACH TO LANGUAGE AND COMPUTATION.

To provide the reader with further motivation, an overview of the paradigm introduced in the report follows:

i. The historical development of human language has not been optimal, for it's use as a template for formal and programming languages. Tree syntax is common to all natural languages, and is defective in that as used in a conventional manner, it can structurally only directly link one part of speech, to at most one more complex part of speech. Conventional tree syntax cannot indicate the sharing of subexpressions of an expression, or directly describe many to many relationships. Subexpression repetition is required, which discards a potential logarithmic compression in the size of the representation. Such trees exhibit a high degree of variability, where any individual branch may be arbitrarily long, requiring a complex parsing phase before semantic processing. Trees alone cannot directly express spatial information indicating data transfers and resource allocation within a computational environment, or which subexpressions may be semantically processed simultaneously. Non-spatial languages are unsuitable as parallel programming languages, because their compilation involves the solution of a combinatorial explosion, which is argued to be one source of the parallel computing crisis. The use of names of constructions, or pointers to locations in a computer memory alone, to access and reuse results of subexpression evaluations, represent a partial solution which discards an opportunity to devise a better general purpose language structure.

ii. The emergence of the non-spatial tree, as the de facto, standard language structure for syntax and semantics, has had serious consequences for our capacity to describe and reason about complex objects and situations. The inability to directly share subexpressions contributes to code bloat in commercial software, disconnected representations of environments, and a kind of linguistic schizophrenia. An unrestricted recursive application of rewriting rules for symbol strings is suboptimal linguistically, in that it is not conducive for describing simultaneous processes. Tree formalisms have deterred the introduction of an explicit notion of time and computation into mathematics.

iii. The problematic nature of subexpression repetition in tree languages has been noticed before, and has given rise to graph/data flow models, such as Term Graph Rewriting, Petri Nets, Semantic Nets, and Dataflow models. But these approaches have not entered the mainstream. Although the basic structure used is that of a graph, they are described in conventional tree-based mathematics, involving the non-spatial transformation of expressions alone, and lack an explicit notion of a computational environment. They are implemented on networks of Turing Machines/processors, do not call for a fundamental rethink of formal models of computation, and rarely call for an alternative computer architecture to the processor network.

iv. An alternative to conventional tree based syntax and semantics has been devised in the form of an a language environment called *interlanguage*. The environment consists of a language based on the notion of the interstring, and an abstract memory and functional unit array, capable of storing elements, and performing operations of some given algebra. Interlanguage allows the sharing of subexpressions to be explicitly represented, with linear cost with respect to the number of subexpressions. The tree form of an interstring is highly regular, requiring only a minimal syntactic analysis phase before semantic processing. Interstrings indicate which subexpressions may be semantically processed simultaneously, and allow resource allocation to be performed implicitly. Interstrings are also suitable for representing data structures with shared parts, and are intended to replace trees and graphs as standard programming structures.

v. The $\alpha$-Ram family of machines are formal models of computation, which have been developed to be the target machines for the compilation of high level programs expressed in an interlanguage. Members of the $\alpha$-Ram family with infinite memories are Turing Computable.

vi. A member of the $\alpha$-Ram family with finite memory, called the Synchronic A-Ram, may be viewed as a formal model underpinning the concept of an FPGA or reconfigurable machine. It supersedes finitistic versions of the Turing Machine and the $\lambda$-Calculus, the current standard models of Computer Science, in it's ability to efficiently support a high level parallel language. There is the prospect of a proper formalisation of parallel algorithmics, a new way of relating operational and denotational program semantics, and novel opportunities for parallel program verification. Massive instruction level parallelism can be supported, storage and processing resources are integrated at the lowest level, with a control mechanism similar to a safe Petri Net marking.

vii. An interlanguage called *Space*, has been designed to run on the Synchronic A-Ram. Space is an easy to understand, fully general purpose parallel programming model, which shields the programmer from low level resource allocation and scheduling issues. Programs are textual rather than graphic, and iteration, data structures, and performance evaluation are supported. Space has a high level sequential state transition semantics, and solves the conceptual problem of how to orchestrate general purpose parallel computation, in a way that has not been achieved before.

viii. The set-theoretical/logical definition of procedures for assembling constructions in mathematics, and the constructions themselves, are normally considered to reside in a universe of discourse, which is neutral and abstract from any computational implementation. A claim is made however, that conventional tree based formalisms in pure mathematics, harbour implicit notions of sequential, asynchronous and recursion oriented computation. Further, a universe of discourse incorporating an explicit parallel computational environment, is amenable to the adoption of parallel forms of reasoning, that bypass an implicitly sequential style in conventional mathematical discourse.

ix. Synchronic Engines are physical architectural models derived from the Synchronic A-Ram and Space, and are composed of large arrays of fully, or extensively connected storage and processing elements. The models suggest optoelectronic, and spin-wave based hardware specifications. If interconnect issues can be overcome, there is a new avenue for developing programmable and efficient high performance architectures.

Without having yet provided detail, the class of Space-like interlanguages, and the associated formal and hardware platforms, which during execution preserve their parallelism and lack of resource contention, constitute a paradigm of parallel computation that will be termed *synchronic computation.*

## 1.4 SPACE.

Space is a programming interlanguage with a functionality comparable to C. Space programming has an applicative style, and bypasses the readability and efficiency issues associated with recursion based, functional style programming. In order to explore design issues arising from the interaction of interlanguage and machine resources, a Synchronic A-Ram simulator has been written, and a substantial software project has resulted in a programming environment called *Spatiale* (*Spatial E*nvironment) being developed. Spatiale is a non-GUI, unix console application written in C, and incorporates a compiler that transforms Space programs into Synchronic A-Ram machine code. The package and documentation are available via links on www.isynchronise.com.

Spatiale is intended to serve as a prototype for Synchronic Engine programming environments. Space would require little adaptation in order to program Synchronic Engines. It is an explicitly parallel, deterministic, strictly typed, largely referentially transparent language, that retains the notion of updateable states. Although the Space programmer is obliged to consider some scheduling and resource allocation issues, these are relatively transparent within the narrow, synchronous and deterministic context of a module, and he is shielded from issues pertaining to pre-defined modules. They have been resolved by earlier composition, leaving the compiler to implicitly perform these tasks at compilation time.

Space modules are not generally intended to retain states between activations. At the current stage of the compiler boot strapping process, a high degree of referential transparency can be attained. It cannot be unequivocably ascribed to Space, because the programmer is obliged to ensure a module resets it's internal values after execution. In addition, memory allocation and reconfigurable interconnect features are required to bridge the gap between a high level program environment and a low level machine. It is envisioned that later versions of Space will have built in support for low level mechanisms, that will guarantee referential transparency for new program modules.

In Space, as well as in the Synchronic A-Ram machine code, more than one simultaneous write to a storage area, and more than one simultaneous call to a processing resource, results in machine error. The error mechanisms do not appear to restrict the expression of deterministic parallel algorithmics. Space modularisation and programming methodology, lead to the avoidance of race conditions and deadlocks, and enhanced software maintainability. The ability to modularise scheduling and resource allocation, and avoid resource contention, gives rise to programming models and architectures, which have decisive advantages over multi-threading for processor networks.

A deterministic Space program with simultaneous sub-programs running in a synchronous (or virtually synchronous) environment, is much easier to understand than a non-deterministic, asynchronous network of Von Neumann processes. Space has the benefits of functional programming, such as modular construction and lack of side effects, despite having updateable states. In addition, there are not the stack related inefficiencies associated with recursive function based computing. In order to provide proof of concept for synchronic computation, a range of massively parallel high level programs have already been successfully run on the simulator with outputs as expected. This has, to the best of my knowledge, never been achieved before with a simulated formal model of computation.

An implementation of synchronic computation onto processor networks is conceivable. Parallel sub-processes could be broken down into coarse grained blocks, and then sequentialised to run individually on a core, in the hope that some parallel speedup is preserved. Unfortunately, this approach would likely lead to low utilization of the panoply of conventional processor resources, and poor performance overall. Fine grained processes would need to synchronise and communicate across non-adjacent cores, resulting in long waits for maintaining cache coherency, and for the interconnection network to transfer results, leaving ALUs idle for many machine cycles. In addition, interlanguages offer no obvious opportunities for exploiting the extensive hardware resources dedicated to supporting speculation, predication, and the elimination of race and time hazards for multiple, out of order instruction issue.

## 1. 5 ORGANISATION OF THE REPORT.

Chapter two justifies the introduction of the interlanguage environment, by comparing the ability of interstrings to represent dataflow and dataflow processing, with trees and graphs, and by providing a critique of historical attempts in Computer Science to deal with the structural defect of tree languages. Chapter three describes the α-Ram formal models of computation, inspired by interstrings. Chapter four defines a programming language called *Earth*, which is close to the Synchronic A-Ram machine code, and allows the definition of the most primitive program modules used in Space. Chapters 5 to 7 present the Space interlanguage itself. Space's type system and program declarations are laid out in chapter 5, and chapter 6 defines the basic interstring language structures, and presents some simple program examples. Chapter 7 covers programming constructs, enabling the description of massive parallelism, along with a range of program examples. In chapter 8, Synchronic Engines are presented. Chapter 9 discusses the relative merits of the standard models compared with α-Ram machines, and gives an outline of how efficiently simulable models offer new opportunities for unifying logic and mathematics with foundational computer science.

**References.**

## Chapter 2

## TREES, GRAPHS, AND INTERSTRINGS.

### 2.1. TREES, AND THE SINGLE PARENT RESTRICTION.

The syntax of every natural language expression is a tree structure, whose nodes represent various parts of speech, and whose leaves are words in the language. Fig 2.1 depicts a typical Chomskian interpretation of the syntactic structure of a sample English sentence, with a single relative clause.

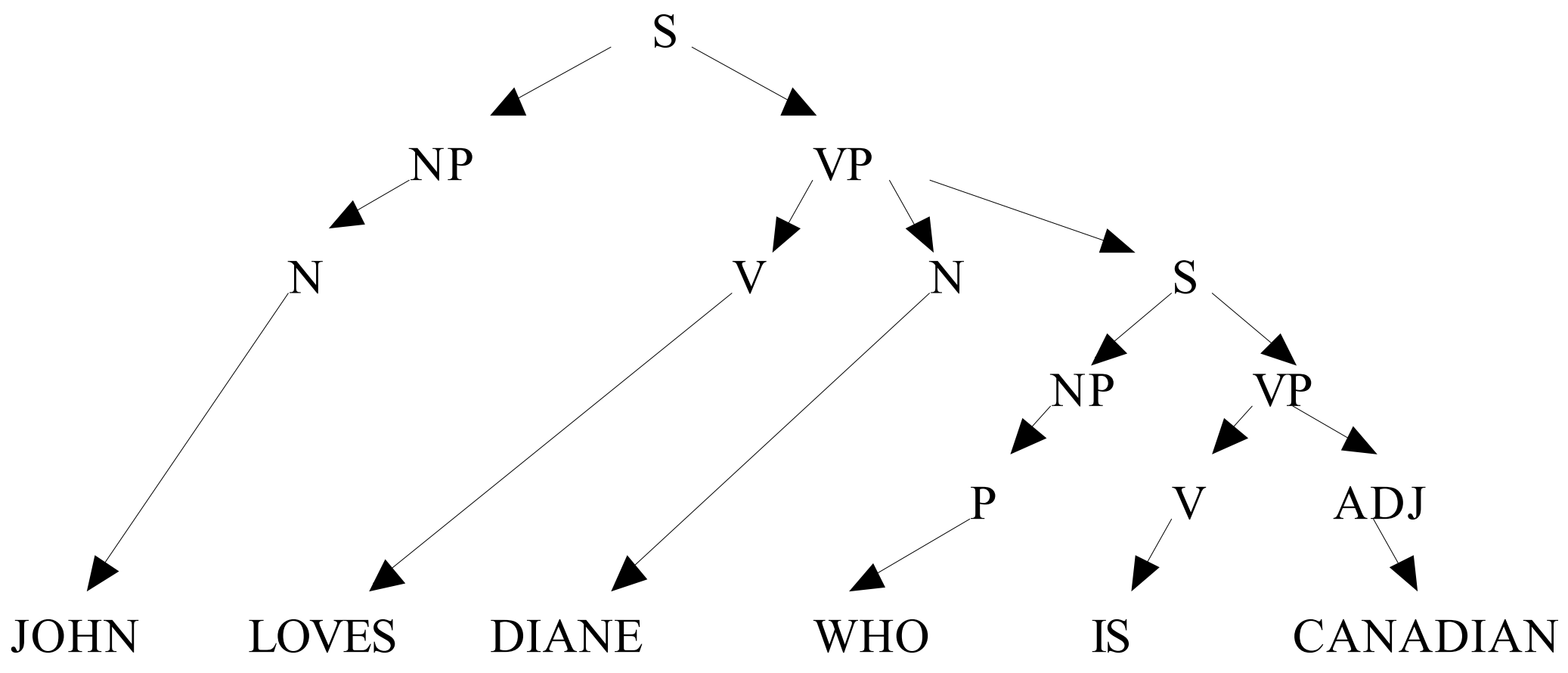

Figure 2.1 Syntax tree for "John loves Diane who is Canadian."

The parts of speech are connected by directed arrows, and have a semantic aspect, because they are named as nouns (N) and noun phrases (NP) suggesting objects, and verbs suggesting relations. The precise relationship between grammar and meaning in natural language is a mystery, because the manner in which language is generated and analysed by humans is not well understood. Current psycholinguistic experimentation allows the examination of neural activity of the brain's language areas at low resolution during sentence comprehension, using electroencephalography and functional magnetic resonance imaging. It is not yet known in detail what leading semantic structure a syntactic structure is initially transformed into by human cognition, or how semantic representations are processed, or even whether syntactic and semantic processing are so entwined as to render a distinction between the two meaningless from a cognitive viewpoint [1] [2] [3]. These issues are more likely to be settled, when high resolution techniques become available that can track individual axonic and dendritic state changes of very large numbers of neurons.

It can be said that what differentiates syntax trees from mathematical objects as directed acyclic graphs, is that by definition each node is restricted to having at most one arrow directed towards it. In other words, each node or part of speech in the tree, is restricted to being an immediate child (or subpart) of at most one parent (or more abstract) part of speech. In fig 2.1, the noun part of speech N whose leaf is "Diane", is a child of only the verb phrase (VP) part of speech of the sentence's main clause. This linguistic aspect concerning the construction of expressions in tree languages, will be referred to as the single parent restriction (SPR).

A language system possessing SPR, limits a part of speech describing an object, from participating directly in more than one predicate or relationship. In normal discourse, pronouns[6] or some form of naming of complex objects is used, to ameliorate the effects of SPR. It will be argued that the circumventions are inadequate, and that conventional tree languages are unsuited for describing many-to-many relationships between objects, which are ubiquitous in practice. In 2.2 it is explained that SPR is an arbitrary, sub-optimal linguistic attribute, because it imposes unnecessary limits on the size and complexity of what can be described[7]. SPR also has impact on the efficiency with which expressions can be processed.

There seems to have been an unstated assumption in the historical development of formal and logical languages that SPR is not problematic. Context free grammars, Backus Naur statements, or inductive rules for First Order Predicate Logic (FOPL) are used to define formal tree languages. Expressions are generated by the free recursive application of a succession of primitive rewriting steps, and generally speaking, any branch of the syntax tree may be arbitrarily long, resulting in trees with high structural variability. In 2.4.2 I review how the inductively defined, semantic interpretation of a FOPL functional term, generates a tree structure of evaluations within the semantic domain. Strictly speaking the λ–calculus has no semantic interpretation, but the application of $\beta$-reduction, the main mechanism for evaluating tree-based λ-expressions, always results in another λ-expression It is common for the SPR to be preserved in the semantic interpretation of expressions of formal tree languages.

[6] It would be preferable for "Diane" to directly participate in the noun phrase (NP) of the relative clause as well, rather than having to employ the pronoun "who", which refers to "Diane". The utility of pronouns in performing their role is strictly limited. Work on the semantic level is required to disambiguate the identity of the pronoun's referent. If a pronoun is not in close representational proximity to it's referent, it is difficult or impossible to identify the referent, entailing that they may not appear in adjacent sentences, let alone larger pieces of text.

[7] English's SPR may have prevented a more concise presentation, but it is conceded that SPR has not been a barrier to expressing the report's contents.

It is likely that SPR has been carried over into leading semantic representations in human cognition, even if later processing produces semantic networks (discussed in 2.3) not subject to the SPR, where an object node may participate in many relationships. The circumstantial evidence for this is the semantic aspect of parts of speech, suggesting trees are significantly involved on the semantic level, and the fact that a non-trivial amount of work is required to transform a tree into a non-tree network or graph like form. If this is the case, then a further sub-optimal outcome ensues for natural languages, as well as for formal languages whose semantics preserve the tree aspect of their syntax. In 2.2, it is explained how a semantic tree obscures which subexpressions may be semantically processed simultaneously.

To arrive at a spatial version of natural language, would seem to require a nontrivial protocol that names storage and processing resources in the minds of communicators. Perhaps it was inevitable that non-spatial language was the first general purpose communication system to have evolved for humans.

### 2.1.1 Overview of the chapter.

This chapter touches on a number of disciplines, and sets the scene for the rest of the report. It might assist the reader to provide an overview of what follows. An argument is presented that the size of the description of an algebraic dataflow using tree expressions with high structural variability, is logarithmically inefficient with respect to the depth of a dataflow, and is not conducive to the simultaneous semantic manipulation of subexpressions. A structurally variable tree can also problematic as a basis for a data structure, because the content of such a tree can only be accessed by following an irregular chain of links between nodes, which can impede the speed and simultaneity with which the contents of the data structure can be accessed and processed.

Trees with high structural variability are occasionally useful, but are not suited as standard, general purpose structures for the description of computation, and for high performance programming languages[8]. Attempts to ameliorate the defects of tree languages with fixes and extensions are wasteful, because greater productivity ensues from considering alternatives. The directed acyclic graph has been suggested as an alternative, but it is problematic to derive a language rewriting system for graphs. They have not led to a practical new paradigm of computer architecture, or a successful high level parallel programming model. A more suitable approach for describing shared components and dataflow is the *interlanguage environment*, which consists of a language based on *interstrings*, which are strings of strings of short expressions, and an abstract memory and abstract functional unit array, capable of storing elements and performing the operations of some algebraic semantic domain.

Interstrings are trees, in which only a small subset of branches may be arbitrarily long, and coupled with an abstract memory, exhibit an structure intermediate between syntax and semantics, which allows subexpressions to participate in more than one more complex subexpression. They also facilitate the simultaneous semantic processing of subexpressions. Formal models of computation able to represent and process interstrings, and to exhibit characteristics of the interlanguage environment, should therefore be considered.

To achieve this, the formal model should be *synchronically oriented*, which encompasses three characteristics: i). an act of computation is always associated with a hardware resource (a defining characteristic of spatial computation referred to in 1.1), ii). a clocked environment, in which there is a facility to activate a selection of hardware resources simultaneously, and finally iii). there should be a high degree of, or full connectivity between hardware resources.

A mathematically rigorous foundation for an interlanguage environment for describing algebraic dataflows is given. Results are not formally justified, excepting one about the representational inefficiency of a tree based function series, and a theorem that asserts for any tree expression in the language of functional terms in predicate logic, there exists a semantically equivalent interstring/memory pair. At this stage all that is required are hints for a better format for a parallel language, and for the design of low level parallel computation models, which are described in the next chapter.

[8] A future paper will discuss a novel means of migrating mathematics into a computational environment. It will be argued in 9.6 that an additional consequence of SPR, is that conventional tree grammar has complicated the introduction of time and computation into mathematical formalisms, which in turn has disproportionately focused interest on sequential, recursion oriented, and asynchronous models of computation.

## 2.2. DATAFLOW EXPRESSED AS TREES AND GRAPHS.

To devise a better language system for describing objects generally, macro-type dataflow is considered. Dataflow is ubiquitous as a semantic structure, and is ideal for understanding the potential for implementing parallel computation efficiently. Although dataflow lacks a direct ability to iterate or perform selection, it can be highly expressive. FOPL formulas have a quantifier-free prenex form, and prenex formulas are essentially dataflows. Consequently any proof containing a sequence of FOPL formulas may also be presented as a sequence of dataflow representations.

Given a semantic domain such as the algebra $\mathbf{A} = (A,+,-,\times,\div)$, consider the following expressions for the functions $f,g\colon A^3 \to A$, which are representations of the SPR language $D$(A)[9], defined by the Backus-Naur statement $\langle\text{EXP}\rangle := c \mid v \mid (\langle\text{EXP}\rangle\ op\ \langle\text{EXP}\rangle)$ , where $v$ is drawn from some denumerable alphabet of variables $V$, c is drawn from the set $A$, and $op$ is from the finite set of operation symbols in $\mathbf{A}$ :

$$\left(\left((x+y)+(y\times z)\right) \times \left((x+y)-(y\times z)\right)\right) \quad (f)$$

$$\left(\left((x+y)+(y\times z)\right) \times \left((x+y)-(y\times z)\right) + \left((x+y)+(y\times z)\right) \div \left((x+y)-(y\times z)\right)\right) \quad (g)$$

The use of brackets in the Backus-Naur definition permits a unique parsing of (*f*) and (*g*), which yield the trees in figure 2.1(a) and figure 2.2(a) respectively. A rigorous description of a system of labelled, directed acyclic graphs intended to describe dataflow, is somewhat involved. It will suffice for present purposes, is describe it as a set of vertices *Vert*, a set of directed edges between vertices $E \subseteq Vert \times Vert$, and optional functions for labelling vertices and edges with symbols e.g. $h_1\colon Vert \to V \cup \{+,-,\times,\div\}$ to mark vertices with variables and operations, and $h_2\colon E \to N$ to indicate the index of a functional input edge. Further, for any vertex $v$, there is no directed path of length greater than zero that starts and ends on $v$. Each function may be depicted as a directed acyclic graph $\langle Vert, E, h_1, h_2 \rangle$ in 2(b) and 3(b), where for readability $h_2$ is not shown.

[9] As it would not assist the presentation here, no separation between syntax and semantics will occur until section 2.4, which contains a semantics related theorem.

To evaluate the function applied to values of inputs given 2(a) or 3(a), one might iterate a search through the tree, applying parent operations to pairs of children whose values are known, removing the children, the edges to the parent node, and the parent node itself, and replacing them with the result of the arithmetic operation etc. until the final result is known. The evaluation of a function as a graph structure, as in 2(b) or 3(b), is commonly known as the Circuit Value Problem. The procedure in the latter case is slightly different, because only those child nodes and edges can be removed which are not associated with another parent.

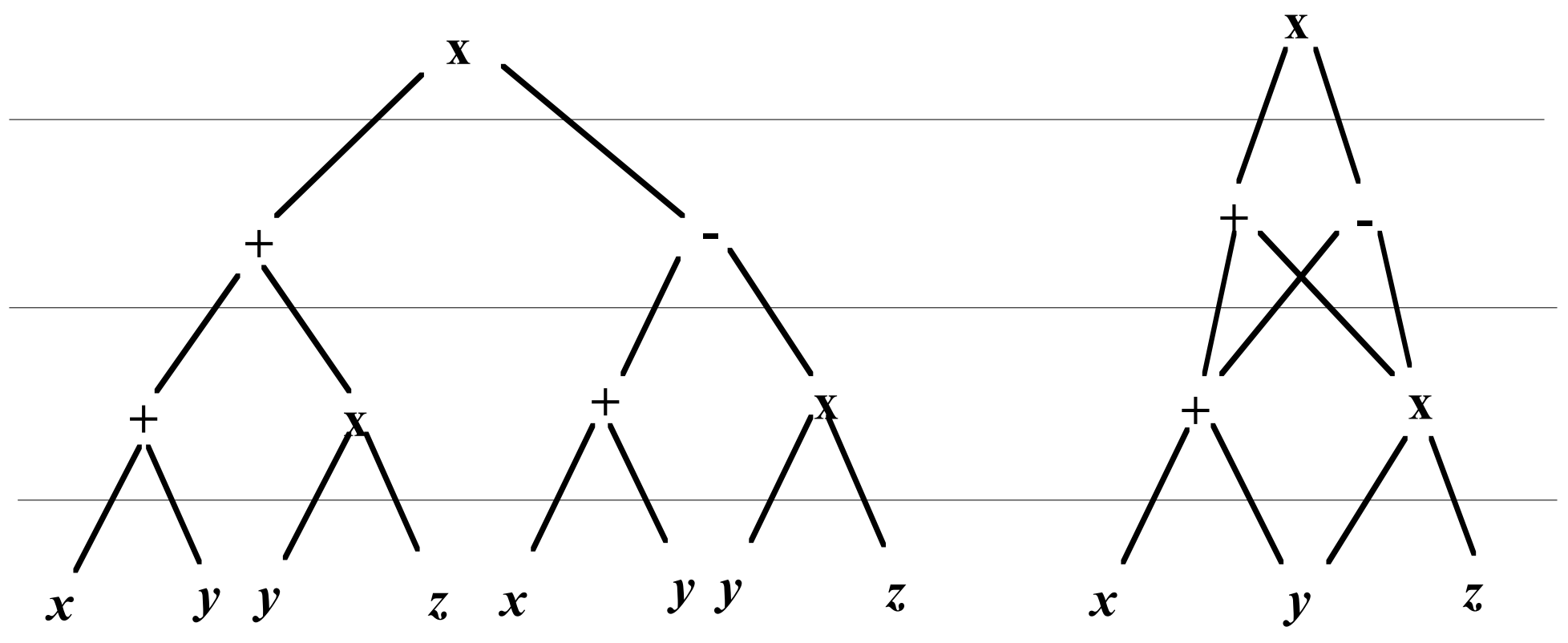

Figure 2.1. $f$ described as (a) a tree, and (b) a graph

The figures suggest that serious problems ensue from $D$(A)'s SPR, from both syntactic and semantic points of view.

i. For the representations (*f*) and (*g*) in the language *D(A)*, and the tree representations 2.1(a), and 2.2(a), the SPR results in a considerable repetition of subexpressions. The addition of one layer in the depth of the dataflow results in the length of (*g*) being approximately twice the size of (*f*), and the same can be said of the corresponding tree forms. Yet the sizes of *f* and *g*'s graph forms differ by only two nodes and four edges. The example suggests that sizes of surface and tree representations may approximately double with the addition of each extra layer, resulting in an exponential growth with respect to the number of layers.

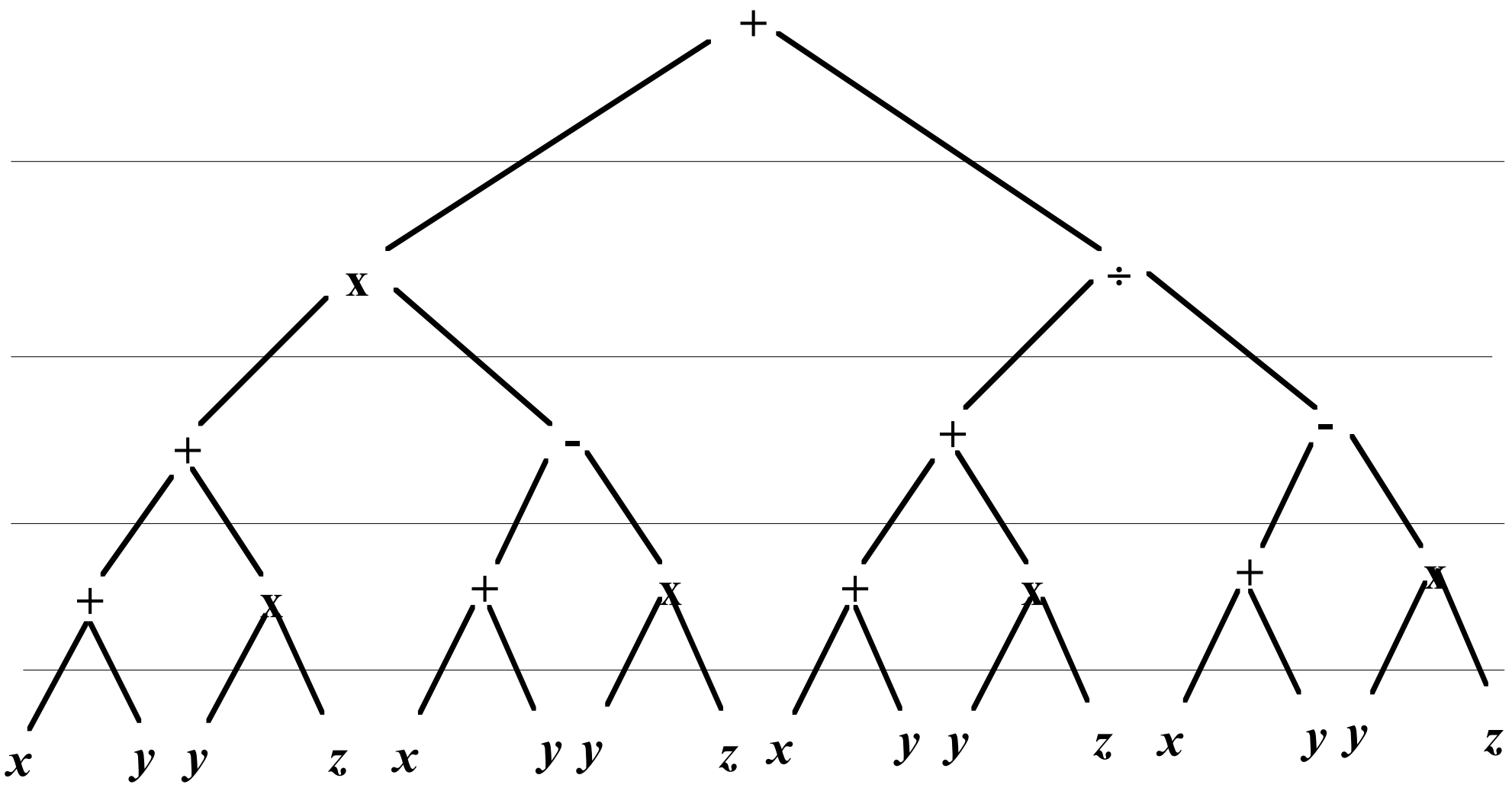

Figure 2.2(a). *g* described as a tree

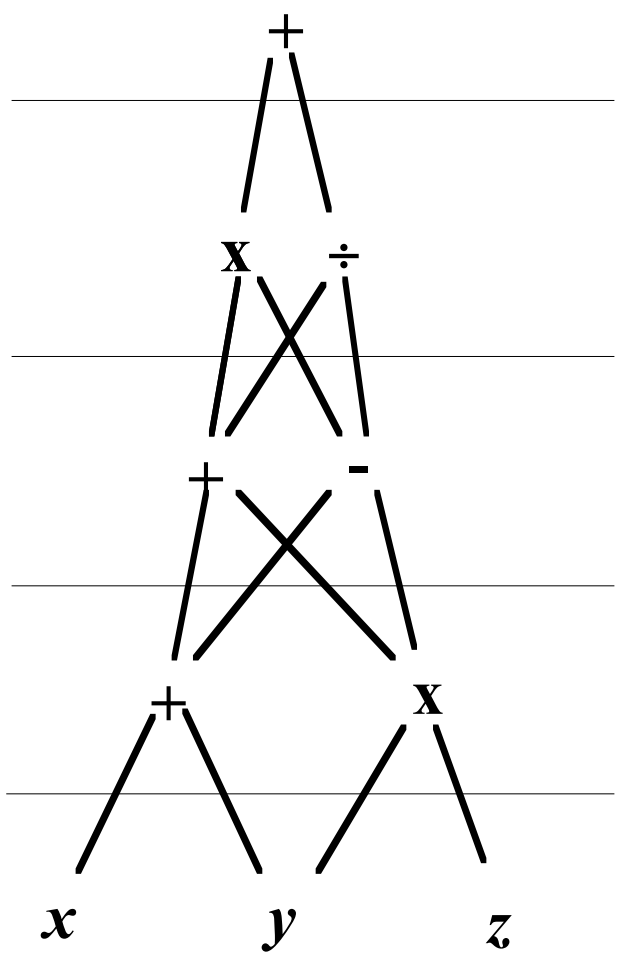

Figure 2.2 (b). *g* described as a graph

ii. Dataflows expressed as graphs do not exhibit the SPR. The graph format in 2.1(b) and 2.2(b) avoids subexpression repetition entirely, by allowing subexpressions to participate in more than one parent expression.Compared to the tree format, the graph seems to exhibit a linear growth with respect to dataflow depth, suggesting that a potential logarithmic compression in the size of representations is available.

iii. From a semantic perspective, subexpression repetition also imposes unnecessary computational cost, by having to repeat identical computations. The graph versions, by sharing subexpressions, avoid the need to repeat computations.

iv. Subexpressions with the same degree of bracket nesting, associated with the layering which appears in the diagrams, may be computed at the same time (assuming all operations execute in unit time), but the tree, in neither syntactic nor semantic forms, can directly identify which these are. But there is a similar problem with the graph as well, because the layering in a structure like $\langle Vert, E, h_1, h_2 \rangle$ is also not explicit. Additional processing, or a new structure would be required in both cases, to co-associate subexpressions/nodes with the same degree of bracket nesting, which places a further burden on language processing.

In 2.5 an alternative representation for dataflow is presented, which can bypass SPR and can also explicitly indicate dataflow layering. In the next section, a series of functions modified from the Fibonacci series are described, which clearly exhibit the exponential growth of tree descriptions with respect to the number of layers in the dataflow.

### 2.2.1 EXPONENTIAL GROWTH IN TREE DESCRIPTION OF FUNCTION SERIES DERIVED FROM FIBONACCI SERIES.

The Fibonnacci series is defined by the recursive system of equations on the left. Consider a derived function taking further variables $x$ and $y$, $xyfib$: $N^3 \rightarrow N$ on the right:

$$\begin{aligned} fib(0) &= 0 & \qquad xyfib(0,x,y) &= x \\ fib(1) &= 1 & \qquad xyfib(1,x,y) &= y \\ fib(n) &= fib(n-1) + fib(n-2) & \qquad xyfib(n,x,y) &= xyfib(n-1,x,y) + xyfib(n-2,x,y) \end{aligned}$$

The *xyfib* definition recursively defines for *n* a tree representation, expressing a dataflow with no shared subexpressions, where *n-1* is the depth of the dataflow. For example, $xyfib(5,x,y) = \left(\left(\left((x+y)+y\right)+(x+y)\right)+\left((x+y)+y\right)\right)$ has a dataflow depth of 4. The tree syntax of this expression is depicted in Fig 2.3(a), and it's corresponding graph form with no subexpression repetition is depicted in Figure 2.3.

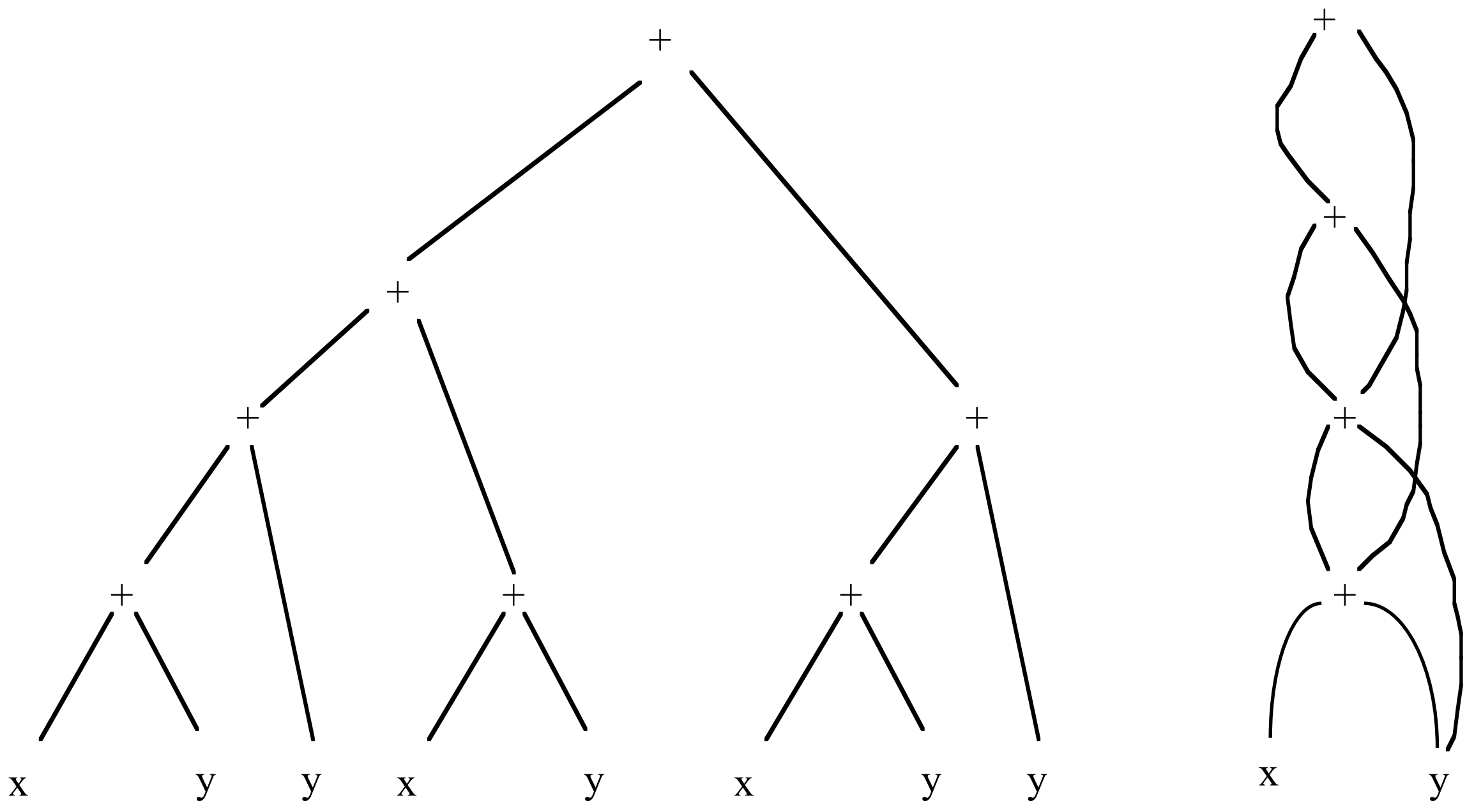

Fig 2.3 Tree expression (a), and graph expression (b) for *xyfib(5,x,y)*.

The linear function *addgraph* gives the number of additions in the graph expression, and is defined by :

$$addgraph(0) = 0$$
$$addgraph(n) = n - 1$$

The function *addtree* gives the number of additions in the tree expression, and may be inductively defined as follows:

$$addtree(0) = 0$$
$$addtree(1) = 0$$
$$addtree(n) = add(n-1) + add(n-2) + 1$$

It is well known that the ratio between one number and its predecessor in the Fibonacci series approaches the golden ratio $\left(\frac{1+\sqrt{5}}{2}\right)$, and the series therefore increases at a well-defined exponential rate with respect to *n*. Consequently the function *addtree* also increases at an exponential rate, because for $n \geq 4$, $addtree(n) > fib(n)$. Doubtless there are many other examples of function series whose tree description size grows exponentially with respect to dataflow depth, whilst a compacted graph description size grows linearly.

The obvious computational solution to the exponential explosion in the size of tree descriptions, is to employ pointers to addressable locations in the memory of a computational device with updateable states, to be used as temporary storage locations for later reuse of subexpression evaluations. An attempt to generalise this approach, and an explanation of why it is a only a partial solution, is discussed in section 2.4.1

## 2.3 Work related to addressing the SPR.

The problem of subexpression repetition in tree languages was noticed decades ago, and has directly and indirectly inspired considerable activity in computer science. Research has focused on the idea of transforming graphs rather than trees, within a number of different contexts. With one exception, theoretical approaches for graph related transformations are not co-presented with a parallel hardware model, other than the processor network. Some brief notes and references follow on the major responses that have been proposed, which can be grouped into four areas; Semantic Nets, Petri Nets, Dataflow Models, and Parallel Graph Rewriting in Functional Programming. Term graph rewriting is considered as an aspect of the last category. The latter two approaches provide a fully programmable model of computation.

### 2.3.1 Semantic nets.

Semantic Nets (SN) emerged from early Artificial Intelligence [4] [5] [6], and were an attempt to develop a graphed version of FOPL, in the hope that alternative ways of encoding knowledge about the real world would result, together with new forms of deduction and querying of information. Nodes in an SN represent objects, and multiple links are allowed between nodes, which represent relationships, and in addition are allowed between collections of nodes. Therefore SPR as it relates to the SN equivalent of a part of speech, need not arise.

One major benefit ascribed to SN is that objects that are found to be identical are obliged to have the same node. This allows all facts pertaining to an object to be accessible by only having to reference links from that node, which has the additional benefit that not having to repeat nodes, makes semantic networks smaller and easier to process.

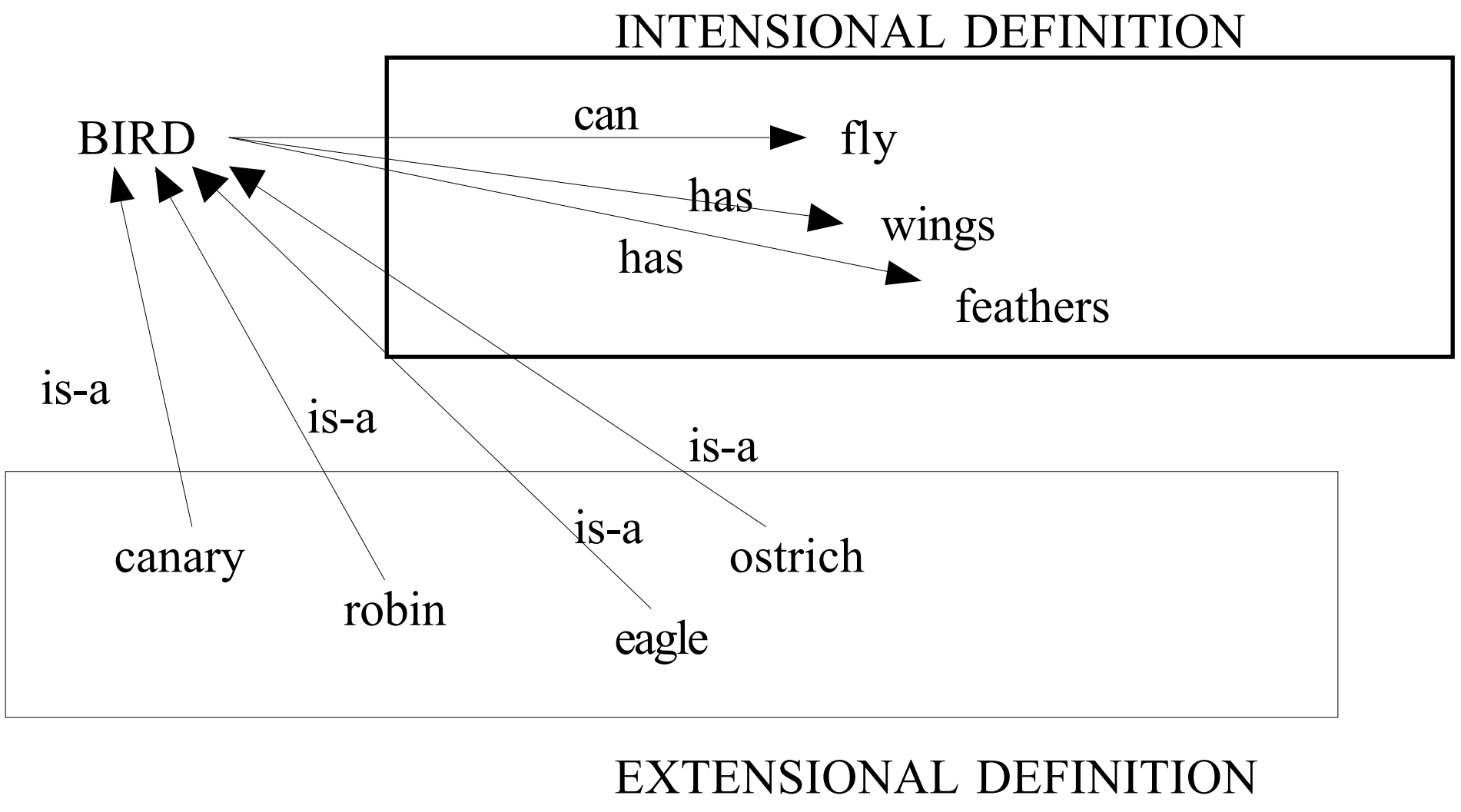

Figure 2.6 Example of a Semantic Net.

Another feature in SN is that a distinction can be made between certain types of facts or predicates. For example, in figure 2.6, there is a difference between what are described as intensional (has an attribute) and extensional (is an instance of) predicates. The most important type of SN deduction involves pattern matching on the graph structure of nets, which involves identifying patterns or fragments of nets, which may be replaced by others according to pre-defined transformation rules, resulting in new facts being attached to a node. Unfortunately pattern matching is computationally expensive, because it has complexity of subgraph isomorphism, which is known to be NP-complete [7], placing a significant practical limitation on it's deductive power. In 2.3.4, it will be discussed how the complexity of subgraph isomorphism can also affect parallel graph reduction and term graph rewriting.

Woods in [8] described ambiguities in SNs, indicating they do not have a clear semantics, imposing so few restrictions on the types of allowable nets and pattern matching rules, that little practical benefit can arise. Although there have been some notable exceptions which are almost direct pictorial representations of FOPL formulas, including the LOGIN system [9], and Sowa's conceptual graphs [10] [11].

The main benefit of the SN model seems to have been as a pictorial aid in inspiring new approaches and implementations for AI systems. The main criticism is that ambiguous semantics and pattern matching, prevented the emergence of novel forms of deduction more efficient than predicate logic based reasoning, such as resolution for Horn clauses. Lately, however, interest has been rekindled by the emergence of the Semantic Web [12] [13].

### 2.3.2 Petri nets

Petri Nets (PN) possess the exotic combination of being both synchronous and non-deterministic [14] [15], and were introduced to describe, and establish safety conditions for non-deterministic systems composed of nodes/modules with many to many relationships, e.g. concurrent programs, and workflow management. A PN is a graph consisting of a set of place nodes *P*, a set of transition edges *T*, and a flow relation $F \subseteq (P \times T) \cup (T \times P)$. The PN description has the potential to avoid the SPR, because the flow relation may be represented as an adjacency list or matrix, avoiding the need to repeat node names, in order to express many to many relationships between places and transitions. The state of a net is a marking $\mu{:}P \rightarrow N$, which associates a number of tokens with each place. A transition can optionally choose to fire if all its input places are marked at least once, resulting in a new marking in which one token is subtracted from the input places, and one token is added to the output places. A run consists of the marking moving across the totality of places in a non-deterministic manner, in a sequence of globally clocked steps, and proceeds until no further firing of a transition is possible.

A PN is not intended to be a fully programmable model of computation, because it is not modular in it's standard presentation, which limits the ability to define complex systems. In addition, because a transition *non-deterministically* fires when *all* of it's input places are marked, it is not clear how a PN could be used to define the simplest computational structures such as logic gates, boolean circuits, and adders.

A curious aspect of a PN is that it is equipped with a global clock, and yet a component transition may be non-deterministic, suggesting that the internal environment of the transition is unavoidably asynchronous. But if a global clock is available at the most general level, why not within the transition? Withstanding these issues however, the important concept of an evolving marking/set of tokens, found it's way into the control mechanism of the next category of models, and into that of the α-Ram family.

### 2.3.3 DATAFLOW MODELS

The Dataflow Model (DM), developed in [16] [17] and [18], offered some desirable characteristics for parallel computing. The approach acknowledged the primacy of dataflow, the need to consider alternatives to the Von Neumann building block, and had the benefit of describing dataflows as graphs rather than trees, at least at some stage of the compilation process. DM is not a low level model of computation, being understood to be a network of Turing Machines, Finite State Machines, or Von Neumann machines. At some level of abstraction, a DM program is described as a labelled graph structure bearing some similarity to a program flowchart.

Program control relies on a notion of a network marking, consisting of a number of data tokens, not dissimilar to a PN marking, being associated with the inputs (places) of a functional node, capable of executing some predefined function, or primitive, built in arithmetic operation. As with PNs, a functional node (transition) can be activated, usually deterministically, when all of it's input places have at least one data token. Similar to a conventional program flowchart, a data token arriving at an if-type node is routed down the appropriate wire after the selection criteria has been applied to the token's associated data. DM is the only approach with a potentially novel hardware aspect, since nodes on the hardware level can be restricted to being arithmetic-logic units with some control circuitry, rather than full Von Neumann machines.

An attractive aspect of the DM was that the parallelism in the system is implicit. Because DM node executions/functional evaluations are generally not explicitly activated, operations may be triggered as soon as operands are available, and the expression of which node executions should be performed in parallel is not required.

The scheduling of when a functional node fires is further determined by whether the DM is *data driven* or *demand driven.* Functional nodes in a data driven model such as the Static Model [19] are always busy waiting for inputs. The activation of a data driven node proceeds, precisely when all input operands are available, signalled by the arrival of tokens. Functional nodes in a demand driven model such as the Kahn-MacQueen model [20] only wait for inputs, when they are "busy" i.e. when they have a received a request for their output, from a "downstream" functional node needing inputs. The activation of a busy demand driven node proceeds, when all input operands are available, signalled by the arrival of tokens.

It is not easy to generalise about DMs, because the field fractured to address differing limitations of data-driven and demand driven approaches. The limitations involved issues in hardware implementation and programmability With the exception of Static Dataflow, the requirement that a number of tokens may sit on top of a node, leads to a hardware issue in the need for a buffer for every node, and the potential problem of buffer overflow[10].

Johnston et al in an influential survey [21] describes software and some hardware related issues with dataflow models: the token system of program control cannot easily support iteration and data structures for dataflow programming languages, without adding significant complexity and potential deadlocks to programs. This places a serious limitation on their programmability. The survey also argues there is a mismatch between the fine grained parallelism of dedicated dataflow architectures, and the coarse grained nature of many problems, which map better onto Von Neumann machines.

Given the combination of hardware and software issues, the prospects of a general purpose DM-based architecture seem poor. The alternative of mapping a DM to a processor network programming model is problematic, not only because a good one does not exist. Johnson's survey also explains how the implementation of a dataflow cycle imposes a high overhead; busy waits, phases to identify tokens, node execution, and the generation of new tokens[11]. A synchronous, deterministic form of DM based on systolic spatialism, referred to in 1.1, oriented to Digital Signal Processing, and sometimes termed the *datastream* model, is the only DM that might be described as being currently available as hardware [22] [23]. Datastream computing has been suggested as a basis for general purpose model by Hartenstein [24], and this claim is questioned in [25] and [26].

There are further reasons why the DM failed to live up to the early expectations that it would replace the Von Neumann paradigm. It is natural to characterize dataflow as a composite entity, because it consists of a network containing functional nodes, whose internal workings need not be defined in the program. To focus on dataflow as a non-primitive, mid-level object however, can obscure the possibility that in order to construct a viable dataflow oriented model of computation, more fundamental areas also need to be considered:

i. As already argued, there is a case for a new language environment for describing dataflow, beyond formalisms derived from conventional tree languages and graphs.
ii. An alternative model of computation to the Turing Machine network is needed, able to support the language environment, to model the internal working of functional nodes, and to encompass their relationship to the higher levels of abstraction of a dataflow program. The model should suggest new architectures, and better accommodate iteration, data structures, scheduling and resource allocation.

[10] Static Dataflow pays for this advantage by not being able to support data parallelism.

[11] Some dataflow architectural concepts had a recent appearance in the Explicit Dataflow Graph Execution Model in TRIPS [27].

### 2.3.4 Parallel graph rewriting and functional programming.

Functional Programming, together with a predicate calculus analogue called Logic Programming, formed the declarative paradigm, in which all iteration in program control is absorbed into recursive mechanisms[12]. Functional Programming languages (FPLs) are tree languages based on a formalism called the λ-calculus, introduced by Alonzo Church in the early twentieth century [28] and [29], that was presented as a general mathematical theory of computable functions. Graph rewriting is a technique that has been used for improving the efficiency of FPL sequential and parallel implementations [30], and [31].

The λ-calculus defines computable functions as λ-terms, defined by the simple BN statement $\langle\lambda - term\rangle ::= v \mid (\lambda v\langle\lambda - term\rangle) \mid (\langle\lambda - term\rangle\langle\lambda - term\rangle)$, where $v$ is drawn from some denumerable alphabet of variable symbols. The statement is powerful enough to enable any number, discrete object or computable function to be coded, somewhat inefficiently, as a λ-term. A λ-defined computable function can be applied to a λ-defined input, and evaluated according to a few *reduction rules*, which transform one λ-term into another.

The reduction rules are applied one at a time to those subexpressions which exhibit the input structure for some rule, known as *redexes* (reducible subexpressions). As will be discussed later in the section, parallel graph rewriting has been used partly to deal with subexpression repetition, and partly as an attempt to overcome this inbuilt sequential aspect of λ-evaluation. The pure λ-calculus cannot represent nor assign variables, resulting in a formalism with no updateable states, and implements iteration as a recursive function definition, involving a special λ-term called a Y-combinator.

[12]The declarative approach was proposed as the next great programming revolution in the 80s/90s, and it's failure was in sharp contrast to the widespread industry takeup of object oriented programming.

For the purposes of describing dataflow, the λ-calculus is in the category of a tree formalism, and is subject to the problems covered in the previous section. An attempt to implement λ-evaluations as a computer simulation, unsurprisingly results in the customary explosion of storage requirements and execution time, as lamented by Wadsworth in his PhD thesis on page 136, and in Plasmeijer in [32]. Moreover, there is no unique sequence for applying reduction rules to multiple redexes in a λ-term. The most efficient evaluation strategy, called applicative order evaluation, can have the undesirable property of not halting when a less efficient strategy, called normal order evaluation, does. Unfortunately normal order evaluation may also lead to repetitive argument evaluation.

Despite these problems, considerable effort has gone into making the λ-calculus the foundation of a computing paradigm, because it has been perceived as a means of facilitating program verification, and of unifying mathematics with computer science. Variations on the λ-calculus, fixes, and increasingly complex extensions to the theory, have resulted in a range of FPLs and compilers. The benefits ascribed to FPLs are considered in turn, with critiques:

i. **No side effects.** Side effects are software bugs associated with imperative languages (C, Fortran, Cobol etc..), which occur when subroutines or functions are allowed to modify global variables. These variables can result in a function/sub-routine retaining and changing states between calls, thereby delivering different outputs from identical inputs, at different stages of program execution. By eschewing variables and updateable states of any kind, FPLs are not subject to side effects. My view is that this is throwing out the baby with the bath water. In 2.4, it is argued that the absence of updateable states deprives a language environment of an effective mechanism for bypassing the SPR, for indicating which functions may be instructed to execute simultaneously, and for allocating computational resources. Updateable states do not in themselves cause side effects, and the complete abandonment of variables is unnecessary, because it is possible to program in imperative languages with a methodology that eschews global, or semi-global variables. The methodology can avoid global variables, and functions retaining states, by extending the list of input and output arguments handed down to each (sub)function.

ii. **Succinctness of functional definitions.** Succinctness ensues from two sources. The SPR in FPLs is to some extent mitigated by a fix called a *where* construction, which is a facility that dilutes the principle of not allowing function definitions within functions, by permitting an internal definition. However, nesting of *where* constructions is forbidden, perhaps because it is felt that nesting would be complex and unnecessary, and because avoiding subexpression repetition would require some form of the dreaded updateable state. Consequently a function definition can efficiently incorporate a dataflow with shared subexpressions, with only one or two layers. In order to describe a dataflow with sharing of depth $2k$ efficiently, the programmer is obliged to introduce $k$ conceptually irrelevant functions, which obscures the overall structure of the definition[13]. The second source of succinctness is the use of recursion and pattern matching to express all forms of iteration, which can often encrypt an entire iterative loop into one line of code. Recursion may be to the taste of some programmers, but many find it much less readable than iteration. Moreover, non-tail recursive function definitions can lead to exponential running times, whilst tail recursive definitions require the introduction of further variables, making programs more difficult to understand.

iii. **Referential Transparency.** The notion of referential transparency (RT) is related to the notion of side effects, and applies to λ-terms, FPL programs, and to the language *D(A)*. RT is the attribute where the replacement of a functional subexpression by it's value, is always independent of it's surrounding context. In *D(A)*, RT results from the unremarkable fact that algebraic operators have no internal state, entailing that every call of an operator with the same inputs, always results in the same output. From the point of view of program verification, RT is useful in allowing equivalent function definitions to be substituted for each other, allowing the derivation of new equations from old ones. (RT is of course undesirable for programs implementing memory allocation, explaining why memory allocation is dealt with implicitly by lists in FPLs.) It is suggested by proponents of parallel FPLs, that because RT entails that the order in which sub-expressions on the same layer are evaluated may be arbitrary, functional programs are inherently suited to parallel execution. But why is this obvious? It is more relevant to say that RT is only one aspect of parallelism; there are many other linguistic, theoretical, and implementational factors to consider as well. In a related argument, it is also proposed that RT is conducive to the optimization of FPLs, because it is a foundation for subexpression elimination, and the graph rewriting approach to parallelizing FPLs. I argue that lack of subexpression repetition should be an inherent feature of languages, rather than the result of compiler optimization. The effectiveness of graph rewriting is considered later in this section. Finally, it is worth noting that because suitably devised programming methodologies lack side effects, imperative languages can also be referentially transparent.

[13] It is onerous to be restricted to only two levels of nesting/brackets, because it can result in the granularity of a functional concept not being captured within a single function definition.

iv. **The Introduction of Computation into Mathematics.** A significant benefit ascribed to FPLs, is the facility with which assertions about recursive function definitions can be proved. The brevity and inductive aspect of recursive definitions, make them attractive for the formulation of inductive proofs, facilitating the verification of program correctness. The Curry-Howard isomorphism is a means of relating proofs in various, tree based logics, to various extensions of the tree based λ-calculus [33] [34] [35], and can be extended to relate proofs to FPL programs. But it has been argued here that tree formalisms limit the size of possible descriptions with shared subexpressions, and cannot explicitly indicate which semantic transformations may occur simultaneously. Consequently the size and complexity of the rules that FOPL formulas can describe is restricted, as is the potential parallelism and complexity of FPL programs and λ-expressions. Along with the doubtful status of the λ-calculus and extensions as useful models of computation in their own right, these factors place a question mark on the practical utility of the Curry-Howard isomorphism. This topic is discussed further in 8.4.

With the possible exception of program verification, the FPL advantages described above may be duplicated or superceded by imperative languages. The recursive nature of FPLs, and the storage/execution time issues arising from the SPR, had negative implications for runtime efficiency on sequential machines. To attempt to overcome this problem, parallel graph reduction introduced by Wadsworth, and term graph writing [36] [37], and more recently uplinks and λ-DAGs [38] were proposed. The defective tree structure of a λ-term with subexpression repetition, is transformed into a graph form without subexpression repetition. Reduction rules for λ-terms have to be reinterpreted as transformation rules for graphs, resulting in a new kind of graph based calculus. It is possible for a number of redexes to be present in a graph, able to be transformed simultaneously, in a process known as parallel graph reduction. A parallel graph compiler and runtime environment, must identify where these redexes are located by a process of pattern matching., and find a way of allocating work to a network of processors.

Concurrent Clean is a parallel FPL, dating from the early 90s. In [39], Plasmeijer observed that “an implementation (of FPLs) on parallel hardware is much more difficult than on sequential hardware, mainly because parallel machines (processor networks) are hard to program anyway”. Attempting to meet the challenge, Concurrent Clean uses annotations to mark subexpressions, which the programmer deems suitable for the compiler to parallelize. The compiler transforms tree based Clean programs into graphs, using RT as a basis for eliminating subexpression repetitions, and employing a range of sophisticated transformation techniques to optimise graph representations, and to improve runtime execution. The compiler relies on a work distribution model called ZAPP, in order to allocate work to a network of processors. Clean was a good example of the best achievable performance of it’s time, that could be expected from the technique of applying parallel graph rewriting to the execution of FPL programs on processor networks. There are four main obstacles to achieving good performance by parallelizing FPLs:

i. Rules of graph transformation are non-spatial and purely textual. There is no suggestion of a graph rewriting machine, closely related to graph based software, which might bypass the need for pattern matching. An implementation will be subject to software and runtime performance overheads, arising from the translation of mathematical structures into low level data structures, and from the implementation of graph transformations as subroutines, that are suitable for the chosen hardware.
ii. Pattern matching on graphs has the complexity of subgraph isomorphism, which is NP-complete [7]. Although there are techniques such as graph unwinding to more easily locate reducible expressions, their use is restricted to certain graph formats, and not scalable in a parallel context.
iii. For a processor network implementation, the lack of a good programming model means that a graph compiler must interface either with existing sub-optimal models, or deal directly with the hardware.
iv. The lack of layering in high level graph (and tree) languages, limits the extent to which they can express parallelism explicitly, entailing that additional mechanisms must be introduced.

Not surprisingly, the development of a mainstream FPL approach to parallel processing has not emerged, since the concepts were introduced over 18 years ago[14]. It has been argued that imperative languages are able to match or supercede FPL advantages, other perhaps than in the field of verification. How susceptible a Space program is to the verification process, has yet to be fully determined, but it's conventional sequential state transition character (see 6.3), referential transparency, and lack of side effects, are promising attributes. If a good verification framework for Space is achievable, then the inherent obstacles to parallelizing FPLs, cast doubt on the existence of a future growth path for the paradigm.

### 2.3.5 SUMMARY OF RELATED WORK

Trees alone are not suited to describe dataflow. Fixes and add-ons that are are employed to ameliorate the difficulties arising from lack of layering information for simultaneous semantic processing, and exponential overheads resulting from subexpression repetition in syntactic and surface representations, are inadequate. The issues that have been raised concerning graphs at the end of 2.3.4, and the notable abscence of a graph oriented, general purpose computer architecture, support the case that an entirely new approach for describing dataflow is needed.

## 2.4 HOW TO REPRESENT AND COMPUTE A DATAFLOW?

The conventional mathematical representation of a dataflow is as a labelled directed acyclic graph, whose most direct translation from the set theoretical definition to a data structure, would consist of an undifferentiated list of tuples with labelling information. The data structure would require transformation (into say an adjacency list or adjacency matrix), to make it suitable as input for a program for a parallel model of computing. It would be preferable to devise a mathematical structure that is optimised for parallel computation, requiring little or no transformation. The next section explains why a naive attempt to employ pointers to addressable locations in the memory of a computational device, which are used as temporary storage locations for the reuse of subexpression evaluations, will not suffice.

[14] At the time of writing, commercial applications are not listed on the Concurrent Clean website.

### 2.4.1 FIRST ATTEMPT AT EXPRESSING DATAFLOW WITHOUT SUBEXPRESSION REPETITION.

Consider an algebra $\mathbf{A} = \langle A,+,-,\div,..\rangle$ with no relations and only binary operations on the set $A$. A means of describing dataflow-type functions of the form $f{:}A^n \to A^m$ is described, composed out of the elementary operations of A. Define a memory function $\sigma{:}N \to A$, which is an updateable state, and can be viewed as a countably infinite collection of memory cells, each of which can store any value from A. Let $\Sigma = \{\sigma \mid \sigma{:}N \to A\}$. A function $g{:}\Sigma \to \Sigma$ is called a state transformer. Now the function $write{:}\Sigma \times A \times N \to \Sigma$ may be defined, which becomes a state transformer, if the latter two arguments are supplied. *Write* replaces the contents of the $n$th memory cell with the value $a$ :

$$write(\sigma,a,n) = \{\langle n,a\rangle\} \cup \{\langle i,x\rangle \in \sigma \mid i \neq n\}.$$

An *extended function* is a means of expressing an algebraic operation on the contents of memory locations. Let $g$ be an binary function in the algebra **A**. Then the extended function $g^{\sim}$ is defined as the function $g^{\sim}{:}\Sigma \times N^3 \to \Sigma$, where

$$g^{\sim}(\sigma,i_1,i_2,k) = write\left(\sigma,\ g(\sigma(i_1),\sigma(i_2)),\ k\right).$$

The application of the function $g^{\sim}$ is similar to the execution of an assembly language instruction (e.g. addR1R2R3), where the contents of the memory cells $i_1$, and $i_2$ are read, fed into the function $g$, and the result stored in memory cell $k$. A way of addressing the location of where a functional evaluation is stored (and read) has been introduced. Addressable memory enables the result of a functional evaluation to be available for as long as necessary as input for other evaluations, thus providing a way of avoiding subexpression repetition.
$f$ may now be expressed as:

$$f(x,y,z) = \sigma'(7), \text{ where}$$

$$\sigma' = \times^{\sim}\left(-^{\sim}\left(+^{\sim}\left(\times^{\sim}\left(+^{\sim}\left(write\left(write\left(write(\sigma,x,0),y,1\right),z,2\right),0,1,3\right),1,2,4\right),3,4,5\right),3,4,6\right),5,6,7\right) \qquad (f)'$$

Expressions like *(f)'*, that are composed of a series of applications of extended and write functions, will be called an *extended definition*. Informally, let *E(A)* be the language composed of all extended definitions using a sequence of nested extended functions applied to a sequence of nested write functions from the algebra A. The method for constructing extended definitions from a tree structure describing a dataflow like 2(a) , would be to:

i. Transform the tree to a dataflow graph 2(b) which eliminates subexpression sharing.
ii. Apply a succession of writes to load the inputs into an initial segment of memory, and then, working from the lowest layer upwards, apply in any sequence the associated extended functions with appropriate input values for each layer's algebraic operations.

The transformation of (*f*) into (*f*)' adds a one off cost of inserting inputs and extracting the output, a constant overhead of symbols per algebraic operation in the expression, and has rather reduced readability. But it has clearly done away with the need to repeat subexpressions[15]. The potential logarithmic compression in dataflow expression size is therefore available, using a tree syntax, which has less variability in it's structure. Although subexpression repetition has been avoided, there is no obvious way of introducing a sequence of time steps beyond identifying them with the linear sequence of functional applications, which obscures which algebraic operations might be performed simultaneously. Moreover, an extended definition makes no reference to possible resource allocation of algebraic 'functional units'. The formalism is a partial solution because it cannot make explicit information required for the efficient orchestration of parallel computation, leaving difficult combinatorial issues for a compiler and runtime environment.

The limitations of extended definitions, and the non-inductive nature of the method for constructing extended definitions, together do not justify the effort of separating syntax and semantics, which would enable the proofs of theorems, such as the plausible statement that for every expression $X$ in *D(A)*, there exists an extended definition $Y$ in *E(A)*, which is semantically equivalent to $X$. The next attempt at describing dataflow, however, does.

[15] The scheme also accommodates the ability to efficiently describe a function with more than one output. The function $h{:}\mathrm{A}^3 \rightarrow \mathrm{A}^2$ , defined using bracket notation as

$$h(x,y,z) = \left\langle ((x+y)+(y\times z)) \times ((x+y)-(y\times z))\ ,\ ((x+y)-(y\times z)) \right\rangle$$

is expressed as $h(x,y,z) = \left\langle \sigma'(7), \sigma'(6) \right\rangle$, with $\sigma'$ defined as before.

## 2.4.2 INTERLANGUAGE ENVIRONMENT: A SECOND ATTEMPT AT EXPRESSING DATAFLOW

A tree structure with restricted variability called the interstring is presented, along with a computational environment in which a notion of time, an array of memory cells, and an array of functional units, that allows a circumvention of the problem of indicating which subexpressions are shared, and which may be simultaneously evaluated. The interstring can also express data transfers and allocation of resources for primitive algebraic operations, which is important for attaining the goal of a spatially oriented programming language. The interstring is a program which transforms the memory, whose output is always stored in a designated cell.

The idea is to design a tree syntax, where the degree of freedom that an individual branch may be arbitrarily long, is restricted to certain child branches of the rightmost branch, as in figure 2.5. Instead of dataflow expressions being trees with a high degree of structural variability, requiring a complex parsing phase, interstrings have a highly regular structure (strings of strings of simple expressions). A theorem is presented, that for any arithmetic-style expression with high structural variability, there exists a semantically equivalent pair of an abstract memory, and interstring with low structural variability.

An interstring exhibits an intermediate structure, which enables the equivalent of a child part of speech to participate in more than one parent part of speech, and only requires simple and readily parallelizable forms of syntactic/semantic processing. To compare this approach for describing dataflow with tree languages, a syntax and semantics is defined for a tree dataflow language L(V,F,C) derived from a standard Tarskian treatment of 2-ary functional terms in FOPL.

(1) Syntax for the language L(V,F,C).

Basic Symbols

i. parentheses ( )
ii. variables $x,y,z,......$(Countable set V)
iii. 2-ary function symbols $Id,+,\times,\div,-,...$ (Countable set F)
iv. constant symbols $a,b,c,...$ (Countable set C)

The set of terms L(V,F,C) is the smallest set satisfying these conditions:

i. any variable is a term,
ii. any constant symbol is a term,
iii. if $f$ is a function symbol, and $t_1$ and $t_2$ are terms, then $f(t_1,t_2)$ is a term.

(2) Semantics.

A model for L(V,F,C) is a pair M = < D, I >, where

i. D is a non-empty set, called the domain of M.
ii. I is a function such that $\forall c \in C( \mathrm{I}(c) \in \mathrm{D} )$, and $\forall f \in \mathrm{F}, \mathrm{I}(f)$ is a function $\mathrm{D}^2$ to D, and there is a function name $Id \in F$, reserved for a binary identity function on D.

An assignment in a model M is a function $A: V \rightarrow D$.

$[[t]]_{M,A}$ is called the denotation of the term t in M relative to A.

$\forall t \in \mathrm{L}(\mathrm{V,F,C})\ [[\mathrm{t}]]_{\mathrm{M,A}} =$

I(c), if t is a constant symbol.

A(v), if t is a variable,

$I(f)([[\mathrm{t}_1]]_{\mathrm{M,A}},[[\mathrm{t}_2]]_{\mathrm{M,A}})$, if t $= f(t_1,t_2)$ and $f \neq Id$.

$[[\mathrm{t}_1]]_{\mathrm{M,A}}$ , if t $= Id(t_1,t_2)$

Given a model M = < D, I > and an assignment A, any syntactic element in L(V,F,C) may be mapped to a semantic element in D. The inductive definition generates a tree structure of semantic evaluations for a functional term. As a by-product, the definition also yields a tree expression within the semantic domain, similar to the way the syntactic inductive definition does for functional terms. The SPR in L(V,F,C) has been transferred into the semantic interpretation.

A system called the *interlanguage environment* is now presented, using the same symbol sets V, F, C, for describing dataflows, which is considerably more complex than L(V,F,C). The interlanguage environment does however, overcome the problems of the previous approach, and gives clues for the design of novel parallel language systems, and formal models of computation.

(3) Definition of an interlanguage environment I(V,F,C).

Given the sets V, F, C as before, an interlanguage *environment* I(V,F,C) is a subset of the cartesian product $\Psi \times \mathrm{Y}$, defined as follows:

i. $\psi$ is a finite memory of cells, each of which stores an element of $\mathrm{V} \cup \mathrm{C}$ . The memory is attached to a notional array of *k* abstract functional units (FUs), as depicted in figure 2.3. The array has a "wireless" connection system, for copying the contents of cells into others. Formally $\psi$ is a function $\psi\colon N_{3k} \rightarrow \mathrm{V} \cup \mathrm{C}$ , and is called a *k*VC-memory. Let $\Psi = \{\psi \mid k > 0, \text{ and } \psi\colon N_{3k} \rightarrow V \cup C\}$.

*ii.* *Y* is an interstring, which is a structure belonging to a subset of $\left((F \times N)^{*} \cup (N \times N)^{*}\right)^{*}$

The expression in fig 2.4 is a simplified depiction of an interstring, and represents an abstract program for the 2VC-memory, consisting of an alternating sequence of columns of pairs of symbols. The columns are separated by double colons " :: ", and terminated by a colon and semi-colon ' :; ". The construction of an interstring for a *k*VC-memory involves two types of columns: *An alpha column* is a string of pairs of function symbols and integers, represented as $f(j)$, where $f(j)$ is understood to express the activation of the *j*th functional unit, $1 \le j \le k$, by taking as the first and second inputs the contents of the 3*j*+1 and the 3*j*+2 memory cells respectively, and to write the result of applying the notional algebraic operation associated with the "opcode" symbol *f* into the 3*j*+3 memory cell. Any number of functional units may be activated in the array, but there is a restriction that no attempt is made to activate the same functional unit twice. Formally an alpha column $\alpha \in A_k \subset (F \times N)^{*}$, where $A_k$ is defined as follows. Let $length(\alpha)$ be the length of the string $\alpha$. The expression $\partial_p(\alpha) = \langle f,i \rangle$ means that the tuple $\langle f,i \rangle$ appears as the *p*th tuple in $\alpha$, where $1 \le p \le length(\alpha)$. Further $\partial^1_p(\alpha) = f$, $\partial^2_p(\alpha) = i$, if $\partial_p(\alpha) = \langle f,i \rangle$. The set of $\alpha$ columns is defined as:

$$A_k = \left\{ \alpha \in (F \times N)^{*} \;\middle|\; \forall p,q \left( \begin{array}{l} (1 \le p,q \le length(\alpha)) \Rightarrow \\ \left( (1 \le \partial^2_p(\alpha) \le k) \wedge (1 \le \partial^2_q(\alpha) \le k) \wedge \right. \\ \left. \exists i\left( (\partial^2_p(\alpha) = i) \wedge (\partial^2_q(\alpha) = i) \Rightarrow (p = q) \right) \right) \end{array} \right) \right\},$$

A *beta column* is a string of pairs of integers, represented as $n \rightarrow m$, each of which describes an instruction that copies "wirelessly" the contents of cell *n*, into cell *m.* The copies are understood to all occur in one step. Multiple reads from a cell are permissible , but multiple copies to the same cell are disallowed. It is also permissible for a cell to appear as both source and destination in a beta column, because a copy cycle consists of a read phase followed by a write phase. A beta column $\beta$ is a finite string of tuples in $N \times N$, and is member of a set $B_k \subset (N \times N)^*$, where $B_k$ is defined as follows. Let $length(\beta)$ be the length of the string $\beta$ of integer pairs, and the expression $\chi_p(\beta) = \langle i,j \rangle$ means that the tuple $\langle i,j \rangle$ appears as the $p$th tuple in $\beta$, where $1 \le p \le length(\beta)$. Further $\chi_p^1(\beta) = i$ , $\chi_p^2(\beta) = j$ if $\chi_p(\beta) = \langle i,j \rangle$. Then the set of $\beta$ columns is defined as:

$$B_k = \left\{ \beta \in (N \times N)^* \;\middle|\; \forall p,q \left( \begin{array}{l} \left(1 \le p,q \le length(\beta)\right) \Rightarrow \\ \left( \left(1 \le \chi_p^1(\beta), \chi_q^1(\beta) \le 3k\right) \wedge \left(0 \le \chi_p^2(\beta), \chi_q^1(\beta) \le 3k\right) \right) \\ \exists i \left( \left(\chi_p^2(\beta) = i\right) \wedge \left(\chi_q^2(\beta) = i\right) \Rightarrow (p = q) \right) \end{array} \right) \right\}$$

An interstring $Y$ of dimension *k,* denoted $Y^k$*,* where *k* is a non-negative integer*,* may now be defined as a member of the set construction $(A_k \cup B_k)^*$ , where by convention an interstring is non-null and composed of an alternating series of alpha and beta columns, which takes the form $Y^k = \langle \alpha_0, \beta_0, \alpha_1, \beta_1, \ldots \alpha_{n-1}, \beta_{n-1} \rangle$, $\alpha_i \in A_k$, $\beta_j \in B_k$, where $n \ge 1$. The terminating column always contains a single element, of the form $\langle\langle t,0 \rangle\rangle$, where $t \bmod 3 = 0$. For ease of representation, $Y^k$ will be depicted as $\alpha_0 :: \beta_0 :: \alpha_1 :: \beta_1, \ldots :: \alpha_{n-1} :: \beta_{n-1} :;$ , with a pair of double colons separating the columns, and a colon/semi-colon pair terminating the expression. A sample interstring's syntactic tree structure is depicted in fig 2.5.

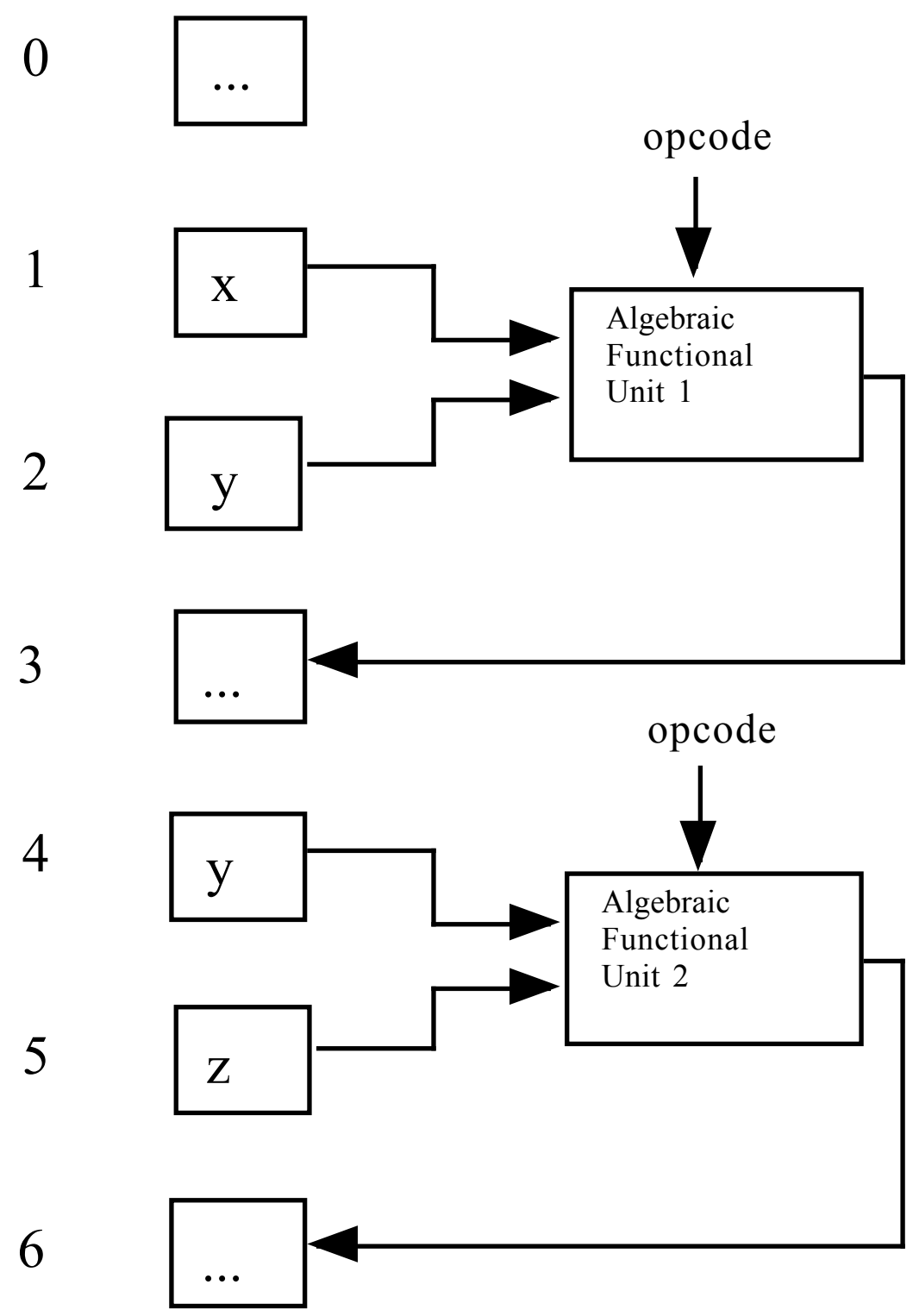

.Figure 2.3. A 2VC-memory with abstract functional unit array.

$$
\begin{array}{ccccccccccccc}
+(1) & :: & 3\to 1 & :: & +(1) & :: & 3\to 1 & :: & \times(1) & :: & 3 \to 0 & ::; \\
\times(2) & & 6\to 2 & & -(2) & & 6\to 2 & & & & & \\
 & & 3\to 4 & & & & & & & & & \\
 & & 6\to 5 & & & & & & & & &
\end{array}
$$

Figure 2.4 Simplified depiction of an interstring program, as a string of string of symbol pairs, which computes $f(x,y,z)=(((x+y)+(y\times z)) \times ((x+y)-(y\times z)))$, with the result stored in cell #0.

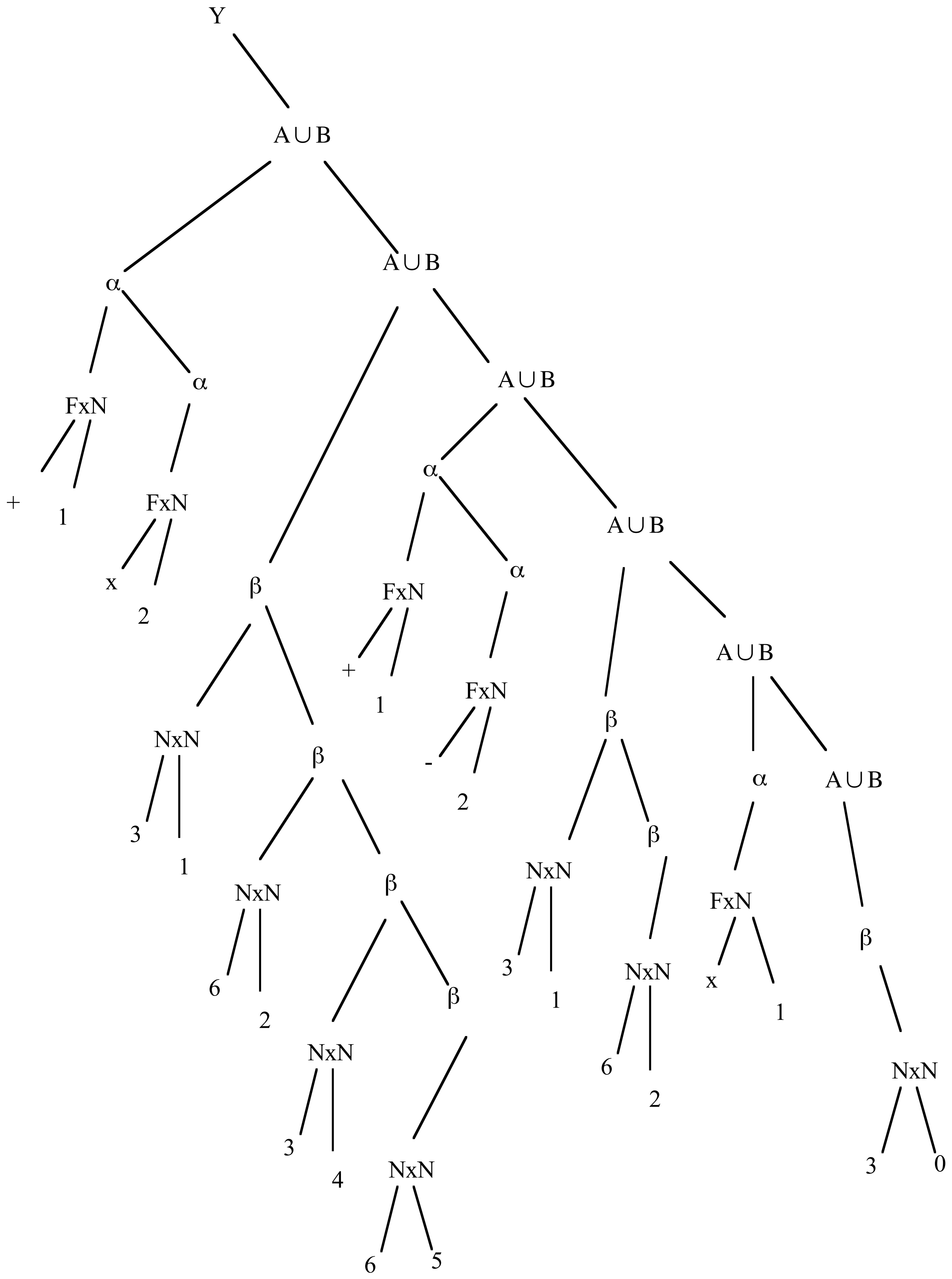

Figure 2.5 Tree syntax of interstring in figure 2.4, based on the set definition, rather than the "::" construction.

The interlanguage environment I(V,F,C) may now be defined:

$$\text{I(V,F,C)} \ = \ \left\{ \left\langle \psi, Y \right\rangle \in \Psi \times \text{Y} \,\middle|\, k > 0,\ \psi \text{ is a } k\text{VC memory, and } Y \text{ is of dimension } k \right\}$$

For the purpose of relating L(V,F,C) and I(V,F,C), the execution of the interstring program $Y^k$ on a $k$VC-memory, given M and A, is understood to yield a single output from a dataflow, which is stored in the memory's zeroth cell. The definition of Y as $\left(A_k \cup B_k\right)^*$, where string elements are understood to alternate between $A_k$ and $B_k$, rather than the more mathematically concise $\left(A_k \times B_k\right)^*$, is to simplify the theorem's proof.

(4) Semantics for I(V,F,C).

The interlanguage environment I(V,F,C) has been defined as the set of pairs of *kVC* memories and interstrings of dimension *k*, according to the definitions of the constructions $\Psi$,Y, and the given sets V,F,C. Given an element $\left\langle \psi, Y \right\rangle \in \text{I}\left(\text{V,F,C}\right)$, a model $\text{M} = \left\langle \text{D,I} \right\rangle$, and an assignment A for L(V,F,C) , the term $\left\langle\left\langle \psi, Y \right\rangle\right\rangle_{\text{M,A}}$ is called the denotation of $\left\langle \psi, Y \right\rangle$ in M relative to A, which will be defined presently.

A function of the form $\sigma : N_{3k} \rightarrow \text{D}$ is called a *kD-memory*. The denotation of a $k$VC-memory $\psi$ in M relative to A, is the $k$D-memory $[[\psi]]_{\text{M,A}} : N_{3k} \rightarrow \text{D}$:

$$[[\psi]]_{\text{M,A}} \ = \ \left\{ \left\langle i,\ [[t]]_{\text{M,A}} \right\rangle \,\middle|\, \left\langle i,t \right\rangle \in \psi \right\} \qquad \text{(i)}$$

Given M and A, the abstract array of FUs of a $k$VC-memory $\psi$ may now be translated into a concrete domain array of FUs for the *kD*-memory, where an FU may be instructed to perform any binary operation $\text{I}\left(f\right) : \text{D}^2 \rightarrow \text{D},\ f \in \text{F}$. A $k$D-memory and domain FU array, constitutes a simple model of computation called the *intermachine*, with interstrings as programs. An interstring *Y* applied to $[[\psi]]_{\text{M,A}}$, produces a sequence of $k$D-memories, by executing an alternating succession of alpha and beta columns individually, from left to right. A notion of time may therefore be introduced, where an intermachine time step or machine cycle is identified with the execution of either an alpha or beta column element. The number of elements in *Y*'s outermost string, is represented by the expression *length*(*Y*).

For $Y = \alpha_0 :: \beta_0 :: \alpha_1 :: \beta_1, ... :: \alpha_{n-1} :: \beta_{n-1} ::$, $length(Y) = 2n$, and an intermachine executes $Y$ in $length(Y)$ cycles. For the purpose of establishing the theorem below, the interstring is understood to compute only a single output from a dataflow, which is stored in the memory's zeroth cell.

The assembly of the pair $\langle \psi, Y \rangle$ in I(V,F,C) is analogous to the construction of a functional term $X$ in L(V,F,C), although as we will see, not every interlanguage element represents a functional term or meaningful dataflow. The execution of the interstring program $Y$ on $[[\psi]]_{M,A}$ , and the reading the contents of the zeroth memory cell, is analogous to the semantic evaluation of the functional term in L(V,F,C), given M and A.

Given a model $M$, the function $cycle_M$ is applied to an integer $h$, a $k$D-memory and an interstring of dimension $k$, and generates a new D-memory state, after the first $h$ of an interstring's columns have been processed. We define $cycle_M$ as follows.

Let the set of $k$D-memories $\Sigma_k = \{ \sigma \mid \sigma : N_{3k} \rightarrow D \}$, then the functions $apply_M : \Sigma_k \times A_k \rightarrow \Sigma_k$ and $copy_M : \Sigma_k \times B_k \rightarrow \Sigma_k$, where $\alpha \in A_k$, $\beta \in B_k$, are defined:

$$apply_M(\sigma,\alpha) = \left( \bigcup_{(\delta_p(\alpha) = \langle f, i \rangle)} \langle 3i, \mathrm{I}(f)(\sigma(3i-2), \sigma(3i-1)) \rangle \right) \cup \left\{ \langle j, x \rangle \in \sigma \;\middle|\; j \neq 3(\delta_p^2(\alpha)) \right\}$$

where $1 \leq p \leq length(\alpha)$ (ii)

$$copy_M(\sigma,\beta) = \left( \bigcup_{(\chi_p(\beta) = \langle i, j \rangle)} \langle j, \sigma(i) \rangle \right) \cup \left\{ \langle k, l \rangle \in \sigma \;\middle|\; k \neq \chi_p^2(\beta) \right\}$$

where $1 \leq p \leq length(\beta)$ (iii)

$apply_M$ and $copy_M$ generate sets rather than multisets, because the definitions of $A_k$ and $B_k$ do not allow the union iterators to generate repetitions, and because in each case both set constructions in the outer union of the expression, are disjoint.

We define $ca_M : \Sigma_k \times (A_k \cup B_k) \rightarrow \Sigma_k$ to be

$$ca_M(\sigma,\gamma) = apply_M(\sigma,\gamma) \quad \text{if } \gamma \in A_k$$
$$copy_M(\sigma,\gamma) \quad \text{if } \gamma \in B_k \qquad \text{(iv)}$$

The notion of a memory being modified by an interstring after $h$ intermachine cycles[16] is captured by the partial function $cycle_M : N \times \Sigma_k \times (A_k \cup B_k)^* \rightarrow \Sigma_k$.

For a non-null interstring $Y = \langle \gamma_0, \gamma_1, \ldots \gamma_{m-1} \rangle$,

$$cycle_M(h, \sigma, \langle \gamma_0, \gamma_1, \ldots\ldots \gamma_{m-1} \rangle)$$
$$= ca_M(\ldots\ldots(ca_M(\, ca_M(\sigma,\gamma_0),\gamma_1))\ldots\ldots,\gamma_{h-1}) \quad \text{for } 1 \le h \le m.$$

and is undefined for $h > m$. (v).

The denotation of $\langle \psi, Y \rangle$ in M relative to A may now finally be given

$$\langle\langle \psi, Y \rangle\rangle_{\mathrm{M,A}} \equiv cycle_M(\, length(Y),\, [[\psi]]_{\mathrm{M,A}},\, Y\,)\,(0) \qquad \text{(vi).}$$

In expression (vi), "(0)" expresses the reading of the contents of the zeroth cell of the memory construction $cycle_M(length(Y),\, [[\psi]]_{\mathrm{M,A}},\, Y)$. [17]

**Theorem.**

Let $X$ be a functional term in L(V,F,C). Then there is an element $\langle \psi, Y \rangle$ of I(V,F,C), such that for every model M, and every assignment A, $[[X]]_{\mathrm{M,A}} = \langle\langle \psi, Y \rangle\rangle_{\mathrm{M,A}}$.

---

[16] $cycle_M$ might be given a recursive definition, given that it processes strings of variable length. In 8.5.1, it is argued that recursive functions are not conducive to parallel computation. With the long term aim of introducing parallel computation into mathematical discourse in mind, recursion and least fixed point semantics are avoided, by using "iterative" definitions.

[17] Unlike functional terms in FOPL, the interlanguage environment can be modified to describe a function with more than one output without subexpression repetition, by adding an extra element of an "output string" of integers $m_1 m_2 . m_r$ representing cell numbers, to the environment: $\Psi \times \mathrm{Y} \times N^*$. Instead of reading only the output of the zeroth memory cell, we might have

$\langle\langle \psi, Y, \langle m_1 m_2 . m_r \rangle\rangle\rangle_{\mathrm{M,A}} \equiv \langle \sigma(m_1), \sigma(m_2), .. \sigma(m_r) \rangle$, where $\sigma = cycle_M(\, length(Y),\, [[\psi]]_{\mathrm{M,A}},\, Y\,)$.

**Proof.** By structural induction on functional terms in L(V,F,C). Recall $Id \in F$ is reserved as a special element, such that for every M = < D, I >, $I(Id)$ is the binary identity function on D

**Base Case.** $X \in V \cup C$.

Let $\psi{:}N_3 \rightarrow V \cup C$ be a *1VC*-memory, where $\psi(i) = X$, $0 \le i \le 3$. $[[X]]_{M,A} = d$, for some $d \in D$, therefore $[[\psi]]_{M,A} = \{ \langle i,d \rangle \mid 0 \le i \le 3 \}$ by (i). $\psi(0)$ already contains the value $[[X]]_{M,A}$, but the inductive step requires a non-null interstring $Y$ for the base case, so let $Y = \langle \langle\langle Id,1 \rangle\rangle, \langle\langle 3,0 \rangle\rangle \rangle$.

$\langle\langle \psi,Y \rangle\rangle_{M,A}$ = $cycle_M\left(2, [[\psi]]_{M,A}, \langle\langle\langle Id,1\rangle\rangle,\langle\langle 3,0\rangle\rangle\rangle\right)(0)$ by (vi) and subst. for $Y$ and $length(Y)$.

$ca_M\left( ca_M\left([[\psi]]_{M,A}, \langle\langle Id,1\rangle\rangle\right), \langle\langle 3,0\rangle\rangle \right)(0)$ by (v).

$ca_M\left( apply_M\left([[\psi]]_{M,A}, \langle\langle Id,1\rangle\rangle\right), \langle\langle 3,0\rangle\rangle \right)(0)$ by (iv).

$ca_M\left( [[\psi]]_{M,A}, \langle\langle 3,0\rangle\rangle \right)(0)$ by (ii), alpha column with *Id* preserves $[[\psi]]_{M,A}$

$copy_M\left( \{ \langle i,d \rangle \mid 0 \le i \le 3 \}, \langle\langle 3,0\rangle\rangle \right)(0)$ by (iv) and subst. for $[[\psi]]_{M,A}$

$\{ \langle i,d \rangle \mid 0 \le i \le 3 \}(0)$ by (iii), beta column preserves cells with same value.

$= d = [[X_{M,A}]]$.

**Inductive Case.** $X = f(t_1,t_2)$

By the base case hypothesis, there are *VC*-memories $\psi_1$, $\psi_2$, and non-null interstrings $Y_1$, $Y_2$ such that $[[t_1]]_{M,A} = \langle\langle \psi_1,Y_1 \rangle\rangle_{M,A}$, $[[t_2]]_{M,A} = \langle\langle \psi_2,Y_2 \rangle\rangle_{M,A}$ for all M and A. The inductive step requires the construction of $\psi_3$, $Y_3$, such that $[[f(t_1,t_2)]]_{M,A} = \langle\langle \psi_3,Y_3 \rangle\rangle_{M,A}$.

The step involves conjoining the two base case memories into a larger *VC*-memory, a column by column appending of the base case interstrings with numeric offsets added to elements of $Y_2$, that will preserve the calculations of $[[t_1]]_{M,A}$, and $[[t_2]]_{M,A}$, removing the final beta columns, and appending alpha and beta columns to the resulting interstring to calculate $[[f(t_1,t_2)]]_{M,A}$, and transferring the value to the zeroth cell of the larger *VC*-memory. For some non-zero natural numbers *k,l*, $\psi_1{:}N_{3k} \rightarrow V \cup C$, $\psi_2{:}N_{3l} \rightarrow V \cup C$.

Let $\psi_3 : N_{3(k+l)} \rightarrow V \cup C$, where $\psi_3(i) = \begin{cases} \psi_1(i) & \text{if } 0 \le i \le 3k \\ \psi_2(i-3k) & \text{if } 3k+1 \le i \le 3k+3l \end{cases}$.

Let $Y_1 = \langle \gamma_0, \ldots, \gamma_{n-1} \rangle$, where $\gamma_i \in A_k \cup B_k$, and $Y_2 = \langle \mu_0, \ldots, \mu_{m-1} \rangle$, where $\mu_i \in A_l \cup B_l$. For simplicity assume $n \ge m$. Define $Y_3 = \langle \lambda_0, \ldots, \lambda_{n-1}, \lambda_n, \lambda_{n+1} \rangle$ as follows.

1). For $1 \le i \le m-2$, if $i$ is odd, $\gamma_i \in A_k$, $\mu_i \in A_l$, and for some series of function symbols $f_w$, $g_x$ in *F*, and some series of natural numbers, $u_y$, $v_z$

$\gamma_i = \langle\langle f_1, u_1\rangle, \langle f_2, u_2\rangle \ldots \langle f_s, u_s\rangle\rangle$, $\mu_i = \langle\langle g_1, v_1\rangle, \langle g_2, v_2\rangle \ldots \langle g_t, v_t\rangle\rangle$. Then construct $\lambda_i$ by adding an offset of *k* to the second element of the $\mu_i$ pairs, and append the result to $\gamma_i$ as follows:

$\lambda_i = \langle\langle f_1, u_1\rangle, \langle f_2, u_2\rangle \ldots \langle f_s, u_s\rangle, \langle g_1, v_1 + k\rangle, \langle g_2, v_2 + k\rangle \ldots \langle g_t, v_t + k\rangle\rangle$, where $\lambda_i \in A_{k+l}$

2). For $1 \le i \le m-2$, if *i* is even, $\gamma_i \in B_k$, $\mu_i \in B_l$, and for some series of natural numbers $u_y$, $v_z$,

$\gamma_i = \langle\langle u_1, u_2\rangle, \langle u_3, u_4\rangle \ldots \langle u_{s-1}, u_s\rangle\rangle$, $\mu_i = \langle\langle v_1, v_2\rangle, \langle v_3, v_4\rangle \ldots \langle v_{t-1}, v_t\rangle\rangle$. Then construct $\lambda_i$ by adding an offset of 3*k* to both elements of every pair in $\mu_i$, and appending the result to $\gamma_i$ as follows:

$\lambda_i = \langle\langle u_1, u_2\rangle, \langle u_3, u_4\rangle \ldots \langle u_{s-1}, u_s\rangle, \langle v_1 + 3k, v_2 + 3k\rangle, \langle v_3 + 3k, v_4 + 3k\rangle \ldots \langle v_{t-1} + 3k, v_t + 3k\rangle\rangle$

where $\lambda_i \in B_{k+l}$. Here the series $u_y$, $v_z$ and their lengths are understood to be different from the case when *i* is odd.

3). If $m \ne n$, then for $m-1 \le i \le n-2$, $\lambda_i = \gamma_i$. [18]

4). To complete the definition of $Y_3$, $\lambda_{n-1} = \langle\langle 3,1\rangle, \langle 3k+3, 2\rangle\rangle$, $\lambda_n = \langle\langle f, 1\rangle\rangle$, and $\lambda_{n+1} = \langle\langle 3,0\rangle\rangle$.

Lemma: $cycle_M\left((n-2), [[\psi_3]]_{M,A}, Y_3\right)(3) = [[t_1]]_{M,A}$

and $cycle_M\left((n-2), [[\psi_3]]_{M,A}, Y_3\right)(3k+3) = [[t_2]]_{M,A}$

[18] The construction of $Y_3$ does not implement the elimination of subexpression sharing. It may easily be achieved in steps 1, 2 and 3, by searching for and deleting duplicates of function terms and operands in each alpha and beta column pair, assuming each function/operands combination appears as early (i.e. in the leftmost interstring column) as possible, without affecting the semantics of the dataflow.

Sketch Proof: $\psi_3$ has been assembled so that the contents of $\psi_1$ have been copied into the first $3k+1$ cells of $\psi_3$, and the contents of $\psi_2$, excluding the zeroth cell, have been copied into the remaining cells of $\psi_3$. $Y_3$ has been constructed as a column by column appending of the base case interstrings $Y_1$ and $Y_2$, with numeric offsets added to elements of $Y_2$. In the first $n-2$ cycles, the execution of $Y_3$ preserves the evaluation of $Y_1$ alpha and beta elements on the $\psi_1$ segment of $\psi_3$, without interfering with the evaluation of $Y_2$ alpha and beta elements on the $\psi_2$ segment of $\psi_3$, where the latter evaluation may have completed in an earlier cycle.

$cycle_M\left((n-2),\ [[\psi_3]]_{M,A},\ Y_3\right)(3) = [[t_1]]_{M,A}$ by Lemma

$\Rightarrow cycle_M\left(n-1,\ [[\psi_3]]_{M,A},\ Y_3\right)(1) = [[t_1]]_{M,A}$, by (iii), (iv) and (v).

$cycle_M\left((n-2),\ [[\psi_3]]_{M,A},\ Y_3\right)(3k+3) = [[t_2]]_{M,A}$ by Lemma

$\Rightarrow cycle_M\left(n-1,\ [[\psi_3]]_{M,A},\ Y_3\right)(2) = [[t_2]]_{M,A}$, by (iii), (iv) and (v).

$\Rightarrow cycle_M\left(n,\ [[\psi_3]]_{M,A},\ Y_3\right)(3) = [[f(t_1),t_2]]_{M,A}$ by (ii), (iv), and (v).

$\Rightarrow cycle_M\left(n+1,\ [[\psi_3]]_{M,A},\ Y_3\right)(0) = [[f(t_1),t_2]]_{M,A}$ by (iii), (iv), and (v) □

The theorem says that any functional term in L(V,F,C), is semantically equivalent to some element of I(V,F,C). This demonstrates that any tree expression, which can exhibit high structural variability, can be coded by a simple finite memory, and an interstring whose tree structure has very restricted structural variability. Although the syntax and semantics for I(V,F,C) is more complex, they provide a foundation for a parallel computing framework, that is absent from L(V,F,C), and similar tree formalisms for expressing dataflow.

If a function name and FU number pair $\langle f,k \rangle$ in an alpha column, is associated as a "part of speech", then the ability of a beta column to describe the copying of the contents of memory cells into other cells, enables the "part of speech" to be available as a component for a plurality of more complex "parts of speech". Consequently, in this restricted context of functional terms, a means is provided to avoid the SPR and subexpression repetition. Bypassing the SPR has resulted in a collapse of the complexity of the syntax of an interstring considered as a tree, compared with the syntax of conventional tree expressions.

If the result of a functional evaluation is intended to be input for use in a pair later than the next alpha beta pair, the result may either be left for as long as necessary in the output cell of an FU, which is intentionally not activated again until the result is no longer needed, or the result may be copied into temporary storage in some other memory cell.

Interstrings have three superior characteristics as a format compared to conventional tree based expressions of dataflow.

i. Although as presented they do not guarantee the absence of subexpression repetition, they provide a means for avoiding it entirely.
ii. The interstring supports scheduling and resource allocation, by directly indicating in a alpha column which functional evaluations and in which FUs, may be performed simultaneously in each time step.
iii. The interstring specifies data management by explicitly listing data transfers between machine locations in a beta column.

The algebraic array has been wired somewhat inefficiently, since the output of a FU could be written into one of the inputs, thus having one FU per 2 instead of 3 cells. This alternative array can also reduce the size of an interstring's copy columns, because outputs can remain in place as inputs for the evaluations in next function column. But the 3 cell arrangement has the advantage of improving interstring readability.

An element $\langle \psi, Y \rangle$ may not actually describe a dataflow in a coherent way, because beta columns in $Y$ may write to cells that are never used as inputs or outputs, and functional units in alpha columns may be activated, which have not received inputs since their last activation. A formal language theory of interstrings would be desirable, capable of describing the incremental construction of well formed, readable dataflows expressed as elements of $\langle \psi, Y \rangle$, lacking subexpression repetition, for an efficiently wired algebraic array. [19] To achieve this, the theory would encompass some kind of compilation phase to transform readable interstrings into semantic versions of interstrings, which would constitute machine code for the intermachine.

[19] It would also be desirable to explore the consequences of extending the interlanguage environment to cover the equivalent of predicate terms in predicate logic, and this is dealt with in a future paper.

The description of a system for constructing well-formed interstrings is somewhat more involved than a Backus Naur definition. The interlanguage environment's primary purpose, is to give clues as what might be required from a parallel-oriented low level formal model of computation. The presentation of an interstring grammar within the context of the interlanguage environment would not be worthwhile at this stage, because as will be discussed in the next section, it is not fully programmable. The presentation of an interstring grammar will be deferred until a programmable environment for algebras encoded as abstract datatypes and virtual functions/modules, has been developed.

### 2.5 PROSPECTS FOR A FULLY PROGRAMMABLE SCHEME BEYOND DATAFLOWS.

With the introduction of the interlanguage environment, it has been demonstrated that an interstring is superior to the conventional use of tree based strings with high structural variability at describing general dataflow, and the first attempt at describing dataflow with a memory in 2.3, by eliminating subexpression repetition, by indicating which algebraic operations may be simultaneously evaluated, and by explicitly allocating functional unit resources and data transfers. However, the interlanguage environment and semantics beg many questions, because it is not obvious how to introduce a generally programmable formalism:

i. Is the interstring stored only "on the page", like the instruction table of a Turing Machine, or if in some other fashion, in what format?
ii. If L(V,F,C) and I(V,F,C) are extended to include a set of relation symbols, how are the true/false results of relations computed and stored? How can more complex functions than macros be built up using relations and if statements?
iii. How can data structures and iteration be implemented?
iv. How might interstrings and interstring programs themselves be treated as inputs to functions?

Whilst there is no immediate way in which the interlanguage environment suggests answers, it does hint at what might be required from a low level, parallel oriented formal model of computation, capable of supporting a high level, fully programmable environment. The intermachine, however limited in programmability, has the following attributes:

- An act of computation is always associated with a particular hardware resource (for generality not specifically a functional unit), and the result of any computation is always stored in an addressable location.
- The machine has a clocked environment, in which there is a facility to activate an arbitrary selection of hardware elements in one time step.
- Full connectivity between hardware elements in constant time, where connections occupy a minimal, or at most sub-quadratic amount of space.

A model of computation that accommodates these characteristics will be called *synchronically oriented*[20]**.** Synchronic orientation inspired the design of the α-Ram family of models, which is the subject of the next chapter. Space is the first fully programmable language based on interstrings, with a functionality similar to C. It is designed to run on a model of computation called the Synchronic A-Ram, and is presented in chapters 5-7.

## 2.6 Concluding comments.

It is difficult to imagine how SPR and subexpression repetition can be dealt with, without the use of addressable storage locations. A tree language, co-presented perhaps with an transducer-type automata for testing the syntactic structure of expressions, but without consideration for addressable memory, and at least an abstraction of a computer environment for semantic processing, is not sufficient to adequately describe shared subexpressions. The inability to directly share subexpressions leads to disconnected representations of dataflow, many-to-many relationships, and environments in general, generating a kind of linguistic schizophrenia. Given these considerations, it is hard to avoid the following conclusion; the conventional tree form with high structural variability, is wholly unsuited as a general purpose syntactic structure for formal languages whose semantic interpretations are dataflows, including high performance programming languages.

An abstract machine and memory environment designed to relate the processing of syntactic expressions into semantic objects, can bypass severe restrictions on the size and complexity of the dataflows that can be described, and on how efficiently semantic processing can take place. The semantics for such an environment has no need to explicitly represent graphs, nor to incorporate pattern matching, and as will be demonstrated, has the potential to inspire innovatory models of computation, and novel parallel architectures.

[20] The Turing machine and the λ-calculus, the standard models of computer science, are not synchronically oriented. A cell of a Turing Machine tape is not directly addressable or reachable in constant time, and a λ-expression has no explicit updateable state, beyond the expression itself.

There is another problematic aspect of tree and graphs, if they are used as a basis for data structures to describe objects other than dataflows, such as partial orders and other mathematical structures. The internal content of such data structures can only be accessed by following the chain of links between nodes, which can impede the speed and simultaneity with which the data structure is processed. Trees in particular are normally inductively assembled, somewhat arbitrarily providing a justification for recursive procedures, in order to process them. In a future paper, it is argued that recursion itself, is not conducive to parallelism, and should only be introduced in rare, inherently sequential contexts, when it's use does not impact on performance.

In 5.6, an interstring-like data structure called a sub-module map, is used to represent the relational structure between sub-routines for a Space program, that provides a more direct way of accessing the content of the relational structure. This will facilitate a more efficient and parallelisable implementation of the Space compiler. Interlanguages for data structures differ from the interlanguage environment presented in this chapter, in that many to many relationships are expressed between data structure elements, and do not require a denotational environment, in the form of an explicit memory function or functional unit array.

Ultimately an interstring is defined using sets and is a tree construct, and a language of interstrings could be defined by a context free grammar, perhaps equipped with guards of some kind, to limit expressions to legal alpha and beta columns. It could be argued that the interstring is not as novel as is being suggested, because a tree parser of some kind is still required to process interstrings. But this argument does not take into account the highly regular structure of interstrings, and the potential parallelism of the processes that manipulate them. Instead of a dataflow expression being a tree with wide structural variability that needs a complex parsing phase, an interstring's syntax is simple with a restricted structural variability, and whose components are easy to access simultaneously.

As part of a boot strapping project, a compiler for the Space language has been written in C to run on a simulation of a highly parallel, formal model of computation called the Synchronic A-Ram. In chapter 6, Space programs are described as a series of interstring-like structures. The interstrings are stored as a data structure incorporating the simplified form (strings of strings of fixed length expressions), in conjunction with information regarding the length of each sub-interstring. The module that processes the data structure will not require a recursive descent parser, or a parser at all as such, merely an ability to directly locate sub-interstrings of an interstring.

The various stages for compiling interstrings have extensive opportunities for parallelization, that are not present in the compilation stages of tree based programming languages. The parallelized compilation of Space programs, would constitute a form of parallel rewriting, that is in contrast to the sequential activity of transducer type automata operating on tree expressions [40].

It has been suggested that the construction of an interstring grammar, and by implication a formal theory of interlanguages, has to be deferred until a suitable synchronically-oriented, formal model of computation has been assembled, that can fully support interlanguages. In a future paper, it will be argued that one member of the α-Ram family of formal models presented in the next chapter, called the Synchronous B-Ram, can fulfill that role. The equivalent of an automata that accepts or rejects expressions from a language of interstrings, would be a large Synchronous B-Ram program.

**References**.

# Chapter 3
# α-RAM FAMILY OF MODELS.

## 3.1 INTRODUCTION.

α-Ram models have varying degrees of synchronic orientation. The most synchronically oriented presented here, provide an efficient platform for interlanguages, that can exploit their parallel friendly features. In turn, the design and implementation of the first fully programmable form of an interlanguage, was influenced by the definitions of α-Ram machines. They are low level models of computation, in that the most primitive machine steps are on the level of switches and single bit manipulations, and do much less work than a Turing Machine step, or a λ-calculus beta reduction. But for following reasons, it will be argued that they are superior candidates to be the basis for complexity theory, and to be the standard models of computer science, rather than the Turing Machine or the λ-calculus.

i. They are lower level, and yet can support high level models of computation (programming languages), without the severe complexity costs imposed by the standard models (see chapter 9). This feature opens the door to couching mathematical discourse in terms of α-Ram programs and data structures, and to introducing explicit notions of time and computation into discrete mathematics.
ii. A forthcoming paper will set out how a simple account of the semantics of programming languages is potentially available, helping to clarify the relationship between denotational and operational semantics. This would provide some new approaches for the formalization of sequential and parallel algorithmics, and generate new opportunities for parallel program verification.
iii. They suggest promising new parallel architectures which allow implicit scheduling and resource allocation for pre-defined modules, coupled with enhanced programmability.
iv. Possessing registers containing data and instructions, they also happen to bear a closer resemblance to the machines we are familiar with. The Sequential A-Ram is more suitable as a theoretical basis for the Von Neumann model, than the Turing Machine, and the Synchronic A-Ram has some claim to be a theoretical basis for FPGAs.

Two categories of machine are proposed, the A-Ram and the B-Ram. The A-Ram has a finite array of registers and is simulable, whose size is limited only by available disk space.

The B-Ram has an infinite set of finite arrays of registers, and is of interest from the perspective of computability and comparison with the standard models. This chapter presents only two instances from each category, but a wide variety of state transformation functions and machine definitions, are conceivable within the $\alpha$-Ram framework.

The main focus of interest will be a model called the Synchronic A-Ram, as it seems promising for deriving high performance computing architectures. It is not the intention of this chapter to be a mathematical treatise, but a rigorous definition of the machines will be provided to fix their semantics, together with less rigorous arguments pertaining to their computability.

The formal descriptions of the $\alpha$-Ram machines rely on notions of sets and functions, the natural numbers, some standard logical notation, and the successor and predecessor functions for natural numbers. For brevity the arithmetic notation "$i+3$" will be used instead of "$succ(succ(succ(i)))$", and any use of "$i+j$" in the definitions will only occur within a context where there is an upper bound for $j$[21].

## 3.2 A-RAM.

The A-Ram has a memory block consisting of an array of registers, each containing an array of locations storing an element of the set $\{0,1\}$, and a marking which represents a subset of registers, indicating which instructions (one per register) are to be executed in situ, in the next machine cycle[22]. The machine state is the pair of the memory block and the marking. An A-Ram also has a state transformation function, which is usually based on an machine error detection scheme and some instruction set. A machine run consists of the state transformation function being repeatedly applied to the machine state either indefinitely, or until a special termination condition arises. An A-Ram process may therefore be seen as a sequence of pairs of memory blocks and markings.

[21] Much less compact descriptions of $\alpha$-Ram machines are possible, which do not require numbers or successor/predecessor functions, and logical notation, but are beyond the scope of the current work.

[22] The ability of the A-ram model to execute instructions "in situ", has something in common with the Process In Memory concept, found in associative models such as the ASC [6]. But program control is handled differently, data structures are not limited to the structure code format, and there is no ALU machinery associated with each register.

Unlike the formalisation of standard models, the definition of the A-Ram includes an explicit notion of working memory. It is non-generic in that it can neither represent or execute indefinitely large programs, nor represent indefinitely large inputs and outputs. It does however, permit the integration of data and instructions into one memory structure within a low level formal model of computation. A fully generic extension of this machine will later be considered, called the B-Ram. Section 3.5.1 sketches a machine which manipulates symbol sets larger than {0,1}. Formally, an A-Ram is a tuple $\langle p,\sigma,\mu,\eta \rangle$ :

1. Offset. $p \geq 3$ is an integer and is called the offset. Let $n = 2^p$. $n$ is the number of *elements* in a *register*, where each element stores a member of the set $\{0,1\}$. ($p$ is the minimum number of elements required to encode the position of an element in a register, using the alphabet of $\{0,1\}$ .)

2. Memory. The memory block is a function $\sigma: N_{2^{(n-p-2)}} \times N_n \rightarrow \{0,1\}$ which takes a register index $x$, an element index $y$, and delivers the contents of the $y$th element of the $x$th register.

register *0* :

| | | | |
|---|---|---|---|
| $\sigma(0,n-1)$ | $\sigma(0,n-2)$ | ......... | $\sigma(0,0)$ |

register *1*

| | | | |
|---|---|---|---|
| $\sigma(1,n-1)$ | $\sigma(1,n-2)$ | ......... | $\sigma(1,0)$ |

⋮

register $2^{(n-p-2)}-1$

| | | | |
|---|---|---|---|
| $\sigma(2^{(n-p-2)}-1,n-1)$ | $\sigma(2^{(n-p-2)}-1,n-2)$ | ......... | $\sigma(2^{(n-p-2)}-1,0)$ |

Fig 3.1 Diagrammatic representation of the memory map $\sigma$.

Although it is possible to formulate the memory block as $\sigma: N_{2^{(n-2)}} \rightarrow \{0,1\}$, the proposed definition will simplify matters in describing the state transformation function. Let $\Sigma = \{ \sigma \mid \sigma: N_{2^{n-p-2}} \times N_n \rightarrow \{0,1\} \}$ be the set of all possible memory blocks.

3. Marking. The marking $\mu \subseteq (N_{2(n-p-2)} - \{0\})$ represents a non-null subset of the registers in the memory block, or the special termination value $\varnothing$. A marked register contains an instruction, which is obliged to execute in the next machine cycle. The zeroth register cannot be marked and is reserved for machine status bits. Let $\Pi = P(N_{2^{n-p-2}} - \{0\})$ be the set of all markings.

4. State transformation function. The state transformation function is a partial function $\vartheta: \Sigma \times \Pi \rightarrow \Sigma \times \Pi$. A run of the A-Ram commences with an application of $\vartheta$ to $\langle \sigma, \mu \rangle$, which constitutes the first machine cycle. Thereafter, a succession of cycles proceeds, where the input to $\vartheta$ is the output from the previous cycle, either indefinitely, or until a tuple is produced which is undefined for $\vartheta$. The run is then understood to terminate, and the user may determine the success or type of failure of the run, by examining the bits of register 0. A least fixed point treatment of state transformation is eschewed, as part of an approach in avoiding recursion.

Although the notion of a marking has been borrowed from Petri Nets, the manner in which α-Ram markings participate in transforming machine states is deterministic. Within a deterministic context, the A-Ram offers a precise characterization of the distinction between sequential and parallel processes; any marking of the former may only contain a maximum of one element, whereas at least one marking of the latter has to be a non-empty, non-singleton set.

Within a framework that links determinism with clocked environments, such as CCS, CSP, or Gurevich's Abstract State Machines [1], a parallel A-Ram process would of course be characterized as sequential (even though it might involve many program modules running at the same time). Two formulations of the state transformation function are now presented, which yield sequential and parallel versions of the A-Ram.

### 3.2.1 Sequential A-RAM

There is a similarity in program control between a formal version of the Von Neumann model, such as the RAM [2] and a Sequential A-Ram, in that a program counter may associated with a marking, which always contains a maximum of a single register, and the marking always advances to the next instruction (register) unless it contains a jump or selection instruction. The fetch and execute cycle of the RAM may be compared with the execution of the Sequential A-Ram's single instruction, determined by the marking, occurring in situ.

#### 3.2.1.1 Informal description of Sequential A-RAM.

The Sequential A-Ram is an A-Ram, whose state transformation function $\eta'$, is based on an instruction set composed of four primitive instructions, two of which are write instructions modifying the content of a register, and the other two are concerned with the modifying the marking (program control). An instruction fits into a single register, and a marking may only ever contain one register.

Let (x,y) designate the yth bit of xth register, where x is called the destination cell, and y is called the offset. There are 4 assembly language type instructions, which are described in Figure 3.1

| opcode | Instruction (assembly language) | action |
|---|---|---|
| 00 | **wrt0 x y** | Write 0 into (x,y). |
| 01 | **wrt1 x y** | Write 1 into (x,y). |
| 10 | **cond x y** | Examine contents of (x,y).<br>If zero, jump to instruction in next register.<br>Else, jump to instruction in next but one register. |
| 11 | **jump x** | jump to instruction in register x. |

Fig 3.2 Sequential A-Ram Instruction Set.

If the offset length $p$=5, then a register contains 32 bits, then **x** will be represented as a 25 bit binary integer, and y as a five bit binary integer, with the following instruction format:

| opcode | destination cell (x) | offset (y) |
|---|---|---|
| bits 30-31 | bits 5-29 | bits 0-4 |

With such an arrangement, the memory block has 33,554,432 registers. Register 0 is a special register which is reset for each run.

The Sequential A-ram instruction set may be characterized as partially synchronically oriented, because (i) there are instruction formats for indicating where in memory a bit is to be read or written to, and (ii) the most primitive state transformation is associated with an instruction in an individual register.

A marking is invisible to the user, so to indicate that a run is active, the bit (0,0) is set. A run begins by an application of $\eta$ to $\langle \sigma, \{1\} \rangle$, executing the instruction in register 1, which is always **wrt1 0 0**. In general, if the instruction is a write instruction, then the new marking contains the next register, indicating that that instruction is to be executed in the next cycle. Otherwise, the memory block is unaltered, and the new marking is evaluated according to whether the instruction is a jump or a cond.

Instructions of the form **wrt1 0 y**, where **y** $\neq$ 0**,** are reserved by an error detection mechanism, and if marked result in machine error.

A run is successful one cycle after executing the special instruction **wrt0 0 0**, which resets (0,0), thereby halting the run, and indicating to the user that the run has succeeded. The user is informed when a run fails, if the machine detects an error condition, which empties the marking, and sets an error bit in register 0. A run halts iff (0,0) is reset or one of the error bits is set. Leaving aside the undecidability of the Halting Problem for the moment, there are a total of five notional outcomes with any Sequential A-Ram run:

i. The run goes on forever.
ii. The run succeeds.
iii. A *Halt Fail occurs* if the last register in the memory block (register 33,554,431 in the above example) is activated for execution, but is not the instruction **wrt0 0 0**. In this case the machine ignores the instruction, and writes a 1 to (0,1), and then halts.
iv. A *Cond Fail* occurs if the last but one register in the memory block (register 33,554,430 in the above example) is activated for execution, and is a cond instruction. In this case the positive consequent of the cond is undefined, the machine ignores the instruction, writes a 1 to (0,2), and then halts.
v. *Error Fail.* Programs may not write to the designated error bits. The marking of any wrt instruction to any bit in register zero, other than the zeroth bit, is illegal.

The Sequential A-Ram has the key features described by [3], required for a model to be generally programmable, namely sequencing (provided by sequencing of instructions), selection (provided by the cond instruction), and iteration (provided by cond and jump instructions). It will be claimed that a Sequential A-Ram with a sufficiently large memory block can implement any given program or function, which possesses upper bounds for the sizes of the input, the temporary work space required, and the output.

A sample assembly language program appears in Fig 3.3, which increments an unsigned 4 bit integer stored in bits 0-3 of register 27, and sets bit 0 of register 28 if there is an overflow. The non-data part of the program occupies 26 registers and takes 5 to 14 cycles to halt.

For a value of 0111 stored in bits 0-3 of register 27, a run of the machine would halt successfully after executing the instruction/register sequence 1, 2, 4, 7, 8, 10, 13, 14, 16, 19, 20, 21, 23, 24.

```
0     ....               // (0,0) is set iff Sequential A-Ram is busy
1     wrt1 0 0           // Signal the Sequential A-Ram is busy
2     cond 27 0          // test bit zero
3     jump 5             // bit zero is zero, so set bit and halt
4     jump 7             // bit zero is one, so reset bit and test next bit
5     wrt1 27 0
6     wrt0 0 0     // Signal the Sequential A-Ram has halted successfully
7     wrt0 27 0
8     cond 27 1          // test bit one
9     jump 11
10    jump 13
11    wrt1 27 1
12    wrt0 0 0
13    wrt0 27 1
14    cond 27 2          // test bit two
15    jump 17
16    jump 19
17    wrt1 27 2
18    wrt0 0 0
19    wrt0 27 2
20    cond 27 3          // test bit three
21    jump 23
22    jump 25
23    wrt1 27 3
24    wrt0 0 0
25    wrt1 28 0
26    wrt0 0 0
27    ....         // bits 0-3 of register 27 store value to be incremented
28    ....         // bit 0 of register 28 stores overflow bit
```

Fig 3.3: Sequential A-Ram program for incrementing 4-bit integer.

### 3.2.1.2 Formal description of sequential A-RAM.

A sequential A-Ram is an A-Ram $\langle p,\sigma,\{1\},\eta'\rangle$, where $\eta'$ is defined as follows.

Let $i \in N_{2^{n-p-2}}$ be a variable pointing to a register.

i. $\sigma_y(i)$ is the integer represented by the bits in the offset cell. $\sigma_y(i) = \sigma(i,p-1) * 2^{p-1} + \sigma(i,p-2) * 2^{p-2} + \ldots\sigma(i,1) * 2 + \sigma(i,0)$. In other words, $\sigma_y(i)$ is the integer represented by the bit vector $\langle\sigma(i,p-1),\sigma(i,p-2),.\sigma(i,0)\rangle$, with the least significant bit on the right.

ii. $\sigma_x(i)$ is the integer represented by the bit vector $\langle\sigma(i,n-3),\sigma(i,n-4),.\sigma(i,p)\rangle$, with the least significant bit on the right. $\sigma_x(i)$ is the integer represented by the bits in the destination cell.

iii. $\sigma_z(i)$ is the integer represented by the bit vector $\langle\sigma(i,n-1),\sigma(i,n-2),.\sigma(i,0)\rangle$, with the least significant bit on the right. $\sigma_z(i)$ is the integer represented by the contents of the whole register viewed as a binary number.

Recall $\Sigma = \left\{ \sigma \mid \sigma: N_{2^{n-p-2}} \text{ x } N_n \rightarrow \{0,1\} \right\}$. A write function $\omega$: $\Sigma$ x $N_{2^{n-p-2}}$ x $N_n$ x $\{0,1\} \rightarrow \Sigma$, which modifies the content of one cell of one register in $\sigma$.

$$\omega(\sigma,i,j,b) = \{\langle i,j,b\rangle\} \cup \{ \langle x,y,z\rangle \in \sigma \,|\, (x \neq i) \vee (y \neq j) \}$$

Recall $\Pi = P\left(N_{2^{n-p-2}} - \{0\}\right)$. In Fig 3.4 a (partial) state transformation function $\eta' : \Sigma \text{ x } \Pi \rightarrow \Sigma \text{ x } \Pi$ can now be described for the Sequential A-Ram. For brevity the cases have to be evaluated in order from top to bottom. The value of the function is followed on the right by it's associated condition, expressed as a conventional logical formalism.

$$
\eta'(\sigma,\{i\}) = \begin{cases}
\langle \omega(\sigma,0,0,0),\ \varnothing \rangle, & \text{if} \quad \sigma_z(i) = 0 & (1) \\
\langle \omega(\sigma,0,1,1),\ \varnothing \rangle, & \text{if} \quad \left(i = 2^{n-p-2} - 1\right) \wedge \left(\sigma_z(i) \neq 0\right) & (2) \\
\langle \omega(\sigma,0,2,1),\ \varnothing \rangle, & \text{if} \left(i = 2^{n-p-2} - 2\right) \wedge \left(\sigma(i,n-1) = 1\right) \wedge \left(\sigma(i,n-2) = 0\right) & (3) \\
\langle \omega(\sigma,0,3,1),\ \varnothing \rangle, & \text{if} \quad \left(\sigma_x(i) = 0\right) \wedge \left(\sigma_y(i) \neq 0\right) & (4) \\
\langle \omega(\sigma,\sigma_x(i),\sigma_y(i),\sigma(i,n-2)),\ \{i+1\} \rangle, & \text{if} \quad \sigma(i,n-1) = 0 & (5) \\
\langle \sigma,\ \{i+1\} \rangle, & \text{if} \ \left(\sigma(i,n-2) = 0\right) \wedge \left(\sigma(i,n-1) = 1\right) \wedge \left(\sigma(\sigma_x(i),\sigma_y(i)) = 0\right) & (6) \\
\langle \sigma,\ \{i+2\} \rangle, & \text{if} \ \left(\sigma(i,n-2) = 0\right) \wedge \left(\sigma(i,n-1) = 1\right) \wedge \left(\sigma(\sigma_x(i),\sigma_y(i)) = 1\right) & (7) \\
\langle \sigma,\ \{\sigma_x(i)\} \rangle, & \text{if} \ \left(\sigma(i,n-2) = 1\right) \wedge \left(\sigma(i,n-1) = 1\right) & (8)
\end{cases}
$$

Figure 3.4 Definition of the Sequential A-Ram's state transformation function $\eta'$.

To assist the reader, the various conditions correspond to the following:

1. Successful halt instruction **wrt 0 0**.
2. Halt fail.
3. Cond fail.
4. Error fail.
5. wrt0 or wrt1 instruction.
6. Negative consequent of cond instruction.
7. Positive consequent of cond instruction.
8. Jump instruction.

A run commences with an application of $\eta$ to $\langle\sigma,\{1\}\rangle$, which constitutes the first machine cycle. If a tuple of the form $\langle\sigma,\varnothing\rangle$ is generated by $\eta$, then the run halts because

$\langle \sigma,\varnothing \rangle$ is undefined for $\eta$ for all $\sigma$.

**Thesis**. Let $c,d \in N$, and $f : N \to N$ be a Turing-computable partial function, which has the upper bound *c*. Let *n* be a valid input for $f$, and $n \le d$, for some constant *d*. Then there exists a Sequential A-Ram $\langle p,\sigma,\mu,\eta' \rangle$ which computes $f(n)$.

The Sequential A-Ram is a finite construction, and cannot be fully Turing-computable, The thesis claims, however, that a Sequential A-Ram exists which can compute $f(n)$ if we are given $f$ with an upper bound *c*, and an input no larger than *d*. To prove this thesis one might provide a scheme for producing a large enough Sequential A-Ram with a program part that implements the action of the instruction table for the Turing Machine which describes $f$, and a data part that has space for encoding *n* and $f(n)$. A hand waving argument will be presented in Appendix A, that the Sequential B-Ram is fully Turing-computable.

### 3.2.2 SYNCHRONIC A-RAM.

Although the Synchronic A-Ram is intended to be relevant to massively parallel computing, it is presented here as a purely mathematical model, in order to fix it's semantics. In 8.3, it is shown that a physical version of the model is subject to significant hardware and communication resource overheads.

#### 3.2.2.1 INFORMAL DESCRIPTION OF THE SYNCHRONIC A-RAM.

The Synchronic A-Ram's state transformation function is called $\eta$, where the main difference in program control compared to that of the Sequential A-Ram, is that in lieu of having a single element marking, which is in effect a program counter pointing to a single instruction, and a fetch and execute type cycle, there is instead a marking which is a subset of register names holding instructions, to be simultaneously executed in situ. The Synchronic A-Ram may be classified as a CREW (Concurrent Read Exclusive Write) formal model of computation.

It is a CREW model, because a viable high level programming approach has been found which can usefully exploit simultaneous reads, but does not require the need to use or accommodate simultaneous writes, to the same storage location. The machine is parallel, in that the machine has a global clock, and instead of only one instruction being active per cycle, there may be many.

The activation of more than one register is allowed by adding an offset y to the jump instruction. Thus **jump x y** activates registers x through to x+y in the next cycle. The instruction set is described in Fig 3.5

| opcode | Instruction (assembly language) | action |
|---|---|---|
| 00 | **wrt0 x y** | Write 0 into (x,y). |
| 01 | **wrt1 x y** | Write 1 into (x,y). |
| 10 | **cond x y** | Examine contents of (x,y).<br>If zero, mark instruction in next register for execution in next cycle.<br>Else, mark instruction in next but one register. |
| 11 | **jump x y** | In the next cycle, mark all instructions from instruction x, up to an including instruction x+y for execution in the next cycle. |

Fig 3.5 Synchronic A-Ram Instruction Set.

If $p = 5$, a register contains 32 bits, then x will be represented as a 25 bit binary integer, and y as a five bit binary integer, allowing a single jump to mark up to a maximum of 32 instructions for activation in the next cycle.

| opcode | destination cell (x) | offset (y) |
|---|---|---|
| bits 30-31 | bits 5-29 | bits 0-4 |

To describe which (potentially non-contiguous) subset of registers should be active in the next cycle, the Petri net notion of a marking is borrowed, which could either be a subset of the set of integers representing registers, or a mapping of integers representing registers to the set $\{0,1\}$. The former will be chosen to simplify the definition of the A-Ram's transformation function.

If a marking contains more than one register, then there is the undesirable possibility of a new marking being generated which is a multiset, the equivalent of an unsafe marking in Petri Net theory. This introduces unnecessary complexity into the definition and operation of the machine, and in 3.2.2. there is a error condition called a *Marking Fail* designed to detect multiset markings.

Register 0 is a special register which is reset for each run. The first cycle of a run simultaneously executes the instructions in registers 1 and 2. The instruction in register 1 is always **wrt1 0 0**,in order to indicate to the user that the machine is busy. The instruction in register 2 is usually **jump x y**, in order to proceed with the main body of the program.

If a Sequential A-Ram performs a write instruction, then the succeeding instruction is normally executed in the following cycle, whereas this is not true of the Synchronic A-Ram. After the first cycle, any instruction(s) will be activated only by a cond or jump from the previous cycle. With many instructions being active per cycle, a run may exhibit chaotic and undefined behaviour in a number of undesirable ways. The state transformation function therefore incorporates an error detection scheme, that halts in failure if such behaviour occurs.

A run is successful one cycle after a marking containing only the special instruction **wrt0 0 0**, thereby halting the run and indicating to the user that the run has finished in the manner envisaged by the programmer. A run fails if the machine detects an error condition, which empties the marking, thereby halting the run, and sets an error bit in register 0, which indicates to the user that the run has failed. The run has halted iff (0,0) is reset or one of the error bits is set. There are three possible outcomes with any Synchronic A-Ram run:

i. The run goes on forever.
ii. The run succeeds.
iii. The run fails and halts in the cycle after one generating one of nine errors.

### 3.2.2.2 ERRORS.

The Synchronic A-Ram's error conditions listed here are intended to eliminate the possibility of undefined or nonsensical behaviour, and certain types of resource contention. The removal of parallel-related contentions at the most primitive level of machine activity, contributes to eliminating contention in higher level parallel programming. Some of the error conditions are only included for the sake of completeness to cover rare cases, whilst others have been of assistance in the debugging of machine code programs, and of a high level language compiler. The bits of register 0 are reserved for machine status and error flags. Instructions of the form **wrt1 0 y**, where **y** ≠ 0**,** are reserved by the error detection mechanism, and cannot be used in any program. The marking is invisible to the user, but obviously visible to a simulation. The machine communicates to the outside world regarding errors by setting various bits in register 0. The purpose of each error is categorised after the description.

Types of Errors.

1. *Marking Fail.* A legal marking cannot be a multiset. (In Petri Net parlance, each cycle in a run should have a safe marking.) If a register is activated more than once in the same cycle, then the marking is emptied and (0,1) is set. (Precludes Live fail.) Contention.
2. *Write Fail.* The Synchronic A-Ram performs simultaneous writes, but is not a concurrent write machine in the conventional sense. If at least two instructions write to the same location (x,y) in the same cycle, then the marking is emptied and (0,2) is set. This error precludes no other types of error occurring in the same cycle. Contention.
3. *Halt Fail.* The final marking should always only activate the special halt instruction. If **wrt0 0 0** is not the only instruction in a marking, then the marking is emptied and error bit (0,3) is set. (Precludes Live fail and Marking fail.) This error helps to ensure the module halt is meaningful, and as intended.
4. *Live Fail.* A program should not unexpectedly generate an empty marking. If the marking becomes empty (without the halt instruction having been activated in the previous cycle), then the marking is emptied and (0,4) is set, and the machine halts in failure. (Precludes Halt fail and Marking fail.) This error helps to ensure the module halt is meaningful, and as intended.

5. *Cond Fail.* If a cond instruction in the last but one register in the memory block is activated for execution, then the positive consequent of the cond is undefined. The machine ignores the instruction and writes a 1 to (0,5). Completeness.
6. *Consequent Fail.* A cond instruction may also be misused if any two of the triplet of a cond instruction, it's negative consequent instruction, and positive consequent instruction are in a marking. If this occurs the marking is emptied and (0,6) is set. Compiler debugging.
7. *Active Fail.* An instruction should not have its contents modified in the same cycle in which it is active. If the marking contains a write to a bit within a marked register, then the marking is emptied and (0,7) is set. Contention.
8. *Jump Fail.* A jump instruction should not attempt to mark a register index which is illegal or does not exist. If the destination cell points to the zero register, or if the destination cell plus the offset exceeds the total number of registers in the memory block ($2^{n-p-2}$), then the marking is emptied and (0,8) is set. Completeness.
9. *Error Fail.* Programs may not write to the designated error bits. The marking that includes a wrt instruction to any bit in register zero, other than the zeroth bit, is illegal, and (0,9) is set. Completeness.

A Synchronic A-Ram machine cycle may be viewed as having a read phase, followed by a write phase, similar to master-slave type registers in sequential digital circuits. Registers are able to read and write to the same bit in an unmarked register in the same cycle, where the write executes after the read. Therefore instructions **cond a b**, and a **wrtx a b** can occur in the same marking, where the cond instruction reads (a,b), before it is overwritten by the wrt instruction.

It should be stressed that the proposed error detection scheme imposes a substantial computational cost for every machine cycle, not only for a simulation, but also for hardware based on the concept. However, it is believed possible to prove that all types of errors may be eliminated with a suitably designed language, compiler and programming methodology, thereby eliminating the need for the scheme's implementation. Such a proof is beyond the scope of this report. The main benefit of the scheme to date was to assist in the debugging of Spatiale 1.0, and thus far, it can be reported that a wide range of compiled (massively parallel) programs, producing outputs as expected, have not yet generated the error detection mechanisms that have been implemented.

### 3.2.2.3 Threads in space.

The language Space will later be presented, in which a collection of high level subprograms/processes may be activated simultaneously. One may also create machine code with program segments being active simultaneously. In the context of conventional multi-threaded computing, the simplest form of thread is normally understood to involve the activation of a sequence of single primitive machine instructions[23]. A Space program may be written, that only invokes high level sub-programs sequentially with respect to the module's level of abstraction. But Space processes (and Synchronic A-Ram processes generally) are different to threads, in that the finest grain sub-processes involve a sequence of markings, which in general, individually contain a plural collection of primitive instructions.

There is a consequently a fundamental difference between a Von Neumann thread and a Space thread. Having made this qualification, the term *thread* is retained to describe a Synchronic A-Ram active program segment or module. When a number of Space threads at the same level of abstraction are active, there is always one in particular called a *carry thread*, which transfers control to the next stage of the program.

### 3.2.2.4 Program example.

Without a high level language being immediately available, we must contend with machine code whose parallelism can be difficult to understand. When machine coding, care must be taken to ensure that (i) the output of a thread is ready before before moving on to the next stage of the program, thus avoiding the generation of various errors, and (ii) that an attempt to signal a successful halt cannot occur when another thread might still be active (Halt Fail). Additional jump instructions may be used to provide shorter or longer exit sequences from threads as required.

The program in Figure 3.6 is a 4-input AND gate, whose implementation involves a pair of 2-input AND gates running simultaneously. The code is not particularily efficient, because at this simple level a program that tests four bits in sequence is shorter and can be faster for some inputs, but it does exemplify a simple form of Synchronic A-Ram multi-threading .

[23] The ubiquity of pipelining and multiple instruction issue in processors, does complicate this issue somewhat.

The code for a 2-input AND gate must test the two bits separately, because there is no single member of the instruction set which can test two bits. The first 2-input AND gate program module incorporates the carry thread. If the result of the first AND gate is positive, the result of the second AND gate is tested to give the final result. The four bit input is stored in bits 0-3 , the result of the 2nd AND gate in bit 4, and the final output in bit 5 of register 23.

If the input quartet of bits is 1011, where the rightmost bit is stored in (23,0), then the run produces the marking sequence: $\{1,2\},\{3,4\},\{5,10\},\{7,11\},\{9\},\{15\},\{16\},\{20,21\},\{23\},\varnothing$.

The input 1110 gives $\{1,2\},\{3,4\},\{5,10\},\{18,12\},\{19,14\},\{20,21\},\{23\},\varnothing$.

A sequential digital circuit can be implemented as Synchronic A-Ram program, that semantically respects the circuit's parallelism. A range of programs are described in the next chapter.

| Line number | Instruction | Comment |
|---|---|---|
| 1 | wrt1 0 0 | // indicate machine is busy |
| 2 | jump 3 1 | // jump to initiate both AND gates |
| 3 | jump 5 0 | // activate 1st AND gate module, which is the carry thread |
| 4 | jump 10 0 | // activate 2nd AND gate |
| 5 | cond 24 0 | // test input bit 0 |
| 6 | jump 18 0 | // input bit 0 is zero, so jump to long exit sequence |
| 7 | cond 24 1 | // now test input bit 1 |
| 8 | jump 19 0 | // input bit 1 is zero, so jump to short exit sequence |
| 9 | jump 15 0 | // input bits 0,1 are one, so test result of 2nd AND gate |
| 10 | cond 24 2 | // test input bit 2 |
| 11 | wrt0 24 4 | // input bit 2 is zero, so write 0 into result bit of 2nd AND gate, |
| 12 | cond 24 3 | // now test input bit 3 |
| 13 | wrt0 24 4 | // input bit 3 is zero, so write 0 into result bit of 2nd AND gate |
| 14 | wrt1 24 4 | // input bit 3 is one, so write 1 into result bit of 2nd AND gate |
| 15 | cond 24 4 | // result of 1st AND gate was one, test result of 2nd gate |
| 16 | jump 20 1 | // result of 2nd gate was zero, hence write final result of zero |
| 17 | jump 21 1 | // result of 2nd gate was one, hence write final result of one |
| 18 | jump 19 0 | // long exit sequence |
| 19 | jump 20 1 | // short exit sequence |
| 20 | wrt0 24 5 | // write final result of zero |
| 21 | jump 24 0 | // jump to halt |
| 22 | wrt1 24 5 | // write final result of one |
| 23 | wrt1 0 0 | // instruct machine to successfully halt |
| 24 | // reg 24 stores input bits 0-3, 2nd AND gate result in bit 4, and final result in bit 5 | |

Figure 3.6: 4-input AND gate involving two 2-input AND gates running simultaneously.

### 3.2.3 FORMAL DESCRIPTION.

A Synchronic A-Ram is an A-Ram $\langle p,\sigma,\{1,2\},\eta\rangle$, where $\eta$ is defined as follows. Recall $\Sigma = \left\{ \sigma \mid \sigma: N_{2^{n-p-2}} \text{ x } N_n \to \{0,1\} \right\}$, $\Pi = P\left(N_{2^{n-p-2}} - \{0\}\right)$, and let $\mu \subseteq \Pi$. The function $\eta$ will be defined by a set of error conditions followed by a definition for a legal input, which will split the function $\eta(\sigma,\mu) = \langle \eta_1(\sigma,\mu),\eta_2(\sigma,\mu)\rangle$, where $\eta_1{:}\Sigma\text{x}\Pi \to \Sigma$, and $\eta_2{:}\Sigma\text{x}\Pi \to \Pi$. To simplify the definition of the errors, the cases have to be evaluated in order from top to bottom, rather than simultaneously per machine cycle.

This makes little practical difference to the debugging of a program for a simulation specified by the definition, because the debugging of one error will not suppress the existence of other errors within the same cycle, unless they have same root cause, in which case it does not matter. The other errors will still be detected in subsequent runs, allowing all errors to be dealt with eventually. Note that a machine always halts in the first cycle that an error occurs, allowing no further dependent errors to be generated which might appear in subsequent cycles A run commences with an application of $\eta$ to $\langle \sigma,\{1,2\}\rangle$, which constitutes the first machine cycle. If a tuple of the form $\langle \sigma,\varnothing\rangle$ is generated by $\eta$ and $\sigma(0,5) = 0$ and $\sigma(0,0) = 1$ (live fail), then the error bit (0,5) will be set, and the run will halt in the next cycle. $\eta$ is undefined for for all other tuples of the form $\langle \sigma,\varnothing\rangle$, which terminate the run. To assist the reader, the various conditions in Figure 3.7 correspond to the following:

1. Success.
2. Marking fail
3. Write fail.
4. Halt fail.
5. Live fail.
6. Cond fail.
7. Consequent Fail
8. Active fail
9. Jump fail
10. Error fail
11. Non-halting legal cycle.

$$
\eta(\sigma,\mu)=
\begin{cases}
\langle \omega(\sigma,0,0,0),\ \varnothing\rangle, & \exists! i\in\mu\ (\sigma_z(i)=0) \qquad (1)\\[2ex]
\langle \omega(\sigma,0,1,1),\ \varnothing\rangle, & \mu \text{ is a multi set.} \qquad (2)\\[2ex]
\langle \omega(\sigma,0,2,1),\ \varnothing\rangle, & \exists i,j\in\mu \begin{pmatrix} (i\neq j)\wedge(\sigma(i,n-1)=0)\wedge(\sigma(j,n-1)=0) \\ \wedge(\sigma_x(i)=\sigma_x(j))\wedge(\sigma_y(i)=\sigma_y(j)) \end{pmatrix} \qquad (3)\\[2ex]
\langle \omega(\sigma,0,3,1),\ \varnothing\rangle, & \exists i,j\in\mu\ ((i\neq j)\wedge(\sigma_z(i)=0)) \qquad (4)\\[2ex]
\langle \omega(\sigma,0,4,1),\ \varnothing\rangle, & (\mu=\varnothing)\wedge(\sigma(0,0)=1)\wedge(\sigma(0,5)=0) \qquad (5)\\[2ex]
\langle \omega(\sigma,0,5,1),\ \varnothing\rangle, & \exists i\in\mu \begin{pmatrix} (\sigma(i,n-1)=1)\wedge(\sigma(i,n-2)=0) \\ \wedge\ (i=\ 2^{n-p-2}-2) \end{pmatrix} \qquad (6)\\[2ex]
\langle \omega(\sigma,0,6,1),\ \varnothing\rangle, & (\exists k\in 2^{n-p-2})(\exists i,j\in\mu) \begin{pmatrix} ((\sigma(k,n-1)=1)\wedge(\sigma(k,n-2)=0)) \\ \wedge \begin{pmatrix} ((i=k+1)\wedge(j=k+2))\ \vee \\ \left((i=k)\wedge\begin{pmatrix}(j=k+1)\vee \\ (j=k+2)\end{pmatrix}\right) \end{pmatrix} \end{pmatrix} \qquad (7)\\[2ex]
\langle \omega(\sigma,0,8,1),\ \varnothing\rangle & \exists i,j\in\mu\ ((i\neq j)\wedge(\sigma(j,n-1)=0)\wedge(\sigma_x(j)=i)) \qquad (8)\\[2ex]
\langle \omega(\sigma,0,9,1),\ \varnothing\rangle & \exists i\in\mu \begin{pmatrix} (\sigma(i,n-1)=1)\wedge(\sigma(i,n-2)=1) \\ \wedge\ \left(\left((\sigma_x(i)=0)\vee\left(\sigma_x(i)+\sigma_y(i)\geq 2^{n-p-2}\right)\right)\right) \end{pmatrix} \qquad (9)\\[2ex]
\langle \omega(\sigma,0,10,1),\ \varnothing\rangle & \exists i\in\mu((\sigma_x(i)=0)\wedge(\sigma_y(i)\neq 0)) \qquad (10)\\[2ex]
\langle \eta_1(\sigma,\mu),\ \eta_2(\sigma,\mu)\rangle & \qquad (11)
\end{cases}
$$

Figure 3.7 Definition of the Synchronic A-Ram's state transformation function.

The function $\eta_1{:}\Sigma x\Pi \rightarrow \Sigma$ is now defined.

$$\eta_1(\sigma,\mu) \;=\; \left( \bigcup_{\sigma(i,n-1)=0} \left\langle \sigma_x(i),\sigma_y(i),\sigma(i,n-2) \right\rangle \right)$$

$$\cup \; \left( \sigma - \left\{ \langle p,q,r \rangle \,\middle|\, \exists i \left( (i \in \mu) \wedge \left(\sigma(i,n-1)=0\right) \wedge \left(\sigma_x(i)=p\right) \wedge \left(\sigma_y(i)=q\right) \right) \right\} \right) \;.$$

$\eta_2 : \Sigma \mathrm{x} \Pi \rightarrow \Pi$ is now defined. If $i,j \in N$, $0 < j \leq 2^p - 1$, let $[i,0] \equiv i$ and $[i,j] \equiv i, i+1, .. i+j$, then

$$\eta_2(\sigma,\mu) \;=\; \left( \bigcup_{(\sigma(i,n-1)=1) \wedge (\sigma(i,n-2)=0) \wedge \left(\sigma\left(\sigma_{x(i)}, \sigma_{y(i)}\right)=0\right)} i+1 \right)$$

$$\cup \; \left( \bigcup_{(\sigma(i,n-1)=1) \wedge (\sigma(i,n-2)=0) \wedge \left(\sigma\left(\sigma_{x(i)}, \sigma_{y(i)}\right)=1\right)} i+2 \right)$$

$$\cup \; \left( \bigcup_{(\sigma(i,n-1)=1) \wedge (\sigma(i,n-2)=1)} \left[ \sigma_x(i),\sigma_y(i) \right] \right) \;.$$

In addition to having an updateable and addressable memory in a clocked environment, and having any act of computation always associated with a specific set of registers, the Synchronic A-Ram can activate multiple program segments simultaneously in one machine time step. By also supporting full connectivity between registers in unit time, with connections themselves occupying no space in the machine, it may be characterised as fully synchronically oriented. A notion of propagation delay for the A-Ram is discussed in 3.6.

The definitions of the state transformation function involve a number of operations, which would normally be considered to be performed serially by the reader. This is somewhat at odds with the claim that Synchronic A-Ram instructions execute "in parallel" and "in situ". But the block diagram for a Synchronic A-Ram register in 8.3, makes plain that a truly physically parallel Synchronic A-Ram could be in theory be built. A claim is made in 9.6 that conventional mathematical discourse is non-spatial, and implicitly harbours a sequential notion of computation. If a notions of spatial and parallel computation were to be introduced into mathematical discourse, then a definition of the state transformation function could be given, that would more obviously respect the claim of parallelism and in situ operation.

## 3.3 B-RAM.

The Turing Machine [4] represents indefinitely large programs, by employing an indefinitely extendable state set and instruction table. The instruction table is not explicitly stored in some memory, and exists only "on paper". Turing solved the problem of processing indefinitely large inputs, by proposing an infinite denumerable sequence of tape squares, a cursor capable of pointing to any square (without quite being a number), which can be moved leftwards or rightwards one square at a time. I purloin a more extensive helping of the same trick in deriving a Turing-computable version of the A-Ram, and assume a cursor is essentially a numeric address, or part of a co-ordinate numerical address of a memory location.

The B-Ram memory consists of an infinite sequence $\sigma_0, \sigma_1, \sigma_2, ....$ of A-Ram type memory blocks. Each register in *each* memory block has a cursor, which is capable of pointing to any memory block $\sigma_i$, $i \geq 0$. This arrangement may seem excessive, but it will enable the B-Ram to be Turing-computable and also easily programmable. There is a simplification of the definition of the B-Ram's marking, if we only wish to consider a sequential machine, but with the general synchronic case, the B-Ram marking will have to be capable of referring to multiple blocks, and have to describe a set of subsets of the memory block register sets. A B-Ram is a tuple $\langle p, \sigma, \mu, \rho, \zeta \rangle$.

1. Offset. $p \geq 3$ is a positive integer and is called the offset length. Let $n = 2^p$. $n$ is the number of *elements* in a *register*, where each element stores a member of the set $\{0,1\}$. ($p$ is the minimum number of elements required to encode the position of an element in a register, using the alphabet of $\{0,1\}$ .)

2. Memory. $\sigma$ is a memory composed of an infinite sequence of memory blocks $\sigma_0, \sigma_1, \sigma_2, ....$, each containing $2^{(n-p-2)}$ registers, where $\sigma : N \times N_{2(n-p-2)} \times N_n \rightarrow \{0,1\}$ $\sigma$ takes a block index $i$, a register index $j$, and an element index $k$, and delivers the contents of the $k$th element of the $j$th register of the $i$th memory block. Let $\Sigma = \{ \sigma \mid \sigma : N \times N_{2^{n-p-2}} \times N_n \rightarrow \{0,1\} \}$ be the set of all possible memory blocks.

3. Marking. The marking is the subset

$$\mu \subseteq \left( N \times P\left(N_{2(n-p-2)}\right) - \left\{ \langle 0,M \rangle \middle| (0 \in M) \wedge \left(M \subseteq N_{2(n-p-2)}\right) \right\} \right)$$

The zeroth register of the zeroth memory block cannot be marked and is reserved for machine status bits. The set $\mu$ represents a possibly infinite set of subsets of memory block registers. Another characterization of a marking could be as a partial function $\mu: N \rightarrow P\left(N_{2(n-p-2)}\right)$. Let the set of all markings be

$\Pi = P\left( N \times P\left(N_{2(n-p-2)}\right) - \left\{ \langle 0,M \rangle \middle| (0 \in M) \wedge \left(M \subseteq N_{2(n-p-2)}\right) \right\} \right)$. (If I only wished to consider sequential B-Rams, then I could get away with defining $\Pi = N \times N_{2(n-p-2)}$ ).

4. Cursor function. $\rho$ is called the cursor function, and maps every register in every memory block to a memory block index. $\rho: N \times N_{2(n-p-2)} \rightarrow N$.

Let $\mathrm{P} = \left\{ \rho \mid \rho: N \times N_{2(n-p-2)} \rightarrow N \right\}$ be the set of all possible cursor configurations.

5. State transformation function.

The state transformation function is a partial function $\zeta : \Sigma \times \Pi \times \mathrm{P} \rightarrow \Sigma \times \Pi \times \mathrm{P}$ A run of the B-Ram commences with an application of $\zeta$ to $\langle \sigma,\mu,\rho \rangle$, which constitutes the first machine cycle. Thereafter, a succession of cycles proceeds, where the input to $\zeta$ is the output from the previous cycle, either indefinitely, or until a tuple is produced which is undefined for $\zeta$. The run is then understood to terminate, and the user may determine the success or type of failure of the run, typically by examining the contents of register 0 of $\sigma_0$ of the final cycle.

Two formulations of the state transformation function are now considered, which yield sequential and parallel versions of the B-Ram.

### 3.3.1 SEQUENTIAL B-RAM.

#### 3.3.1.1 INFORMAL DESCRIPTION OF SEQUENTIAL B-RAM.

In common with the Sequential A-Ram, only one instruction (of only one memory block) in the Sequential B-Ram memory may be marked. In addition to the four Sequential A-Ram instructions, there are an extra two instructions which manipulate register block pointers, all depicted in Fig 3.8. If it is stated that an instruction points to a memory block, it is meant that the register in which the instruction resides points to a memory block.

At the beginning of each run, every register cursor in every memory block points to the same memory block as the register itself resides in: $\rho(i,j)=i$, for $i \geq 0$ and $0 \leq j \leq 2^{n-p-2}-1$.

| opcode | instruction (assembly language) | action |
|---|---|---|
| 000 | **wrt0 x y** | Write 0 into (x,y) of memory block pointed to by instruction. |
| 001 | **wrt1 x y** | Write 1 into (x,y) of memory block pointed to by instruction. |
| 010 | **cond x y** | Examine contents of (x,y) of memory block pointed to by instruction.<br>If zero, jump to instruction in next register of the same memory block as the instruction.<br>Else, jump to instruction in next but one register. |
| 011 | **jump x** | Jump to instruction in register x of memory block pointed to by instruction. |
| 100 | **mvrt x** | Move cursor of register x of memory block pointed to by instruction to the right. |
| 101 | **mvlt x** | Move cursor of register x of memory block pointed to by instruction to the left.. |

Fig 3.8 Sequential B-Ram Instruction Set.

The Sequential B-Ram begins a run by activating the instruction represented by bits in register 1 in memory block $\sigma_0$. It then executes the instruction in the succeeding register of the same memory block, and so on, unless instructed to perform a conditional instruction, or perform a jump to an instruction in a possibly different block, or finishes a run when instructed to write 0 to (0,0) in memory block $\sigma_0$. If an instruction is activated at the end of any memory block $\sigma_i$, which is not a jump or a halt, then there is a machine error.

If $p = 4$, then the register has 16 bits, and x will be represented as a 9 bit binary integer, and y as a 4 bit binary integer, with the following instruction format:

| opcode | destination cell (x) | offset (y) |
|---|---|---|
| bits 13-15 | bits 4-12 | bits 0-3 |

With such an arrangement, a Sequential B-Ram memory block has 512 registers. Register 0 of $\sigma_0$ is a special register which is reset for each run. A Sequential B-Ram instruction may be referencing a bit, or a register, in a different memory block. The Sequential B-Ram begins a run by executing the instruction in register 1 of $\sigma_0$, which is always **wrt1 0 0** pointing to $\sigma_0$, to indicate to the user that the machine is busy, then the succeeding instruction, and so on, unless instructed to perform a jump or a cond. The run ends successfully one cycle after any cycle executing the instruction **wrt0 0 0** pointing to $\sigma_0$. A run fails if the machine detects an error condition, which empties the marking, thereby halting the run, and sets an error bit in register 0, which indicates to the user that the run has failed. There are five possible outcomes with any run:

i. The Sequential B-Ram runs indefinitely.
ii. The Sequential B-Ram succeeds, and halts if the instruction **wrt0 0 0** pointing to $\sigma_0$, is executed.
iii. The Sequential B-Ram causes a *Halt Fail* if the last register in the array (register 511 in the above example) in some memory block is activated for execution, but is not wrt0 0 0 pointing to $\sigma_0$. The machine ignores the instruction, empties the marking, writes a 1 to (0,1) in $\sigma_0$.
iv. The Sequential B-Ram causes a *Cond Fail* if the last but one register in some memory block (register 510 in the above example) is activated for execution, and is a cond instruction. In this case the positive consequent of the cond is undefined, the machine ignores the instruction, empties the marking, writes a 1 to (0,2).

v. The Sequential B-Ram causes a *Cursor Fail* if a mvlf instruction is applied to a register in some memory block, which is already pointing to $\sigma_0$. The machine ignores the instruction, empties the marking, writes a 1 to (0,3).

A sample assembly language program in Fig 3.9 will now be considered, which increments an indefinitely large integer. The integer is stored as a succession of bits in the (0,0) cells of the succession of memory blocks after $\sigma_0$. The (0,1) bit of a block will be set to indicate if the block contains the final, most significant bit, and reset otherwise. To avoid redundancy, it is assumed the most significant bit of the integer is set. This scheme for representing an integer is wasteful of memory space, but it considerably simplifies the program, which otherwise would have to implement finite incrementers for destination and offset cells, integer comparators tests for reaching the end of a memory block etc. The program only occupies 26 registers (the same as a 4-bit Finite Sequential Ram incrementer!).

It is claimed that the Sequential B-Ram can implement any program or function, and encode any data required for computation, ie the Sequential B-Ram is Turing-computable. Note that in order to run the program a second time, all the cursors of the registers marked (c) would have to be rewound to point to $\sigma_0$.

3.3.1.2 FORMAL DESCRIPTION OF SEQUENTIAL B-RAM.

Let $\Pi' = \left\{ \left\{ \left\langle i,\{j\} \right\rangle \right\} \;\middle|\; i \in N, j \in N_{2(n-p-2)} \right\}$ be the set of all B-Ram markings composed of singleton tuple referring to a single instruction in a single memory block. (If I only wished to consider sequential B-Rams, then I could define $\Pi' = N \times N_{2(n-p-2)}$ )

Recall $\Sigma = \left\{ \sigma \mid \sigma: N \times N_{2^{n-p-2}} \times N_n \rightarrow \{0,1\} \right\}$ is the set of all possible memory blocks, and $P = \left\{ \rho \mid \rho: N \times N_{2^{n-p-2}} \rightarrow N \right\}$, is the set of all possible cursor configurations.

i. $\sigma_x(i,j)$ is the integer represented by the bit vector $\left\langle \sigma(i,j,n-3), \sigma(i,j,n-4), .. \sigma(i,j,p) \right\rangle$ and is the integer represented by the destination cell x of the register $j$ in the $i$th memory block.

ii. $\sigma_y(i,j)$ is the integer represented by the bit vector $\langle \sigma(i,j,p-1),\sigma(i,j,p-2),..\sigma(i,j,0)\rangle$, and is the integer represented by the offst cell y of the register $j$ in the $i$th memory block. $\sigma_z(i,j)$ is the integer represented by the bit vector $\langle \sigma(i,j,n-1),\sigma(i,j,n-2),..\sigma(i,j,0)\rangle$

iii. $\sigma_z(i,j)$ is the integer represented by the contents of the whole register viewed as a binary number.

| Line | Instruction | Comment |
|---|---|---|
| 0 | .... | //cell (0,0) of register 0 indicates if Sequential B-Ram is busy |
| 1 | wrt1 0 0 | // Signal the Sequential B-Ram is busy, next, begin loop |
| 2 | mvrt 10 | // move cursors of instructions marked (c) to next block |
| 3 | mvrt 13 | |
| 4 | mvrt 16 | |
| 5 | mvrt 18 | |
| 6 | mvrt 20 | |
| 7 | mvrt 21 | |
| 8 | mvrt 24 | |
| 9 | mvrt 25 | |
| 10 | cond 0 1 | //(c) test if bit in current memory block is final |
| 11 | jump 13 | // bit is non-final, so test next bit |
| 12 | jump 20 | // bit is final, which is always reset |
| 13 | cond 0 0 | //(c) test non-final bit |
| 14 | jump 16 | // jump to set bit,and halt |
| 15 | jump 18 | // or jump to reset bit, and repeat loop |
| 16 | wrt1 0 0 | //(c) set bit of non final bit |
| 17 | wrt0 0 0 | // halt |
| 18 | wrt0 0 0 | //(c) wrt0 into (0,0) of (next) memory block |
| 19 | jump 2 | |
| 20 | wrt0 0 1 | // (c) rewrite final bit status of current block |
| 21 | wrt0 0 0 | // (c) reset bit |
| 22 | mvrt 23 | // move onto next block to write final bit |
| 23 | mvrt 24 | // move onto next block to indicate final bit |
| 24 | wrt1 0 0 | //(c) |
| 25 | wrt1 0 1 | //(c) |
| 26 | wrt0 0 0 | // halt |

Fig 3.9: Sequential B-Ram program for incrementing indefinitely large integer.

The write function is defined $\omega$: $\Sigma$ x $N$ x $N_{2(n-p-2)}$ x $N_n$ x $\{0,1\} \rightarrow \Sigma$, which writes the bit value $b$ into the $k$th cell of the $j$th register in the block $\sigma_i$.

$$\omega(\sigma,i,j,k,b) = \{\langle i,j,k,b\rangle\} \cup \{ \langle x,y,z,w\rangle \in \sigma \mid (x \neq i) \vee (y \neq j) \vee (z \neq k) \}$$

The cursor-write function is defined $\kappa$: P x $N$ x $2^{n-p-2}$ x $N \rightarrow$ P, which instructs the cursor of a register $j$ in the block $\sigma_i$ to point to $n$.

$$\kappa(\rho,i,j,l) = \{\langle i,j,l\rangle\} \cup \{ \langle x,y,z\rangle \in \rho \mid (x \neq i) \vee (y \neq j) \}$$

Then the state transformation function $\xi'$ : $\Sigma$ x$\Pi'$ x P $\rightarrow \Sigma$ x$\Pi'$ x P for the Sequential B-Ram is defined in Fig 3.10, evaluating the special cases in sequence:

The conditions correspond to the following:

1. Success.
2. Halt fail.
3. Cond fail.
4. Cursor fail.
5. Either wrt0 or wrt1 instruction.
6. Negative consequent of cond instruction.
7. Positive consequent of cond instruction.
8. Jump instruction.
9. Mvrt instruction.
10. Mvlt instruction.

At the beginning of a run, every cursor points to it's own block, ie, $\rho(i,j) = i$, for $i \geq 0$ and $0 \leq j \leq 2^{n-p-2} - 1$. A run of the Sequential B-Ram commences with an application of $\xi'$ to $\langle\sigma,\rho,\{\langle 0,\{1\}\rangle\}\rangle$, which constitutes the first machine cycle. Thereafter, a succession of cycles proceeds, where the input to $\xi'$ is the output from the previous cycle, either indefinitely, or until $\xi'$ yields a tuple of the form $\langle\sigma,\rho,\{\langle 0,\varnothing\rangle\}\rangle$. In the latter case the function $\xi'$ may no longer be applied, the run terminates, and the user may determine the success or type of failure of the run by examining the bits in register zero.

$$
\xi'\left(\begin{array}{l}\sigma,\rho,\\ \{\langle i,\{j\}\rangle\}\end{array}\right)=
\left|\begin{array}{lll}
\langle \omega(\sigma,0,0,0,0),\ \rho,\ \{\langle 0,\varnothing\rangle\}\rangle, & (\sigma_z(i,j)=0)\wedge(\rho(i,j)=0) & (1)\\
\langle \omega(\sigma,0,0,1,1),\ \rho,\ \{\langle 0,\varnothing\rangle\}\rangle, & (j=2^{n-p-2}-1)\wedge((\sigma_z(i,j)\neq 0)\vee(\rho(i,j)\neq 0)) & (2)\\
\langle \omega(\sigma,0,0,2,1),\ \rho,\ \{\langle 0,\varnothing\rangle\}\rangle, & \left(\begin{array}{l}(j=2^{n-p-2}-2)\wedge(\sigma(i,j,n-1)=0)\\ \wedge(\sigma(i,j,n-2)=1)\wedge(\sigma(i,j,n-3)=0)\end{array}\right) & (3)\\
\langle \omega(\sigma,0,0,3,1),\ \rho,\ \{\langle 0,\varnothing\rangle\}\rangle, & \left(\begin{array}{l}(\sigma(i,j,n-1)=1)\wedge(\sigma(i,j,n-2)=0)\\ \wedge(\sigma(i,j,n-3)=1)\wedge(\rho(\rho(i,j),\sigma_x(i,j))=0)\end{array}\right) & (4)\\
\langle \omega(\sigma,i,\sigma_x(i,j),\sigma_y(i,j),\sigma(i,j,n-2)),\ \rho,\ \{\langle i,\{j+1\}\rangle\}\rangle, & \left(\begin{array}{l}(\sigma(i,j,n-1)=0)\\ \wedge(\sigma(i,j,n-3)=0)\end{array}\right) & (5)\\
\langle \sigma,\ \rho,\ \{\langle i,\{j+1\}\rangle\}\rangle, & \left(\begin{array}{l}(\sigma(i,j,n-1)=0)\wedge(\sigma(i,j,n-2)=1)\\ \wedge(\sigma(i,j,n-3)=0)\wedge(\sigma(i,\sigma_x(i,j),\sigma_y(i,j))=0)\end{array}\right) & (6)\\
\langle \sigma,\ \rho,\ \{\langle i,\{j+2\}\rangle\}\rangle, & \left(\begin{array}{l}(\sigma(i,j,n-1)=0)\wedge(\sigma(i,j,n-2)=1)\\ \wedge(\sigma(i,j,n-3)=0)\wedge(\sigma(i,\sigma_x(i,j),\sigma_y(i,j))=1)\end{array}\right) & (7)\\
\langle \sigma,\ \rho,\ \{\langle \rho(i,j),\{\sigma_x(i,j)\}\rangle\}\rangle, & \left(\begin{array}{l}(\sigma(i,j,n-1)=0)\wedge(\sigma(i,j,n-2)=1)\\ \wedge(\sigma(i,j,n-3)=1)\end{array}\right) & (8)\\
\langle \sigma,\kappa(\rho,\rho(i,j),j,\rho(i,j)+1),\{\langle i,\{j+1\}\rangle\}\rangle, & \left(\begin{array}{l}(\sigma(i,j,n-1)=1)\\ \wedge(\sigma(i,j,n-2)=0)\\ \wedge(\sigma(i,j,n-3)=0)\end{array}\right) & (9)\\
\langle \sigma,\kappa(\rho,\rho(i,j),j,\rho(i,j)-1),\{\langle i,\{j+1\}\rangle\}\rangle, & \left(\begin{array}{l}(\sigma(i,j,n-1)=1)\\ \wedge(\sigma(i,j,n-2)=0)\\ \wedge(\sigma(i,j,n-3)=1)\end{array}\right) & (10)
\end{array}\right.
$$

Figure 3.10 Definition of the Sequential B-Ram's state transformation function

#### 3.3.1.3 TURING COMPUTABILITY OF SEQUENTIAL B-RAM.

A proof that the Sequential B-Ram is Turing-computable is desirable, in order to establish it as a fully generic, general purpose model of computation. Given that a Sequential B-Ram's definition allows us to write programs and manipulate and transfer data, across an infinite series of memory blocks, intuitively it should be plausible that the Sequential B-Ram is Turing-computable. One way of proving this would be to:

1. Describe a sequential high level, modular language for Sequential B-Ram, call it D-language
2. Devise a specification for compiling D-language into Sequential B-Ram machine code, implement the specification, and prove the implementation satisfies the specification.
3. Write a general TM simulation program in D-language, which simulates a TM by processing data structures representing the tape, and the TM.
4. Prove the D-Language program is a true simulation of the TM.

Unfortunately, the infrastructure is not in place to execute this plan. At the time of writing, neither Sequential B-Ram simulator, high level language nor compiler have been completed. (A major part of the research to date has been the development of an interlanguage environment for the Synchronic A-Ram.)

A Sequential B-Ram simulation of a one way, one tape Turing machine with the alphabet $\{0,1,\#\}$ has been written in psuedo code, and implemented as a machine code program consisting of 268 lines. The machine code is nontrivial, and a proof of it's correctness has not been attempted. At this stage of development, the reader is invited to satisfy himself of the Sequential B-Ram's Turing Computability, by examining various levels of description of the simulation program in Appendix A. The appendix will be easier to understand if the reader has been introduced to the language *Earth*, to be described in the next chapter.

Landin [5] defined a device called the SECD machine for evaluating λ-expressions. A Sequential B-Ram simulation of the SECD is feasible, but would be significantly more complex than a TM simulation program.

### 3.3.2 SYNCHRONIC B-RAM.

For the sake of completeness, a synchronic version of the B-Ram is presented, but without giving program examples. This model may assume practical relevance for scaling the Synchronic Engine beyond the size limits imposed by the largest viable wafer-scale systems. If the reader's interest is not directed to machines with indefinitely large amounts of memory and parallelism, he may prefer to move on to the next chapter.

#### 3.3.2.1 INFORMAL DESCRIPTION OF SYNCHRONIC B-RAM.

The machine is a natural extrapolation of the Synchronic A-Ram and Sequential B-Ram concepts. A B-Ram marking is called infinite if it is composed of an infinite set of tuples. Having adopted the style of transformation functions described so far, a necessary condition for a B-Ram run to generate an infinite marking, is for the initial marking to be infinite. In order to avoid this complication and keep the set of all Synchronic B-Ram programs denumerable, the machine is restricted to having a finite initial marking.

The main difference between a Synchronic and a Sequential B-Ram, is that a marking may be a non-singleton finite set referring to multiple blocks, so the marking could take the the following form, where $i_x$ is a label referring to $x$th memory block:

$$\mu = \left\{ \left\{ \left\langle i_0, \left\{ j_{00}, j_{01}, j_{02}, j_{03}, \ldots j_{0k_0} \right\} \right\rangle \right\}, \left\{ \left\langle i_1, \left\{ j_{10}, j_{11}, j_{12}, j_{13}, \ldots j_{k_1} \right\} \right\rangle \right\}, \ldots\ldots\ldots \right\}$$

where $i_x \in N$, and $0 < j_{yz}, k_w < 2^{n-p-2}$.

At the beginning of each run, every register cursor in every memory block points to the same memory block as the register itself resides in: $\rho(i,j) = i$, for $i \geq 0$ , and $0 \leq j \leq 2^{n\text{-}p\text{-}2} - 1$. The instruction set is described in figure 3.5 Register 0 is a special register which is reset for each run.

A run begins by simultaneously executing the instruction in registers 1 and 2 of memory block $\sigma_0$ in the first cycle. The instruction in register 1 of memory block $\sigma_0$ is always **wrt1 0 0**,in order to indicate to the user that the machine is busy. The instruction in register 2 of memory block $\sigma_0$ is a jump, in order to proceed with the main body of the program.

The run ends successfully one cycle after any cycle having the marking consisting only of an instruction **wrt0 0 0** pointing to $\sigma_0$. A run fails if the machine detects an error condition, which empties the marking, thereby halting the run, and sets an error bit in register 0, which indicates to the user that the run has failed.

| opcode | Instruction (assembly language) | action |
|---|---|---|
| 000 | **wrt0 x y** | Write 0 into (x,y) of memory block pointed to by instruction. |
| 001 | **wrt1 x y** | Write 1 into (x,y) of memory block pointed to by instruction. |
| 010 | **cond x y** | Examine contents of (x,y) of memory block pointed to by instruction.<br>If zero, jump to instruction in next register of the same memory block as the instruction.<br>Else, jump to instruction in next but one register. |
| 011 | **jump x y** | In the next cycle, mark all instructions from register x of memory block pointed to by instruction, up to an including register x+y of the same memory block for execution in the next cycle. |
| 100 | **mvrt x** | Move cursor of register x of memory block pointed to by instruction to the right. |
| 101 | **mvlt x** | Move cursor of register x of memory block pointed to by instruction to the left., if it can. |

Fig 3.11 Synchronic B-Ram Instruction Set.

There are three possible outcomes with any Synchronic B-Ram run:

i. The run succeeds.
ii. The run fails and halts in the cycle after one generating one of ten errors.
iii. The run never terminates.

3.3.2.2 SYNCHRONIC B-RAM ERRORS.

The bits of register 0 are reserved for machine status and error flags. Instructions of the form **wrt1 0 y**, where y≠0 are reserved by the error detection mechanism, and cannot be used in any program. The machine indicates errors by setting various bits in register 0.

Types of Errors.

1. *Marking Fail.* A legal marking cannot contain a tuple containing a multiset. If some instruction in some memory block is activated more than once in the same cycle, then the machine empties the marking and (0,1) of $\sigma_0$ is set. The machine halts unsuccessfully in the next cycle. Precludes Live fail.

2. *Write Fail.* The machine is not a concurrent write machine. If at least two instructions in any two memory blocks write to any location (x,y) in any memory block in the same cycle, , then the machine empties the marking and (0,2) of $\sigma_0$ is set. This error precludes no other types of error occurring in the same cycle.

3. *Halt Fail.* The final marking should always be a singleton with only the special halt instruction **wrt0 0 0** pointing to $\sigma_0$ . Otherwise, the machine empties the marking and the error bit (0,3) of $\sigma_0$ is set. Precludes Live fail and Jump fail.

4. *Live Fail.* A program should not unexpectedly generate an empty marking. If the marking becomes empty (without a Live fail having been activated in the previous cycle), then (0,4) to $\sigma_0$ is set. Precludes Halt fail and Jump fail.

5. *Cond Fail.* If a cond instruction in the last but one register of any memory block is activated for execution, then the positive consequent of the cond is undefined. The machine empties the marking and sets the bit (0,5) of $\sigma_0$.

6. *Consequent Fail.* If any memory block, any two of the triplet of a cond instruction, it's negative consequent instruction, and positive consequent instruction are in a marking. If this occurs the marking is emptied and (0,6) to $\sigma_0$ is set.

7. *Active Fail.* An instruction should not have its contents modified in the same cycle in which it is active. If a write instruction to a cell of some register of any memory block occurs in a marking, in which the same register of the same memory block is also marked, then the machine empties the marking, and (0,7) to $\sigma_0$ is set.

8. *Cursor Fail*. A register may not have it's cursor moved the left of the zeroth block. If a mvlf instruction is applied to a register in some memory block, which is already pointing to $\sigma_0$, then the machine ignores the instruction, empties the marking, and (0,8) to $\sigma_0$ is set.

9. *Jump Fail.* In a marking, a jump instruction should not attempt to mark the illegal zeroth register of the zeroth memory block, or the register index of some block which does not exist (i.e. the destination cell plus the offset exceeds the index $2^{n-p-2}-1$ of the last register of a block ), then the marking is emptied and (0,9) is set.

#### 3.3.2.3 State transformation function for synchronic B-RAM

Recall $\Sigma = \left\{ \sigma \mid \sigma: N \times N_{2^{n-p-2}} \times N_n \rightarrow \{0,1\} \right\}$ is the set of all possible memory blocks, and $\mathrm{P} = \left\{ \rho \mid \rho: N \times N_{2^{n-p-2}} \rightarrow N \right\}$, is the set of all possible cursor configurations, and $\Pi = P\left( N \times P\left(N_{2(n-p-2)}\right) - \left\{ \langle 0, M \rangle \,\middle|\, (0 \in M) \wedge \left( M \subseteq N_{2(n-p-2)} \right) \right\} \right)$ is the set of all markings $\mu$. Recall the $o$ functions:

i. $\sigma_x(i,j)$ is the integer represented by the bit vector $\langle \sigma(i,j,n-3), \sigma(i,j,n-4), .. \sigma(i,j,p) \rangle$ and is the integer represented by the destination cell x of the register $j$ in the $i$th memory block.

ii. $\sigma_y(i,j)$ is the integer represented by the bit vector $\langle\sigma(i,j,p-1),\sigma(i,j,p-2),..\sigma(i,j,0)\rangle$, and is the integer represented by the offst cell y of the register $j$ in the $i$th memory block.

iii. $\sigma_z(i,j)$ is the integer represented by the bit vector $\langle\sigma(i,j,n-1),\sigma(i,j,n-2),..\sigma(i,j,0)\rangle$. $\sigma_z(i,j)$ is the integer represented by the contents of the whole register viewed as a binary number.

Recall the write function $\omega$: $\Sigma \text{ x } N \text{ x } N_{2(n-p-2)} \text{ x } N_n \text{ x } \{0,1\} \rightarrow \Sigma$, which writes the bit value $b$ into the $k$th cell of the $j$th register in the block $\sigma_i$.

$$\omega(\sigma,i,j,k,b) = \{\langle i,j,k,b\rangle\} \cup \{\ \langle x,y,z,w\rangle \in \sigma \mid (x \neq i) \vee (y \neq j) \vee (z \neq k)\ \}$$

Also recall the cursor-write function $\kappa$: $\text{P x } N \text{ x } 2^{n-p-2} \text{ x } N \rightarrow \text{P}$, which instructs the cursor of a register $j$ in the block $\sigma_i$ to point to $n$.

$$\kappa(\rho,i,j,l) = \{\langle i,j,l\rangle\} \cup \{\ \langle x,y,z\rangle \in \rho \mid (x \neq i) \vee (y \neq j)\ \}$$

Then the state transformation function $\xi : \Sigma \text{ x} \Pi \text{ x P} \rightarrow \Sigma \text{ x} \Pi \text{ x P}$ for the Synchronic B-Ram is defined as follows, evaluating the special cases in sequence:

$$
\xi(\sigma,\mu,\rho) = \begin{cases}
\left\langle \omega(\sigma,0,0,0,0), \left\{\left\langle 0,\varnothing \right\rangle\right\}, \rho \right\rangle, & (\exists i \in N)\left(\exists j \in 2^{n-p-2}\right) \\
& \left(\left(\mu = \left\{\left\langle i,\{j\}\right\rangle\right\}\right) \wedge \left(\sigma_z(i,j) = 0\right) \wedge \left(\rho(i,j) = 0\right)\right) \quad (1) \\
\left\langle \omega(\sigma,0,0,1,1), \left\{\left\langle 0,\varnothing \right\rangle\right\}, \rho \right\rangle, & (\exists i \in N)\left(\exists M \subseteq 2^{n-p-2}\right) \\
& \left( \left(\left\langle i,M \right\rangle \in \mu\right) \wedge \left(M \text{ is a multiset}\right) \right) \quad (2) \\
\left\langle \omega(\sigma,0,0,2,1), \left\{\left\langle 0,\varnothing \right\rangle\right\}, \rho \right\rangle, & (\exists i,j \in N)\left(\exists M_1,M_2 \subseteq 2^{n-p-2}\right)(\exists k \in M_1)(\exists l \in M_2) \\
& \begin{pmatrix} \left(\left(\left\langle i,M_1\right\rangle \in \mu\right) \wedge \left(\left\langle j,M_2\right\rangle \in \mu\right)\right) \\ \wedge \left((i \neq j) \vee (k \neq l)\right) \wedge \left(\sigma(i,k,n-1) = 0\right) \\ \wedge \left(\sigma(j,l,n-1) = 0\right) \wedge \left(\sigma(i,k,n-2) = 0\right) \\ \wedge \left(\sigma(j,l,n-2) = 0\right) \wedge \left(\rho(i,k) = \rho(j,l)\right) \\ \wedge \left(\sigma_x(i,k) = \sigma_x(j,l)\right) \wedge \left(\sigma_y(i,k) = \sigma_y(j,l)\right) \end{pmatrix} \quad (3) \\
\left\langle \omega(\sigma,0,0,3,1), \left\{\left\langle 0,\varnothing \right\rangle\right\}, \rho \right\rangle, & (\exists i,j \in N)\left(\exists M_1,M_2 \subseteq 2^{n-p-2}\right)(\exists k \in M_1)(\exists l \in M_2) \\
& \begin{pmatrix} \left(\left(\left\langle i,M_1\right\rangle \in \mu\right) \wedge \left(\left\langle j,M_2\right\rangle \in \mu\right)\right) \\ \wedge \left((i \neq j) \vee (k \neq l)\right) \wedge \left(\sigma_z(i,k) = 0\right) \end{pmatrix} \quad (4) \\
\left\langle \omega(\sigma,0,0,4,1), \left\{\left\langle 0,\varnothing \right\rangle\right\}, \rho \right\rangle, & \left((\mu = \varnothing) \wedge \left(\sigma(0,0,0) = 0\right) \wedge \left(\sigma(0,0,5) = 0\right)\right) \quad (5) \\
\left\langle \omega(\sigma,0,0,5,1), \left\{\left\langle 0,\varnothing \right\rangle\right\}, \rho \right\rangle, & (\exists i \in N)\left(\exists M \subseteq 2^{n-p-2}\right)(\exists k \in M) \\
& \begin{pmatrix} \left(\left\langle i,M \right\rangle \in \mu\right) \wedge \left(k = 2^{n-p-2} - 2\right) \\ \wedge \left(\sigma(i,k,n-1) = 0\right) \wedge \left(\sigma(i,k,n-2) = 1\right) \\ \wedge \left(\sigma(i,k,n-3) = 1\right) \end{pmatrix} \quad (6)
\end{cases}
$$

Figure 3.12 First part of definition of the Synchronic B-Ram's state transformation function

$$
\xi(\sigma,\mu,\rho)=
\begin{cases}
\left\langle \omega(\sigma,0,0,6,1),\ \{\langle 0,\varnothing\rangle\},\ \rho\right\rangle, & (\exists l\in N)\left(\exists M\subseteq 2^{n-p-2}\right)\left(\exists k\in 2^{n-p-2}\right)(\exists i,j\in M) \\
 & \left(\begin{array}{l}\left(\begin{array}{l}((\langle l,M\rangle\in\mu)\wedge(\sigma(l,k,n-1)=0)\\ \wedge(\sigma(l,k,n-2)=1)\wedge(\sigma(l,k,n-3)=0)\end{array}\right)\\ \wedge\left(\begin{array}{l}((i=k+1)\wedge(j=k+2))\ \vee\\ \left((i=k)\wedge\left(\begin{array}{l}(j=k+1)\vee\\ (j=k+2)\end{array}\right)\right)\end{array}\right)\end{array}\right) \qquad (7) \\
\left\langle \omega(\sigma,0,0,7,1),\ \{\langle 0,\varnothing\rangle\},\ \rho\right\rangle, & (\exists i,j\in N)\left(\exists M_1,M_2\subseteq 2^{n-p-2}\right)(\exists k\in M_1)(\exists l\in M_2) \\
 & \left(\begin{array}{l}((\langle i,M_1\rangle\in\mu)\wedge(\langle j,M_2\rangle\in\mu))\\ \wedge\ ((i\neq j)\vee(k\neq l))\wedge(\sigma(i,k,n-1)=0)\\ \wedge(\sigma(i,k,n-2)=0)\wedge(\rho(i,k)=\rho(j,l))\\ \wedge(\sigma_x(j,l)=i)\end{array}\right) \qquad (8) \\
\left\langle \omega(\sigma,0,0,8,1),\ \{\langle 0,\varnothing\rangle\},\ \rho\right\rangle, & (\exists i\in N)\left(\exists M\subseteq 2^{n-p-2}\right)(\exists k\in M) \\
 & \left(\begin{array}{l}((\langle i,M\rangle\in\mu))\wedge(\sigma(i,k,n-1)=1)\\ \wedge(\sigma(i,k,n-2)=0)\wedge(\sigma(i,k,n-3)=1)\\ \wedge(\rho(\rho(i,k),\sigma_x(i,k))=0)\end{array}\right) \qquad (9) \\
\left\langle \omega(\sigma,0,0,9,1),\ \{\langle 0,\varnothing\rangle\},\ \rho\right\rangle, & (\exists i\in N)\left(\exists M\subseteq 2^{n-p-2}\right)(\exists k\in M) \\
 & \left(\begin{array}{l}(\sigma(i,k,n-1)=0)\wedge(\sigma(i,k,n-2)=1)\\ \wedge\ (\sigma(i,k,n-3)=1)\ \wedge\\ \left(\begin{array}{l}(\sigma_x(i,k)=0)\vee\\ ((\sigma_x(i,k)+\sigma_y(i,k))\geq 2^{n-p-2})\end{array}\right)\end{array}\right) \qquad (10) \\
\left\langle \xi_1(\sigma,\mu,\sigma),\ \xi_2(\sigma,\mu,\sigma),\ \xi_3(\sigma,\mu,\sigma)\right\rangle & \qquad (11)
\end{cases}
$$

Figure 3.13 Second part of definition of the Synchronic B-Ram's state transformation function

The conditions correspond to the following:

1. Success.
2. Marking fail.
3. Write fail
4. Halt fail.
5. Live fail.
6. Cond fail.
7. Consequent fail
8. Active fail.
9. Cursor fail.
10. Jump fail.
11. Non-halting legal input.

I proceed with the definitions of the component functions for the final case of non-halting legal inputs:

1. $\xi_1 : \Sigma x \Pi x \mathrm{P} \rightarrow \Sigma$.

$$\xi_1(\sigma,\mu,\rho) = \left( \bigcup_{\exists k \exists M \begin{pmatrix} (\langle k,M \rangle \in \mu) \\ \wedge\ (\sigma(k,i,n-1)=0) \\ \wedge\ (\sigma(k,i,n-2)=0) \end{pmatrix}} \langle k, \sigma_x(k,i), \sigma_y(k,i), \sigma(k,i,n-3) \rangle \right)$$

$$\cup \left\{ \langle p,q,r,s \rangle \in \sigma \;\middle|\; \forall k \forall M \forall i \begin{pmatrix} ((\langle k,M \rangle \in \mu) \wedge (i \in M)) \Rightarrow \\ \begin{pmatrix} ((\sigma(k,i,n-1) \neq 0) \vee (\sigma(k,i,n-2) \neq 0)) \\ \vee (\rho(k,i) \neq p) \vee (\sigma_x(k,i) \neq q) \\ \vee (\sigma_y(k,i) \neq r) \end{pmatrix} \end{pmatrix} \right\}.$$

2. $\xi_2 : \Sigma \times \Pi \times \mathrm{P} \to \Pi$.

If $i,j \in N$, $0 < j \leq 2^p - 1$, let $[i,0] \equiv i$ and $[i,j] \equiv i, i+1, .. i+j$. The definition will make use of a partial function $f : N \times \Sigma \times \Pi \times \mathrm{P} \to P\left(2^{n-p-2}\right)$, where $f(i,\sigma,\mu,\rho)$ is the subset of $2^{n-p-2}$ associated with a block index $i$, if it is to appear in $\eta_2(\sigma,\mu.\rho)$.

$$\xi_2(\sigma,\mu,\rho) = \left\{ \langle i, f(i,\sigma,\mu,\rho)\rangle \;\middle|\; \exists k\, \exists M\, \exists j \left( \begin{array}{c} (\langle k,M\rangle \in \mu) \wedge (j \in M) \\ \wedge \left( \begin{array}{l} \left( \left( \begin{array}{l} \wedge\, (\sigma(k,j,n-1) = 0) \\ \wedge\, (\sigma(k,j,n-2) = 1) \\ \wedge\, (\sigma(k,j,n-3) = 0) \end{array} \right) \wedge (k = i) \right) \\ \vee \left( \left( \begin{array}{l} \wedge\, (\sigma(k,j,n-1) = 0) \\ \wedge\, (\sigma(k,j,n-2) = 1) \\ \wedge\, (\sigma(k,j,n-3) = 1) \end{array} \right) \wedge (\rho(k,j) = i) \right) \end{array} \right) \end{array} \right) \right\}$$

where
$$f(i,\sigma,\mu,\rho) = \left( \bigcup_{\exists M \left( \begin{array}{l} ((\langle i,M\rangle \in \mu) \wedge (j \in M) \wedge (\sigma(i,k,n-1)=0) \\ \wedge\, (\sigma(i,k,n-2)=1) \wedge (\sigma(i,k,n-3)=0) \\ \wedge\, (\sigma(\rho(i,k),\sigma_{x(i,k)},\sigma_{y(i.k)})=0) \end{array} \right)} k+1 \right)$$

$$\cup \left( \bigcup_{\exists M \left( \begin{array}{l} ((\langle i,M\rangle \in \mu) \wedge (j \in M) \wedge (\sigma(i,k,n-1)=0) \\ \wedge\, (\sigma(i,k,n-2)=1) \wedge (\sigma(i,k,n-3)=0) \\ \wedge\, (\sigma(\rho(i,k),\sigma_{x(i,k)},\sigma_{y(i.k)})=1) \end{array} \right)} k+2 \right)$$

$$\cup \left( \bigcup_{\exists M \left( \begin{array}{l} ((\langle p,M\rangle \in \mu) \wedge (q \in M) \wedge (\sigma(p,q,n-1)=0) \\ \wedge\, (\sigma(p,q,n-2)=1) \wedge (\sigma(p,q,n-3)=1) \\ \wedge\, (\rho(p,q)=i) \end{array} \right)} [\sigma_x(p,q), \sigma_y(p,q)] \right) .$$

3. $\xi_3 : \Sigma \times \Pi \times P \rightarrow P$.

$$\xi_3(\sigma,\mu,\rho) = \left( \bigcup_{\exists i \exists M \exists j \left( \begin{array}{l} (\langle i,M\rangle \in \mu) \wedge (j \in M) \wedge (\sigma(i,j,n\text{-}1)=1) \\ \wedge (\sigma(i,j,n\text{-}2)=0) \wedge (\sigma(i,j,n\text{-}3)=0) \\ \wedge (\rho(i,j)=p) \wedge (\sigma_x(i,j)=q) \end{array} \right)} \langle p,q,\rho(p,q)+1\rangle \right)$$

$$\cup \left( \bigcup_{\exists i \exists M \exists j \left( \begin{array}{l} (\langle i,M\rangle \in \mu) \wedge (j \in M) \\ \wedge (\sigma(i,j,n\text{-}1)=1) \wedge (\sigma(i,j,n\text{-}2)=0) \\ \wedge (\sigma(i,j,n\text{-}3)=1) \wedge (\rho(i,j)=p) \\ \wedge (\sigma_x(i,j)=q) \end{array} \right)} \langle p,q,\rho(p,q)-1\rangle \right)$$

$$\cup \left\{ \langle p,q,r\rangle \in \rho \;\middle|\; \forall i \forall M \forall j \left( \begin{array}{l} ((\langle i,M\rangle \in \mu) \wedge (j \in M)) \Rightarrow \\ \left( (\sigma(i,j,n-1) \neq 1) \vee (\sigma(i,j,n-2) \neq 0) \right. \\ \left. \vee (\rho(i,j) \neq p) \vee (\sigma_x(i,j) \neq q) \right) \end{array} \right) \right\} .$$

At the beginning of a run, every cursor points to it's own block, ie, $\rho(i,j) = i$, for $i \geq 0$ and $0 \leq j \leq 2^{n-p-2} - 1$. A run of the Synchronic B-Ram commences with an application of $\xi$ to $\langle \sigma, \{\langle 0,\{1,2\}\rangle\} \rho, \rangle$, which constitutes the first machine cycle. Thereafter, a succession of cycles proceeds, where the input to $\xi$ is the output from the previous cycle, either indefinitely, or until $\xi$ yields a tuple of the form $\langle \sigma, \{\langle 0,\varnothing\rangle\} \rho, \rangle$. In the latter case the function $\xi$ may no longer be applied, the run terminates, and the user may determine the success or type of failure of the run by examining the bits in register zero of memory block $\sigma_0$.

## 3.4 Computability relationships between members of the α-Ram family.

If we accept the Sequential B-Ram is Turing-computable, then it is obvious that the Synchronic B-Ram is also Turing-computable, because any program for the latter may be readily transformed into a program for the former, by sequentialising instructions generating non-singleton markings. Register space allowing, the same relationship holds for the A-ram machines. Fig 3.14 presents a table describing the computability relationships within the α-Ram family.

There may not be enough registers to store the transformation of a large Sequential A-Ram $\langle r,\sigma,\{1\},\eta'\rangle$ programs into a Synchronic A-Ram $\langle r,\sigma,\{1,2\},\eta\rangle$ program, unless the size of the latter's offset is increased. The same may be said in reverse.

If $p$ is fixed, then B-ram programs cannot in general be transformed into A-Ram programs.

| | Sequential A-Ram $\langle p,\sigma,\{1\},\eta'\rangle$ | Synchronic A-Ram $\langle p,\sigma,\{1,2\},\eta\rangle$ | Sequential B-Ram $\langle p,\sigma,\{1\},\rho,\xi'\rangle$ | Synchronic B-Ram $\langle p,\sigma,\{1,2\},\rho,\xi\rangle$ |
|---|---|---|---|---|
| Sequential A-Ram $\langle p,\sigma,\{1\},\eta'\rangle$-computable | Yes, if $p > q$ | Yes, if $p > q$ | Yes | Yes |
| Synchronic A-Ram $\langle p,\sigma,\{1,2\},\eta\rangle$-computable | Yes, if $p > q$ | Yes, if $p > q$ | Yes | Yes |
| Sequential B-Ram $\langle p,\sigma,\{1\},\rho,\xi'\rangle$-computable | No | No | Yes | Yes |
| Synchronic B-Ram $\langle p,\sigma,\{1,2\},\rho,\xi\rangle$-computable | No | No | Yes | Yes |

Figure 3.14 Summary of computability relationships.

## 3.5 Extending the α-RAM concept.

The next two subsections discuss extensions, that demonstrate the α-Ram framework is flexible, and capable of incorporating notions of larger symbol sets and propagation delay.

### 3.5.1 Larger symbol sets

Larger symbol sets are now considered, where each memory cell can store an element of $\{0,1,.m-1\}$. Taking the Sequential A-Ram as an example, there would be two modifications to the instruction set. A further *m-2* wrt instructions would be needed, to be able to write the extra symbols into a register cell. The instruction **cond x y** in register $i$, would jump to register $i+j+1$ if $\sigma(x,y)=j$, resulting in the set depicted in fig 3.15:

| opcode | Instruction (assembly language) | action |
|---|---|---|
| 0 | **wrt0 x y** | Write 0 into (x,y). |
| 1 | **wrt1 x y** | Write 1 into (x,y). |
| 2 | **wrt2 x y** | Write 2 into (x,y). |
| . . . . | . . . . | . . . . |
| m | **wrt(m-1) x y** | Write m-1 into (x,y). |
| m+1 | **cond x y** | Examine contents of (x,y).<br>If zero, jump to instruction in next register.<br>Else if one, jump to instruction in 2nd next register.<br>Else if two, jump to instruction in 3rd next register.<br>.<br>.<br>Else if m-2, jump to (m-2)th next register<br>Else, jump to (m-1)th next register. |
| m+2 | **jump x** | jump to instruction in register x. |

Fig 3.15 Sequential A-Ram Instruction Set for machine with symbol set $\{0,1,.m-1\}$.

Formally, an extended A-Ram is a tuple $\langle m, p, \sigma, \mu, \eta \rangle$ :

1. Symbol set size. $m$ is the number of elements in a finite symbol set.

2. Offset. $p \geq 3$ is an integer and is called the offset. Let $n = m^p$. $n$ is the number of *elements* in a *register*, where each element stores a member of the set $\{0,1,..m-1\}$. ($p$ is the minimum number of elements required to encode the position of an element in a register, using the alphabet of $\{0,1,..m-1\}$ .)

3. Memory. Let $a = \overline{\log_m(m+2)}$. The memory block is composed of $m^{(n-p-a)}$ registers, containing $n$ cells, each capable of storing an element of $\{0,1,..m-1\}$. It is a function $\sigma$: $N_{m^{(n-p-a)}}$ x $N_n \rightarrow \{0,1,..m-1\}$ which takes a register index $x$, an element index $y$, and delivers the contents of the $y$th element of the $x$th register. Let $\Sigma = \left\{ \sigma \mid \sigma: N_{m^{(n-p-a)}} \text{ x } N_n \rightarrow \{0,1,..m-1\} \right\}$ be the set of all possible memory blocks.

4. Marking. The marking $\mu \subseteq \left(N_{m^{(n-p-2)}} - \{0\}\right)$ represents a possibly non-contiguous subset of the registers in the memory block, or the special termination value $\varnothing$. The zeroth register cannot be marked and is reserved for machine status bits. Let $\Pi = P\left(N_{m^{n-p-2}} - \{0\}\right)$ be the set of all markings.

5. State transformation function. As before, the state transformation function is a suitably revised version of the function $\eta'$ : $\Sigma$ x $\Pi \rightarrow \Sigma$ x $\Pi$. The number of cells required to represent the opcode is $\overline{\log_m(m+2)}$. The extended A-Ram would have the instruction format below[24]. I leave the definition of the state transformation function, and the definition for an extended B-Ram, to the reader.

| opcode | destination cell (x) | offset (y) |
|---|---|---|

cells (n-a-1) to (cell n-1) cells $p$ to cells (n-a-2) cells 0 to ($p$-1)

[24] For $m > 2$, there will be unused opcodes. Although code for an extended A-Ram $\langle m, p, \sigma, \mu, \eta \rangle$ will to some extent be compressed because one cond statement can make a selection from $m$ alternatives, extra code will be needed to cover the extra cases in the symbol set for any program.

### 3.5.2 INTRODUCING PROPAGATION DELAY.

The α-RAM machines as presented are physically unrealistic, in that transmission of data or marking information between any two bits or registers occurs in unit time, regardless of a machine's memory size. It is feasible however to introduce a metric function for any two locations in memory, and outline a notion of propagation delay in the definition of state transformation functions. The A-Ram $\langle p,\sigma,\mu,\eta \rangle$ is considered. As the A-Ram is finite, it is reasonable to suppose that there is a maximum distance of say $k$ units between any two registers in the machine. A distance function may then be defined as a function on integers $d$: $N_{2^{n-p-2}}$ x $N_{2^{n-p-2}} \rightarrow \left(N_{k+1} - \{0\}\right)$, where the minimum distance between registers will be taken as 1, in order to make it convenient to associate distance with the number of cycles for a marked instruction to execute. The function $d$ might be based on a conception of the memory block as a one-dimensional, or a multidimensional array of registers in space. An instruction in register $i$ referencing register $j$, is normally understood in previous definitions of the state transformation function, to affect the memory block in the next cycle, i.e. one cycle. We may now understand the instruction to require $d(i,j)$ machine cycles to take effect, where $1 \leq d(i,j) \leq k$ .

Consider a new definition of the set of markings $\Pi^{\sim} = \left(P\left(N_{2^{n-p-2}} - \{0\}\right)\right)^{k}$, where a marking is now a vector with $k$ elements. Each element is a (potentially empty) subset of registers in the memory block, and represents a different propagation stage of markings in flight from source to destination registers. An instruction that is marked in any cycle requiring $d(i,j)$ cycles to take effect, will be added to vector element $k - d(i.j)$ in the cycle's output marking. The final $k$th vector element contains instructions that will arrive at, and affect the memory block $\sigma$ in the next cycle. Members of preceding vector elements are right shifted in each cycle to the succeeding vector elements to reflect propagation.

A modified state transformation function $\vartheta^{\sim} : \Sigma \text{ x } \Pi^{\sim} \rightarrow \Sigma \text{ x } \Pi^{\sim}$ may now be conceived. $\vartheta_1^{\sim} : \Sigma \text{ x } \Pi^{\sim} \rightarrow \Sigma$ is similar to a conventional state transformation of the $\sigma$ memory block, in that only the final vector element plays a role; in effect $\vartheta_1^{\sim} : \Sigma \text{ x } \Pi_k^{\sim} \rightarrow \Sigma$ , where $\Pi_k^{\sim}$ is the rightmost vector element of $\Pi^{\sim}$. The definition of $\vartheta_2^{\sim} : \Sigma \text{ x } \Pi^{\sim} \rightarrow \Pi^{\sim}$ however, must take into account a conventional presentation of $\vartheta_2$ , together with the right shifting of members of vector elements, to reflect propagation delay. The full definition of the state transformation function is left to the reader.

The B-Ram can be given a similar treatment, with the proviso that the distance function may take into account the difference in cursor integer values of the two instructions, as well as their relative displacement between locations in the memory block:

$$d: N \times N \times N_{2^{n-p-2}} \times N_{2^{n-p-2}} \rightarrow (N - \{0\})$$

The marking for a B-Ram with propagation delay would be a string rather than a vector of instruction markings, belonging to the construction

$$\left( P\left( N \times P\left(N_{2(n-p-2)}\right) - \left\{ \langle 0, M \rangle \middle| (0 \in M) \wedge \left(M \subseteq N_{2(n-p-2)}\right) \right\} \right) \right)^{*}$$

**References.**

# Chapter 4
# EARTH.

## 4.1. A LANGUAGE FOR PRIMITIVE PARALLEL PROGRAM MODULES.

An interstring was defined in chapter 2, to be an element of the set construction $(A_k \cup B_k)^*$. From now on, the term will be used more loosely to refer to what was described in Chapter 2 as the simplified form of an interstring: a string of strings of simple expressions, where the simple expressions belong to what will be termed a *base language*. A base language is a tree language, whose set of expressions is finite. The base language's syntax is devoid of any inductive or recursive rule, resulting in syntax trees having a maximum depth, and strings having a maximum length[25]. The term *interlanguage* will henceforth be used informally, to refer to a language whose expressions are interstrings in some base language.

The Space interlanguage represents a modular and functional-style approach to programming, where the semantics of a program are described in terms of interstrings and A-ram states, rather than the intermediate forms of the λ-calculus used for FPLs [1] [2]. The target machine for compilation is the Synchronic A-Ram $\langle 5,\sigma,\{1,2\},\eta\rangle$[26], generating a memory block with 33,554,432 32-bit registers, which occupy 128 MBytes of disk space. Machine size is adequate for a massively parallel program with potentially tens of thousands of sub-programs running simultaneously. Recall a $\langle 5,\sigma,\{1,2\},\eta\rangle$ instruction has the following format:

| opcode | destination cell (x) | offset (y) |
|---|---|---|
| bits 30-31 | bits 5-29 | bits 0-4 |

The next chapter will explain how a Space program module's code consists of a series of numbered interstrings, whose base expressions contain references to other modules (sub-programs), called *sub-modules*, drawn from a library pre-defined by the user. In common with other structured programming approaches, the set of library program modules has a partial order based on the sub-module inclusion relationship. The sub-modules occupying the bottom layer of the partial order, are those modules that do not include references to other sub-modules. They are composed in a language called *Earth*, rather than Space. Earth is a primitive parallel programming language, equipped with a *replicator* mechanism employing a control variable, for repeating code segments. In addition to VHDL and Verilog array type definitions of logic blocks and simple data transfer patterns between blocks, Earth replicators allow the concise definition of more complex repetitive data transfer patterns. Earth code is close to the level of $\langle 5, \sigma, \{1,2\}, \eta \rangle$ machine code, and is the subject of this chapter.

[25] If the base language had a recursive rule, it is arguable that a high structural variability tree syntax is a significant part of the language environment, in which it is the interstring that is supposed to encode dataflow (see 4.7).

[26] The restriction on the initial marking will be slightly relaxed to accommodate program modules that rewrite their own code, and have two initial markings.

Earth and Space modules have a *level*, which is an unsigned integer representing the distance of the module from the bottom of the partial order (depth of module composition). Earth modules are all level zero. The level of a Space module is equal to the maximum of the levels of the module's sub-modules, plus one.

Earth is powerful enough for modules to duplicate the functionality of complex sequential digital circuits, with a high degree of circuit parallelism. The size and depth of a sequential digital circuit, is proportional to the size and cycle completion time respectively of an Earth implementation, providing the programmer matches the module's parallelism to the circuit's parallelism. Subject only to the constraint imposed by memory block size, the programmer is perfectly at liberty to do this, given the highly parallel nature of the Synchronic A-Ram. Earth modules can readily implement primitive operations performed by devices such as *n*-input logic gates, register shifters, demultiplexers, incrementers, adders, etc.. Earth can also implement many of the arithmetic-logic functional units found in processor cores, and a module can generate thousands of lines of machine code[27].

To compose and understand a non-trivial level zero module, only using the notational style of **wrt0 x y**, **wrt1 x y**, **cond x y**, and **jump x y**, can be difficult and error-prone. Earth therefore has features which facilitate the composition and readability of level zero modules. An Earth module begins with a list of declarations, some of which refer to storage, followed by code, whereas a compiled Earth module begins with machine code, followed by registers holding the module's storage locations. If storage were to occupy an variable amount of the memory block registers from memory address 1, then the initial program marking in the succeeding code would also be obliged to be variable. The decision of having the compiled code followed by storage rather than the other way round, was made in order to have fixed initial markings of $\{1,2\}$ for non meta-modules (see 2. above), and $\{1,2\}$ and $\{3,4\}$ for meta-modules.[28]

[27] Appendix B describes a serial style 32 bit adder module, which consists of 138 lines of non-iterated code, and a barrel shift module, whose iterated code compiles into 2,190 machine instructions.

Unlike Space, Earth is not an interlanguage. Earth code without replicator expressions, consists of a simple sequence of machine code like instructions, and the issue of using interstrings does not arise. Replicator expressions however, can exhibit a nested structure, which could be represented as an interstring, rather than as a tree. A decision was taken to use tree syntax/brackets to express replication, because interstrings would have added another labelling system to Earth code, in addition to an existing one, and because code subexpression repetition seemed unlikely[29].

The Spatiale environment runs in a Unix terminal, and allows the user to add a representation of a Space or Earth module to the library, where the module is given a numerical index. Spatiale compiles a library module, and loads the code into the A-Ram memory block. The user is prompted for program inputs in stdin, the $\langle 5,\sigma,\{1,2\},\eta \rangle$-simulator runs the code, and the program's outputs are printed to stdout.

---

[28] The order also facilitated the development and debugging of the Earth compiler, by having compiled text file line numbers matched with A-ram memory block addresses.

[29] The decision in retrospect seems wrong, because interstrings would render a future Earth compiler written in Space, more parallelisable. In addition, interstrings may have sooner identified a problem encountered with the implementation of nested replication discussed in 4.4.7. The Space language itself is not affected by the latter problem, whose main effect is to impose more work on the Earth programmer, and will be fixed in the next version of the Spatiale environment.

## 4.2 DECLARATIONS, AND LEVEL ZERO TYPES.

Earth has a non-extendable set of fixed types, and declarations which can name the location in memory of a typed storage entity. This frees the programmer from having to remember the numeric memory block addresses of storage entities, or having to recalculate them when adding extra lines of code. As with modules, types in Earth and Space have a level, indicating depth of type composition. An Earth type occupies up to the maximum of the whole of a single register, and is designated to be level zero. In Space there is an extensible library of types, where a new type may be derived by forming a construct whose members are pre-defined, existing types. A Space type's level is the maximum of it's member type levels, plus one.

An Earth module begins with declarations, which include a list of named storage entities, of level zero types such as BIT, BYTE, WORD, and REGISTER, followed by some code. The storage declarations state the *interface category* of a named storage entity, i.e. whether it is *input*, *output*, both input and output (*ioput*), or internal storage marked *private*, which cannot be accessed by higher level modules[30]. The final TIME declaration concerns the number of cycles the module takes to complete. Individual Earth declarations have to occupy a single line of text. Beginning with an example, they take the general form:

1. `NAME: myearthmodule;` The name of the module, which is new to the library, unless module editing is intended. (Obligatory)
2. `META: 2;` The meta-status, which refers to special category of Earth modules called *meta-modules*, which can modify their own code, and have two initial markings. Meta-status is covered in more detail in 4.6. (Optional).
3. `BITS: busy private, overflow output;` The BITS declaration consists of a series of names, coupled with interface categories, and takes the form
   BITS: *firstbitname interface_category*, *secondbitname interface_category*,...., *finalbitname interface_category*; . These declarations refer to storage entities occupying a single bit. A BIT declaration is obligatory and always includes a special bit called "busy", which indicates when the module is active. [31] A succession of up to 32 BIT declarations would be assigned to a succession of bits in the same register by the compiler. (Obligatory).

[30] Interface categories are used by Spatiale to check that a Space module does not refer to private storage in sub-modules, and also to prompt the user for module inputs, and print module outputs when running a module.

[31] A busy bit is required for each module, because a module's sub-modules cannot all use the A-Ram's busy bit at (0,0).

4. `BYTES: input0 input, input1 input;` The BYTES declaration consists of a series of names with interface categories, which refer to storage entities occupying a byte. A declaration of up to four BYTE names would be assigned to the same register. (Optional)
5. `WORDS: ioput0 ioput;` The WORDS declaration consists of a series of names with interface category, which refer to storage entities occupying a word, i.e. 2 bytes, allowing two words per register. (Optional)
6. `REGS: output0 output, output1 output;` The REGS declaration consists of a series of names with interface category, which refer to storage entities occupying a whole register. (Optional)
7. `OFSTS: offset0 ioput;` Along with the next two types, the OFSTS declaration refers to a rarely used storage entity occupying a fixed segment of a register, which eases the description of self-modifying code. In this case the offset is the rightmost 5 bits (bits 0-4). A succession of offset names will occupy successive registers. (Optional)
8. `DSTNS: destination ioput;` The DSTNS declaration refers to the destination cell, which occupies bits 5-29. (Optional)
9. `BITAS: bitaddress ioput;` The BITAS declaration refers to the bit address region of a register, able to specify a single bit within the A-Ram memory block. It is the combined destination cell and offset, which occupies bits 0-29. (Optional)
10. `TIME: 4-12 cycles;` The final TIME declaration states the minimum and maximum number of cycles the module takes to complete. If the module is too complex for these values to be reliably stated, then by convention the declaration `TIME: 0-0 cycles;` is made. (Obligatory).

Types defined in clauses 7,8, and 9 are rarely used. Declarations must always begin with NAME, and end with TIME, otherwise they may be in any order.

### 4.3 CODE.

The readability of Earth code is enhanced by replacing the numeric destination cells of Synchronic A-Ram instructions, referring to memory block addresses, with sugared notation. The naming of storage locations allows the destination cells of **cond**, **wrt0**, and **wrt1** instructions to be expressed by *destination names*, which consist of a storage entity name, together with an *index* if the storage entity occupies more than a bit. The *i*th bit of a storage location of type BYTE, for example, is referred to by *bytename.i* , where $0 \le i \le 7$. Thus we can have instructions such as **cond bitname**, or **wrt0 registername.31**. The Earth compiler calculates the numeric destination cell and offset of destination names, and substitutes them in the code where necessary. The naming of storage locations is of assistance in the goal of ensuring that Earth modules do not retain unwanted states between activations(see 4.6).

The destination cell of the **jump** instruction is named as a number relative to the program, called a *relative jump number*, rather than an absolute numeric memory block address. This frees the programmer from having to recalculate jump destination cells and resubstitute them in the code where necessary, when adding extra lines of code. Relative jump number are implemented by labelling a line of code with some positive integer, called a *linename*, similar to the line numbering system used in conventional assembly languages, and FORTRAN.

The mechanism for replicating code will be covered in 4.4. If the code has no replicator expressions, then there is no obligation to have linenames increasing , or increasing by equal increments, as one proceeds through the code. If there are replicator expressions however, then the programmer is obliged to have linenames increasing by single integer increments, in order to ensure that linenames with an replicator component are handled correctly by the current version of the compiler.

Code is terminated by the dummy instruction **endc**, and may be annotated by comments preceded by the double slash '//'. We are now ready to give our first trivial example in figure 4.1, which is a five bit incrementer with overflow.

Recall a Synchronic A-Ram run begins by activating the first two lines of code in the module. The code has a repetitive pattern, which can be compressed into an replicative structure, to be presented in 4.4. Alternatively, there is the possibility of modifying code during runtime, enabling the re-use of code segments. As will be seen in 4.3 and 4.5, the latter approach is often not worthwhile so close to the machine code level, because of the space and time overheads involved in rewriting so many bits in destination cells and offsets of particular instructions.

```
NAME: inceq5bit;
BITS: busy private, overflow output;
OFSTS: ioput ioput;
TIME: 4-12 cycles;

        wrt1 busy
        cond ioput.0            //test first bit
        jump 1 2               // '0' found, set bit and exit
        jump 2 1               // '1' found, reset bit, and test second bit etc..
1       wrt1 ioput.0
        wrt0 overflow
        jump 11 0
2       wrt0 ioput.0
        cond ioput.1
        jump 3 2
        jump 4 1
3       wrt1 ioput.1
        wrt0 overflow
        jump 11 0
4       wrt0 ioput.1
        cond ioput.2
        jump 5 2
        jump 6 1
5       wrt1 ioput.2
        jump 11 0
        wrt0 overflow
6       wrt0 ioput.2
        cond ioput.3
        jump 7 2
        jump 8 1
7       wrt1 ioput.3
        jump 11 0
        wrt0 overflow
8       wrt0 ioput.3
        cond ioput.4
        jump 9 2
        jump 10 2
9       wrt1 ioput.4
        jump 11 0
        wrt0 busy
10      wrt0 ioput.4
        wrt1 overflow                // ioput was initially 31
        jump 11 0
11      wrt0 busy
        endc
```

Figure 4.1 5 bit incrementer with overflow.

## 4.4 BRACKETED AND ABSOLUTE ADDRESSING.

We introduce *bracketed addressing* of relative jump numbers, in order to facilitate the writing of meta-modules, and other modules which rewrite their own code. Bracketed addressing allows another way of expressing destination cells for cond and wrt instructions, as relative jump numbers in square brackets, which the compiler translates into the final absolute memory block addresses. For example, the instructions **cond [3] 0** and **wrt0 [3] 0**, will test and reset respectively, the zeroth bit of that register holding the line of compiled code associated with the linename '3'.

Earth also has *absolute addressing*, which allows the destination operands of cond and wrt instructions, to be described as positive integers between 1 and 33,554,431, referring to the absolute numeric address of a memory block register. The destination operand of an Earth jump instruction, has no obvious need for absolute addressing, and is therefore always a relative jump number[32]. Absolute addressing increases the risk of side effects, and should be avoided (see 4.6). A module that inverts four bits and does not modify it's own code, can be written in 26 lines, requiring only 13 cycles to complete. In order to illustrate module rewriting and bracketed addressing, we present module in figure 4.2 called *Negate4bits*, which is less efficient. *Negate4bits* increments the offsets of a single cond triplet of instructions, which perform bit inversion, thereby allowing the same triplet to be reused for all four bits of the input. [33]

## 4.5 REPLICATIVE STRUCTURE.

In sequential languages, iteration involves the repetition of the execution of a code segment. In a spatial environment, iteration can be implemented by either the sequential or simultaneous execution of repeated segments of code. A textual mechanism that can replicate segments of code, in which certain values are modified in each replication, would be useful in a level zero language[34]. The treatment presented here is not entirely satisfactory, but does demonstrate that a powerful form of Earth code replication is viable. Code with nested replication is described by nesting curly brackets, which consequently exhibits a rudimentary tree syntax.

[32] The operands of absolute and bracketed addressing must be constant integers (see 4.4.1)

[33] The code relies on the knowledge that the compiler allocates the first byte declaration to the rightmost 8 bits of a register.

[34] In chapter 7, it will become apparent that code replication in Space is more powerful than loop unrolling.

```
NAME: negate4bits;
BITS: busy private;
BYTES: ioput ioput;
TIME: 26-26 cycles;

        wrt1 busy
        jump 1 1      // jump to main loop
1       jump 5 0

2       cond ioput.0          // inverter
3       wrt1 ioput.0
4       wrt0 ioput.0

5       jump 6 0
6       cond [2] 0            // increment offsets of instructions 2,3, and 4
        jump 7 3
        jump 8 3
7       wrt1 [2] 0
        wrt1 [3] 0
        wrt1 [4] 0
        jump 1 1              // repeat loop
8       wrt0 [2] 0
        wrt0 [3] 0
        wrt0 [4] 0
        cond [2] 1
        jump 9 3
        jump 10 3             // '11' found, so exit
9       wrt1 [2] 1
        wrt1 [3] 1
        wrt1 [4] 1
        jump 1 1              // repeat loop
10      wrt0 [2] 1
        wrt0 [3] 1
        wrt0 [4] 1
        jump 11 0
11      wrt0 busy

        endc
```

Figure 4.2 Self modifying 4 bit inverter.

The mechanism is called an *replicative structure*, which may be informally depicted as:

<*leftlimit;replicator;rightlimit*>{
*Earth construct 1*
*Earth construct 2*
.
.
}

A *replicator* is a single alphabetic character, which has a similar role to that of a control variable name for the C *for* structure. *Leftlimit* and *rightlimit* are non-negative integers (or arithmetic expressions with replicators as arguments, see 4.4.1), representing the initial and final values of the control variable, where *leftlimit* < *rightlimit*. The code inside the structure is a sequence of *Earth constructs*, where a construct is either an instruction (with optional linename), or another replicative structure. A succession of incremented[35] control values are generated by the compiler, which creates a succession of copies of the Earth construct sequence, where all instances of *replicator*, are replaced by the control value. The result of the compiler applying a replicative structure to code, is termed *replicated code*. We now consider a simple example. Figure 4.3(a) and figure 4.3(b), depict replicative and replicated code respectively for an 4-bit input AND gate, which tests bits serially.

```
NAME: seqand8;
BITS: busy private, output output;
BYTES: input input;   // leftmost 4 bits not used
TIME: 3-0 cycles;

       wrt1 busy
<0;i;3>{
       cond input.i
       jump 1 1
}
       jump 3 1
1      wrt0 output
       jump 2 0
2      wrt0 busy
3      wrt1 output
       jump 2 0
       endc
```

Figure 4.3(a) Replicative 4 bit AND gate.

```
       wrt1 busy
       cond input.0
       jump 1 1
       cond input.1
       jump 1 1
       cond input.2
       jump 1 1
       cond input.3
       jump 1 1
       jump 3 1
1      wrt0 output
       jump 2 0
2      wrt0 busy
3      wrt1 output
       jump 2 0
       endc
```

Figure 4.3(b) Replicated code for 4 bit AND gate.

[35] The arithmetic function which modifies the control value, is restricted to integer incrementation. Experience thus far suggests that it is sufficient for Earth modules.

### 4.5.1 THE NUMEX: A STANDARD FORMAT FOR REPLICATIVE ARITHMETIC EXPRESSIONS.

Replicating Earth requires a means of expressing leftlimits, rightlimits, linenames, relative jump numbers, and offsets, as arithmetic expressions with replicators as arguments. Rather than use a tree or interstring grammar to represent expressions equivalent to (*integer*+*replicator*) or (*integer*+(*integer***replicator*)), the current version of Earth uses a standard representation called a *numex*. The numex is used, because only a limited range of arithmetic expressions are needed, and because certain constant integer operands have a special role, and have to be identified for the correct replication of linenames.[36] There are 8 individual formats with obvious interpretations, which are described in the following non-inductive Backus-Naur definition:

*numex* ::= *replicator*| *integer* | (*integer***replicator*) | (*integer*+*replicator*)
| (*integer*+*integer***replicator*) | (*integer*+*replicator1*+*replicator2*)
| (*integer-replicator*) | (*replicator1*+*integer***replicator2*)

The expressions are evaluated during replication as follows:

1. *integer.* Evaluated as the *integer* itself.
2. *replicator.* Evaluated as the control value of the *replicator.*
3. (*integer***replicator*). Evaluated as the multiplication of *integer,* and of the control value of *replicator*
4. (*integer*+*replicator*) . Evaluated as the addition of *integer* and the control value of *replicator.*
5. (*integer1*+*integer2***replicator*). Evaluated as (*integer1* + (*integer2***replicator*) ).
6. (*integer*+*replicator1*+*replicator2*). Evaluated as (*integer* + *replicator1* +*replicator2* ).
7. (*integer-replicator*). Evaluated as the subtraction of the control value of *replicator*from *integer.*
8. (*replicator1*+*integer***replicator2*). Evaluated as (( *replicator1* + (*integer***replicator2*) ).

The first *integer* operand of types 5, 6 and 7, is called the *leading number*, and has a role in expressing linenames and relative jump numbers in nested replication. The operand assists the reader in understanding the relationship between linename/relative jump number numexes in different replicative expressions, within a module's code. *Replicative numexes* are those that includes an replicator(2-8).

[36] See 4.6 for a discussion of why the numex might not be the best approach.

### 4.5.2 REPLICATION FOR LINENAME NUMEXES.

The present compiler obliges a module with replicative linenames, to have leading numbers monotonically increasing by single integer increments, in order that replicated linenames are handled consistently. As will be seen in 4.3.3, the numex's format allows leading numbers to be related between structures.

If a structure has a non-replicative linename, i.e. one that is a constant integer, then that integer appears only once in the first cycle of replicated code, and no additional linenames are generated. But if a structure does have a replicative linename numex, then a succession of new line numbers will be generated. The issue of how this change affects other linenames in the code then arises. For example, the following structure generates 25 linenames ranging from 1 to 25, resulting in a net addition of 24 new linenames to the module.

```
<0;j;24>{
(1+j)   cond input.j
        jump (10+2*j) 0
        jump (11+2*j) 0
}
```

A structure's linenames' greatest leading number is called the *floor*, and in the above case is '1'. To implement replication, all of the module's linenames, and instructions which mention relative jump numbers, whose leading number is greater than the floor (iterated and non-iterated), have to be incremented by 24, otherwise the module's linenames become inconsistent. The increment is known as the *linename increment*, and under nromal circumstances, is also applied to leading numbers greater than the floor, of instruction destination operands with an replicative component, within the same replicative structure.

### 4.5.3 DASH MODIFIER FOR THE REPLICATIVE STRUCTURE.

There is often a need however, for the linename increment not to be applied to replicative jump numbers, within the *same* structure. In order to describe a succession of jumps, which implements a carry thread (see 3.3.1.3), or a wait for a code segment to complete before the next stage of a program, we might have:

```
1       jump 2 0
2       jump 3 0
3       jump 4 0
4       jump 5 0
.
.
32      jump 33 0
33      some instruction
```

To deal with this kind of example, a replicative structure has a modifier, called a *dash-modifier*. A dash or hyphen is placed before the first curly bracket, in order to be able to generate jump sequences as above, and other desirable pieces of code. The dash modifier instructs the compiler not to increment the leading number of instruction destination operands within the structure before replication. Thus the above jump sequence may be described as:

```
<0;k;31>-{
(1+k) jump (2+k) 0
}
2       some instruction
```

Note that the leading number in the destination operand of the structure's jump instruction, matches the linename of `some instruction`. We are now in a position to give less trivial module examples, that further illustrate Earth's ability to implement sequential digital circuits with circuit parallelism.

### 4.5.4 32-BIT INVERTER

The module in figure 4.4 inverts a register's bits in parallel. A sequence of 32 jumps is generated, to activate 32 cond triplets, one per bit of the input register.

```
NAME: bitwiseinverter32;
BITS: busy private;
REGS: input input, output output;
TIME: 4-4 cycles;

        wrt1 busy
        jump 1 1

1       jump 2 0
        jump 4 31


<0;k;2>-{           // dash modifier used to implement exit sequence
(2+k) jump (3+k) 0
}
3       wrt0 busy

<0;i;31>{
4       jump (5+i) 0
}

<0;i;31>{

(5+i)   cond input.i
        wrt1 output.i
        wrt0 output.i
}

        endc
```

Figure 4.4  32 bit parallel inverter.

### 4.5.5 32-INPUT OR GATE IMPLEMENTED SERIALLY.

It is not uncommon for serial implementations of very simple operations such as logic gates on the Synchronic A-Ram, to have favourable space and time complexity characteristics, compared with parallel implementations. The module `seqor32` in figure 4.5 compiles to 101 lines, and completes in 3-66 cycles, compared with 140 lines and 23-30 cycle completion times for `paror32`, described in 4.4.6. The sequential module will actually run faster if some bit is set within the rightmost ten bits of the input register. In contrast to sequential circuit parallelism, the benefits of Synchronic A-Ram parallelism are manifested, when the boolean function to be implemented is of a minimum size. There is little benefit to be had parallelising logic gates with inputs of 8 or less.

```
NAME: seqor32;
BITS: busy private, output output;
REGS: input input;
TIME: 3-66 cycles;

       wrt1 busy
<0;i;31>-{  // dash modifier used to describe OR operation.
(1+i) cond input.i
      jump (2+i) 0
      jump 4 1
}
2      jump 3 1
3      wrt0 output
4      jump 5 0
       wrt1 output
5      wrt0 busy
       endc
```

Figure 4.5 32 bit, serial OR gate.

### 4.5.6 32-INPUT AND GATE, WITH 4 PARALLEL 8-INPUT AND GATES .

The module *parand32* in figure 4.6, performs a 32-input AND operation on the contents of the input register, and compiles into 96 lines of code. A nested structure implements four 8-input AND gates, which are simultaneously activated to test the four bytes in *input*, whose results are written into the first four bits of *temp*. Meanwhile, a carry thread waits until for 8-input AND gate completion, and then tests the content of *temp*. Note that the numex format allows the leading number of the destination operand in instruction number 1, `jump (4+j) 0`, to be matched with the linename `(4+j)` in the structure describing the four, 8-input AND gates. A sequential version of a 32-bit AND gate, compiles to 69 lines of code, and completion times of 3-34 cycles.

### 4.5.7 LINENAMES, AND RELATIVE JUMP NUMBERS WITH TWO REPLICATORS

To write a 32-input OR gate with four 8-input OR gates active simultaneously, there is a need for linenames with two replicators. This is because we cannot duplicate the way the cond instruction with linename `(4+j)` in *parand32*, was able to select a second consequent through instruction positioning, without having to explicitly mention a linename[37]. The module *paror32* in figure 4.7 performs a 32-input OR operation on the contents of the input register, and compiles into 140 lines of code.

```
NAME: parand32;
BITS: busy private, output output;
BYTES: temp private;
REGS: input input;
TIME: 14-17 cycles;

       wrt1 busy
       jump 1 4      // jump to jumps to 8-input ANDs, and carry thread
<0;j;3>{             // generate jumps to activate the four 8-input AND gates
1      jump (4+j) 0
}
<0;k;8>-{ // dash modifier used to generate carry thread to wait for AND gates to finish.
(2+k) jump (3+k) 0          // carry thread
}
3 jump 5 0
<0;j;3>{      // nested replication generates 4, 8-input AND gates.

       <(8*j);i;(7+8*j)>{
(4+j)         cond input.i
              wrt0 temp.j
       }
       wrt1 temp.j
}
 <0;i;3>{                 // checks results of the 4, 8-input AND gates
5      cond temp.i
       jump 6 1
}
       jump 7 1
6      wrt0 output
7      jump 8 0
       wrt1 output
8      wrt0 busy
       endc
```

Figure 4.6  32 bit, byte parallel, AND gate.

A nested structure implements four 8-input OR gates, which are simultaneously activated to test the four bytes in *input*, whose results are written into the first four bits of *temp*. Program control is essentially the same as in *parand32*. Another module that requires linenames with two iterators, is the *barrelshift* module, which implements 32 bit register barrel shift in 8 cycles, and appears in Appendix A.

[37] If the cond instruction were redefined to select the second consequent if the bit to be tested were found to be reset, then paror32 would be the easier module to define, and parand32 the harder.

Incrementation seems necessary for some examples with linenames having leading numbers, which are less than the outermost structure's floor. There is a need for linename incrementation within inner replication structures, even if they are dashed. The issue of how to handle linename incrementation for multi-replicator linenames was not resolved in time for the current version of Spatiale, but a solution has been arrived at, and will appear in the next implementation. The present compiler can cope with the examples in this report.

```
NAME: paror32;
BITS: busy private, output output;
BYTES: temp private;
REGS: input input;
TIME: 22-29 cycles;
       wrt1 busy
       jump 1 4

<0;j;3>{
1      jump (8+9*j) 0  // there are 9 relative jump numbers included in each or8 block
}

<0;k;16>-{
(2+k) jump (3+k) 0
}

3 jump 4 0

<0;i;3>-{  // dashed replication used to implement 4 input OR gate.
(4+i) cond temp.i
      jump (5+i) 0
      jump 7 1
}
5      jump 6 1
6      wrt0 output
7      jump 8 0      // to avoid halt fail
       wrt1 output
8      wrt0 busy
<0;j;3>{
       <0;i;7>-{
(8+j+i)        cond input.(i+8*j)
               jump (9+j+i) 0
               wrt1 temp.j
       }
(9+j)          wrt0 temp.j
}
       endc
```

Figure 4.7 32-input, byte parallel OR GATE.

The lack of interstrings to describe nested replication also appears to have complicated matters. The way in which multi-replicator linename incrementation takes place was difficult to understand, partly because of the recursive nature of the compiler's procedures, required to process the recursive data structures that were devised to represent nested replication. A non-recursive interstring representation would have simplified matters, facilitating a non-recursive implementation of compiler functions which de-iterate code.[38] Another complicating factor was the restricted arithmetic range of the numex.

In the meantime, Earth is restricted to having only one such replicative structure with two replicator linenames, appearing at the end of a module. Two replicator linenames (at least), will be fully treated with in the next version of Spatiale.

It is worth noting that the destination numex of the jump instruction in line 1 has to be incremented by 9 in each cycle of the replication, in order to match the total number of new linenames, that are generated in each outermost cycle of the terminating replication structure. No immediate way has been found to automate this linkage, which does not impose on the programmer that much.

### 4.6 META-MODULES FOR PROGRAMMABLE COPY.

We encountered the `negate4bits` module in fig 4.2, which modified it's own code. We now consider meta modules, which can modify a segment of their own code, and can then be separately instructed to execute that modified segment, by having a choice of two initial markings. Meta-modules have roles in re-using code segments, and implementing high level programming features (see 6.12). They can also transform the code of a module, so that it is oriented to the faster processing of an instance of a module's input.

A meta module has two phases. The first phase performs a compiler-like task of modifying code, and has the initial marking $\{1,2\}$. There is something of an analogy between the first phase, and the re-configuration of a Field Programmable Gate Array, in order to process certain program types more efficiently. The second phase of a meta-module, executes the modified code, and has the initial marking $\{3,4\}$. A meta module requires a second busy bit for the second phase, which is always called *mbsy*. The compiler will complain if it is missing, and a meta declaration has been made.

[38] No such problems were encountered when implementing a more complex form of replication for Space, using an interstring to describe replications of interstrings.

The `META: 2;` declaration states that the module is a meta module, and that the initial marking of the second role is composed of the instruction with the linename ‘2’, and the succeeding instruction. By convention, the second phase’s initial marked instructions are always the third and fourth code lines in the program, where the third line is labelled with the linename ‘2’.

An important application for meta modules, is the implementation of code which can transfer the contents of arbitrary memory block bit locations, into other arbitrary bit locations. If we wish to write a code segment which copies the contents of $(x_1, y_1)$, into $(x_2, y_2)$, which are known before run time, then the following triplet would obviously suffice:

cond $x_1$ $y_1$
wrt0 $x_2$ $y_2$
wrt1 $x_2$ $y_2$

If however, the bit addresses are not determined before runtime, then a module would have to receive them as inputs, and copy them into the relevant destination and offset operands of the triplet’s instructions, so that the actual bit copying could take place. We will call such an operation *programmable copy*.[39] The maximally parallel implementation of a programmable copy for a single bit which appears in figure 4.8, involves simultaneously writing to all 90 bit- address bits in the cond triplet. The module is somewhat costly, and compiles into 379 lines of code. A maximally parallel implementation of programmable copy for all 32 bits in a register appears in appendix A, and compiles into 5,249 lines of code!

In order for Space to apply operations to arrays of values, there is an extensive need for programmable copy operations, because the source and target locations involving an array value in an replicative context, vary at run time. The fact that such modules are costly might not bode well for a real implementation. In chapter 8, we discuss how the Synchronic Engine design is predicated on the technical availability of very large numbers of fast re-configurable interconnects. Each reconfigurable interconnect would, in effect, be a programmable register copy. Providing the non-trivial technical issue is resolved, the cost of implementing programmable copy here, would not be transferred to Synchronic Engines.

[39] There is no simple analogue in sequential digital circuits for the programmable copy. An ability to copy the contents of an flip flop located arbitrarily within a circuit, into another arbitrarily located flip flop, would require a fully interconnected network spanning the entire area of the circuit.

The `progcopybit` module accepts a bit address as a 30 bit value, rather than separating the bit address into destination operand and offset. It is the instructions with linenames 3,4, and 6 which perform the second phase's bit copying, whose bit addresses are rewritten by the first phase of the module. A meta module called *progcopyreg*, which can programmable copy the contents of a whole register into another, appears in appendix A.

```
NAME: progcopybit;
META: 2;
BITS: busy private, mbsy private;
BITAS: source input, target input; // inputs refer to nonmeta function only.
TIME: 4-7 cycles;                  // the meta function requires no inputs or outputs

1       wrt1 busy                  // first phase: lines 3,4, and 6 are modified
        jump 7 2
2       wrt1 mbsy
3       cond 0 0                   // second phase, perform bit copy
        jump 4 1
        jump 6 1
4       wrt0 0 0
        jump 5 0
5       wrt0 mbsy
6       wrt1 0 0
        jump 5 0

7       jump 14 0 // exit sequence
        jump 8 29
        jump 10 29

<0;i;29>{
8       jump (9+i) 0
}


<0;i;29>{
(9+i)   cond source.i
        wrt0 [3] i
        wrt1 [3] i
}

<0;i;29>{
10      jump (11+3*i) 0
}

<0;i;29>{
(11+3*i)   cond target.i
           jump (12+3*i) 1
           jump (13+3*i) 1
(12+3*i)   wrt0 [4] i
           wrt0 [6] i
(13+3*i)   wrt1 [4] i
           wrt1 [6] i

}

<0;i;3>-{
(14+i)  jump (15+i) 0
}
15      wrt0 busy

        endc
```

Figure 4.8 Programmable bit copy.

## 4.7 MODULE STATES, AND REFERENTIAL TRANSPARENCY.

Earth has features that can modify declared and undeclared storage, which can result in a module retaining a state between activations. This has the potential to generate side effects, and degrade referential transparency of Space modules employing Earth sub-modules.

There is one context in which Earth modules are intended to retain states between activations, where side effect are unlikely. The first phase of a meta-module imbues it with a state, which is retained in a run of the Synchronic A-Ram, until the first phase is activated again. But this is a specific exception, and the programmer is always notified by the environment, if he has selected a meta-module for inclusion in a Space module. In general, in order to improve the referential transparency of Space programs, it is the responsibility of the programmer to ensure that Earth modules do not retain states between activations. There are at least four possible sources of side effects:

i. Private Storage. Storage whose interface category has been declared to be private, will naturally not be overwritten when a Space module is writing inputs to an Earth module, which has been previously activated. A methodology whereby the halting of an Earth module is preceded by a resetting of all bits declared to be private, will eliminate this kind of state retention. A future enhancement can easily automate this solution.

ii. Failure to initialise all of the inputs to Earth modules, within Space modules. This issue pertains to Space programming methodology.

iii. Non-meta modules with bracketed addressing. The normal way in which the contents of registers are updated, requires a reference to declared storage. The `negate4bits` module in 4.3, and the `adder32` module in appendix A.1, are examples of non-meta modules that rewrite their own code using bracketed addressing, in order to reuse code segments. The rewritten code constitutes a form of non-declared storage, where bracketed addressing can result in the module retaining a state, such as the first two bits in lines 2,3, and 4 of `negate4bits`. It is essential in these cases that the Earth programmer ensures all non-declared storage is reset before module termination. A future enhancement can automate this solution, because the affected bits of bracketed lines are explicitly referenced .

iv. Absolute addressing. In addition to the above, an alternative way of updating registers without a reference to declared storage involves absolute addressing. There is a case for removing this facility altogether from Earth.

v. Input/output crossover. It is conceivable that the Earth programmer declares storage to be an output, when it is in fact treated by the module as input or ioput. Consequently the module will appear to the Space programmer to be subject to side effects, in the sense that identical so-called 'inputs' will generate differing outputs. This kind of error involves an obvious programming mistake, and in any case the compiler could be enhanced to read code, to ensure output storage is not read by cond instructions.

## 4.8 FURTHER ENHANCEMENTS.

The decision to employ numexes to describe arithmetic expressions with replicator arguments, was motivated partly by a desire to avoid tree grammar, and partly to exploit the concept of a leading number embedded within a numex, which could assist the programmer in relating links between replicative structures. The downside is that the programmer is restricted to the narrow range of arithmetic functions associated with the numex.

Replicator arithmetic expressions tend to represent small, shallow dataflows with little subexpression repetition. Rather than using numexes, or interstrings and their accompanying machinery, it would have been better to use a conventional tree grammar (i.e. conventional bracketed arithmetic expressions) to represent arithmetic functions involving replicators. In this particular context, tree syntax would have been more flexible and succinct, and it's limitations have little significance. The issue of having a leading number could have been resolved by restricting a bracketed expression to having only one constant integer operand within the outermost bracketed expression, for convenience located in the leftmost position after the first bracket, to be identified as the leading number.

Consequently the next compiler version will dispense with the numex, and will incorporate a recursive descent parser to process small, tree based, replicative arithmetic expressions. Allowing a tree expression to describe the incrementor function of structures, is an acceptable incursion of tree syntax, into the base expressions of an interstring describing the replication of Earth code. Dataflows are shallow, and hence tree structural variability is limited. As part of a boot strapping process, an Earth compiler written in Space would result in being able to avoid tree grammar in code, by allowing primitive incrementor functions to be defined in Earth without the use of replication, enabling them to be subsequently used as incrementor functions for replicative structures.

Given that Earth data structures are never larger than a register, there is a limit to how much nesting of replicative structures is possible, or useful. But the ability of the compiler to handle linename incrementation for multi-replicator linenames in nested structures, needs to be improved. The improvement will be assisted by employing interstrings to describe nested replication, and by replacing numexes with tree expressions.

# Chapter 5
# SPACE DECLARATIONS.

## 5.1 INTRODUCTION TO SPACE.

The Synchronic A-Ram is a finite, fully programmable and highly concurrent model of computation. It is claimed that it can support any finite model with acceptable complexity overheads, whilst respecting the model's implicit or explicit parallelism, including procedural, object oriented, declarative, non-deterministic message passing, or shared memory programming approaches[40]. The Space interlanguage was designed to take advantage of the interstring's ability to efficiently represent and execute dataflows, and of the Synchronic A-Ram's simple broadcast style of communication. Modules are composed in a synchronous, deterministic fashion, exhibiting a small state transition system at the highest level of abstraction. Space is function oriented and strictly typed, and has nested iterative and replicative structures, as well as a basis for supporting performance evaluation.

In chapter 9, it is argued that the Turing Machine and λ-calculus have inferior complexity characteristics to the Synchronic A-Ram, with respect to the implementation of addressable memory and the parallel composability of program modules. These features and the interstring's lack of the Single Parent Restriction, permit a Synchronic A-Ram simulation on conventional computers of very large parallel programs within reasonable timeframes. Why bother constructing a compiler for Space, with a mathematical model of computation as the target? Firstly, an extremely simple semantics is made available for a high level parallel language. Secondly, if a compiler were written for a derived or coarser grained machine, then various technical solutions might be dependent on the idiosyncrasies of that machine, and insights into fundamental issues in deterministic parallelism might be skewed, or missed entirely. The compilation and execution of Space programs, serve to inform the design of high performance architectures derived from the Synchronic A-Ram. The Space interlanguage is also intended to serve as a prototype for languages for high performance architectures.

[40] The state transformation function of the A-Ram is capable of expressing machine models with a variety of instructions sets, and broadcast and non-broadcast communication styles, that could fulfill the same role.

No attempt has yet been made to implement abstract data types and higher order functions in Space, before less advanced language features are shown to be working reliably. At this stage of the compiler boot strapping process, a high degree of referential transparency in program modules can be attained, but cannot be guaranteed. Later versions of Space will be able to support advanced language features, and include support for guaranteeing referential transparency. Subject to room being available in the memory block, and a future enhancement relating to memory allocation, Space can describe any high level, massively parallel algorithm, and therefore constitutes a general purpose programming model.

In chapter 8, a range of possible hardware specifications based on the Synchronic A-Ram concept are discussed, called Synchronic Engines. Space would require little adaptation, to run on Synchronic Engines. The runtime environment associated with Space programs, does not involve the dynamic allocation of processing entities. Storage and processing resources are integrated at the lowest machine level, and sub-programs are always linked with machine resources. Consequently, all scheduling and resource allocation of processing entities can be finalised at program composition and compilation time, rather than at run-time.

With a programming methodology that eschews critical regions and contention for resources, Space is not in general subject to the bugs associated with non-deterministic systems, such as deadlocks, livelocks, and race conditions.

## 5.2 Resource allocation and the runtime environment.

In common with an Earth program module, a Space module is described in the spatial style as a hardware functional unit, and has names for the memory locations that hold the module's inputs, outputs, and internal storage. The module has declarations, one relating to typed storage entities, and another to pre-defined library modules, called *submodules*. With some qualifications[41], a submodule does not retain a state between calls, or deliver differing outputs to identical inputs, making Space for the most part referentially transparent. In the role of a submodule, a module represents a category or class of processes, and has instances. A submodule cannot simply be called as an abstract software entity divorced from hardware. A submodule instance must be called, whose label represents a link to machine resources[42]. In keeping with the spatially oriented approach, an act of computation is always associated with some machine resource. Resource allocation in a spatial environment may be accomplished implicitly, whilst ascending layers of abstraction.

If a maximum of only one call to a submodule class is ever made at any stage in a module's execution, then only one instance of that class needs to be declared. But to describe SPMD parallelism for example, multiple instances of a class have to be declared, usually in the form of a labelled array. Each submodule instance will occupy it's own region in the stretch of A-Ram memory, that will hold the module's compiled code.

Although the Space programmer is obliged to consider explicit scheduling of submodules when composing a module, he is shielded from these and allocation issues pertaining to all sub sub-modules, because they have been automated by earlier composition, leaving the compiler to perform these tasks before run-time. The fact that the programmer only has to consider parallelism within the narrow confines of a single module, makes potential contention issues so obvious, that they are easy to avoid. In any case, the programming methodology described in chapters 6 and 7, makes it straightforward to avoid contention for submodules. Good modularisation, combined with an appropriate methodology, considerably facilitate the construction of safe and reliable parallel code.

---

[41] Apart from meta-modules and memory allocation modules, Space and Earth submodules are not intended to retain states between activations (see 4.6 ). Ensuring modules do not retain states appears to be straightforward, providing the programmer is alerted to the most obvious sources of side effects. At the current stage of compiler bootstrapping however, it is the programmer's responsibility to ensure that this is the case.

[42] A similar approach is taken in VLSI schematic capture, and in hardware description languages like VHDL.

The Synchronic A-Ram's full connectivity, gives rise to a benefit for the parallel programmer. In Space, multiple acts of copying storage contents from one location to another may be called for, without the programmer having to consider explicit allocation of communication channels, because they are always available locally.

The static allocation of processing entities, entails that overheads imposed by run-time scheduling and allocation are avoided. This is achieved without cost, because there is no distinction in the model between processing and storage resources. However, in an architecture that involves a memory hierarchy with processing and non-processing layers, there is the prospect that unnecessarily low utilization of processing resources would arise by ignoring dynamic allocation.

A segment of the memory block, called the *storage block*, is set aside from code to exclusively store data, and in the current implementation occupies the last quarter of memory block (containing 8,388,608 registers), beginning at register 25165824. The purpose of the storage block, is partly to store a library of instances of typed data entities for the convenience of the user, and partly to function as a heap for any runtime requirements for more storage. Register 25165824 is set aside to store the next free register in the storage block. The implementation of parallel memory allocation has to wait for a future stage in the process of compiler boot strapping. Memory allocation modules can be written in Space (see 6.13.3), but simultaneous activation of more than one memory allocation sub-module in the current environment would result in a write fail, and other errors during runtime.

The next version of the programming environment will add an extension to the compiler, which will build an accompanying piece of code called a *sidecar*, to run in parallel with the module. The sidecar will handle all memory allocation and garbage collection for the module and sub-modules, and will be written in Space. The sidecar is non-trivial, and would have to dynamically and without conflict, send memory addresses from the heap quickly, to (potentially very many) memory allocator sub-modules simultaneously, and collect garbage from many sources simultaneously. The sidecar will be a good application for demonstrating the advantages of simultaneous resource allocation to multiple recipients in a synchronous deterministic environment, over asynchronous non-deterministic approaches.

Some inductively defined data structures are allowed, but recursive modules are disallowed in Space, because of the space and time overheads that recursion imposes on Synchronic A-ram processes, and on parallel systems in general. The lack of procedural recursion[43], and the fact that Space sub-routines have pre-assigned machine resources, results in there being no requirement for a stack.

## 5.3 General form of a space module.

To add a new Space module to the library, the user must insert the module's text into a text file called *program.dat*, which is stored in the Spatiale program directory. The file begins with declarations, some of which state storage entities, which produce data structures used in code generation, and for checking the syntactic and the semantic consistency of the module's code. The module's text terminates with code that is comprised of a series of numbered interstrings, that can be attached to replicative, and other types of constructs.

A storage entity of a module of type level *n*, has an interface category, which can indicate whether it is private to the module, and cannot be accessed by a module of level *n+1*, or whether the entity is input or output, and is therefore public, and therefore accessible to a module of level *n+1*.

The naming of the module is succeeded by the obligatory declarations relating to storage and sub-modules, followed by optional declarations, ending with an obligatory declaration referring to cycle completion times for the module. The current implementation of Spatiale requires the order of declarations to be adhered to. The general modular form is displayed in figure 5.1.

The rest of this chapter will present the format of a module's declarations. The following two chapters will describe code.

[43] It is argued in a future paper, that lack of explicit recursion in no way places a barrier to extending Space's functionality, to include an ability to support higher order functions, a novel form of parallel automated theorem proving, and receptivity to formal methods.

```
module myspacemodule{          // module name.

        storage{        //declarations of typed storage entities, obligatory

                        ........

        };


        submodules{    // declarations of predefined modules, obligatory.

                        ........

        };


        contractions{   // declarations of sub-modular contractions, optional.

                        ........

        };


        replications{...};//declaration of replicators and incrementor functions, optional.

        meta{...};     // declaration of meta status, optional.

        metatime: ...// metamodule completion times, obligatory if meta declaration made.

        time: ...      // module completion times, obligatory.


        code{           //  interstring code, obligatory.

                        ........

        };

};
```

Figure 5.1 General form of a Space module.

## 5.4 TYPE SYSTEM AND STORAGE DECLARATIONS.

The Space type system is strict, in the sense that with a few exceptions discussed in 6.6.6, the contents of a storage entity, or of a storage entity associated with a sub-module, can only be copied into the contents of another entity of the same type[44]. Recall that Earth's type set is fixed, and includes entities such as BIT, BYTE, WORD, and REGISTER, all having type level zero. Earth's types are pre-loaded into Space's type library, along with some pre-defined level one types defined in 5.4.2. The Space type set is extensible, where a new type may be derived by forming a construct whose members are pre-defined, existing types. A type's level is the maximum of it's member type levels, plus one.

Similar to the *structure* data entity in C, the current implementation's internal organisation of a Space type may be represented as a tree[45], where the nodes are lower level types, and leaves are level zero types. In the current implementation, up to and including level 50 complex types, may be defined. In 5.4.2, it is explained how Spatiale has a sub-menu in which the user may define new types in a text file, and add them to the type library. Type definitions are presently disallowed in module definitions. The general form of an individual type declaration appears in figure 5.2, followed by an example:

```
newtypename{
        existingtypename1 label1aggregate_construct;
        existingtypename2 label2aggregate_construct;
        .
        .
}

newtype{
        REG register;               // singleton of type REG.
        BYTE bytearray[16];         // array of 16 BYTES.
        REG twodimensionarray[10][2]; // up to three array dimensions are allowed
        WORD wordblockstring<32>;    // string of array blocks of 32 WORDS.
        REG pointer*;               // pointer to a register.
}
```

Figure 5.2 Standard type declaration and example.

[44] When abstract data types are supported in a future version of Space, restrictions will apply to ensure abstract data type compatibility.

[45] Future enhancements will employ an interstring to represent the sub-type hierarchy, which will allow sharing of sub-types between the immediate component types involved a new type composition.

For the time being, the editing of a type is not allowed, because other modules may rely on the type definition, and machinery is not in place to deal with possible conflicting type and module definitions. In the example, *newtype* therefore has to be new to the library.

As with a C structure, the type definition has a series of members. Each member declares an existing type, followed by a label name for that type's instantiation, which must be unique to the definition, followed by the *aggregate construct* associated with the member. There are four kinds of aggregate constructs, which indicate whether the member describes a pointer to, or a singleton of the existing typed data entity, or an array, or a string of arrays of identically typed data entities.

### 5.4.1 AGGREGATE CONSTRUCTS.

Each pre-defined type declared within a module, occupies a fixed number of registers in the module's compiled code, call it $r$. Spatiale calculates the space requirement for a new type definition, by summing the amounts of registers that each member of the definition requires, based on it's type and aggregate construct. The four kinds of constructs are:

i. Singleton. A singleton requires $r$ registers.
ii. Pointer. A pointer only needs a single register, to store the 25 bit address of the first register of a type instance. During runtime a pointer either refers to a storage entity residing in the storage block, or declared as modular storage.
iii. Arrays. Arrays may have a maximum of three dimensions. An array grid composed of $n$ elements of the type, requires ( $r\text{x}n + 4$) registers. An extra four registers are required to store information needed for accessing an array element, whose index is not known until runtime.
iv. Blockstrings. A blockstring is not a string or list in the conventional sense. If a list-like data structure is organised so that an element is only accessible via it's predecessor, then the simultaneous processing of list elements is not possible. Therefore Space supports a hybrid data structure consisting of a string of fixed length blocks, each holding a one dimensional array. A data structure may therefore be indefinitely large by using a chain of blocks linked by pointers, but also permit simultaneous access to at least a portion of it's elements. A string with array block length $n$, requires ( $r\text{x}n + 4$) registers in the storage area of the compiled module's code. Subsequent array blocks reside in the heap, entailing that a module declared with blockstring storage entities, compiles into a constant number of registers. The extra four registers state information needed for accessing elements at runtime, and indicate how full the current block is. Blockstrings are discussed further in 6.1.

### 5.4.2. ADDING A NEW TYPE TO THE LIBRARY.

Spatiale has a sub-menu in which the user may define new types in a text file called *typedef.dat*, which is stored in the Spatiale program directory, and add them to the type library. The definitions are enclosed within the *begintypes* and *endtypes* statements. Space has pre-defined, self-explanatory, level one types which are pre-loaded into the type library. Their type definition appears in figure 5.3.

```
begintypes

        unsigned{
                REG register;
        };

        int{
                REG register;
        };

        char{
                BYTE byte;
        };

        float{
                 REG register;
        };

endtypes
```

Figure 5.3 Example of a typedef.dat file.

The single members of *unsigned*, *int* and *float* are all of type REG, as would be expected. Their role is defined by the modules they are mentioned in, e.g. different modules will be required to multiply two 32-bit unsigned numbers, integers in 2's complement representation, and floating point numbers with mantissa and exponent.

### 5.4.3 IEEE STANDARD FOR BINARY FLOATING-POINT ARITHMETIC. (ANSI/IEEE STD 754-1985)

The Spatiale library will eventually be extended to cover the standard arithmetic logic operations, including arithmetic operations on floating point number types. To suggest that Turing Machine or λ-calculus code to implement floating point arithmetic, be made IEEE 754 compliant might seem inapposite, not least given the difficulty of writing and simulating non-trivial programs for the standard models. The register oriented nature, and complexity characteristics of the Earth and Space languages however, present a realistic option that modules implementing floating point arithmetic operations, can be made to be IEEE 754 compliant.

The need to implement the standard might seem less than pressing, given that the target machine is a simulation of a mathematical model of computation, rather than actual hardware. But IEEE 754 is there to standardise the specification, verification, readability and interoperability of functional unit designs. There is good reason to implement the industry standard into $\langle 5,\sigma,\{1,2\},\eta \rangle$ floating point modules, given that it is straightforward to do so, and that the implementation will aid comparison with hardware design

### 5.4.4. STORAGE DECLARATIONS.

A module's storage declaration is obligatory. It is similar to a type definition, but a storage entity member in the declaration has an extra reference to it's interface category. The set of Space interface categories is identical to Earth's: *input, output, ioput*, and *private*. Only the last category indicates that the storage entity cannot be accessed by a higher level module. Interface categories are used by Spatiale to check that a Space module does not attempt to access private storage in sub-modules, and also to prompt the user for module inputs, and print module outputs when running a module.

Halting the module involves the special instruction "HALT", which makes no explicit reference to a busy bit. When processing the storage declaration, the compiler automatically adds a private busy bit, which is mentioned in the compiler's output, and should be ignored by the programmer.

```
storage{
        existingtypename1 label1aggregate_construct interface_category;
        existingtypename2 label2aggregate_construct interface_category;
        .
        .
}:

storage{
        unsigned input input;  // labels may be the same as an interface category
        int integerarray[16] output;
        REG largetwodimensionarray[1024][32] ioput;
        char charblockstring<16> private;
        float floatpointer* input;
};
```

Figure 5.4 Standard modular storage declaration and example.

## 5.5 SUB-MODULES DECLARATION.

The sub-modules declaration is obligatory. Sub-modules provide the essential ingredient for building a high level program; established subroutines that perform tasks, that have been automated by earlier program composition. Space exhibits strong modularity, similar to that found in OOPLs. The programmer may view pre-defined modules as black boxes, which may be edited or replaced without affecting the semantics of higher level modules in which they participate, providing the following conditions are met:

i. The storage declarations of the non-private, or "interface" entities are not changed. Otherwise higher level modules will refer to storage entity names which may no longer exist, or have been modified.
ii. The new or edited module preserves the original module's output behaviour.

Similar to a type definition, the declaration lists a series of member declarations of submodules, or arrays of submodules. A member declaration has a module class name, and a label name with aggregate construct, where the label represents a link to machine resources. A label name can assist the programmer in remembering any special role for the submodule(s). If a class name is lengthy and un-intuitive, then labelling provides an opportunity to rename the submodule, in order to make interstring code easier to read. Figure 5.5 exhibits the standard submodular declaration, followed by an example. Sub-modular aggregate constructs are restricted to singletons, and arrays with up to three dimensions.

Labels facilitate software maintenance. If a submodule declaration is edited so that a sub-module class is substituted for an alternative, more efficient class with identical interface names, whilst retaining the same label, then no further editing of the module's code is required. Labels have to be unique to the sub-modules declaration. Each labelled instance of a sub-module, or element of a sub-modular array, will occupy it's own region in the stretch of A-Ram memory, that will hold the module's compiled code.

```
submodules{
            module_classname1 label1aggregate_construct;
            module_classname2 label2aggregate_construct;
            .
            .
}:

submodules{

      adder32 adderarray[10];  // array of 10 32-bit adders called "adderarray"
      inceq5bit inceq5bit;  // submodule labels may be the same as module name
      parand32 twodimensionalandarray[3][5];
      parand32 onedimensionalandarray[32];

};
```

Figure 5.5 Standard submodules declaration, and example with level 0 modules from chapter 4.

The same class may appear in different declarations. It is often the case that a submodular array is required for one purpose, whilst an additional submodular array of the same class with a different label is needed for another, where the activations of both arrays are simultaneous, or overlap.[46]

## 5.6 SUBMODULE MAPS.

The submodules declared in the previous declaration are called *immediate* submodules. The declaration determines a tree-like hierarchy with immediate sub-modules at the top, submodules of immediate sub-modules in the next layer, etc.. The hierarchy has nodes which consist of a sub-module's identity, it's instance, and information relating to the sub-module's own instances of sub sub-modules. In order to build machine code, the present compiler generates a data structure called a *submodule map*, that represents the hierarchy. A tree based data structure is not used to represent a submodule map, for the following reasons:

i. The internal content of a tree can only be accessed by following the chain of links between nodes, which impedes the speed and simultaneity with which the content can processed.

ii. A tree will not be able to represent a possible sharing of the immediate submodules' submodules, which the module does not need to activate simultaneously. Sharing leads to a significant contraction of memory block space needed for sub-modules' machine code. The sharing is currently specified by the programmer, in the *contractions declaration*. In a future enhancement the compiler will be able to determine before runtime, which sub-modular resources may be shared.

[46] In cases where they do not overlap, the programmer has the opportunity to reuse the same array for both roles, reducing the scale of resources that the module needs to call upon.

It was mentioned in 4.1, that Spatiale assigns a numerical index called an *index*, to a library module. In the processing of module definition, the compiler also assigns a number *instanceindex* to an instance of a submodule. The submodule map is an interstring-like data structure, whose contents may be accessed in parallel, and can represent the sharing of sub-modular resources. A node in the map's interstring base language is an integer pair (index, instanceindex) representing the submodule's library index and instance, followed by a list of (index, instanceindex) pairs representing the node's immediate submodules called the *argument list*.

The leftmost column of a submodule map lists level zero modules, which have no immediate submodules, and hence an empty argument list. The rightmost column of a submodule map contains only one node, which contains the new index assigned to a new module by the compiler, together with it's argument list of sub-modules. At this stage, the interstring represents a tree of dependencies, because no sub-submodule sharing has yet been introduced.

```
(1,1)   :: (2,1) ( ( 1,1), (1,2) ) :: (7,1) ( (2,1), (6,1) ) :: (9,1) ((7,1), (8,1) ) :;
(1,2)      (6,1) ( ( 3,1), (4,1) )    (8,1) ( (4,3), (6,2) )
(3,1)      (6,2) ( ( 3,2), (4,2) )
(3,2)
(4,1)
(4,2)
(4,3)
```

Figure 5.6 Example of a submodule map.

## 5.6 DECLARATION OF SUBMODULE CONTRACTIONS.

The scheduling of immediate sub-modules may allow no sharing, and hence this declaration is optional. In the above example, the new level 3 module has been given an index of 9. The submodule with index 2, requires two instances of submodule index 1, each instance of sub-module index 6, needs one submodule index 3, and one submodule index 4. The new module's two immediate submodules (7,1) and (8,1), both have a module index 6 in their argument list.

If the programmer can judge from the module's interstring code, that the activation periods of submodules (7,1) and (8,1) never overlap, then there is an opportunity for the module's compiled code to share the submodules of (7,1) and (8,1). The module map resulting from the submodule sharing of (7,1) and (8,1) is displayed in figure 5.7.

```
(1,1)   :: (2,1) ( ( 1,1), (1,2) ) :: (7,1) ( (2,1), (6,1) ) :: (9,1) ((7,1), (8,1) ) :;
(1,2)      (6,1) ( ( 3,1), (4,1) )    (8,1) ( (4,2), (6,1) )
(3,1)
(4,1)
(4,2)
```

Figure 5.6 Example of a contracted submodule map.

The sharing is specified in the contractions declaration, which consist of a series of statements between curly brackets:

```
contractions{
            contractionstatement1;
            contractionstatement2;
            ...
};
```

A contraction statement is either *solitary*, or *ranged*. A solitary contraction describes a single sharing of two particular submodule instances, separated by '~'. In the example in fig 5.6, the modules with indices 7 and 8 are obviously not members of arrays, and their contraction statement would take the form:

```
submodulelabel_with_index7 ~ submodulelabel_with_index8;
```

For the contraction of submodules in arrays, the statement would take the form:

```
submodulelabel1[array_identifier1] ~ submodulelabel2[array_identifier2];
```

To share multiple submodule pairs in two arrays, a ranged contraction statement is used to specify the the upper and lower bounds of the array indices of the two arrays, separated by a hyphen:

```
submodulelabelone[array_identifier1-array_identifier2] ~
                    submodulelabelone[array_identifier3-array_identifier4] ;
```

Note that the statement normally has to be on one line. The following expression pair wise shares the first 16 elements of the `submodulelabelone` array, with sixteen elements in `submodulelabeltwo` array.

```
submodulelabelone[0-15] ~ submodulelabeltwo[8-23];
```

To pair wise share sub sub-modules, the width of the two ranges has to be identical. For arrays of more than one dimension, the programmer must insert the ranges into the appropriate array dimensions.[47]

## 5.8 Declaration of replicators and incremental functions.

A module may have no replicative structures, and accordingly the replicators declaration is optional. Two types of replicative structures in Space are presented in chapter 7. The declaration lists the alphabetical strings used to represent replicator names in Space's replicative structures. Separated by a slash character, the declaration also lists arithmetic functions, which fulfill a similar role to that of the numex in the Earth language. They provide a means of representing array identifiers, leftlimits, rightlimits, and immediate values, as arithmetic expressions with replicators as arguments. From the compiler's point of view, the listing of replicator names and functions is not necessary at this stage, but the declaration serves to limit and regularise the usage of terms in the module's code.

[47] The current implementation restricts ranging to the rightmost array dimension, and will be extended to other dimensions in the next implementation.

Rather than use a tree or interstring grammar to represent replicative arithmetic expressions, the current version of Space uses a standard representation called an *incremental function*. Unlike numexes, incremental functions are one input operand functions, drawn from a fixed set of eight functions, represented by a short alphanumeric string.

This approach was originally taken up to avoid the use of a tree grammar, to avoid the charge that tree expressions are necessary to an interlanguage. But it suffers from the same problem affecting the numex; only a limited range of arithmetic expressions are available. As with Earth, it would have been an acceptable limited use, and more flexible, to employ tree-based bracketed expressions, to represent replicative functions. Consequently the next version of Space will dispense with incremental functions, which will be replaced by conventional arithmetic expressions. For the time being, the programmer is restricted to eight functions on a single replicative unsigned integer argument *i*, which permit an extensive programming range. An *incremental expression* takes the form `replicator/incremental_function_name`. Let the control value of the replicator be *i*. The eight types of expression are exemplified below:

i. "`replicator/id`". Identify function, returns the control value *i*.
ii. "`replicator/inc`". Incrementor function, returns $(i+1)$.
iii. "`replicator/plus2`". Add two function, returns $(i+2)$.
iv. "`replicator/dec`". Decrementor function, returns $(i-1)$. Compiler flags an error for zero input.
v. "`replicator/2*`". Multiply by two function, returns $(2*i)$.
vi. "`replicator/2*+1`". Multiply by two plus one function, returns $((2*i) +1)$.
vii. "`replicator/2^`". Power function, returns $(2^i)$.
viii. "`replicator/div2`". Divide by two function, returns the bit representation of the integer shifted once to the right.

The general form of the declaration, followed by an example, appears below:

```
replications{firstname,secondname,..finalname /firstfunction,.., finalfunction};
```

```
replicators{ i, index, currentvalue / id, inc, 2^};
```

## 5.9 META DECLARATIONS.

Space code consists of a series of interstrings, which are given integer based labels. Normally, program execution begins by activating interstring number 1. Space has a facility for describing high level meta modules. The meta declaration is optional, and takes the following form, and instructs the compiler that the second phase of the meta module is activated by the interstring labelled with *integer*:

```
meta(integer);
```

If a meta declaration is made, then it is obligatory to include the *metatime* declaration, which gives the lower and upper bounds for completion times of the second phase.

```
metatime: shortest_time_integer-longestest_time_integer cycles;
```

```
metatime: 23-23;
```

## 5.10 TIME DECLARATION.

The final declaration is obligatory, and gives the lower and upper bounds for completion times of the module's only phase, or first phase if it has been declared to be a meta module. It takes the form:

```
time: shortest_time_integer-longest_time_integer cycles;
```

The time and meta-time declarations are currently there for the convenience of the user, and are not generally guaranteed to be reliable. If completion times are unknown, then the statements "`time: 0-0;`" or "`metatime: 0-0;`" are made. There is no current support for expressing running times as an arithmetic function of module input values, and of other functions expressing running times for submodules. It is believed that the programming environment however does have the foundations to support this in the future, and a thoroughgoing system of performance evaluation.

The present environment can only offer a possibility of accurately reflecting module completion times, providing a programming methodology is adopted whereby modules are only entered in to the library if they can be proven to terminate, and if there is an upper bound for the size of their inputs. Assuming such a methodology is adopted, even with constant length input, it is not trivial to statically determine module running time, because of the complexity of the factors involved: the compiler implementation, and details of the module's code. But if the composer of an Earth or Space module has the insight to pick a small, judicious range of inputs, which can be run to observe the range of possible behaviours, then completion times can often be reliably determined. The module may then be recompiled to the library, with a more meaningful time declaration.

The issue of completion times for modules with inputs of indefinite size, such as strings, is now discussed. Earth modules have fixed length inputs, and the current approach for representing completion times in Earth, fulfills it's limited role. For a Space module with an input storage entity declared to be a string based type, or a blockstring, an alternative approach is needed.

A module will process a string input by sequentially accessing and processing string or blockstring elements. If a module has only one indefinitely sized input, let the numerical length of the input string or blockstring be $n$. It ought to be possible to express completion times as an arithmetic function of $n$, based on the completion times for the module to process single string elements at different stages of the program. The idea can extended to deal with modules with input types incorporating strings at multiple type levels. The construction of any data type instance of an input, could keep track of the lengths of strings in component sub-types. Time declaration times could then be represented as functions of the input's component string lengths, and the completion times of constant length input submodules. It is believed that applying this approach to code for Synchronic Engines, will yield a powerful and accurate means of determining performance times, that is not possible with parallel programs for processor networks.

# Chapter 6.
# SPACE BASE SET.

## 6.1 INTRODUCTION.

Space's interstrings and program constructs, enable the compact description of readable, massively parallel code. The role of an interstring in Space is somewhat, but not completely analogous to that of a statement in an imperative language, and typically computes the result of a dataflow, and records the results. The final column of an interstring may also test a condition, and/or transfer control to other interstring(s). Program constructs provide a means of embedding high level features into Space, and are described in detail in the next chapter. They fulfill some of the roles of control and repetition structures in sequential imperative languages, and include two different kinds of code replication, while and do-while iterations, and switch selection.

An interstring in a module's code section is called a *base-line*, and a program construct is called a *construct-line*. Examples appear in figure 6.2 and figure 6.3 respectively. The Space compiler has a phase that processes and removes construct-lines, leaving behind only base-lines. A construct-line represents a subprogram of the module, and after an unpacking phase consists only of baselines, with a single entry and at least one exit point. The way in which construct-lines relate to base-lines, is a departure from the way in which control/repetition structures contain statements in sequential imperative languages. Space lines have a numerically based labelling system, and if readability were not a concern, the system would remove the necessity to place related lines textually adjacent to each other.

A construct-line may have base-lines as it's components, and a construct-line that has other construct-lines as components is called *composite*. The internal organization of a composite construct-line is a tree structure called a *construct tree*, and a simplified example of a composite construct-line is depicted in Figure 6.1

An *independent* construct-line is one which does not appear as a component in another construct-line, and represents a sub-program in the module, at a level of abstraction higher than it's components. A construct tree has high structural variability, and with the possible exception of some future attempt at implementing virtual modules (higher order functions), sub-expression repetition is unlikely. The tree is represented by the compiler as a string of strings data structure, because it is easier to process, and facilitates the identification of opportunities for future parallelization.

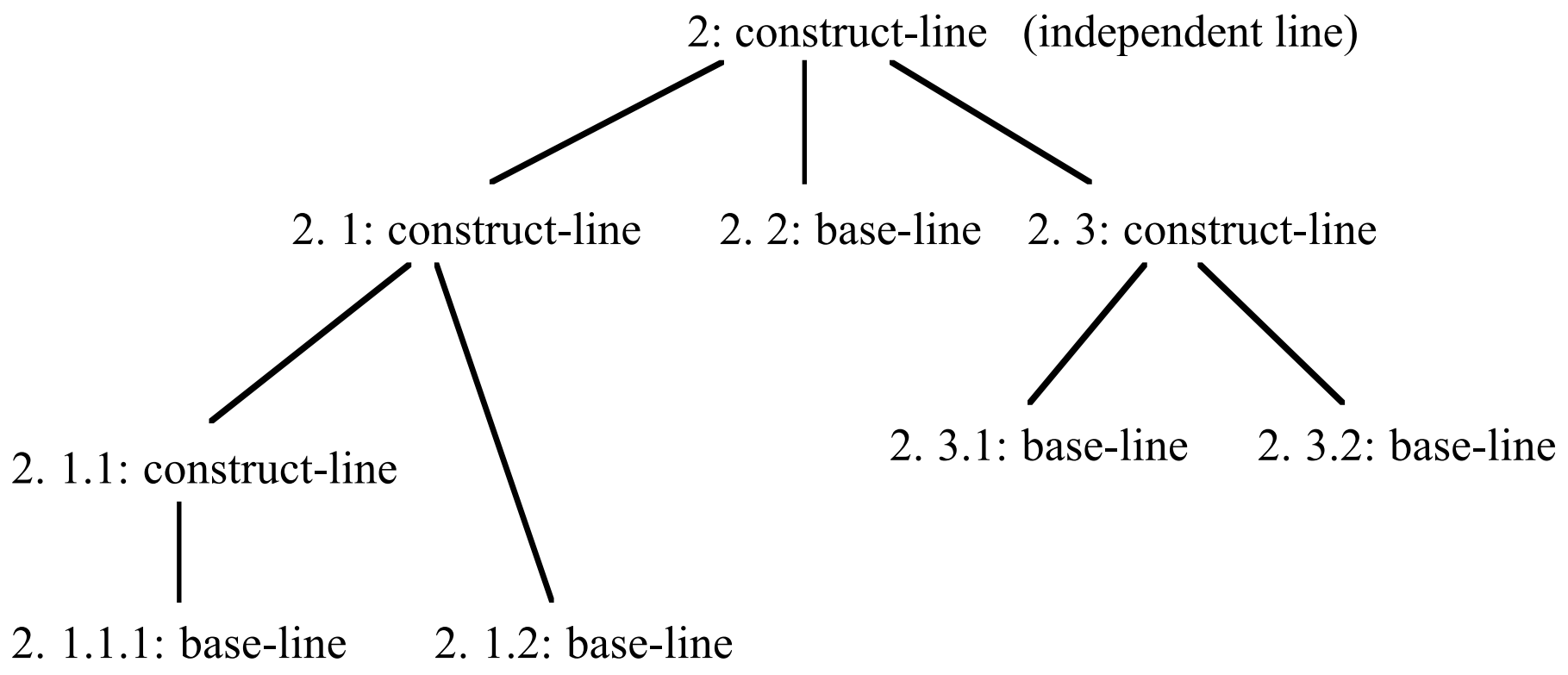

Figure 6.1 Abstraction of a construct tree, with line addresses.

A Space module's storage, submodules, and replicator name declarations, help to establish the base language of the module's base-lines, and the content of construct-lines[48]. A base-line is composed of a sequence of columns of instructions, drawn from a set of instructions called the *base set*, which contains eight elements, and is the main focus of this chapter. A base-line that is not a component of any construct-lines, is also characterised as independent. A line that is a component of a construct-line is called *dependent*.

Both base-lines and construct-lines have a numerically based label, called a *line address*, which is a system where a string of integers is employed for the purpose of expressing component relationships between lines, and is exemplified in fig 6.1. An independent line address is always a single integer, and a dependent line address is an integer string that indicates a path through the layers of the construct tree.

Line addresses are involved in a compilation phase called *expansion*, that processes independent construct-lines and dependent lines, resulting in an output consisting only of independent base-lines. They are used in "if else" and "GOTO" type program control instructions, and provide a means of succinctly referring to high level constructs. Djikstra's warning [1] concerning the dangers of the unrestricted use of the GOTO statement in high level languages, can be applied with excessive zeal in the design of programming languages. The warning is not relevant to the Space interlanguage, because an unexpanded module typically has less than a dozen mentions of the jump instruction, and at every level of abstraction, a program segment has one entry, and at least one exit point[49].

[48] Space could be defined as a context-free language. It was argued in chapter 2 however, that such an approach would not do justice to the inherent parallelism in Space's syntactic and semantic processing. As part of a project to migrate mathematical discourse to α-Ram machines, a formal theory of interlanguages would be presented in terms of α-Ram programs and virtual α-Ram programs. It is an open question whether there is some definitive pairing of grammatical categories of interlanguages and program categories for interlanguage processing, analogous to pairings in the Chomskian hierarchy of languages and automata.

### 6.1.1 BASE-LINES.

```
1: b -> neqz.input  :: _neqz :: cond_neqz.output (3,0) (2,0) :;
   a -> mod.inputa
   b -> mod.inputb
```

Figure 6.2 Independent base-line from the Euclid's algorithm module presented in 6.11.

A Space base-line's topmost line of text has markers, which define the leftmost and rightmost textual cursor positions of individual columns, called the column's *left brace* and *right brace* respectively. A left brace may take the form '*lineaddress*:', or the *column brace* '::'. A right brace may be the column brace '::', the *construct brace* ':>', whose use is defined in chapter 7, or the *line terminator* ':;', A column's instructions must be textually within it's left and right braces, by at least one space character. In Space, a base-line column may only contain instructions of the same type, for reasons that are explained in 6.2 and 6.3.

The placement of instructions in a base-line's column, indicates to the compiler that they are to be all activated in the same Synchronic A-Ram cycle, although they may terminate in different cycles. The simultaneous execution of a column's instructions, is characterised as c*olumn parallelism.* When a base-line is activated, it's first (leftmost) column of instructions are activated. A succeeding column is activated, when all of the instructions in it's predecessor have completed[50]. A base line usually represents a dataflow, and when activated, may terminate either with no further activity, or by jumping to other line(s), or by selecting which line(s) to jump to by testing a condition.

[49] A line address cannot refer to the interior parts of a baseline, and can only refer to relatively high level constructs. The total number of line addresses in pre-expanded module, is typically less than a dozen.

[50] 6.13.3 discusses a future compiler enhancement, that allows asynchronous column parallelism within a globally synchronous environment, in order to match the efficiency of the Dataflow Model's implicit parallelism.

The base-line in figure 6.2 has the line address '1', and is from a module which implements Euclid's algorithm, to be presented in 6.13. The leftmost column is a collection of copy instructions, which transfer the contents of the module's storage entities to a submodule's input storage entities, where '->' indicates the direction of transfer. The subsequent column is called an *activation column*, which in this case is composed of only one instruction. It activates the submodule `neqz`, by placing an '_' before the submodule name `neqz`. The final column has a single *cond* instruction, which fulfills the role of an "if else" control structure. The instruction examines `neqz's` output, and selects from two other line addresses with offsets.

A dataflow of depth greater than one may be expressed, by inserting extra copy and activation columns into a base-line. A base-line's interstring format obviates the need for the brackets in tree languages, used to describe the nesting of expressions for deep dataflows.

### 6.1.2 CONSTRUCT-LINES.

Although construct-lines are covered in the next chapter, an example is given here to give context. The *deep* construct-line in fig 6.3(a) has address '5', and is separated from it's single dependent base-line positioned to the left, with address '5.1' by the construct brace ':>'. The construct-line has a single replicator name '`i`', states a left and right limit, and type of incrementor function, followed by a set of brackets that optionally contain a line address and offset, representing where program control is to be transferred to upon completion of it's code. It expresses a form of 'deep' code replication depicted in 6.3(b), where the base-line is 'deepened', or expanded vertically by the construct-line.

The expansion involves incrementing the control value of the replicator from 0 through to 1023, resulting in base-line columns containing 1024 instructions. The new base-line is given the construct-line's address, and the construct line is discarded. The base-line describes the activation of an array of 1024 adder submodules, and the subsequent transfer of the outputs into the module's output array. Note that the combination of deep construct-line and base-line not only succinctly describes a massive dataflow, but has also implicitly expressed an allocation of machine resources (submodules) to compute the dataflow.

```
5.1: _adder[i] :: adder[i].output -> output[i] :> 5:  deep< i=0; i<= 1023; inc > ()  :;
```

6.3 (a) Dependent base-line with address ‘5.1’, attached to a deep replicator construct-line with address ‘5’.

```
5:   _adder[0]     :: adder[0].output -> output[0]        :;
     _adder[1]        adder[1].output -> output[1]
       .                    .             .
       .                    .             .
     _adder[1023]     adder[1023].output -> output[1023]
```

6.3 (b) Result of expanding deep replicator construct-line.

The way that the dependent base-lines and construct-lines of a composite construct-line are laid out in code, is presented in the next chapter.

### 6.1.3 Co-active parallelism.

At the module’s level of abstraction, a replicated baseline column composed of instructions activating a large array of submodules belonging to the same class, expresses a kind of SIMD parallelism, where the ‘instruction’ (submodule) may represent an arbitrarily complex pre-defined program. A replicated baseline in it’s entirety, can only express a limited form of SPMD parallelism within a module, because there is no explicit program control (see 6.2 and 6.4) involving selection before the final column of a baseline [51].

In order to enable more complex forms of parallel programming, a second source of explicit parallelism in Space, is the ability to simultaneously activate multiple lines. A parallel algorithm implemented in Space, often requires differing forms of replicated code to be active simultaneously. The presence of offsets in program control instructions in base-lines, and certain construct-lines, can instruct a number of subprograms represented by construct lines, and baselines to begin executing simultaneously, that will typically terminate at differing times. Such a set of lines is said to be *co-active*.

Through the use of a facility called *programmable jump* to be described in 7.8.2, co-active parallelism also enables a means during runtime, of varying the range of a submodule array to be activated, in the latter stages of computing a circuit value or parallel prefix tree.

[51] Fully programmable SPMD parallelism at a module’s level of abstraction is achieved through the use of the grow construct, described in 7.7.

The next section explains how co-active parallelism has the potential to simplify a module's state system, in a way that cannot be readily afforded by column parallelism alone

Co-active parallelism permits the multiple activation of base-lines whose copy and activate columns are out of phase, and of construct-lines possessing differing levels of compositional complexity.

## 6.2 CONTAINING PARALLELISM.

There is one perspective from which parallel programming is easier than sequential programming, in that the programmer is freed from being obliged to sequentialise operations, that are naturally conceived as being parallel. Parallel programming is however more often viewed as a less user friendly experience than sequential programming, because in the Von-Neumann network programming model, the coder potentially has to consider massively parallel activity, non-determinism, state explosion, scheduling, allocation, and contention issues. I argue that Space largely avoids these issues.

A Space module contains parallelism at multiple levels of abstraction, that is concealed within the sub-modular hierarchy. Within a module, unconstrained column and co-active parallelism have the potential to generate an undesirable number of states, resulting in the module's behaviour becoming chaotic and difficult to understand. A number of measures are taken to improve readability, and to contain parallelism in Space.

Space does not allow the use of mixed instruction parallelism in a baseline column, i.e. column parallelism has a SIMD character[52]. One of the major themes in this report, is that the basic entity out of which programs are composed, and to which program control is applied, is spatially represented dataflow. Therefore in Space, selection and jumps to other lines, do not appear before the end of a baseline (dataflow), occurring only in the final baseline column.

The deterministic nature of the Synchronic A-Ram, affords a means of eliminating the possibility of the state space expanding uncontrollably. Space is designed so that a module's behaviour may be characterised as a conventional sequential state transition system, where each state is associated with a set of co-active independent lines. The next chapter and program examples in Appendix B, demonstrate that SIMD, M-SIMD, SPMD, MIMD, pipelining, systolic, cellular automata and other kinds of deterministic parallelism, can be implemented by Space modules, characterised as state transition systems. To achieve sequential state transitions, the following programming methodology is adopted, which constrains the way in which the programmer may invoke co-active parallelism.

[52] It is explained in 6.7 how MIMD parallelism may be implemented by an activation column including submodules belonging to differing classes.  Co-active parallelism is another source of MIMD programming.

i. A baseline may not activated, if it has not terminated from a previous activation[53].
ii. One base or construct line in a co-active set, is designated as the *carry line*, and takes as long or longer, to complete than the other lines. The carry line has the role of transferring control to the next state of the program (the next co-active set), at the end of it's execution. The other members of the co-active set are forbidden from activating lines outside the co-active set, either whilst running, or upon their termination.
iii. The co-active sets that may be active at any stage of a module's run, are pre-determined at program composition time.

Space has some synchronisation mechanisms, discussed in 6.10 and 6.11, to assist in establishing that the carry line does not terminate before other members of the co-active set. The mechanisms however, have limited ability to identify the termination of construct-lines, and in the current implementation, the programmer is regrettably obliged to take special care to ensure the carry line is "the last man standing". If this can be achieved, then ensuring that the module's code embodies a sequential state transition system is relatively straightforward.

It is also presently incumbent on the programmer, to check that a module's set of states is connected, i.e. every co-active set can potentially be activated. An example of a state system for the modulus operation on integers is presented in 6.5, and for a serial style adder in 6.13.5 The next chapter contains a range of program examples, which suggest that the above restrictions do not constrain the expression of parallelism. The simplicity of Space's state system also entails that conventional sequential program reasoning techniques, can be applied to modules expressing massively parallel algorithms.

[53] The current compiler implementation cannot statically detect a violation of this or the succeeding restrictions.

A deterministic Space program with explicitly simultaneous sub-modules running in a synchronous environment[54], is easier to understand than an non-deterministic, asynchronous network of Von Neumann programs. Space base-lines and construct-lines clearly depict massively parallel activity, where the scheduling and allocation of sub-processes is implicit, and no allocation of communication channels is required. Resource contention is not a serious issue, and sharing and simultaneous evaluation of dataflow subexpressions is unambiguous.

A Space module describes a relatively small deterministic state transition system, which is conducive to program reasoning. In 6.14, it is argued that there are insignificant, or manageable space and time overheads, imposed by the synchronisation mechanisms required for Space's style of deterministic parallelism.

## 6.3 OVERVIEW OF SPACE COMPILER.

The compiler initially scans the module's declarations, and produces various data structures, which are used for checking syntactic correctness and semantic coherency of the module's code, and for code generation. Code consists of a list of base-lines, with a construct-line, or series of construct-lines, attached to some of base-lines. The internal syntax of base-line instructions, and construct-lines, have no recursive element, and expressions have a maximum length. There is no requirement for extensible, string or list-like data structures to represent them (unlike base-lines, base-line columns, and construct trees).

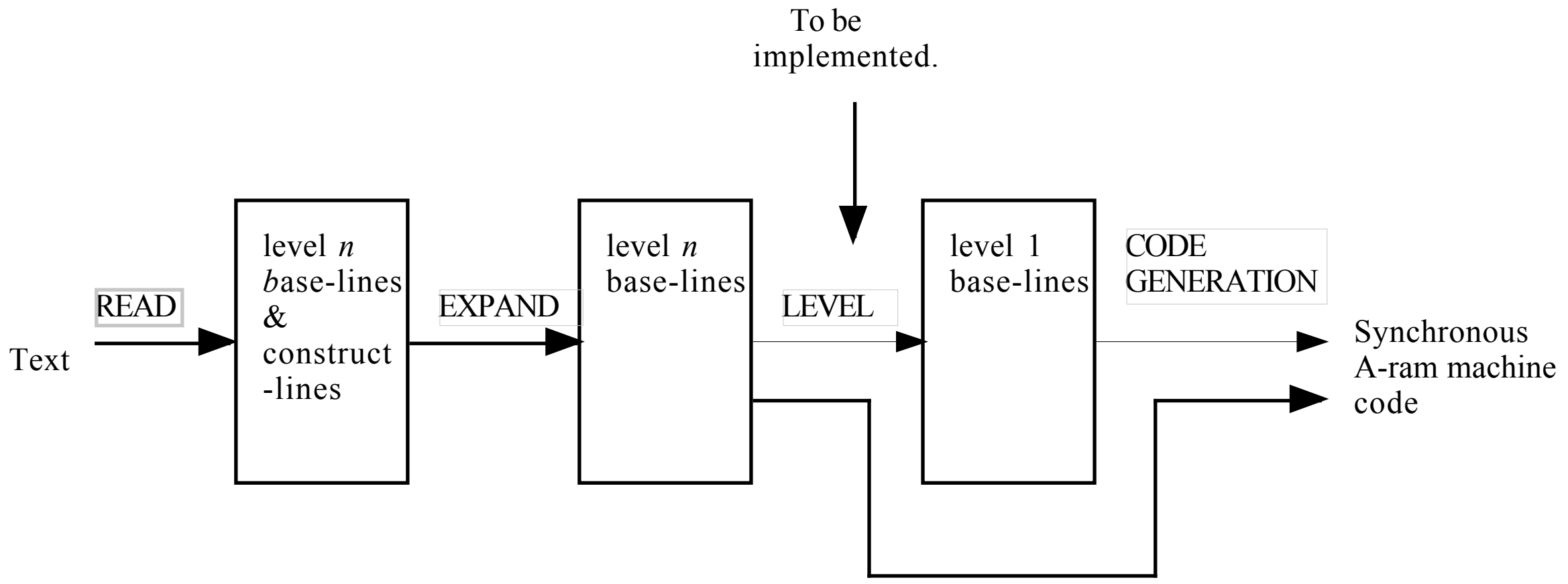

Figure 6.4 Phases in code compilation.

[54] Scaling the Synchronic Engine design may result in asynchronous processing, but this would be concealed from the Space programmer. Modules would be viewed as running in a virtually synchronous environment.

The syntax of (entire) base-lines exhibits low structural variability considered as a tree, because only some of the tree's branches may be indefinitely long. What might be characterized as the single recursive aspect of base-line syntax, relates to appending identically formatted code elements to list-like structures.[55] There are three cases:

i. The number of instructions in a base-line column may be indefinitely large[56].
ii. The number of base-line columns in a base-line may be indefinitely large.
iii. The number of base-lines in a module may be indefinitely large.

In contrast to base-lines, a construct tree can have high structural variability, but the tree is already parsed thanks to the line address system. The limited range and regularity of Space code, entails that it's structure is unambiguous, and many of the conventional compilation phases may be integrated into one phase. Code can be analysed in a topdown manner, without any need for backtracking. There is no requirement for a recursive descent parser, or the overheads associated with predictive parsing [2]. The arrangement of code compilation phases is depicted in figure 6.4. Lexical, syntactic, intermediate code, type checking, and partial semantic analysis phases are completed in one scan of the module's text file, which are integrated into a single *read* phase.

If the maximum level of the module's sub-modules is *n-1*, then the module is level *n*, and the base-lines are also said to be level *n*. The output of the read phase is a complex C data structure called a *program structure*, which stores level *n* base-lines and construct-lines separately. The program structure represents the first stage of intermediate code, with highly nested string components. The program structure is input for the expansion phase, which processes and removes the construct-lines, generating a program structure containing only an expanded list of independent, level *n* base-lines.

The next phase concerns code optimization. The compiler has no need to eliminate subexpression repetition in dataflows, because it is assumed that the programmer has not needlessly introduced repetition into the base-line language, which does not have the single parent restriction[57]. The current version of Spatiale lacks a form of code optimization called

[55] A recursive procedure for the successive processing of the elements of a list-like structure, has no obvious advantage over an iterative procedure, in terms of readability and programmability.

[56] Providing there is room available in the memory block, for the resulting compiled code

[57] Subexpression repetition is not introduced by the read phase, or by the earlier compilation of sub-modules.

*levelling*, which will appear in the next implementation. Levelling translates level *n* code, into semantically equivalent code expressed in lower levels of abstraction, down to and including Earth modules, similar to the hierarchical decomposition of modules made available in HDL programming environments. Levelling will be applied, in order to be able to easily identify and remove all instances of copy and jump propagation at source, generated by high level program composition. This is in contrast to the low level keyhole techniques used in conventional compiler optimization. Until levelling is implemented, the input for the code generation phase is a list of level *n* base-lines.

The code generation phase is analogous to linking, and also handles the production of all code relating to the scheduling of column and line activations. Some semantic analysis occurs in code generation, which is otherwise principally concerned with the direct mapping of base-line instructions into chunks of machine code. Reading, expansion, levelling, and code generation, have extensive opportunities for parallelization, that are not present in the phases of tree based programming languages. Parallelization can be implemented, when a native compiler is written.

### 6.4 Base set.

Eight instruction types are available for use in a base-line column. Mixing instruction types within the same column, would not in itself entail machine error at runtime, but does result in lines which are more difficult to understand, and is not conducive to replicating code and the avoidance of state explosion. To help contain parallelism, and maintain a standard line format and code readability, an individual column may only contain instructions of the same type, and some column types are restricted to being the terminating column of a base-line. The following sections describe instruction formats, which are summarised below.

**i. Copy**. Although no computation occurs in a *copy* instruction, it bears some similarity to an imperative assignment statement. The instruction has two operands, separated by ‘->’, each of which identifies a storage entity appearing either in the storage declaration, an entity appearing in the storage declaration of a submodule’s code, or some immediate value. The instruction copies the contents of the first entity into the second. A copy column may have a number of elements.

ii. **Activate**. An *activate* instruction has only one operand, preceded by an underscore, which identifies a submodule, or single element of a submodular array, appearing in the submodules declaration. The instruction initiates the execution of the submodule's code, usually after the submodule has received an input, or inputs from copy operations. The second phase of a meta submodule may also be activated. A number of activations may appear in a column, where submodules may belong to differing classes.

iii. **Cond**. The cond instruction fulfills the role of the "if..else.." control structure in imperative languages, despite not being the equivalent of a Space programming construct. The first operand is separated from the 'cond' by an underscore, and identifies a one bit storage entity, usually the output of a submodule. The cond instruction inspects the bit, and selects either of the second or third operands pairs, which are are line addresses with offsets. A cond column is always the terminating column of a base-line. Cond columns in pre-expanded code, only contain one element.

iv. **Jump**. A *jump* instruction is analogous to a GOTO command in Fortran and Basic. It has one operand pair of a line address and offset, which specifies a set of lines to be simultaneously activated without selection. A jump column is always the terminating column of a base-line, and may contain a number of jumps.

v. **Skip**. The *skip* instruction has a single line address operand, and is used as a synchronisation mechanism in a (baseline) carry line, to ensure that co-active lines all terminate before the next stage of the program. There may be a number of skips in a column.

vi. **Wait**. A finer grained synchronisation mechanism is the *wait* instruction, where only one wait is allowed per column. The instruction delays the activation of the succeeding baseline column, by the number of machine cycles specified by a single integer operand.

vii. **Subhalt**. As the sequencing of component lines of a composite construct-line is not based on brackets, or on commands appearing on consecutive lines on a page, there is a need to be able to mark where the code within the dependent lines of a construct-line halts. A subhalt column may only have one instruction, and has the construct-line's line address as the single operand. It is always the terminating column of a dependent base-line.

viii. **Halt**. There are two ways of signalling exit points for the program. In addition to a halt instruction for the first or only phase of a module, there is a separate halt for the the second phase of a meta module. A halt column may only contain a single instruction, and is always the terminating column of a base-line.

## 6.5 First Space example: modulus operation on two positive integers.

After the first example of Space code, each instruction is described in detail. A module is now presented in figure 6.5(b) that codes the modulus operation on two non-negative integers. The remainder is obtained somewhat inefficiently, by repeatedly subtracting the divisor from the dividend. The module uses a 32-bit subtractor, which appears in appendix B. The subtractor has non-negative minuend and subtrahend inputs, and a *borrowout* output bit, which is set if the minuend is less than the subtrahend.

```
module modulus{

	storage{
		unsigned dividend input;
		unsigned divisor input;
		unsigned remainder output;
		BIT dividebyzero output; // an error bit is set if divisor b = 0
	};

	submodules{

		sub32 sub;
		paror32 neqz;

	};

	time: 0-0 cycles;

	code{

		1: dividend -> sub.input0 :: _neqz :: cond_neqz.output (5,0) (2,0) :;
		   divisor -> sub.input1
		   divisor -> neqz.input

		2: _sub :: cond_sub.borrowout (3,0) (4,0) :;  // test for overflow

		3: sub.output -> sub.input0 :: jump (2,0) :;    // subtract again

		4: sub.input0 -> remainder :: HALT :;  // recover penultimate sub.output

		5: #1 -> dividebyzero :: HALT :;

	};

};
```

Figure 6.5(b) Modulus operation, where dividend is greater than or equal to divisor.

The main loop is described in lines 2 and 3, which subtracts the subtrahend from the minuend, until the borrowout bit is set. The output is then recovered from the previous cycle's minuend. If $q$ is the integer quotient of the dividend and the divisor, then the loop is invoked $q+1$ times. An error bit `dividebyzero` is set, if the divisor is zero.

The module's state transition system is trivial and is depicted in figure 6.5(a). The code has minimal column, and no co-active parallelism. In this example, each base-line is identified with a state. The only parallelism at the module's level of abstraction, occurs in the copy column of line 1. Each base instruction is now considered in detail.

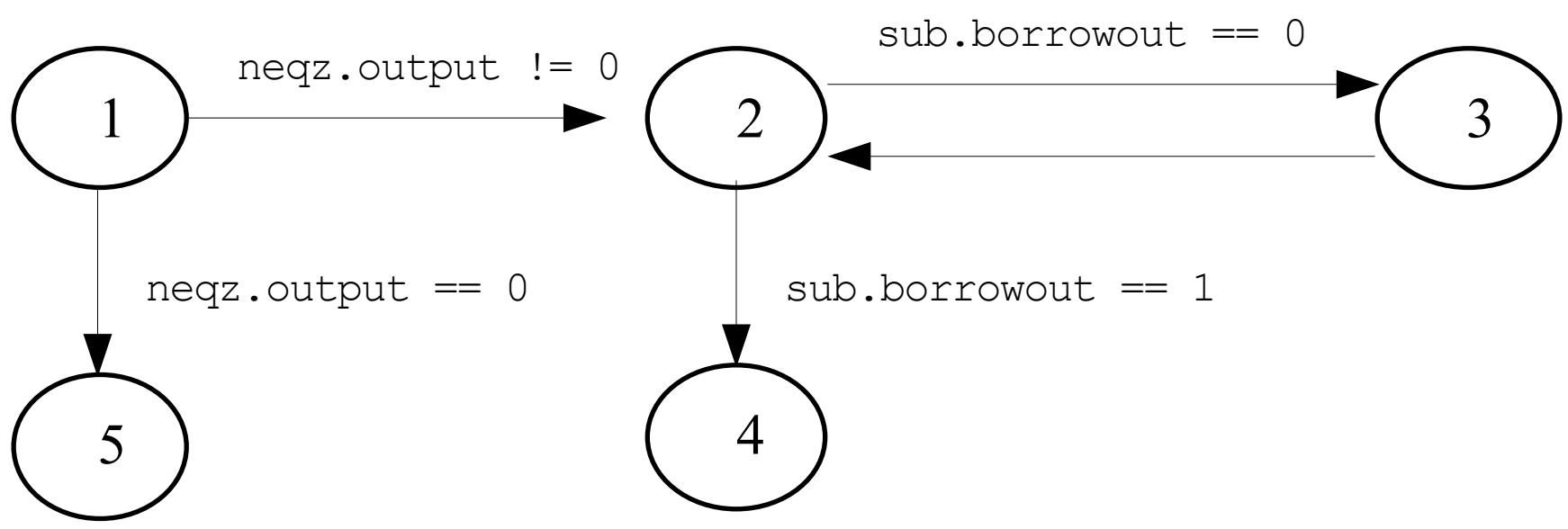

Figure 6.5(a) State transition diagram for modulus program.

## 6.6 COPY.

A copy instruction has the standard format *copy identifier -> copy identifier*, where at least one space character must appear either side of the '->'. The left hand copy identifier distinguishes some part of the memory block, or specifies some constant value, and the right hand copy identifier only distinguishes part of the memory block. The contents of the second identifier are overwritten by the value or contents of the first identifier[58]. Other than submodules writing results to their outputs, this is the only way in which storage can be modified in a Space module. The implementation of a copy column is maximally parallel, and is discussed in 6.14.2. An arbitrary number of copies may appear in a copy column, without any concern for a formal allocation of communication channels.

In order to implement C-type functionality, the language of identifier expressions is somewhat involved, and the description of the copy instruction is considerably more complex than the other instruction formats

[58] The use of the first identifier might therefore be characterised as 'call by value', and the second as 'call by reference'.

The repetition of the same right hand identifier in a copy column is forbidden, because it would result in a type of resource contention. It would copy more than one value into a storage location, generating a write failure at runtime. When the compiler's read phase detects this, it generates an error. The read phase also detects if the same left hand copy identifier appears more than once in a column, and to save space, writes code to read storage once only.

The read phase also tests type compatibility of copy identifiers. Exceptions to type strictness are described (see 6.6.6), that allow the definition of special modules that implement memory allocation, accessing array elements, dereferencing pointers, and the transfer of data between the pre-defined types. There are three categories of identifier:

i. An identifier that refers to an entity in the module's storage declarations, is a *primary identifier*.
ii. An identifier that refers to an entity in a submodule's storage declarations, is a *secondary identifier*.
iii. A *tertiary identifier* encompasses immediate addressing, a size of type operator, constant values, bit segments of level zero types, and the pre-defined level one types.

To maintain good modularity and avoid mixing levels of abstraction, an operand may not refer to any component sub-types of primary and secondary identifiers. This also has the welcome consequence that an identifier expression has a maximum length, and exhibits a limited range of formats. The language of identifiers does not have the recursive aspect of a context free grammar in formal language theory, but could if desired, be defined using Backus Naur statements. I will simply go through the cases. Recall that a storage declaration takes the form in figure 6.6(a), accompanied by examples in 6.6(b).

*typename storagelabel*aggregate_construct interface_category;

6.6 (a) General form of a module's storage declaration.

```
unsigned firstinteger input;
unsigned a input;
int array[16] output;
char charblockstring<16> private;
```

6.6 (b) Examples of storage declarations.

Before the description of primary identifier formats, the aggregate construct of a primary identifier's associated storage declaration has to be considered, followed by a presentation of the formats of a *cell_identifer* and *bit_identifer*. There are four aggregate construct cases.

i. If the `storagelabel` entity has been declared to be a singleton, then no array notation can be included in a primary identifier.
ii. If `storagelabel` has been declared as an array, then an *array_identifer* is required for an element whose identity is known before runtime, consisting of up to three array co-ordinates:[*co-ordinate1*][*co-ordinate*2][*co-ordinate3*], depending on the dimensionality of the aggregate construct. A co-ordinate is either an unsigned integer, or an incremental expression[59]. It needs to be stressed that the current environment has no direct support for accessing array elements, whose co-ordinates are unknown at runtime. This requires the definition of a dedicated module, an example of which is discussed in 6.11.
iii. Space has no direct support for the identification and accessing of elements of blockstrings before and during runtime, which would require the programming of special modules.
iv. A pointer to a typed entity is a single register containing an unsigned integer, representing a memory block address. The contents of a pointer may be copied as if it were a singleton of type `unsigned`. Space has currently no direct support for dereferencing a pointer, which requires a dedicated module for each type, an example of which is described in 6.13.4.

### 6.6.1 BIT IDENTIFIERS AND CELL IDENTIFIERS.

Space can express very fine grained code, because a primary (and secondary) identifier can specify an individual bit or segment of level zero types, and the pre-defined level one types, by using a *bit_identifier*, and a *cell_ identifier* respectively. A bit_indentifier is either an unsigned integer between 0 and 31, or an incremental expression. There are eight cell identifiers:

[59] Recall the definition of an *incremental expression* in 5.7, which mentions a replicator name. During compilation, a control value of the replicator name is taken, and applied to one of eight unary arithmetic functions, that is referred to in the incremental expression.

i. The destination cell of a type occupying a whole register may be identified by using the postfix '`.destn`', which comprises bits 5-29.
ii. The offset cell of a type occupying a whole register may be identified by using the postfix '`.offst`', which comprises bits 0-4.
iii. The bit address segment, of a type occupying a whole register may be identified by using the postfix '`.btadd`', which comprises bits 0-29.
iv. The bytes of a type occupying a whole register may be identified by using the postfixes '`.byte0`', '`.byte1`', '`.byte2`', and '`.byte3`', where '`.byte0`' refers to rightmost byte comprising bits 0-7.
v. The words of a type occupying a whole register may be identified by using the postfixes '`.word0`', '`.word1`', where '`.word0`' `refers` to rightmost word comprising bits 0-15.

### 6.6.2 PRIMARY INDENTIFIERS.

We are now in a position to describe the six formats for primary identifiers. For easier presentation, *`label`* is understood to stand for *`storagelabel`*. Examples of primary identifiers are based on the storage declarations in fig 6.5(b).

| `primary_identifier` formats | Examples of primary identifiers |
|---|---|
| *`label`* | `a, firstinteger` |
| *`label`*`bit_indentifier` | `firstinteger.21, firstinteger.i` |
| *`label`*`cell_indentifier` | `firstinteger.destn, firstinteger.byte3` |
| *`label`*`array_indentifier` | `array[10], array[i], array[i/inc]` |
| *`label`*`array_indentifier.bit_indentifier` | `array[10].i, array[i/inc].0` |
| *`label`*`array_indentifier.cell_indentifier` | `array[i/inc].destn, array[2].word1` |

Storage entities that have been declared to be arrays, cannot be copied in their entirety, using only one copy instruction. An entire has to be copied element by element, a task that is easily accomplished using the deep replicator construct, described in chapter 7.

### 6.6.3 SECONDARY INDENTIFIERS.

Secondary identifiers distinguish storage that is immediate to a submodule, i.e. storage that has been declared in the submodule's storage declarations. Recall that a submodule declaration takes the form in figure 6.7(a), accompanied by examples in 6.7(b).

```
submodulename submodulelabelaggregate_construct;
```

Figure 6.7(a) Standard submodules declaration.

```
inceq5bit inceq5bit[1024];
modulus mod;
parand32 neqz;
parand32 par32array[3][5];
```

Figure 6.7(b) Examples of submodule declaration with level 0 modulea from chapter 4.

A *submodule_identifier*, distinguishes a submodule, or single element of a submodular array, appearing in the submodules declaration. There are only two aggregate construct categories; a submodule is either a singleton, or an element of an array. Consequently there are only two formats for submodule identifiers: `submodulelabel`, and `submodulelabelarray_indentifier`.

The use of the term *primary_identifier* below, refers exclusively to primary identifiers that are legal, *within* the code of submodules. For easier presentation, `smlabel` is understood to stand for `submodulelabel.` The examples of secondary identifiers are based on the submodule declarations in fig 6.7(b).

| `secondary_identifier` formats | Examples of secondary identifiers. |
|---|---|
| `smlabel.primary_indentifier` | `mod.ainput, neqz.output` |
| `smlabelarray_indentifier.primary_indentifier` | `par32array[1][0].output.15` |

### 6.6.4 ADDRESS OPERATOR.

An address operator &, may be applied to any primary or secondary identifier as a prefix. The compiler will substitute a 25-bit unsigned integer, that represents the address of a multi-register entity's first register in the memory block. If the entity is a bit or register segment, the address operator will return a 30-bit integer for the bit, and a 25-bit integer for a register, respectively. Applying the address operator to bytes and words is currently not supported.

### 6.6.5 TERTIARY INDENTIFIERS.

Tertiary identifiers make up a miscellaneous category of lesser used identifiers.

i. Immediate number operator. The operator # can only be used in the first copy operand, and indicates that what follows is either an incremental function expression, or a constant signed integer, unsigned integer, or float value belonging to the pre-defined library level one types. Thus we can have `#34.2 -> floatvalue` , or `#256 -> integervalue, or #i -> integervalue`. The identifier is automatically given the appropriate type.

ii. Immediate char operator. Characters need a separate immediate operator @, because of the existence of incremental expressions in immediate number expressions. Thus we can have @c `-> charvalue`.

iii. Register direct addressing. This is required to implement the limited form of memory allocation, supporting only sequential allocation in the current version of Space. A register with memory block address *integer*, may be identified by placing the address in brackets: (*integer*). Thus we can have `(25165824) -> adder.input`. The identifier is treated as type REG. The use of direct addressing has to be kept to a minimum, to ensure that the module does not generate side effects (see 6.13).

iv. Size of type operator. Involved in array accessing, and also in memory allocation, the size of type operator $, may be applied to any typename in the type library, regardless of whether it has been declared in storage declarations. The compiler will replace it with the number of registers that the typename occupies. This identifier is treated as type unsigned. Thus we can have `$`*`typename`* `-> adder.input`.

v. Array aggregate parameter. These parameters are used to allow the programming of modules, that allow the runtime accessing of array and blockstring elements. The runtime environment is unaware of parameters relating to an array's dimensions and memory block address, or a blockstring's parameters, unless the compiler makes the data available in the memory block. The compiler will substitute expression `arrayname`[[0]] with an immediate integer representing `arrayname`'s first register in the memory block. Up to three dimensional lengths of the array will be represented by `arrayname`[[1]], `arrayname`[[2]], and `arrayname`[[3]], respectively.

vi. Blockstring aggregate parameter. The compiler will substitute expression `blockstringname`<<0>>, with an immediate integer representing the address of the first element of `blockstringname`. The blockstring's block length is represented by `blockstringname`<<1>>. The blockstring's next block address is represented by `blockstringname`<<2>>. The blockstring's offset is represented by `blockstringname`<<3>>.

The main part of explaining the language of Space's base-lines, has now been completed, and the descriptions of the remaining instructions can use established terminology.

### 6.6.6 EXCEPTIONS TO TYPE COMPATIBILITY.

Normally type compatibility in copy identifier pairs is strictly enforced, and conflicting types result in error at compile time. In order to implement various functionalities however, a limited range of exceptions are made for pointers and address operators, and for certain pairings of level 0 and level 1 types, which allow for example, a byte to be copied into the rightmost bits of an unsigned integer. Current exceptions are listed below:

BIT, BYTE, WORD, BITA, DSTN, OFST -> REG, int, unsigned.
(Source is copied into rightmost bits of the target, and leftmost unused bits of target are reset).

unsigned -> BYTE, WORD
unsigned -> OFST
unsigned -> BIT
(Rightmost bits of source are copied to target, and leftmost bits are discarded.)

BYTE -> char
char -> BYTE

unsigned, REG -> `storagename` with aggregate type *pointer*.
`storagename` with aggregate type *pointer* -> REG, unsigned.

&`storagename` -> BITA, DSTN, REG, unsigned

## 6.7 Activate.

An *activate* instruction has the *submodule_identifier* as it's only operand. The instruction has four formats, which involves placing underscores before the submodule identifier. An activation column may include submodule identifiers belonging to mixed classes, and thereby exhibit MIMD parallelism. In the current implementation, the column that succeeds an activate column, may not generally commence, until all of the submodules in the preceding column have halted on their inputs.

The placement of *two* underscores before the submodule identifier in the topmost element of the activate column, signals to the compiler, that the succeeding column may commence, as soon as only the topmost submodule has halted.

By seeding the topmost activate instruction, with inputs that are known to halt later than the inputs of all of the remaining submodules, the use of double underscores entails that a busy wait mechanism that is costly in terms of time and space, does not have to be written by the compiler. This facility is useful if the activate column is particularily deep. In the next compiler version, the need for double underscoring will be bypassed (see 6.14.3), with the introduction of Dataflow Model type mechanisms to regulate the activation of base-line instructions.

| Activate formats | Examples of activate instructions. |
|---|---|
| `_smlabel` | `_mod, _neqz` |
| `_smlabelarray_indentifier` | `_par32array[1][0], _nceq5bit[i]` |

The presence of the same submodule identifier in an activation column, would result in an attempt to activate the same submodule twice in a machine cycle, and generate a jump failure. The compiler can detect identifier repetition, and signals an error.

## 6.8 COND.

The cond instruction fulfills the role of the "if..else.." control structure in imperative languages. The 'cond' is separated by an underscore from the first operand, which is either a primary or secondary identifier of a copy operand expression, with the restriction that the identifier designates a single bit storage entity. In practice, this is often output bit of a submodule.

The activation of a cond instruction involves the inspection of the bit, and the selection of either of the second or third operands pairs, which take the form (line address, offset). The operand pairs are separated from themselves, and the storage operand, by a single space character. The selection of an operand pair, activates the addressed line, and all line addresses up to and including the addition of the offset to the line address. Offsets may be larger than the register width of 32 bits. Neither line address nor offset can be incremental expressions, and must be constant integers. A cond column is always the terminating column of a base-line. In pre-expanded code, a cond column only contains one element.

### Cond formats

```
cond_primaryidentifier (lineaddress, offset) (lineaddress, offset)

cond_secondaryidentifier (lineaddress, offset) (lineaddress, offset)
```

### Examples of cond instructions.

```
cond_firstinteger.1 (2,0) (3,0)  ,  cond_firstinteger.i (2,7) (3,15)

cond_neqz.output (2,0) (3,0) , cond_par32array[1][i].output.13 (2,7) (3,15)
```

## 6.9 JUMP.

A *jump* instruction is akin to a GOTO command in Fortran and Basic. It has one operand pair of a line address and offset separated by a space character, which specifies a set of lines to be simultaneously activated without selection. Neither line address nor offset can be incremental expressions, and must be constant integers. A jump column is always the terminating column of a base-line, and may contain a number of jumps.

There is only one jump format.

| Jump format | Examples of jump instructions. |
|---|---|
| `jump (lineaddress,offset)` | `jump (2,0) , jump (15,63)` |

## 6.10 SKIP.

In a set of co-active lines, only one line may contain a skip column, which is called the *carry line*. The final column of a carry-line transfers control to the next state in the module's state transition system. A skip instruction has a single line address operand, without offset, which refers to another co-active line, called a *skip line*. The instruction tells the carry line to wait until the skip line has terminated, before proceeding to execute the next column of the carry line. A skip column with more than one skip instruction, will wait for all the skip lines to terminate, before proceeding to the next column.

Each skip line has a designated element in a bit array called *skipbits*, which indicates when the skip line is active during runtime. There are two important restrictions on the use of skip instructions:

i. A skip line cannot terminate with a jump or cond instruction, because it is the carryline containing the skip instruction, which transfers control to the next stage of the program (the next set of co-active lines).

ii. For implementational reasons, the current compiler requires that the only the deep construct-line can be a skip line, and in all other cases has to be a base-line.

There is one skip format.

| Skip format | Example of skip instruction. |
|---|---|
| `skip(lineaddress)` | `skip(2)` |

### 6.11 Wait.

The *wait* instruction is used as a fine grained synchronisation mechanism. It delays the activation of the succeeding baseline column, by the number of Synchronic A-Ram machine cycles specified by a single integer operand. There may only be one wait instruction per column.

| Wait format | Examples of subhalt instruction. |
|---|---|
| `wait(number_of_cycles)` | `wait(2), wait(256)` |

### 6.12 Subhalt and Halt.

A subhalt instruction marks where the code within the dependent lines of a construct-line halts. A subhalt column may only have one instruction 'halt' in lower case characters, and has the construct-line's line address as the single operand. It is always the terminating column of a dependent base-line. There is only one subhalt format.

| Subhalt format | Examples of subhalt instruction. |
|---|---|
| `subhalt(lineaddress)` | `subhalt(2), subhalt(3.1.2)` |

To signal the termination of the entire module, the halt instruction is used. A meta-module has two halt instruction formats, and an ordinary module has one. The halt instruction for the only phase, or first phase of a meta-module, is the expression 'HALT. The halt for the second phase of a meta-module, is the expression '-HALT'. A halt column may only contain a single instruction, and is always the terminating column of a base-line. A halt instruction may appear in more than one place in a module's code.

## 6.13 PROGRAM EXAMPLES.

Now that all of the base instructions have been introduced, program examples can be presented that illustrate their use. With one exception exhibiting co-active parallelism, the modules have trivial state transition systems, and are omitted. The first example is a level 2 implementation of Euclid's algorithm, which employs the modulus program as a sub-module.

Many implicitly supported features of C, have to be explicitly programmed in the current version of Space, and require the use of the programmable copy module, and register direct addressing. Both of these can haphazardly modify declared and undeclared storage, which can generate side effects and threaten referential transparency. It is intended that these facilities will only be used where necessary, to make a bridge between low-level and high level environments, and for memory allocation allowing a full panoply of high level programming features. They will be removed from Space, as soon as an appropriate level of compiler boot strapping has been reached.

To begin substantiating the claim that Space has the functionality of C, it is explained how memory allocation, the de-referencing of pointers, and the accessing of array elements whose indices are unknown at compile time, can be programmed. A sequential adder that employs miscellaneous features introduced in this chapter, concludes the section.

### 6.13.1 EUCLID'S ALGORITHM.

A level two module that implements Euclid's algorithm on two non-negative integers stored in *a* and *b* is presented in figure 6.8, where for simplicity the number stored in *a* is stipulated to be greater than or equal to the number in *b*. The greatest common divisor in *gcd* is obtained, by invoking the main loop described in line 2. The module uses the 32-bit and gate `paror32`, defined in 4.4.7, to serve as a test for not equal to zero, The output *gcd* is recovered from the penultimate cycle's remainder, which continues to reside in `mod.dividend`, obviating the need for temporary storage[60]. If *q* is the integer quotient of *a* and *b*, then the loop is invoked $q+1$ times. The code has no co-active parallelism. The only parallelism at the module's level of abstraction, occurs in the copy columns of lines 1 and 2.

[60] It is characteristic of Space programming, that submodule interfaces take on the role of temporary storage.

```
module euclid{
   storage{
       unsigned a input;  // a must be greater than or equal to b
       unsigned b input;
       unsigned gcd output;
   };
   submodules{
       paror32 neqz;
       modulus mod; // modulus is based on somewhat inefficient implementation
    };
    time: 1615-0 cycles;// min time is shown for a=b=1, max time is a long int, if a>>b=1

    code{

       1: b -> neqz.input    :: _neqz :: cond_neqz.output (3,0) (2,0) :;
          a -> mod.dividend
          b -> mod.divisor

       2: _mod :: mod.remainder -> neqz.input  :: _neqz :: cond_neqz.output (3,0) (2,0) :;
                  mod.remainder -> mod.divisor
                  mod.divisor -> mod.dividend

        3: mod.dividend -> gcd :: HALT :;  // transfer penultimate mod output to gcd

      };
};
```

Figure 6.8 Euclid's algorithm.

### 6.13.2 RETURNING AN ELEMENT OF AN UNSIGNED ARRAY .

At the current stage of compiler boot-strapping, there is no inbuilt support for accessing an element of an array, whose index is unknown at compile time. The `arrayreturn` module is an example in Space of a meta-module, which accesses an element of a one dimensional, unsigned integer array. In 6.6.5, reference was made to a category of tertiary copy identifier called the array aggregate parameter. The compiler replaces the aggregate parameter `arrayname[[0]]` appearing as the first copy identifier, with the address of a register, that contains the number of the first register that the array occupies in the memory block[61]. The module has an input `arrayddress`, whose contents have been copied from an aggregate parameter, using a copy of the form `arrayname[[0]] -> arrayreturn.address.`

[61] Since an array's first register address is known at compile time, it would have been easier in this instance for the compiler's read phase to replace the aggregate parameter directly with the register address. The aggregate parameters of blockstrings however, do need to be updated during runtime.

```
module arrayreturn{

   storage{
       unsigned address ioput;
       unsigned index input;
       unsigned value output;
   };
   submodules{
       adder32 adder;
       progsourcecopyreg pcopy;   // source programmable copy for a single register
   };
   meta(2);
   metatime: 0-0 cycles;
   time: 798-798 cycles;
   code{

      1: address -> adder.input1 :: _adder :: adder.output -> pcopy.source // cont below
         index -> adder.input0
                                   :: _pcopy :: HALT :;

      2: -pcopy :: pcopy.out -> value :: -HALT :;
    };
 };
```

Figure 6.9 Returning an element of a one dimensional unsigned integer array.

The first phase of the module is described in line 1, but has been split into two consecutive textual line segments for ease of presentation[62]. Together with the input array index `i`, the first phase of the module activates the first phase of the programmable copy sub-module, loaded with the appropriate input describing the register address of the `i`th array element. The second phase described in line 2, copies the contents of the `i`th element of the array, into the output `value`. The programming of array return as a meta-module, enables the time-expensive part of array return to be expressed as a first phase, which can be favourably dovetailed with other operations in problems, such as an LUT resolver (see appendix B).

The programming of array return modules for multi-register storage types, is assisted by replicator constructs, and will be deferred until the next chapter. The current inability to virtualise Space modules, contributes to it being necessary to write array return modules for each storage type (as well as each dimensionality of array).

[62] Unfortunately the compiler cannot recognise this style of representing long lines. Having a wide monitor screen assists Space programming considerably.

### 6.13.3 Memory allocation.

Lack of virtualisation and inbuilt support for memory allocation, also entails that modules have to be written for each type. The module `memalloc` uses the tertiary copy identifier category of register direct addressing, and the size of type operator. In 5.2, reference was made to the *storage block*, which is a segment of the memory block, one of whose roles is to act as a heap during runtime. Recall register 25165824 is set aside to store the next free register in the storage block. The contents of register 25165824 are transferred to the output using the register direct address mode, the new next free register value is calculated, and copied back into register 25165824. The code assumes `typename` has been added to the type library.

```
module typenameMalloc{ // name is descriptive only, and is not a compiler reserved name.
   storage{
       typename newpointer* output;
   };
   submodules{
       32adder adder;
   };
   time: 0-0 cycles;
   code{
        1: $typename -> adder.input0   :: _adder :: adder.output -> (25165824) :: HALT :;
          (25165824) -> adder.input1
          (25165824) -> newpointer
    };
 };
```

Figure 6.10 Memory allocation module, which retains a state in register `25165824`.

The module illustrates how memory allocation can be handled in Space programs running on the Synchronic A-Ram, but is problematic, because it cannot be used in parallel. The inclusion of multiple memory allocation submodules in a module, or of multiple submodules containing allocation modules in their submodule maps, might result in the register `25165824` being accessed and written to simultaneously by different submodules. This would generate write failures, side effects, and worse. An adequate treatment of parallel memory allocation, referred to as the *sidecar* in 5.3, will appear in a later implementation.

### 6.13.4 De-referencing a pointer to an unsigned integer.

De-referencing modules have to be written for pointers to each type. The presentation of de-referencing modules for multi-register storage types, is assisted by replicator constructs, and will be deferred until the next chapter.

```
module derefunsigned{

   storage{
       unsigned newpointer* input;
       unsigned value output;

   };

   submodules{

       progsourcecopyreg pcopy;

   };

   time: 0-0 cycles;

   code{

      1: newpointer -> pcopy.source  :: _pcopy :: -pcopy :: pcopy.out -> value :: HALT :;

   };

 };
```

Figure 6.11 De-referencing module for pointer to an unsigned integer.

### 6.13.5. Space version of serial 32-bit adder.

The level one `ADDER32` module in figure 6.12, uses the address operator, meta submodules, cell identifiers, co-active parallelism, activate column parallelism, and the skip instruction. The module is a Space version of the Earth serial-style adder that is described in appendix B. It uses a single fulladder submodule, for all 32 triplets of addend bits, addendum bits, and carry-in bits. The module employs an array of 3 programmable copy bit modules described in 4.5, and a 5-bit incrementer described in 4.2.2. The incrementer is used to modify the offsets of the first phase inputs of the `PCOPYBIT`'s submodules, whose second phase loads bit triplets into the fulladder.

Blank lines appear between constituent lines of line 1, in order to improve readability. Line 1 loads initial inputs into the `PCOPYBIT` submodules' first phases, which are then activated, in preparation for the first cycle of the main loop. Lines 2, 3, and 4 describe the main loop, where lines 2 and 3 are co-active, and line 4 is activated on the completion of the carry line 3. The first column of line 2 transfers the carryout from the previous cycle, and the *i*th bits of the addend and addendum, into the fulladder's inputs. When the fulladder has completed, it's carryout bit is copied into the carryin for the next cycle, and the sum bit is copied into the relevant bit of the module's output.

Whilst line 2 is active, the carryline 3 increments a 5 bit integer, which is then loaded into the inputs of the `PCOPYBIT` submodules' first phases. The carryline then has to wait for line 2 to finish by using the skip(2) instruction, which ensures that the second phase of pcopy[2] in line 2 is activated, before pcopy[2] is reprogrammed, by it's first phase being reactivated in line 4. Line 5 transfers the final cycle's carryout, into the carryout output bit of the module, signalling the end of the program.

The `ADDER32` module exemplifies various Space features, but is almost four times slower than it's equivalent, level zero, Earth module `adder32`. The disparity is due partly to an un-optimized implementation of the Space compiler, in particular lacking the *levelling* phase referred to in 6.3, and partly to overheads incurred by coding infrastructure for submodules, and busy wait mechanisms for submodule activations and column completions.

```
module ADDER32{

     storage{
       unsigned addend input;
       unsigned addendum input;
       unsigned sum output;
       BIT carrybit output;
    };

    submodules{
       FULLADDER fulladder;
       inceq5bit inceq;
       progcopybit pcopy[3];
    };
    time: 2714-2714 cycles;

    code{

        1: #0 -> pcopy[0].input0.offst          :: _pcopy[0]    :: jump (2,1) :;
           #0 -> pcopy[1].input0.offst             _pcopy[1]
           #0 -> pcopy[2].input1.offst             _pcopy[2]

           &addend -> pcopy[0].input0.destn
           &fulladder.a -> pcopy[0].input1
           &addendum -> pcopy[1].input0.destn
           &fulladder.b -> pcopy[1].input1
           &fulladder.s -> pcopy[2].input0
           &sum -> pcopy[2].input1.destn

        2:  -pcopy[0]  :: _fulladder  :: fulladder.cout -> fulladder.cin :: -pcopy[2] :;
            -pcopy[1]

        3:  _inceq :: inceq.ioput -> pcopy[0].input0.offst   :: skip(2) :: jump (4,0) :;
                      inceq.ioput -> pcopy[1].input0.offst
                      inceq.ioput -> pcopy[2].input1.offst

        4: _pcopy[0]  :: cond_inceq.Eq31 (2,1) (5,0) :;
           _pcopy[1]
           _pcopy[2]

        5:  fulladder.cout -> carrybit :: HALT :;

       };

};
```

Figure 6.12 32-bit serial-style adder with fulladder , inceq5bit, and programmable copy bit submodules.

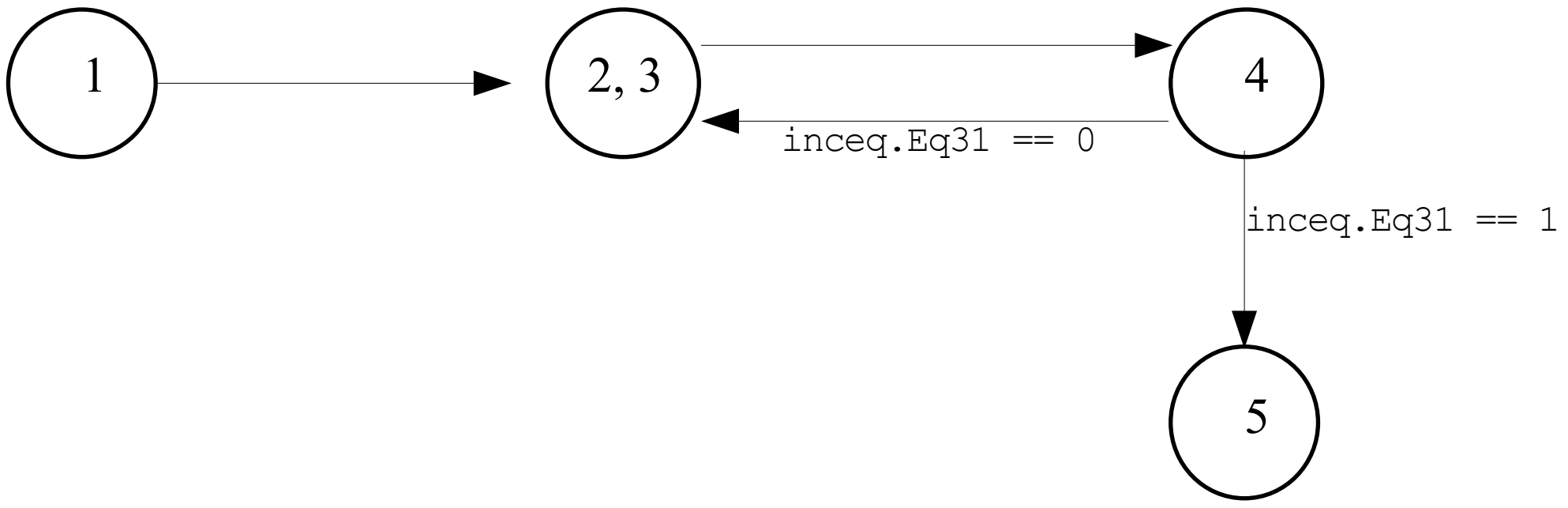

Figure 6.13 State transition diagram for ADDER32 module, exhibiting co-active parallelism.

## 6.13.6. SERIAL 32-BIT MULTIPLIER.

```
module mult32{    // multiplies a pair of 32 bit numbers, and indicates overflow

      storage{

         unsigned inputa input;
         unsigned inputb input;
         unsigned output output;
         BIT overflow output;

      };
      submodules{
         adder32 adder;
         shiftleft32 shiftleft;
         rightshift32 shiftright;
         inceq5 inceq;
      };
      time: 770-24145 cycles;

      code{

   1: #0 -> overflow            :: _shiftright :: cond_shiftright.remainder (2,0) (3,0) :;
      inputa -> shiftleft.ioput
      inputb -> shiftright.ioput

   2: #0 -> adder.input0 :: jump (4,0) :;

   3: inputa -> adder.input0 :: jump (4,0) :;
      inputa -> adder.output

   4: _shiftleft     :: cond_inceq.Eq31 (5,0) (8,0)  :;
      _inceq
      _shiftright

   5: cond_shiftright.remainder (4,0) (6,0) :;

   6: shiftleft.ioput -> adder.input1 :: _adder    //contd.
              :: adder.output -> adder.input0 :: cond_adder.carryout (4,0) (7,0) :;

   7: #1 -> overflow :: HALT :;

   8: adder.output -> output :: HALT :;

   };
};
```

Figure 6.14 32 bit multiplication.

The `mult32` module is a serial implementation of the multiplication of two 32 bit numbers. The submodules include 32 bit register left and right shift modules, with a remainder output to allow inspection of the lost bit. In addition, there is a 5 bit incrementer, and the `adder32` module. In the main loop expressed in lines 4,5 and 6, the second input is right shifted one bit at a time, whilst the first input is left shifted one bit at a time. Whenever a set bit is encountered in the second input, the shifted value of the first input is added to the final result. If an overflow is detected in the adder, the main loop is exited, and the main module's overflow bit is set, otherwise the main loop is invoked 31 times.

## 6.14 SYNCHRONISATION OVERHEADS ASSOCIATED WITH SPACE AND THE SYNCHRONIC A-RAM.

The Space base set, and the language of base-lines have now been introduced. A compilation phase called *expansion* was referred to in 6.3, whose output consists only of independent base-lines, which forms the input for code generation. Even though construct-lines will not be presented until the next chapter, it is possible now to look at synchronisation issues relating to Space, at the machine code level.

It was mentioned in 1.4, that one of the justifications for introducing asynchronism into parallel computing systems, is based on delays associated with centralised program control. A general criticism of globally clocked, parallel computing, exemplified by the BSP programming model for processor networks, is that the approach incurs various synchronisation overheads, that are not present in asynchronous architectures. A large computer with spatially centralised control, will be subject to extended transit times for commands to travel from the control area, to the furthest reaches of the machine.

Space programs are deterministic state transition systems, and program control is conceptually localised. Once the program has been compiled into machine code however, program control is largely spatially distributed, where instructions are processed in situ. Consequently the Space programmer benefits from localised control at program composition time, without having to pay all of the penalty of spatially centralised control at runtime. There are some issues which affect synchronous machines at runtime which appear to be unavoidable, and apply to the Space programs running on the Synchronic A-Ram.

Synchronic A-Ram instructions may read or write to any bit in the memory block in unit time. Therefore synchronisation costs for compiled Space programs are mitigated by the models' high internal connectivity. Nevertheless, there are mechanisms, including the determination of multiple sub-program/thread termination, which need discussion. The term *thread* will temporarily be restricted to mean the process defined either by an instruction in a base-line column, or defined by a base-line. There are three possible sources of overhead, associated with Space's deterministic parallelism.

i. Space and time costs of initiating *n* threads.
ii. Space and time costs of determining the termination of all of *n* threads, which is analogous to the costs of barrier synchronisation in the BSP model.
iii. Overheads incurred by column parallelism in comparison with the Dataflow Model, in particular the restriction that a succeeding base-line column can only be activated, when all of the instructions in it's preceding column have completed

### 6.14.1 MULTIPLE THREAD INITIATION.

The initiation of a single thread requires a single jump instruction, occupying one register in the Synchronic A-Ram $\langle 5,\sigma,\{1,2\},\eta \rangle$. To trigger *n* threads, the compiler builds a *jump tree*, comprising $\overline{\left(\frac{\log_2 n}{5}\right)}$ layers. The final layer is composed of *n* jumps, which activate the threads. Previous layers consist of a series of jump instructions with maximal offsets, which are involved in activating the succeeding layer, the first layer consisting of a single jump. The cycle time to initiate *n* threads in Space is the logarithmic function $\overline{\left(\frac{\log_2 n}{5}\right)}+1$.

The space requirement (number of registers) for the jump tree, is the function *jumptree*(*n*), which is somewhat similar to a convergent geometric series. The function may be expressed as a finite summation series, composed of $\overline{\left(\frac{\log_2 n}{5}\right)}+1$ terms:

$$jumptree(n) = n + \overline{\left(\frac{n}{32}\right)} + \overline{\left(\frac{\overline{\left(\frac{n}{32}\right)}}{32}\right)} + \ldots + 1$$

Since this function is clearly much less than *2n*, it may be concluded that the costs of initiating *n* threads are modest[63].

### 6.14.2 MULTIPLE THREAD TERMINATION.

Jump, halt, and cond columns do not have succeeding columns, and the issue of detecting multiple termination of component instructions does not arise. Copy columns are implemented in a maximally parallel manner, and have minimal execution times known at compile time, dependent on the number of instructions in a column. The scheduling of a succeeding column's activation is therefore straightforward. The remaining column types have variable execution times however, and there is a need for a compiled module segment, called a *barrier*, which can determine multiple thread termination. There are two cases when a barrier is required.

i. In the current implementation, any column succeeding an activation column, cannot be executed, until all of the activation column's submodules have terminated. The compiler writes a barrier, that determines when all of the submodules' busybits have been reset.
ii. A skip column imposes a need to determine when multiple base-lines have ceased activity. The compiler adds a storage entity called *skipbits*, consisting of an array of bits, each bit indicating whether a particular skip line is active. To implement the skip operation, the compiler writes some extra base-line code, and a barrier that determines when all of the skipbits have been reset.

Activation columns, and skip columns resulting from replication, may contain tens of thousands of elements. A barrier is implemented by an array of 8-input and gates. A prefix circuit performs a global boolean operation on the array of input bits. The barrier loops until all busybits/skipbits have been reset, and will generally detect multiple thread termination within a few tens of cycles of it actually happening[64] .

The register size of a barrier for *n* threads is approximately equal to 14*n*, where *n* is less than 1024. The costs of determining thread termination are larger than for thread

[63] A direct comparison with existing parallel systems is not relevant here, because the $\langle 5,\sigma,\{1,2\},\eta \rangle$ machine is a simulated, low level, mathematical model of computation.

[64] It must be admitted that the current implementation is inefficient, and is susceptible to substantial improvement: it can take 75 cycles to determine the simultaneous termination of 1024 threads.

initiation, but are still manageable.

It is possible to bypass the need for the compiler to write a barrier for an activation column, if it is known that the first submodule terminates no later than the column's other submodules. The use of a double underscore before the topmost submodule in an activation column, instructs the compiler to trigger the subsequent column when the first submodule of the column has ceased activity, regardless of when other submodules terminate. The double underscore can be made to work without generating a race hazard, by selecting or loading inputs for the first submodule, which are known to require an execution time greater than or equal to the times for all other inputs.

### 6.14.3 COSTS INCURRED BY COLUMN PARALLELISM IN COMPARISON WITH THE DATAFLOW MODEL.

In the first attempt at implementing the spatial environment, the separation between base-line columns by the characters '**::**', indicates that the instructions in a succeeding column do not begin executing, until all of the preceding column's instructions have finished executing. In the Dataflow Model (DM) described in 2.3.3 however, functional nodes in a dataflow are not organised into columns, but are activated when all input values have been transferred into the node's input locations. This has the benefit that a functional node in a deep and wide dataflow, can be made to activate independently of, and without having to wait for, the completion of other functional nodes, not involved in supplying inputs to the node.

The current implementation will not execute as efficiently as a DM machine, because in a base-line incorporating a succession of copy and activation columns, some activation column's instructions having received all of their inputs, might needlessly wait until all of a preceding activation column's instructions have finished. This will sometimes result in the base-line's execution terminating later than necessary, which will be referred to as *column delay*. The issue is now considered of whether the overall suitability of interstrings for expressing the execution of dataflows on the Synchronic A-Ram, is compromised by column delay.

The elimination of column delay is straightforward, because each base-line column instruction is only used once per base-line activation. No consideration needs to be given to the problem of controlling multiple passes through the base-line dataflow, which affects the Dataflow Model, and is dealt with by multiple tokens, buffers, and either demand or data driven mechanisms. Column delay can be resolved by using a busy wait that signals the completion of the receipt of all of a submodule's inputs, called a *loading wait*, an approach which borrows elements from both demand and data driven mechanisms.

For base-lines that represent dataflows, the base-line activation simultaneously triggers the elements of a copy column followed by an activate column, or simultaneously triggers the elements of an activate column. In the proposed solution, the line activation would also trigger a collection of simultaneously running loading waits for individual submodules in the subsequent activation column, if it exists.

An individual submodule completion in the subsequent column, would then trigger another loading wait for those submodules in the following activation column (if it exists), that accept inputs from the submodule in the previous layer. A loading wait need only be triggered by the termination of one of the previous layer's submodules supplying inputs. Similar to the Dataflow Model, submodules may thus be activated, independently of the completions of other submodules, not involved in supplying inputs to the submodule.

With this approach, the columnar format of base-line interstrings becomes a scheme for representing dataflows, rather than a dogmatic prescription for the scheduling of submodule activations. The advantages of implicit resource allocation/scheduling of submodules, and the explicit parallelism associated with an interlanguage environment can be retained, whilst matching the efficiency of the Dataflow Model's parallelism. In 8.4, it is argued that the Synchronic Engine should include hardware support for loading waits, as well as for the detection of multi-thread termination.

# Chapter 7
# SPACE CONSTRUCT SET.

## 7.1 CONSTRUCTS.

Having laid out the base-line formats, the set of five construct-line formats, called the *construct set*, and the system for composing them, are now presented. The construct set includes two different ways of replicating code, while and do-while constructs, and switch selection. No claim is made that the approach described here, is the most appropriate way of embedding high level features into interlanguages for synchronous models of computation. The intention of this chapter, is rather to present prototype concepts and program examples, which clearly demonstrate that an interlanguage is capable of expressing a wide range of massively parallel programming styles. The next chapter addresses the feasibility of high performance architectures derived from the Synchronic A-Ram, and of a discussion of how a comparison of the performance of Space-derived languages with multi-threaded programming models for processor networks might be arrived at.

The structure of a composite construct-line with it's dependent lines, is a tree whose nodes are construct-lines, and whose leaves are base-lines. A dependent line is said to be an *immediate* line of a construct-line, if that construct line is it's direct parent in the construct tree[65]. With some exceptions, a construct-line format may have as it's immediate lines, a base-line or any other construct-line format.

The line address of a dependent line encodes it's position in the construct tree. Line addresses provides a convenient means of directly expressing the highly variable structure of construct trees, bypassing the need for the compiler to conduct tree parsing. If readability were not a concern, line addresses would also remove the necessity of placing a construct-line's dependent lines adjacent to each other in the module's text.

During expansion, a construct-line generates a potentially massive collection of base-lines. A module's expanded code may be characterized as a larger state transition system, if the module does not employ the *grow* construct[66], to be described in 7.7. An independent construct line represents a sub-program of the expanded code, with at least one entry, and at least one termination point.

[65] The line address of an immediate line, always consists of some integer appended to the line address of it's parent construct line.

Each of the five construct formats specifies a single immediate base-line or construct-line with postfix line address ".1" as the first to be executed, upon entry. Other than the *deep* and *switch* constructs, the termination of the sub-program represented by the construct-line, is signalled by a subhalt instruction. A construct format also features an optional line address and offset pair, called an *egress*, to indicate where program control is to be transferred to upon the construct-line's termination. An egress associated with a dependent construct-line, may be used to transfer control to other dependent lines. It can also be used by independent carrylines to transfer control to another co-active set. I next consider how construct trees are stored and processed by the compiler.

## 7.2 The expansion line.

In contrast with other imperative languages, Space's base-lines (roughly analogous to statements), can be entirely separated from construct-lines (roughly analogous to program constructs), without loosing syntactic information. Base-lines are amalgamated during the read phase into a collection, called the *base-line list*. All of the module's construct lines are gathered into a string of strings of construct-lines, called the *expansion line*. The expansion line integrates the module's construct trees into one interstring-like representation.

Trees are normally inductively defined, which encourages the use of recursive procedures. The representation of an expansion line as a string of strings structure, removes any need for recursion in the processing and compilation of Space's programming constructs. An opportunity is also created to parallelize expansion by accessing columns and column elements simultaneously, which can be exploited when a native version of the Space compiler is written. An expansion line is exemplified in figure 7.1, which assumes that the module contains two independent construct-lines. The first has line address 1, with one dependent construct-line, and the second has line address 2, whose construct tree was exemplified in figure 6.1.

[66] Grow constructs confer greater programmability, but at the costs of enlarging the expanded code's state space, and of interfering with the notion of identifying a state with a set of co-active lines.

| | | | | |
|---|---|---|---|---|
| 1.1: construct-line | :: | 1: construct-line | :: | 2: construct-line :; |
| 2.1.1: construct-line | | 2.1: construct-line | | |
| | | 2.3: construct-line | | |

Figure 7.1 Simplified view of a sample expansion line, where 2.2 is deemed to be a baseline.

The leftmost column elements have base-lines as their only dependents. The rightmost column contains only independent construct lines, the rightmost but one column contains immediate dependents of the elements of the rightmost column, and less complex independent construct-lines. The base-line list and the expansion line are the inputs to the compiler's expansion phase, which involves line address changes, the modification of existing lines, and the creation of new lines.

An iterative procedure is performed, which works through the columns of the expansion line from right to left. In the rightmost column, a independent construct-line undergoes a process in which it is removed, and the dependent construct-lines and base-lines, are transformed into an expanded collection of less complex, construct-lines, and potentially larger base-lines.

When all of the rightmost column's construct-lines have been processed and the column itself can be removed, the same procedure is applied to the new rightmost column. The procedure is repeated until all of the construct-line columns have been removed, generating a potentially vast collection of base-lines.

## 7.3 Construct set.

The construct set contains five elements. The following sections describe the construct-line formats, which are summarised below.

i. **Deep.** The deep construct has a base-line as it's only dependent line. As exemplified in 6.1.2, it defines a vertical replication of base-line code, in which a control variable is modified.
ii. **Grow.** Grow is a powerful construct that can be applied to a multi-line sub-program, and replicates the entire sub-program, in which a control variable is modified. Grow allows fully programmable SPMD parallelism within the module's level of abstraction, and can also be involved when a reduction in the size of the submodular hierarchy is required, specified by the submodules contraction declaration described in 5.6.

iii. **While.** Operates as standard while construct, which tests the contents of a single bit storage entity, and either executes the first line of a multi-line sub-program body of the construct, whose exit is directed to repeat the test, or exits to a specified line address and offset. The entry line is the first immediate line.

iv. **Do-while.** Operates as standard do while construct, which executes the body of construct, then tests a bit and either repeats the body, or exits to a specified line address and offset.

v. **Switch.** Similar to a C language switch construct, it can select between a variety of cases in one step. It may currently only be applied to positive integer cases between the range 0-15.

The first two construct formats are replicative in nature. They include a control variable called a *replicator*, which is modified between a left and right limit, according to an incrementor function. A replicative construct duplicates dependent lines in a particular way, whilst inserting the current control value into every instance of the replicator in the dependent lines.

Recall the replications declaration described in 5.7, which listed various terms to be used in replicative construct lines. The declaration began with the alphanumeric names of the module's replicators. These were followed by a list of the compiler's pre-defined incrementor functions employed that are used in the module's constructs, called *indexical functions.* Further recall an *indexical expression* takes the form `replicator` or `replicator/indexical function`. Indexical expressions are used to describe left and right limits in replicative construct formats, as a function of the control value of the replicator. A *comparator* states a relation between two integers or indexical expressions, and is drawn from the list ">=", "<=", "<", and ">".

Recall that indexical functions, and their associated expressions, take the following forms. Let the control value of the replicator be *i*.

i. "*`replicator`*", or "*`replicator/`*`id`". Identity, returns *i*. The identity function does not need to appear in the replications declaration.

ii. "*replicator/inc*". Incrementor function, returns (*i+1)*.

iii. "*replicator/plus2*". Add two function, returns (*i+2)*.

iv. "*replicator/dec*". Decrementor function, returns (*i-1)*. The compiler flags an error for zero input.

v. "*replicator/2**". Multiply by two function, returns (2**i)*.

vi. "*replicator/2*+1*". Multiply by two plus one function, returns ((2**i) +1)*.

vii. "*replicator/2^*". Power function, returns (2^*i)*.

viii. "*replicator/div2*" Divide by two function, returns the bit representation of the integer shifted once to the right.

## 7.4 DEEP.

The deep construct can express SIMD and a limited form of SPMD style parallelism, by vertically replicating a single baseline, where a control variable is modified between left and a right limit. The format has an egress, which transfers control to the lines specified by the format's rightmost line address and offset, upon completion of the execution of the expanded base-line. It is left as an empty pair of brackets if no egress is to be specified. For ease of presentation, let *ln_add* stand for line address, *rep* for replicator, *ind_fun* for indexical function, *ind_exp* for indexical expression, and *cmp* for the comparator.

In common with all construct formats, the deep construct is to the right of it's first immediate dependent base-line, separated by the construct brace ':>'. The deep construct format takes the following form.

```
ln_add: deep< rep = ind_exp1; rep cmp ind_exp2; ind_fun >  (ln_add,offset)
```

(line address) (left limit) (right limit) (incrementor) (egress)

Examples of deep constructs, attached to dependent base-line.

```
1.1 _adder[i] :: adder[i].output -> output[i] :> 1: deep<i=0; i<=65535; inc > (4,0)

2.1.1: array[j] -> array[value/inc] :> 2.1: deep<value = 31; value >= j/inc; dec> (2.2,10)

7.1 #j/2* -> seededinput[j] :> 7: deep<j=0; j<=1023; inc> ()
```

Note that if a replicative construct-line is dependent on another replicative construct-line (not shown) as in the second example above, then the latter's replicator may appear in the dependent construct-line's left or right limit indexical expressions. I now consider a toy parallel program in Figure 7.2. The module `bigaddition` employs a massive array of submodules belonging to the `adder32` class, which adds two 32-bit integers, and has a maximum running time of 736 cycles. The module completes 65,536 simultaneous additions in 759 cycles. The module seeds the `adder32` submodules with distinct addends and addendums, and outputs the results into a register array. Please note that the current compiler cannot recognise the style of representation below, where for reasons of space, a construct-line appears textually below it's first immediate line.

```
module bigaddition{
  storage{
    REG output[65536] output;
  };
  submodules{
    adder32 adder[65536];
  };
  replications{i / inc, 2*};
  time: 759-759 cycles;

  code{  //

    1.1: #i -> adder[i].input0      :: __adder[i] :: adder[i].output -> output[i]  //cntd
         #i/2* -> adder[i].input1

                 :>  1: deep<i=0; i<=65535; inc > (2,0) :;  // attached to baseline above

     2: HALT :;
   };
};
```

Fig 7.2 A module with a submodular array of 65,536 32-bit adders, each seeded with distinct inputs

The `adder32` submodule is known to have the maximum running time, when given zero value inputs. Recall that the use of a double underscore before the topmost submodule in an activation column, instructs the compiler to trigger the subsequent column as soon as only the first submodule of the column has ceased activity, independently of when the other submodules terminate. As the first adder is given zero inputs, the double underscore can be used to bypass the need to write a barrier to detect the termination of 65, 536 adders, which would be very costly in terms of code and running time[67]. Care must be taken that the topmost submodule does not terminate before the other submodules when using the double underscore, otherwise the activation column may continue to be busy, when it's succeeding column is activated.

[67] Even with the use of the double underscore, `bigaddition` takes approximately half an hour to compile, and two and a half hours to simulate on a single core 1.5 GHz processor.

The construct-line has a single replicator name '`i`', whose expansion is depicted in 7.3, in which the dependent base-line has been vertically replicated. The expansion involves incrementing the control value of the replicator from 0 through to 65,535, generating a baseline, whose first three columns contain 131,072, 65,536 and 65,536 instructions respectively.[68] Since the egress is specified, expansion adds a final jump column to perform the transfer of program control. The new base-line is given the construct-line's address, and the construct line is discarded.

```
1: #0 -> adder[0].input0   :: __adder[0] ::  adder[0].output -> output[0] :: jump (2,0) :;
   #0 -> adder[0].input1       _adder[1]     adder[1].output -> output[1]
   #1 -> adder[1].input0       _adder[2]     adder[2].output -> output[2]
   #2 -> adder[1].input1       _adder[3]     adder[3].output -> output[3]
   #2 -> adder[2].input0       _adder[4]     adder[4].output -> output[4]
   #4 -> adder[2].input1       _adder[5]     adder[5].output -> output[5]
         .                    .                    .
         .                    .                    .
#32767 -> adder[32767].input1  _adder[65535]  adder[65535].output -> output[65535]
         .
         .
#65535 -> adder[65535].input0
#131070 -> adder[65535].input1
```

Fig 7.3 Result of expanding the construct-line in `bigaddition` module.

The deep construct-line and dependent base-line, succinctly describe a massive dataflow, and implicitly allocate machine resources to compute the dataflow. The inclusion of a skip column in the dependent baseline of a deep construct, is currently not allowed. The inclusion of a *wait* column is however, and results in the other columns of the baseline being replicated normally, whilst the wait column is left untouched. *Wait* is used in the next program example.

[68] The double underscore is not copied beyond the first element of the activation column.

## 7.5 MULTI-LINE DEEP CONSTRUCTS.

To program grid based computation in a sequential language, such as a multi-dimensional cellular automata (CA), normally requires the use of two or more control variable names, in order to express the multiple grid co-ordinates of an internal cell of the automata. To implement a CA in Space, a way of vertically replicating a baseline, subject to the modification of multiple replicators is needed.

A natural way of achieving this might be to 'nest' deep constructs, involving the composition of a series of deep constructs. However, because a deep based replication always has a single dependent baseline, nesting is achieved through a *multi-line* deep construct, in order to compress the textual width of code, and improve readability. The multi-line format is similar to the single-line format, where additional internal constructs may be added on consecutive textual lines, placed between the cursor positions occupied by the line address and egress:

```
ln_add: deep< rep1=ind_exp1; rep1 cmp1 ind_exp2; ind_fun1 > (ln_add,offset)
        deep< rep2=ind_exp3; rep2 cmp2 ind_exp4; ind_fun2 >
                                   .
                                   .
                                   .
```

Example of multi-line deep construct, attached to the dependent base-line.

```
3.1: array[j][i/inc] -> adder[j][i].input0   :>   3: deep<i=0; i<=31; inc>  (5,5)  :;
      array[j/inc][i] -> adder[j][i].input1           deep<j=1 ; j<=30; inc>
```

### 7.5.1 A PROGRAM WITH MULTI-LINE DEEP CONSTRUCTS: A SIMPLE CELLULAR AUTOMATA.

The module `cellautomata` in Figure 7.4(a) and Figure 7.4(b) implements a two-dimensional automata using various multi-line deep constructs. The CA has a very simple behaviour, but it's code exemplifies how CA's computations and pattern of internal communications, can be efficiently handled in Space.

Dataflow that can be represented by a single deep construct, such as in fig 7.3, is limited in being able to express complex interactions between replicated components.

`Cellautomata` illustrates one justification for co-active parallelism, because of the cellular automata's requirement for the simultaneous activation of subprograms represented by differing deep constructs.

The `adder2` submodule is a restricted form of `adder32`, where the inputs are two bit positive integers, and the output is a 3 bit integer. The `rem2` submodule acts as modulus two operation, i.e. determines the parity of it's single integer input. In the program example, the user has to supply the initial states of the `array[33][33]` grid, and the number of cycles the automata is to compute.

It is assumed that on the boundary of the array there is a one-cell thick cordon of cells, whose states are not updated. These boundary cells are capable of transmitting their constant state in one direction only, to inner cells. The corner cells do nothing, and it is only the inner 31x31 cells whose states are updated. The new state of an internal cell is given by the arithmetic expression:

$$\text{array}[i][j] = \left(\text{array}[i-1][j] + \text{array}[i][j-1] + \text{array}[i+1][j] + \text{array}[i][j+1]\right) \bmod 2$$

```
module cellautomata{

  storage{
       unsigned array[33][33] input;
       unsigned cycles input;
  };

  submodules{
       adder2 adder[32][32];
       rem2 rem[31][31];
       inceq32 inc;
       equal32 equ;
  };

  replications{i,j / inc, dec};
  time: 0-0 cycles; // time dependent on cycles input, illustrating limitation of
                    // current time declaration.
  code{
              // code appears in figure 7.4(b)
  };

}:
```

Fig 7.4(a) Declarations of a simple cellular automata for updating the internal 1024 cells of a 33x33 array.

```
code{

   1: #0 -> inc.ioput        :: jump (2,2) :;  // initialise incrementer and equaliser
      cycles -> equ.input0

 2.1: array[0][i/inc] -> adder[0][i].input0  :>  2: deep<i=0; i<=30; inc > ()  :;
      array[1][i] ->   adder[0][i].input1

 3.1: array[j][i/inc] -> adder[j][i].input0   :>   3: deep<i=0; i<=31; inc> (5,5)  :;
      array[j/inc][i] -> adder[j][i].input1           deep<j=1 ; j<=30; inc>

 4.1: array[31][i/inc] -> adder[31][i].input0   :>  4: deep<i=1; i<=31; inc> ()  :;
      array[32][i] -> adder[31][i].input1

// top row of computations
 5.1: _adder[0][i] :: adder[0][i].output -> adder[0][i].input0 :: wait(82)    //contd

       :: _adder[0][i] :: adder[0][i].output -> rem[0][i].input  :: _rem[0][i]  //contd

        :: rem[0][i].output -> array[1][i/inc]      :> 5: deep<i=0; i<=30; inc> () :;

 // internal computations
 6.1: _adder[j][i] :: adder[j][i].output -> adder[j/dec][i/dec].input1  //contd
                     adder[j][i].output -> adder[j][i].input0

        :: _adder[j][i] :: adder[j][i].output -> rem[j][i].input :: _rem[j][i] //contd

         :: rem[j][i].output -> array[j/inc][i/inc] :> 6: deep<i=1; i<=30;inc> (11,0) :;
                                                            deep<j=1; j<=30;inc>

 // leftmost column of computations
 7.1: _adder[j][0] :: adder[j][0].output -> adder[j][0].input0 :: wait(82)  //contd

       :: _adder[j][0] :: adder[j][0].output -> rem[j][0].input :: _rem[j][0] //contd

       :: rem[j][0].output -> array[j/inc][1] :> 7: deep<j=1; j<=30; inc> () :;

 // rightmost column
 8.1: _adder[j][31] :: adder[j][31].output -> adder[j/dec][30].input1   //contd

       :> 8: deep<j=1; j<=30;inc> () :;

// bottom row
  9.1: _adder[31][i] :: adder[31][i].output -> adder[30][i/dec].input1   //contd
              :>   9: deep<i=1; i<=31; inc> () :;

  10:   _inc :: inc.ioput -> equ.input1 :: _equ  :;
  11:   cond_equ.output (5,5) (12,0) :;
  12: HALT:;

};
```

Fig 7.4(b) A cellular automata which updates the internal 1024 cells of a 33x33 array.

A cell's new state is computed by a parallel dataflow, composed of two simultaneous additions of its western/northern, and eastern/southern neighbours, whose outputs are then added together and the modulus taken. The output of the `adder2` submodule computing an internal cell's first addition, is copied back into the submodule's first input, and into the second input of it's north-western neighbour in the submodular array (line 6). The module thereby efficiently shares the results of identical additions across the 33x33 cellular array, requiring only a 32x32 `adder2` array, whilst retaining the maximisation of parallelism with respect to the module's level of abstraction.

The module's state transition diagram is depicted in figure 7.5. The majority of the lines deal with boundary conditions. The module starts by loading some initial values into the `inceq32` and `equal32` submodules. The cellular cycle begins with the co-active set comprising lines 2, 3 and 4, which loads the array's states into the inputs of the `adder2` array. Lines 2 and 4 load the top and bottom rows of the `adder2` array respectively. Line 3 is a multi-line construct which loads the internal rows, and has the deepest columns because the baseline is replicated 928 times. It will require a deeper jump tree to initiate copies, and will have the longest execution time. It is therefore given the role of the carry line.

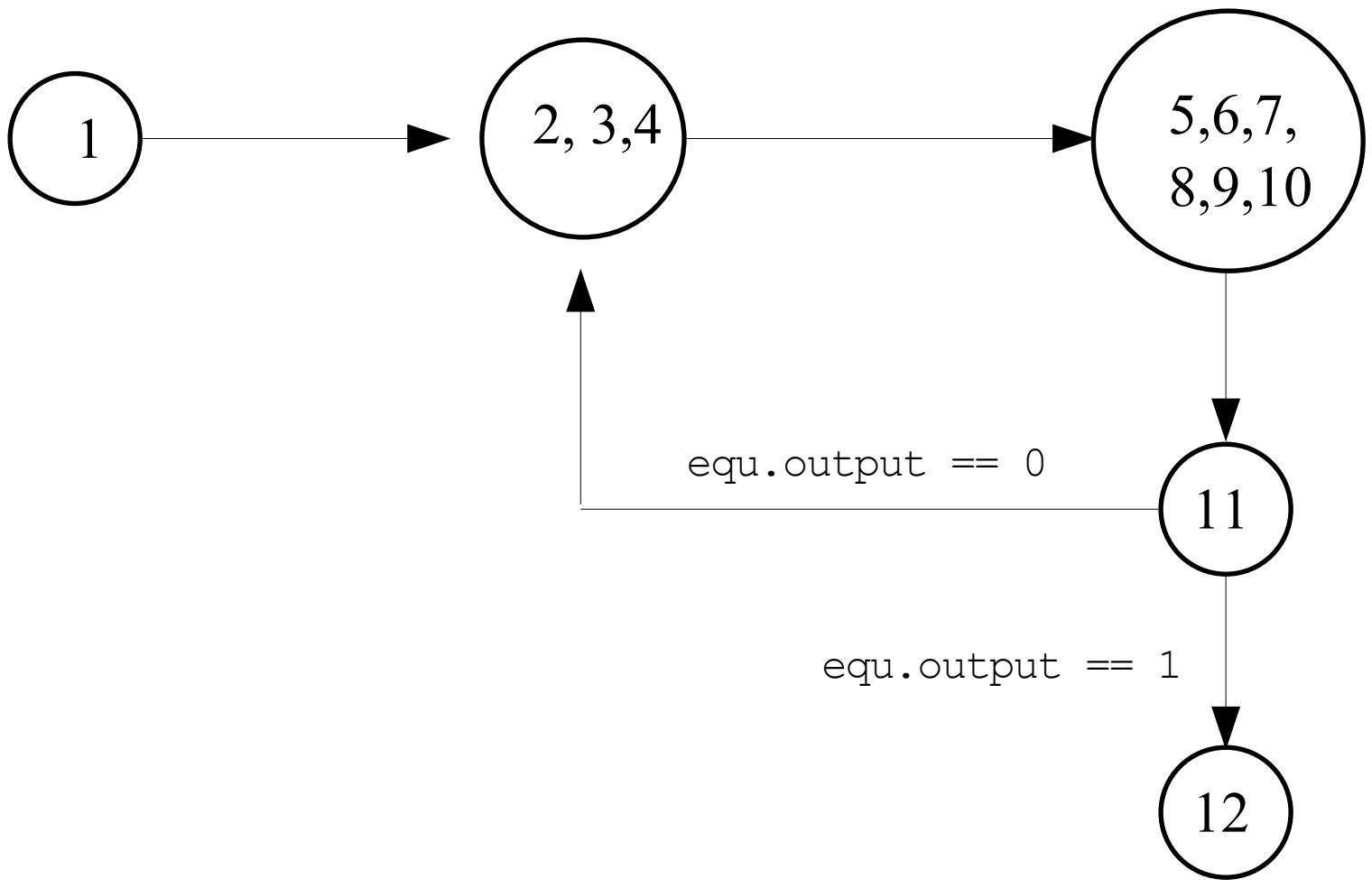

Fig 7.5 State transition diagram for `cellautomata`.

The next co-active set is composed of lines 5, 6, 7, 8, 9, and 10, computes the cell's new states, where line 10 increments a counter representing the number of cycles the automata has completed, and determines the result of comparing it with the input `cycles`. The network of computations and copying of results takes up the other five lines, where four of the lines deal with boundary conditions.

The new value of `array[i][j]` is received from the output of `rem[i-1][j-1]`, whose input has been received from the second addition performed by the submodule `adder[i-1][j-1]`. Line 5 regulates the activity of the leftmost 31 elements of the top row of the `rem2` submodules, and of the `adder2` submodules, which perform two additions. Line 7 regulates the leftmost column of `rem2` and `adder2` submodules with two additions, barring the northwestern corner, which is handled by by line 5. Line 8 regulates the activity of the rightmost 31 elements of the bottom row of the `rem2` submodules, and of the `adder2` submodules, where only one addition is required. Line 9 regulates the bottom column of `rem2` and `adder2` submodules with one addition.

The states of the remaining internal 29x29 array, are computed by line 6, which is the designated carryline, because it is replicated the most at 841 times. Another factor contributing to line 6 terminating later, is the need to implement barriers for 841 threads in columns 1, 3 and 5, whereas the barriers for lines 5 and 7 will terminate much earlier for determining the termination of 29 and 31 threads respectively.

Consequently, after the first column, the execution of line 6's columns will drift out of phase with respect to those of lines 5 and 7. Unfortunately, this has an undesirable consequence that the second activation of an adder in column 4 in lines 5 and 7, will not have received their second inputs from the first adding operations in line 6, unless a synchronisation mechanism is used. Therefore a wait of 82 cycles is specified after the first addition in column 3 of lines 5 and 7, in order to ensure all inputs are in place in time for the second addition activation. The wait instruction is a somewhat crude, low level mechanism, and the next implementation will upgrade the skip instruction to be able to identify the termination of a specific column in a baseline, as well as the termination of a baseline itself.

The next co-active set is comprised of line 11, which tests the output of `equal32` to determines whether the module is to be halted in line 12, or whether another cellular cycle is to be executed.

## 7.6 TWO PROGRAM EXAMPLES: PARALLEL ADDITION OF INTEGER ARRAY, AND MAXIMALLY PARALLEL MATRIX MULTIPLY.

The module `add32array` depicted in fig 7.6, sums all 32 elements of an unsigned input array as a parallel prefix tree in 5 cycles of addition, using an array of 16 adders. In the first line, all of theadders are used, whereas only one adder is needed in the final line. In 7.8, it is explained how the repetitive nature of the code may be avoided, (useful for summing very large arrays) by using grow replication and a special module called the programmable jump.

```
module add32array{ //adds all elements of array containing 32 integers
  storage{
    unsigned A[32] input;
    unsigned sum output;
  };
  submodules{
    adder32 add[16];
  };

  replications{ i / inc, 2*, 2*+1};
  time: 0-0 cycles;

  code{

  1.1: A[i/2*]    -> add[i].input0 :: _add[i] :> 1: deep<i=0;i<=15; inc > (2,0) :;
       A[i/2*+1] -> add[i].input1

  2.1: add[i/2*].output   -> add[i].input0 :: _add[i] :> 2: deep<i=0;i<=7; inc > (3,0) :;
       add[i/2*+1].output -> add[i].input1

  3.1: add[i/2*].output   -> add[i].input0 :: _add[i] :> 3: deep<i=0;i<=3; inc > (4,0)  :;
       add[i/2*+1].output -> add[i].input1

  4.1: add[i/2*].output   -> add[i].input0 :: _add[i] :> 4: deep<i=0;i<=1; inc > (5,0)  :;
       add[i/2*+1].output -> add[i].input1

    5: add[0].output -> add[0].input0 :: _add[0] :: add[0].output -> sum  :: HALT :;
       add[1].output -> add[0].input1

   };
};
```

Fig 7.6 Parallel summation of a 32 element array.

The module `matrixmultiply` in fig 7.7, performs the matrix multiplication of two 16x16 arrays of unsigned integers, in a maximally parallel manner with respect to the module's level of abstraction. (A totally maximally parallel module would have to employ maximally parallel adder and multiplier submodules). A submodular array of 4096 multiplier modules (described in 6.13.6), computes the first layer of the dataflow. The second layer that sums the results of the multiplications, is performed by a submodular array of 256 `add32array` modules.

```
module matrixmultiply{

  storage{
    unsigned A[16][16] input;
    unsigned B[16][16] input;
    unsigned C[16][16] output;
  };

  submodules{
    addarray32 sum[16][16];
    mult32 mult[16][16][16];
  };

  replications{ i,j,k / inc};
  time: 0-0 cycles;

  code{

   1.1: A[i][k] -> mult[i][j][k].inputa :: _mult[i][j][k]  // contd.
        B[k][j] -> mult[i][j][k].inputb

        :: mult[i][j][k].output -> sum[i][j].A[k] :> 1: deep<i=0;i<=15; inc > (2,0)  :;
                                                        deep<j=0;j<=15; inc >
                                                        deep<k=0;k<=15; inc >

   2.1: _sum[i][j] :: sum[i][j].sum -> C[i][j] :> 2: deep<i=0;i<=15; inc > (3,0)  :;
                                                    deep<j=0;j<=15; inc >
   3: HALT :;

  };
};
```

Fig 7.7 Maximally parallel matrix multiply of two 16x16 arrays of unsigned integers.

The compiled module is one of the largest encountered so far, and occupies 10,274,757 registers, or about 32% of the memory block.

## 7.7 GROW.

The *grow* construct allows the programmer to code Single Program Multiple Data parallelism at the module's level of abstraction, without having to pre-define the 'Single

Program' as a module class. The construct is also used, when a reduction in the size of the submodular hierarchy is required, specified by a *ranged* contraction declaration, described in 5.6.

Grow is applied to a multi-line sub-program, in which the first immediate dependent line (i.e. with postfix '.1'), is designated as the first to be executed. Grow replicates the entire sub-program, where each instance of the replicator is replaced by the control value. Grow is used in conjunction with a special submodule called the *programmable jump*, to vary the range of a sub-programs to be activated, to reduce the amount of work required for a parallel prefix tree of computations.

The activation of a grow construct is implemented by code that simultaneously activates the first line of each replicated subprogram. If an egress has been included in the construct, then a means of establishing the termination of all replicated subprograms is needed. The sub-program's termination cannot be signalled by the module's main HALT instruction, therefore the *subhalt* instruction is included in some dependent baseline of pre-replicated code, whose line address operand is the same as it's parent grow construct. In the current implementation, only one instance of the subhalt instruction with the parent's line address, is allowed within the sub-program.

If an egress is present, expansion adds an additional baseline called the *grow termination line*, that is activated along with the first lines. The compiler finds the grow construct's subhalt instruction, and builds a barrier into the grow termination line, which tests for the cessation of all replicated subprograms. The final column of the grow termination line transfers control to the lines specified by the egress. The grow construct has the same format as deep.

```
ln_add: grow< rep = ind_exp; rep cmp ind_exp; ind_fun >  (ln_add,offset)
```

```
1.1: cond_bits[i] (1.2,0) (1.3,0)   :>    1: grow<i=0;i<=7; inc > (2,0) :;
1.2: #1 -> bits[i] :: jump (1.4,0) :;
1.3: #0 -> bits[i] :: jump (1.4,0) :;
1.4: subhalt(1) :;
```

Fig 7.8 Example of a grow construct, attached to dependent lines.

The baselines in fig 7.8 describe a subprogram, which tests an array's bit element `bits[i]`, and inverts it's value. The grow construct specifies that the subprogram be replicated, and executed in parallel, where $0 \le i \le 7$. Upon detecting completion of all replicated subprograms, the grow termination line (not shown) transfers program control to line 2 (not shown).

A grow construct with an egress, may have another grow as a dependent line, providing it has no egress. This is because of a synchronisation issue which became apparent too late in the current implementation cycle, to easily fix. There appears to be no fundamental obstacle in resolving the problem in future implementations, although there are implications for characterising expanded code as a state transition system (see 7.8.3).

### 7.7.1 PROGRAM EXAMPLE WITH GROW AND CONTRACTIONS DECLARATION.

The module `growexample` depicted in fig 7.9, pairwise compares the values of two unsigned input arrays A and B in parallel. The grow construct tests the results of the comparisons in parallel, adds the numbers if `A[i]≤B[i]`, or subtracts `B[i]` from `A[i]` otherwise, and then sends the results to the output array `C[i]`. For each replicated subprogram, either an adder or subtractor is used, but not both.

Given that the adder and subtractor arrays are not used elsewhere in the module, there is an opportunity for the code segments representing `adder[i]` and `sub[i]`, to share any submodules they have in common, without affecting outputs or the meaning of the program. This may be achieved by using the ranged contractions declaration "`adder[0-7] ~ sub[0-7];`". The compiler identifies and assigns the same `inceq5bit` and `progcopybit` modules, as immediate submodules to `adder[i]` and `sub[i]`. The contraction results in a 25% reduction in the space occupied by `growexample`'s compiled submodular hierarchy.

The contraction would be employed erroneously, if both `adder[i]` and `sub[i]` were simultaneously instructed to activate at some stage during runtime, and would result in various machine errors being generated (jump fail, write fail, marking fail etc.)

```
module growexample{

   storage{
     unsigned A[8] input;
     unsigned B[8] input;
     unsigned C[8] output;
   };
   submodules{
     ADDER32 adder[8];
     SUBTRACT32 sub[8];
     compare compare[8];
   };

  contractions{

    adder[0-7] ~ sub[0-7];

  };

  replications{i / inc};
  time: 0-0 cycles;

  code{

     1.1: A[i] -> compare[i].a ::  _compare[i] :>  1: deep<i=0;i<=7;inc> (2,0) :;
          B[i] -> compare[i].b

     2.1: cond_compare[i].aGEb (2.2,0) (2.3,0)   :>    2: grow<i=0;i<=7; inc > (3,0) :;

     2.2: A[i] -> adder[i].addend :: _adder[i] :: adder[i].sum -> C[i] :: jump (2.4,0) :;
          B[i] -> adder[i].addendum

     2.3: A[i] -> sub[i].subtrahend :: _sub[i] :: sub[i].result -> C[i] :: jump (2.4,0) :;
          B[i] -> sub[i].minuend

     2.4: subhalt(2) :;

     3: HALT:;

    };
};
```

Fig 7.9 Module with grow construct, and ranged contraction declaration.

Note that if the grow's subprogram appeared at a lower level of abstraction, encased within a submodule, then the grow construct action could be performed by a deep construct's activation of an array of those submodules. However, it would then not be possible to specify submodule contractions for SPMD programs, because a module may not reference a submodule's submodules.

### 7.7.2 PROGRAMMABLE JUMP.

The Spatiale module library has a special pre-loaded level zero, meta module called the *programmable jump*, with module name `PJUMP`. Like a conventional jump instruction, the `PJUMP` activates a line address, and consecutive lines, up to and including an offset. Unlike a jump instruction, `PJUMP` can be programmed to vary the offset operand during runtime, up to an including a maximum of 31 in the current implementation. Used in conjunction with the grow construct, this facility is useful in reducing the amount of work (total number of A-Ram instructions executed), required to compute a circuit value parallel prefix tree of computations [69]. The module is declared in a special format within the submodules declaration, and takes the form:

```
PJUMP{maximumoffset} PJUMP;
```

Only one programmable jump is currently allowed per module. The maximum offset that `PJUMP` is required to activate, has to be declared within curly brackets after the class name. After having received the offset input as an integer between 0 and 31, the first phase of the metamodule modifies an internal offset and other values, which enable the second phase to perform the programmed jump itself. The activation of `PJUMP`'s first or second phase, appears in a baseline activation column. The baseline instructions have a single line address operand, which numerically specifies the first of the consecutive lines to be triggered. Instructions to activate the first and second phases, take the forms _`PJUMP` (*`lineaddress`*), and -`PJUMP`(*`lineaddress`*) respectively.

The module `addarray32` in fig 7.10 performs the summation of the 32 integer elements of the input array A, as a parallel tree of addition operations with five layers. The module has a submodular array of 16 adders, which are reused for each layer of the prefix tree. The module begins by simultaneously activating lines 2 and 3. Line 2 loads `PJUMP`'s initial offset, and the initial value of a register right shift module, which divides an integer by two, and then activates it's first phase. The carry line 3 loads the input operand array, performs the first wave of 16 additions, and then transfers control to the main loop, comprising lines

[69] Programmable jump does not in itself reduce the space of compiled code, or the running time complexity of a Space module, it merely reduces the total amount of instructions executed and work done. It is envisaged an advanced runtime environment will be able to dynamically reallocate submodules no longer required in runtime, for other purposes, with a view to improve resource utilisation.

4,5, and 6. The use of `PJUMP` normally entails that the set of the module's co-active sets is extended.

Lines 4 divides `PJUMP`'s current offset value by 2, and determines if it is equal to zero. Line 5 is a grow construct, whose address is `PJUMP`'s line address operand. The construct replicates 8 sub-programs, each of which control the passing of two adder's results from the preceding prefix layer into an individual adder.

```
module addarray32{
  storage{
    unsigned A[32] input;
    unsigned sum output;
  };
   submodules{
     adder32 add[16];
     paror32 neqz;
     rightshift32 rightshift;
     PJUMP{8} PJUMP;
  };
   replications{ i / inc, 2*, 2*+1};
   time: 0-0 cycles;
   code{
       1: jump (2,1) :;

       2:  #8 -> PJUMP.offset      :: _PJUMP(5) :;
           #8 -> rightshift.ioput

     3.1: A[i/2*]   -> add[i].input0 :: _add[i] :> 3: deep<i=0;i<=15; inc > (4,0)  :;
          A[i/2*+1] -> add[i].input1

       4: _rightshift :: rightshift.ioput -> PJUMP.offset :: _PJUMP(5) :;
          -PJUMP(5)      rightshift.ioput -> neqz.input      _neqz

       5.1: add[i/2*].output -> add[i].input0     :: _add[i] :: jump(5.2,0) // contd.
            add[i/2*+1].output -> add[i].input1

                                             :> 5: grow<i=0;i<=7; inc > (6,0)  :;

       5.2: subhalt(5) :;

       6: cond_neqz.output (7,0) (4,0) :;

       7: adder[0].output -> sum :: HALT :;
     };
};
```

Figure 7.10 Parallel adder for 32 numbers.

The programmable jump is used to vary the range of sub-programs triggered in each loop cycle, beginning with all 8 subprograms, then 4,2, and finally just the first subprogram. The module could be made to function as desired without the programmable jump. Each cycle of the main loop could be made to activate all eight sub-programs, but the results of some sub-programs would be wasted, and ignored in later layers of the prefix tree.

Whilst the adders are busy, `PJUMP`'s previous offset is shifted one bit to the right, and the result is fed into `PJUMP`'s first phase, and subjected to a test for zero. When the grow construct terminates, the new offset value is tested, and exits the main loop if found to be zero, or initiates another cycle of the main loop.

The activation of `PJUMP`'s second phase in line 4, is unusual because, in effect, a jump begins at the beginning of a line, instead of at the end. Consequently, the principle that all members of a co-active set are initiated simultaneously is slightly eroded, because the grow construct's subprograms are triggered a few cycles after line 4 is activated. This discrepancy is not of great significance, and will be corrected in the next implementation by modifying `PJUMP`'s code, which will enable moving it's activation into a separate line, that activates lines 4 and 5 together.

### 7.7.3 EFFECT OF GROW CONSTRUCT ON THE CHARACTERIZATION OF EXPANDED CODE AS A STATE TRANSITION SYSTEM.

Through the ability to describe SPMD parallelism at the module's level of abstraction, the grow constructs enhances Space's programmability. In conjunction with programmable jump, grow affords a means of reducing the amount of work required to compute circuit values, and parallel prefix trees. Grow also enables a specification of submodule contractions for SPMD programs, that would not be available if the single program were manifested as a submodule in the module's code. These advantages have a downside.

The grow termination line that is added by the expansion phase if an egress has been specified, is always alive whilst control is transferred between states in the replicated sub-programs. This has the consequence of interfering with the notion of identifying a state with a set of co-active lines, and therefore of characterising expanded code with grow constructs, as a (much larger) state transition system, in the manner described in 6.2. It is not easy to discern whether this may have a negative effect on the ability to apply program reasoning to modules with grow constructs. The inclusion of grow constructs, does not of course affect the status of pre-expanded modular code, as a small, sequential state transition system.

## 7.8 WHILE AND DO-WHILE.

The `while` construct is similar to a standard C language while construct, which tests the contents of a single bit storage entity, and either executes a multi-line sub-program body of the construct, whose exit is directed to repeat the test, or exits to a specified line address and offset otherwise. The entry line is `while`'s first immediate component line.

Recall a primary identifier of a copy operand expression refers to an entity in the module's storage declarations, and a secondary identifier refers to an entity in a submodule's storage declarations. The `while` term is separated by an underscore from it's only operand, which is either a primary or secondary identifier. To test a boolean value, there is naturally a restriction that the identifier designates a single bit storage entity. In practice, this is usually a secondary identifier.

```
ln_add: while_primaryidentifier (lineaddress, offset)

ln_add: while_secondaryidentifier (lineaddress, offset)
```

The `dowhile` construct executes the body of construct, then tests the bit and either repeats the body, or exits to a specified line address and offset.

```
ln_add: dowhile_primaryidentifier (lineaddress, offset)

ln_add: dowhile_secondaryidentifier (lineaddress, offset)
```

A module that serially adds the contents of an array of 32 unsigned integers, containing a dowhile construct, appears in fig 7.11. The sequential nature of the module is reflected in the completion time of 25430 cycles. The module would be even more inefficient, if the `arrayreturn` submodule, requiring 798 cycles, were not scheduled to activate simultaneously with the `adder32` submodule.

```
module arrayserialadd{

 storage{
   unsigned A[32] input;
   unsigned sum output;
  };

 submodules{
       adder32 add;
       arrayreturn Areturn;
       inceq5 inceq;
 };

 time: 25430-25430 cycles;

 code{

  1: A[[0]] -> Areturn.address :: _Areturn :: Areturn.value -> add.output :: jump (2,0) :;
      #2 -> Areturn.index         _add       add.output -> add.input1
      #2 -> inceq.ioput
      A[0] -> add.input0
      A[1] -> add.input1

  2.1: inceq.ioput -> Areturn.index :: _Areturn  :: // cont.
        add.output -> add.input0       _inceq
                                       _add

       :: Areturn.value -> add.input1 :: subhalt(2) :> 2: dowhile_inceq.NEq31 (3,0) :;


   3: add.output -> sum :: HALT :;

   };
 };
```

Figure 7.11 Serial addition of array elements using array return module.

## 7.9 SWITCH.

Similar to a C language switch mechanism, this construct can select between a variety of cases in one step. When the switch is activated, the contents of a storage entity are examined, which currently may only be positive integer cases between the range 0-15. The switch construct's compiled code directs program control to the appropriate dependent baseline or construct line. The format is similar to the while construct:

*lineaddress*: switch_*primaryidentifier* (*lineaddress*, *offset*)

*lineaddress*: switch_*secondaryidentifier* (*lineaddress*, *offset*)

When the dependent line has ceased activity, program control is transferred to the construct's egress. In fig 7.12, a simple module provides the user with a choice of arithmetic operations on input integers.

```
module switchexample{

   storage{
     unsigned A input;
     unsigned B input;
     unsigned C output;
     unsigned choice input;
   };

   submodules{
     ADDER32 add;
     SUBTRACT32 sub;
     mult32 mult;
     modulus mod;
   };
   time: 0-0 cycles;

  code{

     1.1_case 0: A -> add.addend    :: _add :: add.sum -> C  :>  1: switch_choice (2,0) :;
                 B -> add.addendum

     1.2_case 1: A -> sub.subtrahend :: _sub :: sub.result -> C :;
                 B -> sub.minuend

     1.3_case 2: A -> mult.inputa :: _mult :: mult.output -> C :;
                 B -> mult.inputb

     1.4_case 3: A -> mod.dividend :: _mod :: mod.remainder -> C :;
                 B -> mod.divisor

     1.5_dflt:   #0 -> C :;

     2: HALT :;

    };
};
```

Figure 7.12 Simple switch program.

The identification of which case a dependent line is associated with, is achieved through replacing the conventional line address format with the following construction:

*lineaddress*_case *caseinteger*:

The term `caseinteger` must currently be a decimal integer between zero and fifteen. If a dependent line is the switch's default case, then the following is used:

*lineaddress*_dflt:

A dependent base line of a switch construct may not have any program control, be it a terminating jump or cond column. Neither may a dependent construct line have an egress specified, i.e. the egress must be an empty pair of brackets. During expansion, the compiler inserts the switch's egress where appropriate, into expanded dependent line constructions.

Expansion removes the switch construct and inserts a new baseline into the baseline list, with the same line address. It consists only of one special *switch* base instruction, which is an invisible member of the base instruction set. This instruction is only of relevance to the code generation phase, and may not be used explicitly by the programmer.

### 7.10 FUTURE ADDITIONS TO THE CONSTRUCT SET.

To improve programmability, further constructs for future implementation are proposed, to support parallel composition of all types of sub-programs, sequential *for* looping, and sub-program definition.

#### 7.10.1 PARA.

The current implementation allows a parallel composition of subprograms with explicit egress, either through the grow construct for SPMD parallelism, or through the use of skip instructions to describe a carryline for a co-active set, composed purely of baselines. There is a case for upgrading the skip instruction to be able to refer to construct lines as well as base lines. An alternative approach to synchronisations involving construct lines, is the *para* construct, which would describe a parallel composition of sub-programs represented by either construct lines or base lines. Expansion would install a synchronisation mechanism, similar to the grow termination line, that tests for the termination of all dependent lines, so that program control could be transferred to an egress. There would be no need for dependent lines to specify an egress. The format would take the form:

*lineaddress*: para (*lineaddress*, *offset*)

### 7.10.2 FOR.

The construct set includes no element to support sequential iteration of a sub-program, such as a *for*-type programming construct in sequential languages. A similar construct might prove useful for those highly parallel Space programs, whose compiled code exceeds the size of the memory block. Such programs could be recast as sequential programs using sequential rather than parallel iteration.

It would be possible to sequentially replicate a sub-program, in a manner similar to loop unrolling, in which a control variable is modified, and insert jumps between replicated segments. But this seems wasteful, given that a single segment of code may be reused, where the control variable is transformed into one of the module's storage entities.

A construct is therefore proposed, in which the control variable is not replicative in nature. Expansion would automatically add a single unsigned storage element to the module, with some control variable name `var`, being initialised to a left limit. As with the while construct, the first dependent line initiates execution of the sub-program, whose termination is signalled by a sub-halt instruction. The control variable is then incremented, and the sub-program is repeated until the right limit has been reached.

```
ln_add: for< var = ind_exp; var cmp ind_exp; ind_fun >  (ln_add,offset)
```

The system of indexical expressions would have to be modified, to support nesting of replicative and non-replicative control variables.

### 7.10.3 PROG.

The utility of the *switch* construct, and the proposed *para* construct, would be enhanced, from being able to have immediate constructs, which represent arbitrary multiline sub-programs. The first line to be executed is `prog`'s first immediate component line. The termination of the multiline subprogram is specified using the subhalt instruction, and program control is then transferred to the egress in the usual manner. The format would take the form:

```
lineaddress: prog (lineaddress, offset)
```

# Chapter 8
# ARCHITECTURES FOR SYNCHRONIC COMPUTATION.

## 8.1 INTRODUCTION TO SYNCHRONIC ENGINES.

The Synchronic A-Ram provided simple semantics for exploring high level deterministic parallelism, and gave rise to Space and the synchronic computation paradigm. A *Synchronic Engine* is a high performance, general purpose physical architecture, possessing mechanisms which support the key aspects of Space or any Space-like interlanguage module's execution. Space programs require hardware features that are not included amongst the mechanisms found in conventional CGAs, running DSP or systolic programs. In common with some CGAs, an ability to simultaneously transfer the results of large numbers of operations to various destinations in the machine, is needed. Space's sequential state transition system, in which high level state changes are often triggered when a collection of multiple baselines or threads terminate, indicates that hardware support for the fast detection of multiple thread termination might also desirable.

A Synchronic Engine's execution of interlanguage code should preserve it's parallelism and lack of resource contention. Further parameters for the design space concern constraints imposed by network infrastructure characteristics, device granularity, data driven ALU activation, mechanisms for maximising runtime resource ultilisation, and timing and communication protocols. The discussion in this chapter is general rather than detailed, the intention is to sketch out design concepts, that can furnish a backdrop for future discussion.

*Heterogeneous* FPGAs [1] and CGAs for embedded applications [2], are composed of a few large arrays of different device types. The former are spatially interleaved in the silicon plane, so that there are always different device types in proximity to each other. Synchronic Engines have a similar arrangement, where the devices perform different kinds of memory access, arithmetic-logic and program control operations. They communicate quickly with nearest neighbours by wire, whereas devices further afield are connected through a separate network, which might be realised by a wire based packet or circuit switched system, or by a wave based technology. Although wave based communication has yet to migrate commercially into the intrachip arena (see 8.2.1), and has to contend with transformations between digital and analogue forms of data constituting an overhead compared with wire based communication, it has potential future advantages by being faster overall over longer distances, by generating less heat dissipation, and by occupying much less area.

Even if the notion of propagation delay outlined in 3.6 were to be introduced, it is explained in 8.3 that the Synchronic A-Ram is unsuitable as a blueprint for a Synchronic Engine, due to the ultrafine granularity of it's instructions, and overwhelming channel requirements. In 8.4, it will be argued that coarser grained functional units, and hardware support for thread synchronisation and reconfigurable register level interconnects, form a more realistic basis for the construction of Synchronic Engines.

Simulations of massively parallel programs on the Synchronic A-Ram tend to result in between one and five per cent of registers in the area occupied by code, being activated as instructions in any machine cycle. The low level of resource usage, is partly due to lack of optimisations in the current compiler, and partly due to the model's ultrafine instruction granularity. Coarser grained machines, and when the nature of the application allows, systolic spatial programming, will provide opportunities for much better resource utilisation.

Mechanisms for runtime reconfiguration of the kind found in FPGAs and coarse grained architectures [1], including hardware support for the context switching of EDGE dataflows found in the TRIPS architecture [3], would further improve performance. Compiler improvements and the optimisations mentioned above, will determine whether synchronic computing can generate acceptable runtime performance on standard industry benchmarks, compared with multithreading on multiprocessors, or systolic computing on CGAs.

In 1.2.2, reference was made to the possibility of synchronising room sized systems to picosecond intervals, given sufficient investment in an optoelectronic system of clocking. For larger systems, or if such an investment is not feasible, a facility to relax the requirement of a global clock is needed, which would open the door to further scaling, In 8.5, GALS machines are described, where special mechanisms described in [4] are made available for information transfer between zones. Massive programs could still be conceived of as globally clocked processes, aiding programmability, but would run asynchronously across zones.

A Synchronic Engine may be a single globally clocked zone, or a GALS machine composed of a collection of similarily sized synchronous zones. The timing regime for data transfer between devices within a synchronous zone, may assume that communication time is constant between any two points, entailing that the longest communication path dictates the minimum time for all communications. If constant time communication, consisting of a constant number of synchronous zone time steps is chosen, then the absolute restriction that data can travel no more than the speed of light at 30 centimeters in a nanosecond, places a physical limit on a zone's diameter.

Alternatively, propagation delays composed of varying numbers of time steps, might be allowed across variable distances within a synchronous zone. If differing propagation times are allowed, then for wave based connectivity, a mechanism is required in the transmitting device to ensure that the arrival of data is synchronised at an appropriate stage of a receiving device's cycle, in order to ensure that the data is directed to and stored properly, in some designated data transfer register. Variable data transfer times will result in some devices occasionally being unnecessarily idle, whilst waiting for the arrival of inputs. Although idle waits have an adverse effect on the efficiency of resource utilisation and performance, the deterministic, synchronous semantics of programs need not be affected. Optimization in code placement on a machine's fabric to take advantage of data locality, could lessen the impact of this issue.

Any Synchronic Engine will include a processing array and a memory hierarchy, where the latter incorporates a memory layer with direct access to the processing array. Spatial or reconfigurable oriented machines executing instructions in situ, employing graph or interstring data structures to represent dataflow, require no repetitive fetch of instructions, and a lesser degree of repetitive access to data, than do Von Neumann machines. Wire based main memory access transfer times, for the transfer of potentially very large data blocks for I/O, and the runtime reconfiguration of device arrays, might be achieved between 50 and 100 nanoseconds.

These considerations provide starting points for the investigation of the synchronic design space. The apparent current unavailability of wave based intra-chip connections[70], allowing space efficient, nanosecond reconfigurable connections between large numbers of logic modules, suggests that maximally efficient Synchronic Engines will not be fully realisable in the short term. There is a sense in which the universe was not designed for Synchronic Engines, because their performance would benefit from instantaneous communication between every storage and ALU unit of a large machine. But for the purposes of outdoing cost-performance ratios for alternative parallel computation models, variable propagation delay could provide an effective connection regime. In the next section, it is discussed how photonic and spintronic interconnects might offer solutions in the mid to long term.

## 8.2 FULL RECONFIGURABLE CONNECTIVITY.

Hartenstein describes the *Reconfigurable Computing Paradox* [6], relating to fine grained FPGAs, which outperform a microprocessor on many applications, and manage to do so in spite of slower clock rates and severe space overheads. Connections between logic blocks in a typical FPGA application, occupy a significant part of chip area, and also require a large amount of storage logic to configure, resulting in only around 10% of an FPGA's logic gates being used for application logic [1]. Complex logic modules such as floating point units are therefore not generally mapped to fine grained FPGAs, but rather to heterogeneous fabrics that contain larger dedicated logic blocks.

The phenomenon is problematic for the use of FPGAs as cost effective general purpose architectures. It arises from their fine granularity, and from a major limitation of silicon; the cost of laying out of wire routes in a plane. Full, one-to-one, non-broadcast connectivity between $n$ logic blocks in linear time requires $O(n^2)$ area complexity using a cross-bar or mesh arrangement, whereas broadcast connectivity needs $O(n^4)$ area. Even the limited bipartite connectivity exhibited by multi-ported memory, has a quadratic area complexity with respect to the number of ports, resulting in for example, a 32 port memory occupying 80 times the area of a single ported memory [7].

[70] DARPA sponsors research in this area that is classified. It is not inconceivable that connection technology needed for efficient synchronic computation is ready now, or will be in the near future.

Reducing the connectivity of a network of processing resources is not in general conducive to general purpose parallel computing, and interlanguage programmability relies in part on direct links being available between any two locations in the target machine. Direct, constant or minimal time connectivity might benefit from a wave based technology, because there is a potential for lower, even $O(n)$ area complexity. Wave based information may be transmitted by electromagnetic, spintronic, or possibly some other form of radiation. Digital information is encoded through the modulation of a wave's amplitude or phase, transmitted through the wave medium by an emitter, and received by a detector and decoded. A multitude of links can be realised in a wave medium or free space, potentially without crosstalk or interference, in a variety of ways. Two approaches for connectivity in synchronic computation are represented pictorially in fig 8.1, which by no means exhaust all of the different styles of wave based communication.

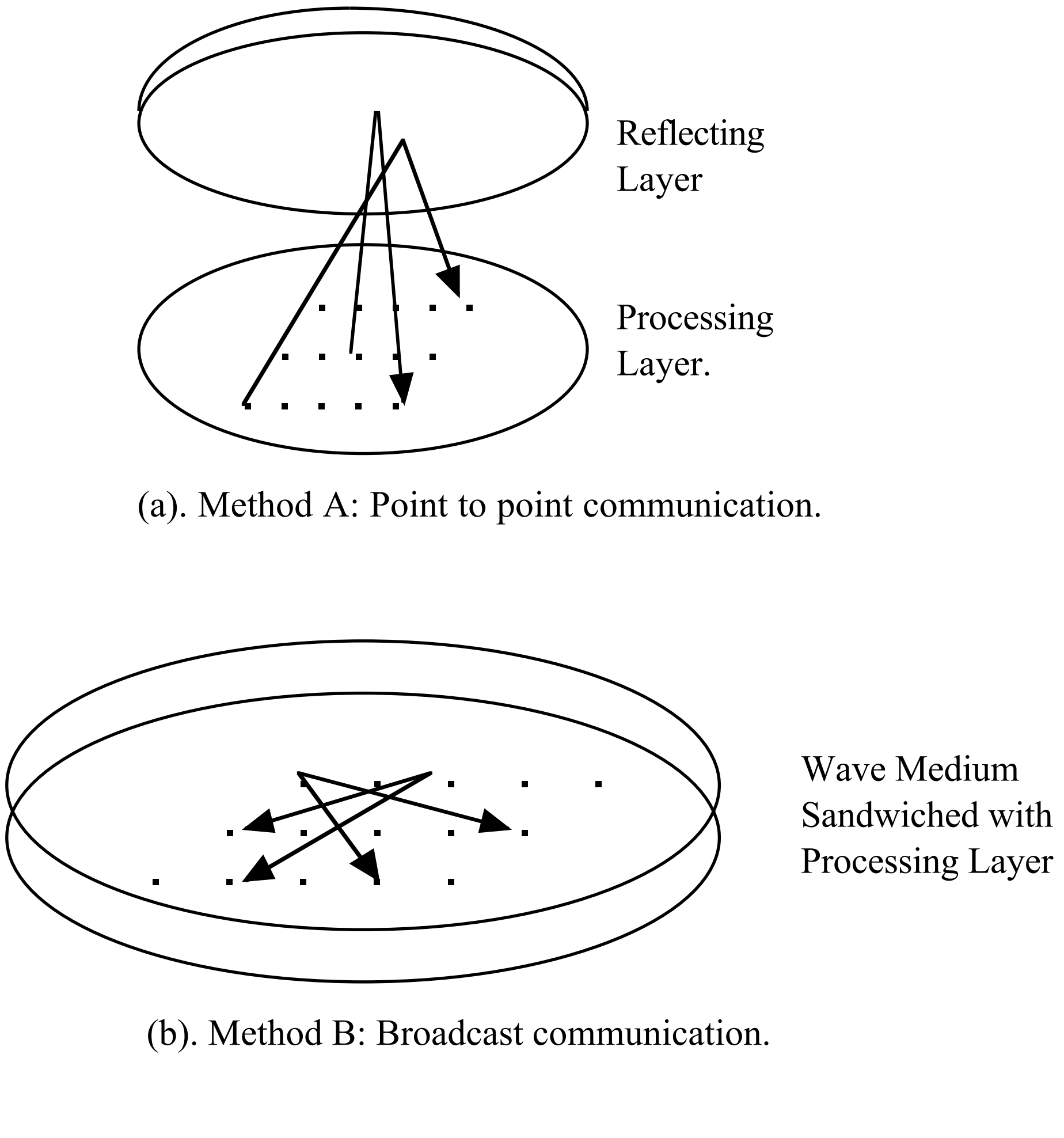

Fig 8.1 Wave based interconnection styles

A. Processing elements are laid out in a plane, and an element incorporates at least one pair of a fixed detector and a pointable emitter or reflector, to aim at most one other element by bouncing a transmission through free space off of a reflecting mirror, or deflecting layer. Propagation times may vary depending on the total flight distance between points. A single emitter is point to point, and does not support broadcast mode.
B. Exclusively employing the technique of wavelength division multiplexing (WDM), each processing element is designated it's own specific wavelength interval, and is equipped with a single wave emitter/detector mechanism, where the wavelength configuration of at least the detector must be adjustable/tunable. The plane of processing elements is sandwiched with a wave medium. Propagation times for messages again may vary, depending on distances between points. Each element may omnidirectionally broadcast on it's own emitting wavelength, through the medium to a plurality of receiving elements, providing the receiving element's detector is tuned to the emitter's wavelength.

Method A is similar to *skywave* propagation, in which radio waves may be sent between two points on the ground, by bouncing them off the ionosphere. Method A may also be seen as an abstraction of the Optical Model of Computation (OMC) [8] [9], which is oriented to PRAM algorithmics. Method A differs from OMC in that processing elements are not restricted to be being Von Neumann processors attached to their own random access memories, the reflecting layer might be planar or curved, and communication times between elements are not in unit time and may vary. Further, all communications succeed because the programming model is assumed to be Exclusive Write.

Method B is analogous to a collection of radio stations omindirectionally broadcasting to multiple different collections of recipients, through the broadcasting medium on differing wavelengths. Ideally, reorientation/retuning, and the transmission of information between any two of potentially tens of thousands of logic blocks or more in both methods, would occur on the scale of a few machine cycles, in the nanosecond range. Longer retuning/reorientation times to access the next interstring block or blockstring, whilst processing a large data set by looping may be tolerable, if retuning and data transfer can be scheduled to coincide with the execution of a loop's interior operations. The next two subsections present brief overviews of recent developments in photonics and spintronics, that may yield interconnect solutions for synchronic computation.

### 8.2.1 PHOTONIC LINKS.

Research in the field of optical links between computational devices, has mainly focused on telecommunication systems for wide and local area networks. Typically these networks need a small number of fixed wavelength or retunable channels supported on fibre optic cables, operating over long distances through the use of repeaters. Telecom networks require large amounts of data on the order of gigabytes and terabytes, being transmitted through each channel per session at speeds of Gbits-Tbits per second.

The architectures presented in this chapter however, need fast connectivity on the inter and intra chip scale, and a very large number of fixed and reconfigurable links, with small amounts of data being transmitted at high speed, on the order of a few bytes. The field of optical interconnections in computer architectures began with the paper by Goodman et al [10]. More recent, supportive surveys [9] [11] [12], suggest that intrachip optical interconnects are becoming feasible, whereas critical surveys argue that technological challenges prevent the economically justifiable introduction of intrachip optical interconnects [13] [14] [15].

The shift to multi-core in the mid noughties, arose out of the failure to maintain frequency scaling in conventional uni-processors. One Method A scheme that would have provided a novel route for maintaining frequency scaling, by adopting (runtime non-reconfigurable) intra-chip optical interconnects, is described in [16] [17]. A multi-scale photonic arrangement is presented, employing microlenses, microprisms, and a curved mirror, in order to effect high bandwidth global communication between logic modules, that purely silicon based approaches cannot duplicate because of area and power constraints[71]. An approach for optically connecting chips in a multi chip module (MCM) is given in [18].

Synchronic Engines are composed of very large numbers (tens of thousands or more) of ALUs distributed across an MCM or wafer scale processing layer. An individual fixed point ALU, may only need to occupy a fraction of a square millimeter, if advanced fabrication technology is used. The Synchronic Engine's efficiency will depend to a large extent on the speed of reconfiguration and transit time of links between ALUs.

[71] It is evident from the historical course of events, that multi-core rolled over more effective approaches for maintaining clock rate and performance improvements.

A high frequency optically based clock distribution scheme is described in [19]. A clocked environment might facilitate the implementation of high speed reconfigurable links in larger Synchronic Engines, because co-ordinating interfaces in a GALS system between asynchronous zones, will impose time and area overheads.

An optical link is reconfigured in Method A, by physically changing the orientation of a single emitter or reflecting mirror, associated with an ALU. In [9], techniques are described where reorientation is achieved through the use of microelectromechanical systems (MEMS), which are micrometer sized moveable structures and mirrors integrated onto silicon chips, capable of a reconfiguration speed in the microsecond range.

An important issue in achieving high density of free space links in method A, using micro-optics, is diffraction. The aperture of an optical emitter will cause diffraction, resulting in the outgoing beam being diverged. In order to avoid cross-talk between a multitude of beams, a receiver must be as big as the size of the arriving, diverged beam. Consequently there is an inverse relationship between link density and the maximum area of the processing layer. I am grateful to Michael J. McFadden for bringing this limiting factor to my attention. He provided a provisional estimate, that a wafer scale processing layer would require an ALU's optical port to be spaced at intervals of at least a few millimeters, constraining the minimum area of an ALU. Increasing link density in order to scale the Synchronic Engines, by reducing ALU size and increasing ALU count, may therefore not be compatible with micro-optics.

In a system that exhibits aspects of both method A and B, but at the macro-optical level, [20] presents the FASTNET scheme, where each of $n$ processing elements is equipped with at least $n-1$ pairs of non-pointable emitters and receivers. Each element pair is dedicated for communication with only one other element. (Further element pairs may be added to increase bandwidth.) A processing element could in theory broadcast information to up to $n-1$ other elements simultaneously. Nanosecond communication times are feasible, and constant time reconfigurability of links between elements is achieved, merely by selecting which transmitter/receiver pair to use.

A drawback is that by having at least one dedicated channel for every uni-directional link between any two nodes, the scheme exhibits $O(n^2)$ area complexity, which limits its use as the only connection mechanism to smaller machine sizes. McFadden further indicates that the additional use of WDM would allow multiple channels per link, and might enable systems with connected zone counts in the low thousands with existing technology. The approach might also be employed as a photonic top layer for a packet or circuit switched network, for interconnecting much larger ALU counts.

Another system from the Lightfleet Corporation [21], called Direct Optical Broadcast Interconnect, may broadcast signals and bounce them off a mirror to effect communication between co-planar elements. Miller in [22] observes that relatively little engineering research has focused on free space intra-chip communications that is characteristic of Method A, and argues that the approach is more suited to regular communication patterns rather than the implementation of reconfigurable connectivity, presumably because of slow speed and power hungry nature of MEMS and other kinds of optical devices. Another factor is that there has been no urgent need in conventional computing models for Method A or B connectivity on the intrachip scale, resulting in fewer engineering solutions being pursued. One notable exception is described in a relatively old paper [23], which presents a scheme for a fully connected system occupying $O(n)$ area, for up to 5000 units. The system has nanosecond retuning of links, and uses a planar waveguide equipped with fresnel lenses, but there seems to have been no follow up.

Copy operations in a Space module which are static and occur between fixed storage locations during runtime, do not require a runtime reconfigurable link. If one were to employ MEMS technology, leaving aside for the moment the issue of link density, then this kind of copy could occur in nanoseconds. Those copy operations that do need to be runtime reconfigurable, would be subject to a microsecond latency. With around one order of magnitude time difference between reconfiguration and ALU execution times, it is questionable whether engineering and programming tricks could significantly mask the reconfiguration latency.

### 8.2.2 SPINTRONIC LINKS.

Spin electronic (spintronic) devices process information encoded into an electron's spin rather than it's charge. Until recently, techniques for the transfer of spin information involved some electron and hence charge transfer. In 2002, Covington et al [24] announced a technique for propagating and detecting spin waves without charge transfer. There are considerable advantages to be had from this approach, because purely spin based transmission of information potentially involves less heat dissipation and device area. Spin waves are generated and detected by a tunable device known as an ACPS line, and propagated along a spin wave medium, realised as a ferromagentic layer sitting on top of a silcon chip.

In 2005, Khitun and Wang [25] announced the transfer of information through the use of encoding data into a spin wave's phase alone, and a technique for implementing logic gates utilising the modification of spin amplitude. The wave character of an electron's spin however, presents an opportunity for implementing wavelength division multiplexing, and full connectivity without cross-talk, on the nanoscale. Eshaghian-Wilner et al in [26], introduced the idea of fully connecting a collection of $n$ processors in a circular arrangement, using $O\left(n^2\right)$ area. Two further interconnection schemes based on a cross-bar and mesh arrangements with $O\left(n^2\right)$ area are presented in [27] [28]. Other aspects of spin-based communication are discussed in [29] [30] [31] and [32].

In theory, spin technology also allows that for $n$ nodes each equipped with an ACPS line, and $n$ frequencies being available, an architecture could operate requiring only O($n$) area for fully interconnecting nodes in broadcast mode (Method B), where $n$ could be in the millions[72]. Further, links could be reconfigured in a nanosecond. Transmission time could either be constant, based on longest distance between two nodes in the connected zone, or be a function of the distance between sending and receiving nodes. Coupling architectures with nano-scale memories is apparently feasible. However, spin information transmission is for short-range, nanoscale interconnects only. Spin waves travel at relatively slow speeds of around $10^{-5}\, m\, s^{-1}$, and their attenuation length at room temperature is 5-10 microns. It is worth bearing in mind that a 64-bit bus will have a width of about 2 microns using 32nm technology. Technologies for achieving complex logic modules at the molecular scale, are surveyed in [33]. The number of processing elements that can be placed in a spin connected zone using current technology, would appear to be limited.

[72] Private communication with Mary Eshaghian-Wilner.

Data held at the molecular scale, will need to be communicated to larger scales, in order to be sent to other molecular devices, to be held in various forms of storage, or transferred to visual displays. Another potential issue concerns how to effect high bandwidth transfer of data held at the molecular scale to and from larger scales.

## 8.3 SYNCHRONIC A-RAM AS AN ARCHITECTURE.

A functional block diagram for an $\langle 5,\sigma,\{1,2\},\eta \rangle$ individual register is presented in fig 8.2, in order to make plain the excessive quantity of connections and address generators that would be required for the execution of the Synchronic A-Ram's four instructions, and to process the state changes specified by active instructions in other registers. (Please consult end of Appendix B if the fig 8.2 is unreadable). The intention is to reveal background, that will inform the construction of Synchronic Engines.

A register has a functionality that would be expensive to realise in silicon, even ignoring the machinery that would be needed for the error detection scheme. The simultaneous read nature of the Synchronic A-Ram entails that a broadcast style of connectivity employing method B would be needed. Considered as a hardware specification, the Synchronic A-Ram is unrealistic for the following reasons:

- The model's instructions are bit-level, and are finer grained than multi-input transistor-based logic gates; the smallest, most efficient computational devices currently available.
- Code for floating point operations requires hundreds of machine cycles to execute, contrasting poorly with the single digit cycle time of silicon based ALUs.
- Runtime reconfigurable interconnection is essential for pointer operations, and iterating over array, string and interstring data structures. The fastest reconfigurable link is the programmable copy operation, which incurs a constant but large space overhead.
- The $\langle 5,\sigma,\{1,2\},\eta \rangle$ has $2^{30}$ bits, each of which requires a distinct communication channel, in order to transmit it's value in the first half cycle, and occasionally to receive a new value in the second half cycle. The $\langle 5,\sigma,\{1,2\},\eta \rangle$ has $2^{25}$ registers, each of which requires 5 distinct communication channels, in order to receive the 5 offset bits of an active jump instruction targeting that register (notwithstanding registers close to the beginning and of the array). A physical $\langle 5,\sigma,\{1,2\},\eta \rangle$ would therefore need on the order of $2^{30} + \left(5 \times 2^{25}\right)$ wavelength intervals in the wave medium.

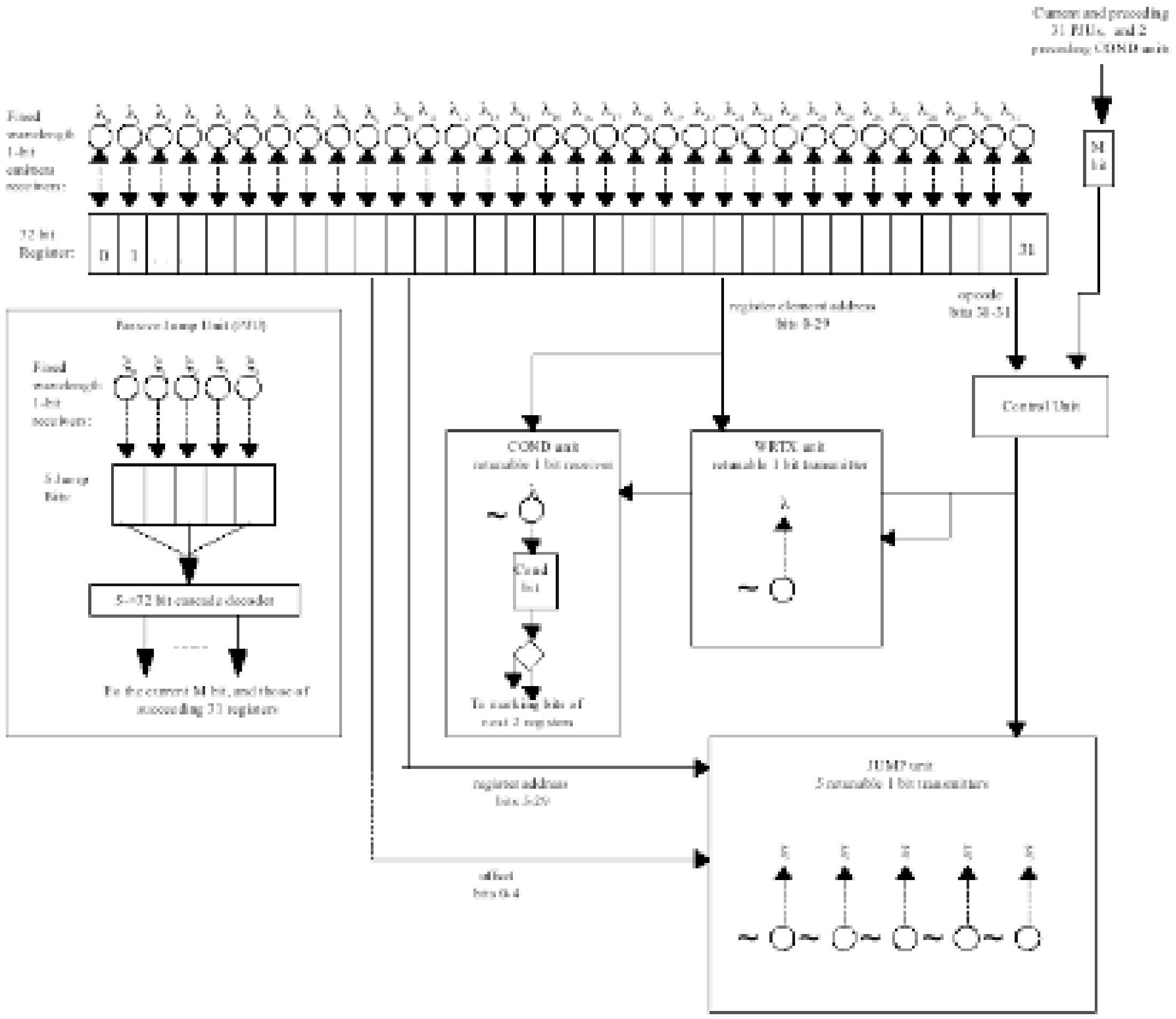

Fig 8.2 Block diagram for a $\langle 5,\sigma,\{1,2\},\eta \rangle$ register.

Each of a $\langle 5,\sigma,\{1,2\},\eta \rangle$ register's 32 bits is attached to it's own emitter and receiver. A bit's emitter is understood to be broadcasting it's value on it's own fixed wavelength into the wave medium in the first half of every cycle. The bit's receiver is only active when another register's instruction writes a zero or one value to it in the second half cycle, on the same fixed wavelength. In addition to it's own 32 storage bits, a register has an additional 7 control bits:

- Marking Bit. The bit indicates whether the register's instruction is to be activated in the next cycle. It receives an OR-ed input from the register's current and preceding 31 registers' Passive Jump Units (see below), and the 2 preceding registers' COND Units (see below). Upon completion of an instruction in the next cycle, the marking bit is reset.
- Cond Bit. If the register is instructed to read the value of some bit in the register array, in order to execute a cond instruction, then the value is stored in the cond bit.
- Jump Bits. If the register is identified by another register's jump instruction's destination cell (bits 5-29), for activation in the next cycle, then the jump instruction's offset bits are received and stored in the register's 5 Jump Bits.

The register has 5 units involved with instruction execution, some of which include fixed wavlength, and retunable emitters and receivers.

- Control unit. When the marking bit is set, the control unit decodes the register's opcode in bits 30-31, and enables the relevant instruction unit.
- WRTX unit. This unit executes the wrt0 and wrt1 instructions. It incorporates an emitter capable of sending a zero or one bit, which is retunable to any of the $2^{30}$ reception wavelengths associated with the register array's bits, specified by the destination bit operand in bits 0-29.
- COND unit. This unit executes the cond instruction. It incorporates a retunable receiver capable of receiving the value broadcast by any storage bit specified by the destination bit operand in bits 0-29, and then selects either the next, or next but one register's marking bit.
- JUMP Unit. This unit executes the jump instruction. It incorporates 5 retunable emitters capable of sending the instruction's offset (bits 0-4) to the Jump Bits of the target register, identified by bits (5-29).
- Passive Jump Unit. This unit includes the Jump Bits. When a jump instruction from another register targets the Jump Bits' wavelengths, the 5 bit value is fed into a 5->32 unary decoder[73], whose output is OR-ed into the current marking bit and the next 31 registers' marking bits.

[73] Instead of setting a single output bit, a *unary* decoder also sets all of the preceding output bits. An input of 010 for a 3->8 unary decoder will therefore set the first three output bits, and reset the remaining 5 bits.

Each register therefore has an extensive collection of 1-bit connection mechanisms, varying subsets of which are required to function simultaneously:

- 37 fixed wavelength receivers.
- 32 fixed wavelength emitters.
- 1 retunable receiver.
- 6 retunable emitters.

Every retunable device, would further require an address generator, which would calculate the wavelength in the relevant wave medium, from the instruction’s operands.

**References.**

# Chapter 9
# FORMAL MODELS FOR COMPUTER SCIENCE.

## 9.1 NEW ROLES FOR FORMAL MODELS.

What roles should a foundational, formal model of computation play in Computer Science? In addition to providing a neutral framework for investigating computability and complexity, it would be desirable if the model could support features needed for high level computation without excessive space and time complexity overheads, including ideally those required for a massively parallel programming environment. It is argued in 9.5, that simulations within feasible timeframes and disk space on a physical machine, of complex programs defined as the model's primitive operations, assist new insights into the relationship between mathematics and computing, and in complexity theory and program semantics.

To simulate a non-spatial, Von Neumann program, Random Access Memory (RAM) is needed, if instruction fetch is to occur in unit time. More generally, RAM allows faster access to internal components of data structures, and affords a means of defining look up tables and for their contents to be accessed in constant time. Addressable memory with constant or at least sublinear access time with respect to memory size, is a pre-requisite for feasibly simulating general purpose computation. An obvious precondition for a parallel computing environment, is the ability to program a parallel composition of pre-existing program modules, without excessive complexity penalties. It is argued in 9.3 and 9.4, that the standard models, the Turing Machine and the λ-calculus, are inherently unable to support RAM and parallel composition efficiently.

Space's construct set, and the successful program runs, have established that the $\langle 5,\sigma,\{1,2\},\eta \rangle$ Synchronic A-Ram is efficient enough to simulate some examples of massive parallelism. Larger programs and inputs can be dealt with by selecting a machine with a larger offset, assuming a compiler for that machine is available, so long as indefinitely large amounts of runtime memory are not required.

A more general argument is now presented, that for non-generic finitistic computation, the model has a superior ability to support high level and parallel computation, than the standard models. For generic computing, the Synchronic B-Ram can run and process indefinitely large programs and inputs across multiple memory blocks. However, the advantages of the Synchronic A-Ram over the Turing Machine do not carry over to the Synchronic B-Ram to the same extent, because register cursors can move only a single memory block position per machine cycle. Further discussion of non-finitistic computation is beyond the scope of this report.

### 9.2 RAM AND PARALLELISM ON THE SYNCHRONIC A-RAM.

The reading of a register bit in a $\langle p,\sigma,\{1,2\},\eta\rangle$, whose address and offset are known before runtime, can be performed in 2 cycles, using code that occupies 3 registers, irrespective of $p$. If the address and offset are known at runtime only, then a programmable copy operation is required, described in 4.6, and appendix B for the $\langle 5,\sigma,\{1,2\},\eta\rangle$. A bit may be read in as few as 8 cycles, using around 380 lines of code, and a register may be read in 15 cycles, with code occupying 5,248 registers. Although runtime access code has a high cost in registers, it is *constant* per bit or register. Providing the A-ram has a large enough offset and memory array, disk space for a simulation is cheap and available.

If a Space module's submodule occupies $k$ registers in the Synchronic A-Ram's memory, then an array of $n$ such submodules occupies $kn$ registers. In 6.14.1, it was explained that to trigger $n$ threads on the $\langle 5,\sigma,\{1,2\},\eta\rangle$, the Space compiler builds a *jump tree*, comprising $\left\lceil \frac{\log_2 n}{5} \right\rceil$ layers, that takes $\left\lceil \frac{\log_2 n}{5} \right\rceil + 1$ cycles to execute. More generally, the cycle time to initiate $n$ threads on a $\langle p,\sigma,\{1,2\},\eta\rangle$ is the logarithmic function $\left\lceil \frac{\log_2 n}{p} \right\rceil + 1$. The space requirement in registers for the jump tree, is the linear function *jumptree*($n$).

$$jumptree(n) = n + \left\lceil \frac{n}{2p} \right\rceil + \left\lceil \frac{\left\lceil \frac{n}{2^p} \right\rceil}{2^p} \right\rceil + \ldots + 1$$

Therefore parallel composition for the current implementation of synchronic computation, has a linear complexity in space, and logarithmic complexity in initiation time for $n$ submodules[74]. The Synchronic A-Ram's complexity characteristics set a good standard, and it is now pertinent to ask of any formal model, how space and time efficient are possible implementations of RAM, and the parallel composition of program modules?

### 9.3 DELTA COMPLEXITY AND THE TURING MACHINE.

The Turing Machine (TM) had an unique role in the historical foundation of Computer Science, and provided a framework for analysing decidability and computability. TM allowed an assessment of memory and time complexities of various low abstraction level problem classes, such as graph reachability, graph isomorphism, clique finding, and the Travelling Salesman Problem. But the model has not had much of a role in higher levels of abstraction. For the purpose of the discussion, a one way infinite, multi-tape Turing Machine will be given a simplified definition, as the tuple M = $(Q,\Sigma,n,\delta)$:

i. $Q$ is a finite non-empty set of states.
ii. $\Sigma$ is a finite set of symbols.
iii. $n \geq 1$ is the number of tapes.
iv. $\delta$: $Q \times \Sigma^n \rightarrow Q \times \Sigma^n \times \{L,R,C\}^n$ is a partial function, and is called the instruction table of M. The $\delta$ function is represented as a list of tuples.

*Delta measure* relates to the size of the TM program's δ function, or instruction table. Although the tables of multi-tape TMs would require larger amounts of bits to encode than single tape machines per tuple, it will be sufficient for present purposes to define the delta measure $d(M)$ of a machine *M*, to be the number of tuples listed in the table. It will be seen that some important TM manipulations involve exponential increases in delta complexity.

There might be a view that delta complexity is not that relevant, and can be dealt with in any case by considering a TM as an encoded input for some fixed Universal Turing Machine (UTM), which would include an explicit representation of the TM's instruction

---

[74] It is hopefully clear that for an address known only at runtime, register access in $\langle p,\sigma,\{1,2\},\eta \rangle$ imposes only a linear space, and logarithmic time overhead with respect to the number of bits in a register $\left(2^p\right)$.

table. It is noteworthy however, that UTMs are rarely considered in complexity analysis. Input TMs have restrictions on their alphabet sets, and on having a single or no more than a fixed number of tapes. UTMs significantly complicate program description and analysis of TM running times and space usage, and the formulation of theorems and proofs.

The arguments that are deployed in this section, might be reframed within the context of the UTM without in my view affecting the conclusions, but would have to consider the degrees of freedom and difficulties involved in writing the UTM and TM encodings. What degree of parallelism should be allowed in the UTM's operations, if parallel module composition is under investigation? A $k$-tape UTM cannot not simulate a $k+1$ tape TM, whilst preserving the TM's parallelism. One might construct a mechanism in the UTM that would yield accurate running times and space usage for a $k+1$ TM, but there is also the UTM's own delta complexity to consider. If the instruction table is subject to exponential growth in order to implement some parallel operations for example, then the UTM becomes un-simulable for practical purposes. In common with the majority of researchers, I find it more convenient to consider TMs alone, but at the same time giving explicit consideration to the delta measure.[75]

### 9.3.1 ADDRESSABLE MEMORY FOR THE TURING MACHINE.

The access of storage in TMs is constrained by tape cursor(s) only being able to move one tape cell per machine step. Suppose some scheme is devised for a finite addressable, register array with $r$ registers composed of a constant number of bits, to occupy a region in the one-way tape, or tapes. We now consider how TMs might be organised, that accept as input the register array, and an integer $1 \le i \le r$, and generate on an output tape, the content's of the $i$th register.

i. An array of $r$ registers composed of a fixed number of bits, could be laid out as a horizontal, numbered sequence of $r$ bit vectors on a single tape. A register access would require the cursor's tape moving in $O(r)$ steps. By implementing a simple loop with a counter, the delta measure may be confined to a constant value, irrespective of $r$.
ii. The register array could be laid out as a vertical sequence of bit vectors on $r$ tapes. Each

[75] It is open to question how a consideration of changes in delta measure in the reductions used in proofs from one problem class to another, would affect our understanding of the standard heirarchy of complexity classes.

cursor in the vertical array of tapes is positioned at the beginning of a register. If a separate machine state is associated with each register, a simple state transition system could be applied to walk across the array, register by register, in $O(r)$ steps. The register is then read one bit at a time, by moving the cursor sequentially across the register's contents. In this case there is a seperate state associated with each register, so the delta measure is $O(r)$.

iii. There is another way for access to a vertical array to be implemented in $O(\log(r))$ time. Assume for simplicity $r$ is a power of 2. Consider an access as a route traversal, considering one bit of $i$ at a time, through a binary decision tree with $2^{r+1}-1$ nodes. Given the tree has $\log(r)+1$ levels, it would take $O(\log(r))$ steps to arrive at the right address state at the tree's bottom layer, and access a register's contents, which is rather better than the first two proposals. Unfortunately, the decision tree requires a separate machine state associated with each tree node and register. There are an exponential number of machine states with respect to $r$, and hence of tuples in the instruction table. Therefore this method of memory access has at least exponential delta complexity.

A hand waving argument has been presented that addressable memory in TMs, requires either linear time, or logarithmic time with exponential delta complexity, with respect to the size of memory. These overheads help explain why TMs' simulable programmability has been confined to low levels of abstraction.

### 9.3.2 PARALLEL COMPOSITION OF TURING MACHINES.

Parallel computing in TMs is achieved through the use of multi-tape machines. In order to help establish multi-tape as a reasonable means of introducing parallel operations into the TM framework, one standard result offered is that the transformation of a *k*-tape machine into a single tape machine, imposes only a polynomial running time overhead. Formally, a given machine *S* where $n > 1$, requiring $f(r)$ steps to halt for input of size *r*, may be simulated on some single tape machine *T* in $f(r)^m$ steps, for some fixed integer *m*.

But in all of the extant proofs, the conversion of a *k*-tape instruction table into a single tape table, involves an exponential increase in the number of tuples, with respect to *n*. Nobody has come forward with a transformation that does not impose an exponential overhead in delta complexity. The space that a UTM encoding of the transformed single tape TM occupies, would certainly be exponentially larger than the encoding of the multi-tape machine, with respect to *n*. TMs ability to access arbitrarily large instruction tables in unit time, explains the polynomial overhead in running time, that would be transformed into exponential overhead, if UTM running times were considered.

Suppose we have a collection of single tape machines $M_i = \langle Q_i, \Sigma, 1, \delta_i \rangle$, where $1 \le i \le k$ Suppose further $M_i$ requires $f_i(r)$ steps to halt for input of size $r$. How might one treat their composition within the TM framework, so that the resulting TM called *MPAR,* runs the *k* machines as parallel threads, and processes *k* inputs at the same time. For simplicity, assume there is no need for *MPAR* to successfully halt by ascertaining when all of the threads terminate, and that *MPAR*'s running time should be equal to the maximum of the set $\{ f_i(r_i) \mid 1 \le i \le k \}$. The solution is to construct a *k*-tape machine, where the *k* inputs are arranged as a vertical array of *k* tapes. *MPAR*'s delta function would take the following form, derived from a cartesian product of the single tape tuple sets.

$$\delta : \left( \prod_{1 \le i \le k} Q_i \right) \times \Sigma^k \to \left( \prod_{1 \le i \le k} Q_i \right) \times \Sigma^k \times \{L, R, C\}^k .$$

By convention, if $\delta(p,\varepsilon) = \langle q,\phi,\varphi \rangle$, let $\delta^1(p,\varepsilon) = q$, $\delta^2(p,\varepsilon) = \phi$, and $\delta^3(p,\varepsilon) = \varphi$. Then $\delta$ for *MPAR* is defined below for all those pairs of tuples $\langle q_1,q_2,..q_k \rangle, \langle \varepsilon_1,\varepsilon_2,..\varepsilon_k \rangle$, where $\langle q_i,\varepsilon_i \rangle$ is defined for some $\delta_i$, $1 \le i \le k$, and some $q_i \in Q_i$, $\varepsilon_i \in \Sigma$.

$$\delta^1(\langle q_1,q_2,..q_k \rangle, \langle \varepsilon_1,\varepsilon_2,..\varepsilon_k \rangle) = \langle \delta_1^1(q_1,\varepsilon_1), \delta_2^1(q_2,\varepsilon_2),..\delta_k^1(q_k,\varepsilon_k) \rangle$$
$$\delta^2(\langle q_1,q_2,..q_k \rangle, \langle \varepsilon_1,\varepsilon_2,..\varepsilon_k \rangle) = \langle \delta_1^2(q_1,\varepsilon_1), \delta_2^2(q_2,\varepsilon_2),..\delta_k^2(q_k,\varepsilon_k) \rangle$$
$$\delta^3(\langle q_1,q_2,..q_k \rangle, \langle \varepsilon_1,\varepsilon_2,..\varepsilon_k \rangle) = \langle \delta_1^3(q_1,\varepsilon_1), \delta_2^3(q_2,\varepsilon_2),..\delta_k^3(q_k,\varepsilon_k) \rangle$$

*MPAR* runs the single tape machines $M_i$ in parallel, by providing a delta tuple for every possible combination of $\delta_i$-defined pairs of single tape states and tape symbols. It is clear from the above definition that $d(MPAR) = \prod_{1 \le i \le k} d(M_i)$, and is therefore exponential with respect to *k*. The only reasonable way of treating parallel composition in TMs exhibits exponential delta complexity. Consequently, it is not realistic to simulate a TM machine, containing even moderately sized parallel compositions.

### 9.4 Complexity considerations relating to the λ-calculus

Upon first attending a λ-calculus course, I queried it's relationship to complexity theory, and was informed the relationship was *orthogonal*. The λ-calculus is not suited for complexity analysis even for low level problem classes, because of the lack of an explicit notion of memory, and the overheads revealed by simulations. If a machine is introduced, sequential high level processes, and even low level sequential processes, are not easy to simulate as a sequence of pure λ-calculus reductions. $\beta$-reduction involves finding each occurrence of a particular variable in the λ-term of length *r* requiring $O(r)$ time, and $O(r)$ space for storing the λ-term, if no optimisations are in place. Optimising the processing of expressions, with potentially massive numbers of variables, would require a dynamic record to be maintained for each variable's positions in the λ-term.

The Single Parent Restriction, and the lack of an explicit memory attendant on the formalism's emphatically non-spatial character, entail that storage cannot be random addressable in constant time. Storage must be represented as a linked list tree structure, or as a binary tree similar to the third TM approach described in 9.2.1. In a linked list, the accessing of the $i$th element requires $i-1$ operations of destructively removing the head of the list, and therefore has a minimum linear time overhead. A binary tree memory of the kind described in (iii) of 9.3.1, has an exponential number of nodes with respect to memory size, which would impact on the length of the λ-term, and the memory required to store it. The inability of the λ-calculus to support memory or memory access efficiently, is fatal for a viable simulation of λ-reductions expressing general purpose, high level processes.

Early attempts to produce compilers for functional programming languages, that employed intermediate forms of the λ-calculus and transformations of tree representations alone, yielded inefficient runtime code [1]. The transformation of a λ-term as a tree of unshared sub-terms, into a graph with shared sub-terms, at least bypassed the representational explosion of describing wide, deep dataflows with multiple shared subexpressions. It was not until compiler writers treated functional program processes as a series of graph transformations, replacing expression reduction in later compiler phases with a graph rewriting mechanism and RAM based memory manipulations, that functional languages became usable, if not efficient compared with imperative languages.

The λ-calculus has no means of explicitly composing pre-existing λ-definitions, which are intended to be reduced in parallel. Graph based calculi raise the prospect of implementing parallel operations on those subgraphs which can be identified as being reducible in the same process step. But the NP-hardness of subgraph isomorphism, the non-spatial, machine neutral character of graph rewriting, and the lack of a target architecture other than the processor network, have entailed that no mainstream parallel implementations have emerged.

### 9.5 SIMULABILITY, COMPUTATION AND MATHEMATICS.

The simulation efficiency of α-Ram models has potential benefits pertaining to the relationship between computer science and mathematics. One reason why TM-based Complexity Theory has generally been restricted to problem classes described at a low level of abstraction, is because the analysis of more complex programs would be obfuscated by the arbitrary complexity overheads of RAM, and especially parallelism in the TM model. The ability of the Synchronic A-Ram to support high levels of abstraction, and to share code and data in one rewritable memory area, suggests there may be a route for unifying complexity theory with the performance evaluation analysis of real world software and hardware systems (see 5.10). A parallel efficient model also provides a theoretical basis for deriving viable high performance architectures and programming environments, and the report promotes Synchronic Engines and interlanguages in this regard.

The denotational description of a Space process is simply the succession of pairs of Synchronic A-Ram memory states and markings. The operational description centers on the idea of a module as a dynamic list of interstrings, which are replenished and consumed during runtime as control passes between states in the module's state transition system. Moreover, a module of level *n*, can in theory be compiled into a semantically equivalent, less abstract module of level *i*, where $1 \leq i < n$ (down to the level of the Earth modules, one level of abstraction away from machine code, and the denotational level).

A future paper will describe how simulation efficient compute model can provide a platform for spatialising logic and mathematics, and thereby realise mathematical argument as a form of computation. Complex structures, procedures, and proofs, could be brought into computational life as abstract data structures, virtual algorithms, and interstring-based logic programs defined in terms of a primitive formal model. Neither the λ-calculus, nor functional programming models for Von Neumann machines would be able to fulfill this role, due to the simulation inefficiency of the former, and the lack of viable example of the latter. The paper will explore in detail a more ambitious agenda than the type theoretical approach, for introducing time and computation explicitly into mathematical discourse. The intention is to allow the use of only enough tree based set theory and informal logic, for the definition of a low level, generic, synchronic model of computation.[76] The Synchronic B-Ram defined in 3.2.2 fulfills this requirement. It is proposed that the following steps are then taken:

i. The construction of a native, high level interlanguage programming environment, able to support virtual programs and abstract data types.
ii. The definition of a prenexed and interstring-based form of predicate logic called *interlogic*, together with a deterministic parallel deduction algorithm that is executable on the model.
iii. Finally, the recasting of the rest of mathematical discourse concerning structures derived from computable algebras, as interstring-based reasoning about virtual programs and abstract data types. Accordingly, a notion of compiling high level discourse into machine language is introduced.

[76] Asynchronous and non-deterministic compute models may be simulated and investigated on such a model, by employing random number generators, and ignoring or failing to take advantage of it's synchrony and determinism. It is not obvious how simulation might be achieved in the opposite direction.

Interlogic will from a novel logico-mathematical basis for alternatives to the deductive and relational database models, oriented to massively parallel deduction/transaction processing. A spatial universe of discourse that may be defined at a sufficiently primitive level, diminishes the possibility of a mathematical system harbouring implicit computational assumptions, a topic that is discussed further in the next subsection. Further, such a universe is amenable to the adoption of "programming methodologies", which may be less susceptible to negative computability and decidability phenomena encountered in conventional, non-spatial accounts of logic and mathematics.

### 9.6 IMPLICIT NOTIONS OF COMPUTATION IN MATHEMATICS.

The set theoretical and logical definition of procedures for assembling algebraic constructions in mathematics, and the constructions themselves, are normally considered to reside in a universe of discourse, which is neutral and abstract from any computational implementation. The intention of this section is to cast doubt on such a perspective. It will be argued that ordinary mathematical discourse exhibits implicit notions of computation, which are rooted in tree based formalisms and data structures with high syntactic variability(see 2.1). In particular, such discourse has a sequential nature, and is asynchronous and recursion oriented.

It was proposed earlier that only enough conventional tree based set theory and informal logic should be used, for the definition of the Synchronic B-Ram model, and then to recast all other mathematical discourse as interlogic based programming on the model. In conventional mathematics, a set construction is a tree expression with high syntactic variability, and is taken to be suitable as a general purpose structure at every level of abstraction. It has been argued however, that tree based data structures are not amenable to parallel operations. Mathematical reasoning exclusively based on set theory and tree based logic, without an explicit idea of computation, encourages a default sequentialism in certain mathematical practices:

i. A list of logically AND-ed comprehension formulas within a set definition, are normally understood to be applied as a sequence of selection filters.
ii. Proof theoretical transformations of logical formulae, are applied one at a time.
iii. Dataflows expressed in tree-based formalisms are normally evaluated in an eager fashion that identifies innermost expressions, and then evaluates them in some random order outwards, sequentially.
iv. Mathematical proofs are considered line by line, and cases in a proof are dealt with one at a time.

A form of automated reasoning can be envisioned that could perform simultaneous or parallel versions of the practices itemised above. The inability of the standard models to support parallelism efficiently may have arisen in part, from a mindset predisposed to sequential style of mathematics. Similarily, a computer scientist working in a universe of discourse that precludes any explicit notion of computation and time, that automatically precludes a notion of global time, is more inclined towards asynchronous approaches to programming environments, than synchronous ones.

The claim that such universes are recursion oriented is now considered. Tree based formalisms (various logics, λ-calculus, process algebras, etc.) do not normally reference a denotational computational environment with a memory. Expressions are generated by the free recursive application of a succession of primitive rewriting steps, and generally speaking, any branch of the syntax tree may be arbitrarily long, resulting in trees with high structural variability. Because trees are recursively assembled, it is natural to use recursive procedures in order to process them. Proofs concerning inductively defined FOPL expressions for example, employ the principle of structural induction.

The ability to describe a recursive algorithm in a programming language might improve it's programmability from one perspective. From the point of view of the runtime environment however, recursive procedures are costly to implement. Each recursive module call requires significant amounts of stack based housekeeping, affecting space requirements and running times. Implementing a recursive call in a parallel context is even more difficult, because it involves the forking of more modules, their allocation to machine resources, and scheduling and co-ordinating their activity. A mathematical procedure that is defined with a high degree of recursion, in an environment without an explicit idea of computation, need not be troubled by considerations of parallelism, or computational cost. But the fact that the recursion appears cost free, in itself encourages the use of recursion. Recursion also encourages a sequential outlook, because parallel forms of recursion are not easy to program or conceptualize.

Can one be confident that the implicitly sequentialised nature of abstract mathematics is transparent, and never obscures promising avenues of development, especially within the context of computer science? Talian and Skillicorn [2] make some interesting comments about modelling the real world with parallel computation.

> "Writing a sequential program often involves imposing an order on actions (in the real world) that are independent and could be executed concurrently. The particular order in which they are placed is arbitrary and becomes a barrier to understanding the program, since the places where the order is significant are obscured by those where it is not."

The implict sequentiality of mathematical constructions may well create a barrier for developing mathematics, for similar reasons.

**References.**

# APPENDIX A.

## SIMULATION OF A UNIVERSAL TURING MACHINE ON THE SEQUENTIAL B-RAM.

### A.1 INTRODUCTION.

A single one way infinite tape Turing Machine is the tuple $M = (Q, \Sigma, I, q_0, \delta, F)$:

i. $Q$ is a finite non-empty set of states.
ii. $\Sigma$ is the finite set of symbols including the blank symbol $\perp$.
iii. $I$ is called the input alphabet, and is a non-empty subset of $\Sigma$, where $\perp \notin \Sigma$.
iv. $q_0 \in Q$, and is called the initial state.
v. $\delta : (Q / F) \times \Sigma \rightarrow Q \times \Sigma \times \{L, R, C\}$ is a partial function, and is called the instruction table of M.
vi. $F$ is a subset of $Q$, and is called the set of final or halting states of M.

A universal Turing Machine program for the Sequential B-ram will be described, which simulates any standard, one way, single tape Turing Machine of the form

$$sM = (Q, \{0,1,\perp\}, \{0,1\}, q_0, \delta, \{success, failure\})$$

A coding scheme translates the delta function for some given sM, and tape, into data structures for the Sequential B-ram. In chapter 4, an assembly sub-language of Space for the Synchronic A-Ram, called Earth was described. For the simulation, we present pseudocode, and a program written in an Earth style language called *B-language*, which can be straightforwardly transformed into assembly and machine code for the Sequential B-Ram. There is currently no compiler for B-language, but procedures for generating the assembly and machine code versions of the program, which runs to 268 lines, are described in A.7. The assembly code is given at the end of the appendix.

There are currently implementations of neither a simulator nor compiler, available for the Sequential B-Ram. The length of the program's machine code, and even of the B-language code, entails that some effort would be required to prove that that it truly simulates the the action of the sM applied to the tape. Such a proof would be sufficient to prove a Sequential B-Ram is Turing-Computable. In the current abscence of a proof, the reader is invited to attempt to satisfy himself that the simulation program is plausible and works, by examining the various levels of code .

Our approach to programming a Sequential B-Ram $\langle p,\sigma,\mu,\rho,\eta \rangle$, is to pick a $p$ sufficiently large to store the executable code in the first block $\sigma_0$, called the *code block*, and to partition the other blocks so as to provide space for a series of indefinitely large data structures. $p = 4$ will provide a memory block with 512 16-bit registers, sufficient for the task in hand. It is possible to write a fixed length program, since the code only has to deal with a data structure describing the delta function of finite size, determined by the sM's number of states. The delta function coding scheme has to allow for the fact the set $Q$ may include an indefinitely large number of states, and therefore has to be able to represent indefinitely large integers.

There is currently no library of routines available for performing basic arithmetic operations on integers on the Sequential B-Ram. Therefore, our program relies on unary representations of integers, simple operations thereon, and a favourable coding of the delta function.

An integer variable storing $k > 0$ will be represented by a bit sequence sandwiched between two '0's, so that $b_0 b_1 b_2 \ldots b_{k+1} =$ 011...10, where the number of '1's is equal to $k$. Since $k$ is indefinitely large, it may well exceed the number of bits in a register or memory block. Therefore, for a given variable, we will pick a location ($i$,x,y) in each memory block $\sigma_i$, $i \geq 1$ to store each bit of $k$, so that $\sigma(i,x,y) = b_{i-1}$. This kind of variable in B-language is called *moveable*, because the same variable name can be invoked to address the same location in differing memory blocks, with the assistance of the cursor of the instruction in which the variable is mentioned. A variable in B-language is *fixed*, if it's storage is confined to one memory block. Moveability and fixedness will be defined more fully within the context of code segments in A.2.1.

Given an sM $=\left(Q,\{0,1,\perp\},\{0,1\},q_0,\delta,\{success, failure\}\right)$ and tape, we are obliged to translate both into Sequential B-Ram data structures. Fig A.1 depicts a memory map of how the data structures are distributed across the infinite series of memory blocks succeeding the code block. Fig A.1 lists the various code segments. (Please consult end of Appendix B if fig A.1 is unreadable).

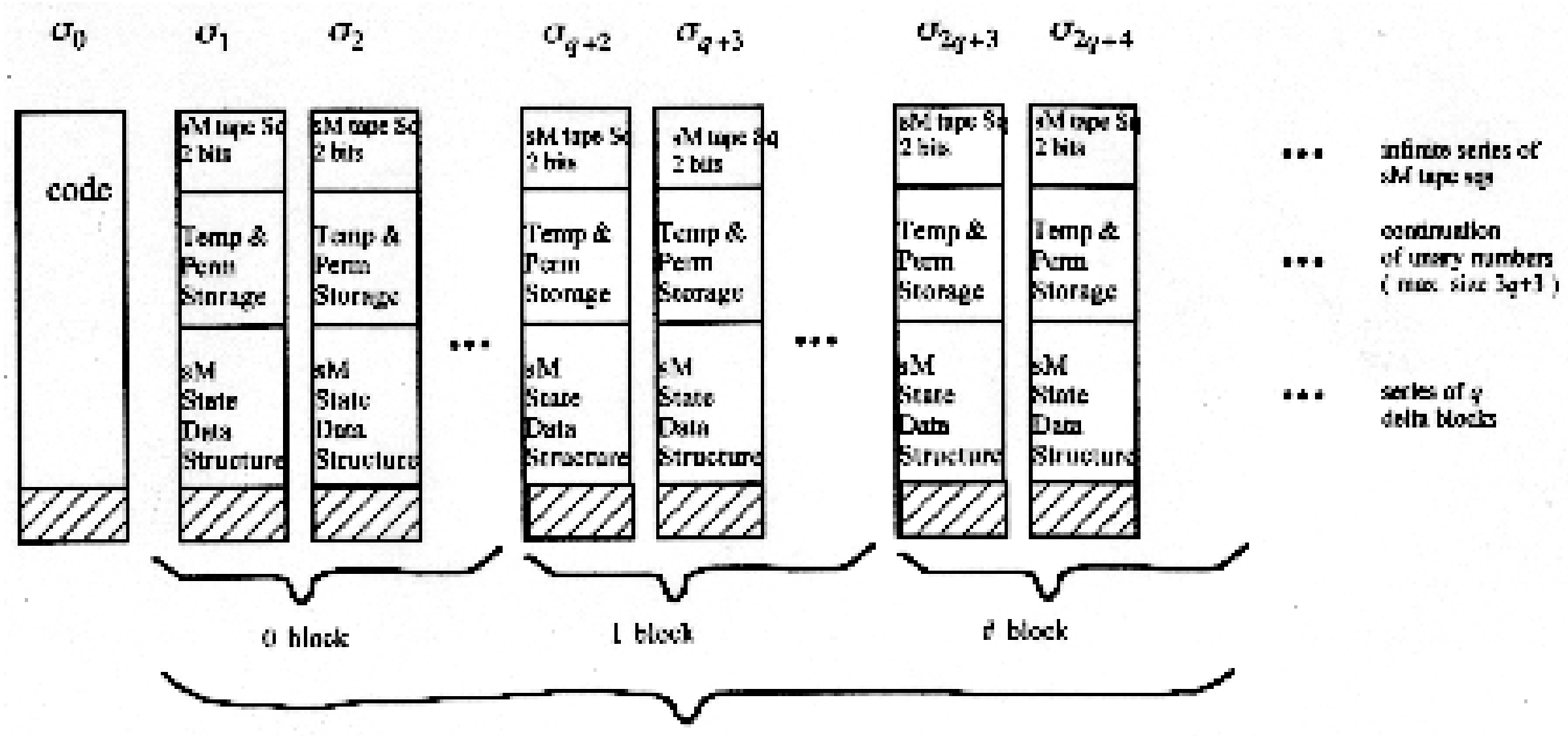

Delta Block (each occupying 3q+3 memory blocks)

Fig A.1 Memory Map for sM simulation program

Let $q=|Q|-1$, be the number of non-failing states in $Q$. The sequential B-Ram's non-code blocks are partitioned into three horizontal areas in fig A.1 with different organizations:

i. sM tape squares. 2 bits will suffice to encode a member of the sM alphabet $\{0,1,\perp\}$. If the sM tape squares are numbered $t_i$, $i \geq 0$, the content of $t_i$ is stored in a moveable 2-bit variable called *tape* in the locations $(i+1,0,1)$ and $(i+1,0,0)$.

ii. Temporary storage for recording copies of current tape square data and current state change information. The latter incorporates a moveable variable, able to store a unary integer of maximum size ($q$-1). This is followed by permanent read only data which represents unary representations of ($q$+1), (2$q$+2), and (3$q$+3).[77]

[77] The program strictly requires only $q$+1, as 2$q$+2 and 3$q$+3 can of course be implemented as repetitions of $q$+1. But extra representations simplify the program somewhat.

iii. The state change information recorded in $\delta$, is stored as a data structure in the form of a series of *delta blocks*, where each delta block occupies (3$q$+3) memory blocks, and encodes all of the information associated with a particular sM state.

### A.2.1 DELTA BLOCKS.

Delta block data occupies only the fourth and/or fifth registers of the memory blocks succeeding the code block. All delta block variables are moveable, the only fixed variables in the program relate to an sM success bit, and temporary storage. Each delta block has a bit located in it's first memory block called *finalstate*, which is set iff the delta block's sM state is the successful terminating state. A delta block is composed of three *tape blocks*; the 0-block, the 1-block, and the #-block, each occupying ($q$+1) memory blocks, which store information corresponding to whether the simulation in a particular sM state, has detected a 0,1, or # in the current sM tape square respectively. [78] The sM's states are given some ordering for the coding scheme, where initial state $q_0$ is allocated the first delta block.

If the tape block's finalstate bit is reset, it incorporates the following information:

i. The delta function is undefined for some state/tape content combinations. A bit read-only moveable variable called *statetapevalid* is set iff the current state/tape content combination is valid.
ii. If statetapevalid is set, then the tape block also records the new tape contents in the 2-bit *newsymbol* variable, and the direction in which to move the sM tape cursor is stored in the 2-bit *tapedir* variable.
iii. If the new sM state is different from the current one, then the tape block also stores the direction in which the new delta block is located in *statedir*. In addition, the unary representation of the number $j$, where $1 \le j \le q-1$, of delta blocks distant to the location of the delta block of the new sM state, is stored as a moveable integer variable *statemove*. This is the only variable name with a kind of nested moveability, in that it references each tape block, and different memory blocks within the same tape block.

[78] A tape block requires $q+1$ memory blocks, because of the wasteful, but easily programmable unary representations as series of single bits in up to $q$+1 memory blocks. By using as little as one bit per memory block, the coding scheme wastes storage on an epic scale, but we are only considering computability here

## A.3 B-LANGUAGE

Language concepts are described which improve readability of assembly code, some of which are unique to the cursor based B-Ram machine. With the exception of mvrt and mvlt instructions, there is a one to one correspondence between B-language instructions and machine code intructions. B-Language code is always confined to the $\sigma_0$ block.

i. Variables occupy some pre-declared continuous segment of a register. They may be fixed or moveable, and are discussed further in A2.2. A variable name replaces the numeric destination cell **x** of instructions **cond x y** and **wrt0 x y, wrt1 x y**. If the variable occupies a single bit, no **y** offset is needed. Otherwise, the offset is an integer referring to the relative position of the bit in the variable's register segment, rather than the absolute register offset. Instead of the usual blank space between destination cell and offset, a B-language instruction is of the form **cond variable_name.y,** or **wrt0 variable_name.y**, or **wrt1 variable_name.y**.

ii. As in current assembly languages, the operand of the jump instruction **jump x** does not refer to an absolute memory block address, but to an integer relative to the B-language program, appearing at the beginning of the instruction line. The actual register address is determined after program composition, during B-language compilation.

iii. The operand of the cursor moving instructions **mvlt x** and **mvrt x** is not an address in the code block, but is instead the name of a *code segment*, which is described in A.2.1.

### A.3.1 CODE SEGMENTS IN B-LANGUAGE.

A *code segment* is a non-contiguous subset of registers in the code block, referred to by a *codesegment_name*, whose register cursors are moved collectively. Code segments only include cond and wrt instructions, and with the exception of the setting/resetting of the B-Ram's busy bit in (0,0,0), all cond and wrt instructions are in some code segment. A statement of the form "mvrt *codesegment_name"* or "mvlt *codesegment_name"* will be shorthand for a list of mvrt/mvlt instructions, which move the cursors of all those instructions/registers associated with the *code_segment_name,* right or left respectively, once only. It is therefore possible to speak of a code segment having a cursor. A stretch of code having a particular function in the program, e.g. processing unary integers, adjusting sM tape content and cursor, implementing state changes etc., will contain a code segment whose cursor has to moved back and forth in order to perform the task. Recall $q = |Q| - 1$, is the number of non-failing states in $Q$. Figure A.2 lists the various code segments.

Two code segments are *linked* if their cursors are aligned and moved together throughout the program, with the exception of parts of a program, where only one code segment cursor is moved with respect to the other. There is one case of linked segments in the program; *shiftstatecode*, and *statecode*. Figure A.3 is a declaration of the code segment names in the simulation program.

Code segments can be used to implement iteration. To perform an iteration $k$ times, a code segment has to track the unary representation $b_0 b_1 b_2 ... b_{k+1} =$ 011...10, from the beginning until the end of the '1's, traversing one memory block for each $b_i$. In order to reuse the tracking code during program execution, it is necessary to "rewind" the code segment's cursor. This is a straightforward piece of housekeeping. The linked *statecode* code segment is concerned with accessing data in a particular delta block and tape blocks, and is linked to the segment *shiftstatecode* is concerned with accessing the unary representation of the number of delta blocks needed to be traversed to access the new state.

| Name of code segment | Description |
|---|---|
| oncerightcode | Miscellaneous code, whose initialisation only requires cursor to be moved once to the right, for main program execution and repetitions of the program. |
| tapecode | Code dealing with testing and rewriting data in sM tape square |
| statecode | Code dealing with testing data in delta block, encoding that part of a delta function relating to one state. |
| shiftstatecode | Code tracking code that deals with state change, which is linked to statecode |
| oneQ1code | First requirement for tracking the number q+1 |
| oneQ2code | Second requirement for tracking the number q+1 |
| twoQ1code | First requirement for tracking the number 2q+2 |
| twoQ2code | Second requirement for tracking the number 2q+2 |
| threeQcode | Only requirement for tracking the number 3q+3 |
| statemovecopycode | Code for making a unary copy of the number of delta block movements required for current state change |

Figure A.2 Code segment names in B-language simulation program with descriptions.

### A.3.2 VARIABLES IN CODE SEGMENTS.

Data referred to by variables in B-Language occupy blocks succeeding the code block. An instruction mentioning a variable is always included in some code segment, in order to have a convenient means of accessing the relevant memory blocks.

At the beginning of a Sequential B-Ram run, the cursor of every register points to it's own memory block. Consequently, cursors of code segments are initialised at the beginning of a B-program, to point away from the code block to access the first instance of data. A code segment cursor is *fixed*, if it's cursor always points to the same memory block throughout program execution after initialisation, and is *moveable* otherwise. Instances of variables in fixed and moveable segments are called fixed and moveable respectively. A read only variable may appear in more than one code segment, e.g. oneQ or twoQ. In any one machine cycle, the same variable name in different code segments may reference potentially different memory blocks, without confusion.

## A.4 DECLARATION OF VARIABLES FOR SIMULATION PROGRAM.

Fig A.3 tables declare variables for the B-language program, and their locations in the Sequential B-ram's memory blocks.

| Variable | Code segments | Cursor | Description |
|---|---|---|---|
| finalstate<br><br>1 bit | statecode | moveable | The finalstate bit is read only. It appears once in each delta block at ($b$,5,7), where $b = 1 + 3n(q+1)$, $0 \le n \le q-1$, and is set iff the delta block's state is a successful final state for the sM. |
| success<br><br>1 bit | once rightcode | fixed | The success bit appears only in (1,0,6). It is only examined after the B-Ram has halted. The success bit is set iff the sM run has succeeded. |
| tape<br><br>2 bits; tape.1 and tape.0 | tapecode | moveable | The tape bits represent the contents of the sM tape. They are located in ($n$,0,0) and ($n$,0,1), for $n \ge 1$.<br>i.e they occur once in every B-Ram block after $\sigma_0$.<br>$\langle 0,0 \rangle \equiv 0$, $\langle 0,1 \rangle \equiv 1$, $\langle 1,0 \rangle \equiv \perp$ |
| tapecopy<br><br>2 bits; tapecopy.1 and tapecopy.0 | once rightcode | fixed | The tapecopy bits are a copy of the contents of the current sM tape square. They are located in (1,0,2) and (1,0,3). |
| oneQ<br>1 bit<br>read only | oneQ1 code<br><br>oneQ2 code | moveable<br><br>moveable | Unary representation of $q$+1, requiring $q$+3 cells stored in locations ($n$,1,0), where $1 \le n \le q + 3$ |
| twoQ<br>1 bit<br>read only | twoQ1 code<br><br>twoQ2 code | moveable<br><br>moveable | Unary representation of 2$q$+2, requiring 2$q$+4 cells stored in locations ($n$,2,0), where $1 \le n \le 2q + 4$ |

Figure A.3(a) Variables with descriptions.

| Variable | Code segments | Cursor | Description |
|---|---|---|---|
| threeQ<br>1 bit<br>read only | threeQ code | moveable | Unary representation of $3q$+3, requiring $3q$+5 cells stored in locations ($n$,3,0), where $1 \le n \le 3q+5$ |
| statetape-valid<br><br>1 bit | statecode | moveable | This bit appears once in each tape square block in each delta block at ($a$,5,0), where $a = 1 + n(q+1)$, $0 \le n \le 3q+2$. It is set iff the delta function is defined for the current sM state tape combination. |
| tapedir<br><br>2 bits,<br>tapedir.1 &<br>tapedir.0 | statecode | moveable | These two read only bits appear once in each tape square block in each delta block at ($a$,5,2) and ($a$,5,1) for every valid state tape combination.They store the direction in which the sM tape cursor should move for current state/tape symbol combination.<br>$\langle 0,0 \rangle \equiv L$, $\langle 0,1 \rangle \equiv R$, $\langle 1,0 \rangle \equiv C$ |
| newsymbol<br><br>2 bits,<br>newsymbol.1<br>&<br>newsymbol.0 | statecode | moveable | These two read only bits appear once in each tape square block in each delta block at ($a$,5,4) and ($a$,5,3) for every valid state tape combination.They store the new tape symbol to be written into the current sM tape square for current state/tape symbol combination.<br>$\langle 0,0 \rangle \equiv 0$, $\langle 0,1 \rangle \equiv 1$, $\langle 1,0 \rangle \equiv$# |
| statedir<br><br>2bits,<br>statedir.1 &<br>statedir.0 | statecode | moveable | These two read only bits appear once in each tape square block in each delta block at ($a$,5,6) and ($a$,5,5) for every valid state tape combination.They store the direction in which the cursors for statecode and shiftstatecode have to be moved, for current state/tape symbol combination.<br>$\langle 0,0 \rangle \equiv L$, $\langle 0,1 \rangle \equiv R$, $\langle 1,0 \rangle \equiv C$ |
| statetape-valid<br><br>1 bit | statecode | moveable | This bit appears once in each tape square block in each delta block at ($a$,5,0), where $a = 1 + n(q+1)$, $0 \le n \le 3q+2$. It is set iff the delta function is defined for the current sM state tape combination. |

Figure A.3 (b) Variables with descriptions.

| Variable | Code segments | Cursor | Description |
|---|---|---|---|
| tapedir<br><br>2 bits, tapedir.1 & tapedir.0 | statecode | moveable | These two read only bits appear once in each tape square block in each delta block at ($a$,5,2) and ($a$,5,1) for every valid state tape combination.They store the direction in which the sM tape cursor should move for current state/tape symbol combination.<br>$\langle 0,0\rangle \equiv L$, $\langle 0,1\rangle \equiv R$, $\langle 1,0\rangle \equiv C$ |
| newsymbol<br><br>2 bits, newsymbol.1 & newsymbol.0 | statecode | moveable | These two read only bits appear once in each tape square block in each delta block at ($a$,5,4) and ($a$,5,3) for every valid state tape combination.They store the new tape symbol to be written into the current sM tape square for current state/tape symbol combination. $\langle 0,0\rangle \equiv 0$, $\langle 0,1\rangle \equiv 1$, $\langle 1,0\rangle \equiv \#$ |
| statedir<br><br>2bits, statedir.1 & statedir.0 | statecode | moveable | These two read only bits appear once in each tape square block in each delta block at ($a$,5,6) and ($a$,5,5) for every valid state tape combination.They store the direction in which the cursors for statecode and shiftstatecode have to be moved, for current state/tape symbol combination.<br>$\langle 0,0\rangle \equiv L$, $\langle 0,1\rangle \equiv R$, $\langle 1,0\rangle \equiv C$ |
| statedir-copy<br><br>1 bit | onceright code | fixed | This bit is a copy of the direction in which the cursors for statecode and shiftstatecode have to be moved. Only one bit is required to code L or R, C is not needed. It is located in (1,0,4). $\langle 0,0\rangle \equiv L$, $\langle 0,1\rangle \equiv R$ |
| statemove<br><br>1 bit | statecode | moveable | Unary representation in each tape block of number of delta blocks that statecode and shiftstatecode have to be moved, in the direction indicated by statedir, in order to point to new delta block. Cells stored from ($a$,6,0) to ($b$,6,0), where<br>$a = 1 + n(q+1)$, $b = 1 + 2n(q+1)$, $0 \le n \le 3q+2$.<br>Statemove has a kind of nested moveability, in that it is the only moveable variable, whose cursor points to different memory blocks within the same tape block. |
| dummy<br>1 bit | onceright code | fixed | Dummy variable in (1,0,15), surplus to requirements, that plays no role in program |
| statemovecopy<br>1 bit | statemove copycode | moveable | Copy of unary representation of statemove, stored in ($n$,4,0), $1 \le n \le q-1$ |

Figure A.3 (c) Variables with descriptions.

## A.5 PREAMBLE TO PSEUDOCODE.

Recall the sM's initial state $q_0$ is stored in the first delta block. After initialising the code segment cursors, the cursors of *statecode* and *shiftstatecode*, which deal with implementing an sMdelta instruction, point to the 0-block of the first delta block. The program's main loop begins by testing the *finalstate* bit, if which indicates if the current state is the terminating state *success*. If it is, the *success* bit is set, and the program terminates. If it is not, the contents of the current sM tape contents are tested, copied into temporary storage, and the code segments *statecode* and *shiftstatecode* are shifted 0, $q$+1 or 2$q$+2 times to point to the 0-block, 1-block or # block respectively, of the current delta block.

The validity of the current state/tape symbol combination are tested in the relevant tape block. If invalid, the SM has halted in the *failure* state, the success bit is reset, and the program terminates. ( The failure state does not require a delta block). If valid, the sM current tape content and cursor are adjusted accordingly. If the new state is different from the current one, it's location is represented by the moveable variable *statedir*, and by the moveable variable *statemove*, which stores the unary number of delta blocks that the cursors of the code segments *statecode* and *shiftstatecode*, have to be shifted by, to access the new state. The code segment *shiftstatecode* tracks and copies of the contents of the *statemove* variable into temporary storage in *statemovecopy* variable, and it's cursor is rewound. The cursors for *statecode* and *shiftstatecode* are then moved 3q+3 times for each unary 1-bit stored in *statemovecopy*, left or right as appropriate.

Finally, the main loop ends by realigning, if necessary, the code segments *statecode* and *shiftstatecode*, to point to the 0-block of the new state, by testing the copy of the previous sM tape square. The first level of pseudocode appears in the next section

## A.5.1 Pseudocode.

Initialise cursors for tapecode, statecode, shiftstatecode, oneq1code, twoq1code, oneq2code, twoq2code, and statemovecopycode blocks of code.

```
Begin loop
        If state is success (finalstate==1)
                Exit successfully

    // Make copy of tape square and move statecode and shiftstatecode to appropriate tape block
        Else if tape square  = #   // tape.1 =1 & tape.0=0
                Make copy of tape square
                Move statecode and shiftstatecode 2q+2 blocks to the right.
                Rewind twoq1code
        Else if tape Square = 1   // tape.1 =0 & tape.0=1
                Make copy of tape square
                Move statecode and shiftstatecode q+1 blocks to the right.
                Rewind oneq1code
        Else if Tape Square = 0        // tape.1 =0 & tape.0=0
                Make copy of tape square only

    // Having accessed the relevant tape block, now perform turing machine step
        If state/symbol combination is valid   // statetapevalid=1

                Copy new tape symbol into tape square.

                        // Make copy of tape direction and move tape cursor if required.
                If tape cursor move  = R
                        Make tape direction copy
                        Move tapecode 1 block to the right.
                Else If tape cursor move = L
                        Make tape direction copy
                        Move tapecode 1 block to the left.
                Else If tape cursor move = C
                        Make tape direction copy only

                        // Move statecode and shiftstatecode to new delta block if required.

                If state change direction != C
                        Make state directioncopy
                        Copy statemove into statemovecopy
                        Rewind shiftstatecode
                        If state direction copy = R
                                Move statecode and shiftstatecode 3q+3 blocks to the right for every '1'  in
                                                                                statemovecopy unary.
                        Else if state direction copy = L
                                Move statecode and shiftstatecode 3q+3 blocks to the left for every '1' in
                                                                                statemovecopy unary.

                        Move statemovecopycode once to the right to realign

                        //  Now realign statecode and shiftstatecode back to '0' tape block if required.
                If tapecopy = #
                        Move statecode and shiftstatecode 2q+2 blocks to the left.
                        Rewind twoq2code
                        Repeat loop
                Else If tapecopy  = 1
                        Move statecode and shiftstatecode q+1 blocks to the left.
                        Rewind oneq2code
                        Repeat loop
                Else If tapecopy  = 0
                        No movement of statecode and shiftstatecode needed.
                        Repeat loop
        Else
                Exit unsuccessfully.
```

## A.6.1 PREAMBLE TO B-LANGUAGE CODE.

Recall the Sequential B-Ram is a sequential machine. Consequently B-Language code has to be interpreted a little differently from Earth. The execution of a wrt, mvlt, or mvrt instruction is always followed by the execution of the next instruction. Jump instructions do not have an offset, and activate only one instruction, numbered relative to the program. The variable name *busy* refers to the B-Ram's busy bit in (0,0,0).

An instruction is declared to be in a code segment by having the code segment name mentioned in brackets after the instruction, and before a comment on the same line, e.g.

```
.
.
.
cond finalstate      (statecode) // ......
.
.
```

## A.6.2 B-LANGUAGE CODE.

```
        wrt1 busy                          // Signal the Sequential B-Ram is busy
        mvrt oncerightcode                 //  code that only needs to be moved once       -14 lines-
        mvrt tapecode               // code dealing with TM tape square, marked (-a-)      -6 lines-
        mvrt statecode              //  code dealing with delta table, marked (-b-)  -8 lines-
        mvrt shiftstatecode         // code tracking code dealing with delta state change  -2 lines-
        mvrt oneQ1code              //first requirement for q+1, has to be moved twice to access first '1'
        mvrt oneQ1code                     //  of '0111...110' of q+1  -2 lines-
        mvrt oneQ2code              //  second  requirement for q+1, has to be moved twice to access first '1'
        mvrt oneQ2code                     // of '0111...110' of 2q+2  -2 ls-
        mvrt twoQ1code              //   first requirement for 2q+2          "            -2 lines-
        mvrt twoQ1code
        mvrt twoQ2code              //  second  requirement for 2q+2         "                -2 lines-
        mvrt twoQ2code
        mvrt threeQcode             //only requirement for 3q+3                           -2 lines-
        mvrt threeQcode
        mvrt statemovecopycode    // statemovecopycode for recording of number of delta block movements -3 ls-
        wrt0 statemovecopy   (statemovecopycode)  //  initialise first '0' bit of " 01111110 "copy of record of
                                    // number of block  movements
        mvrt statemovecopycode    // move again to be able to write first '1' of "011110"

1       cond finalstate     (statecode)         //  test first for final state
        jump 2                                  // repeat loop
        wrt1 success        (oncerightcode)     //indicate success
        wrt0 busy                               // halt

2       cond tape.1          (tapecode)         // examine tape square, test second bit of tape pair
        jump 6                                // must be 0 or 1
        wrt1 tapecopy.1    (oncerightcode)     //performing copy
        wrt0 tapecopy.0    (oncerightcode) // tape= #, so move statecode and shiftstatecode 2q+2 blocks to the right

3       cond twoQ    (twoQ1code)    // now tracking 2q+2 in row 2, must be rewound at the end
```

```
        jump 4
        mvrt statecode              // move statecode and shiftstatecode rightwards
        mvrt  shiftstatecode
        mvrt twoQ1code              // also move right rewind mech
        jump 3
4       mvlt twoQ1code              // rewind twoQ1code
        cond twoQ       (twoQ1code) // rewind mechanism
        jump 5      // rewind complete, now shift forward once to get first 1, then check validity
        jump 4
5       mvrt twoQ1code
        jump 11         //  delta block shift from tape content and rewind completed, check validity

6       cond tape.0     (tapecode)  // we have eliminated #, is tape square 0 or 1?
        jump 10                 // tape is 0, make copy, and no movement of statecode needed, then check validity
        wrt0 tapecopy.1    (oncerightcode)  //    tape = one, copy contents
        wrt1 tapecopy.0    (oncerightcode)  //  tape = 1, hence delta block must be moved q+1 blocks to the right
7       cond oneQ       (oneQ1code) // now tracking q+1 in row 1, must be rewound at the end
        jump 8
        mvrt statecode              // move statecode and shiftstatecode rightwards
        mvrt  shiftstatecode
        mvrt oneQ1code              // also move right rewind mech
        jump 7
8       mvlt oneQ1code              // rewind twoQ
        cond oneQ       (oneQ1code)  // rewind mechanism
        jump 9      // rewind complete, now shift forwad once to get first 1, then check validity
        jump 8
9       mvrt oneQ1code
        jump 11    //  delta block shift from tape content completed, check validity

10      wrt0 tapecopy.1    (oncerightcode) //  tape = 0, so copy
        wrt0 tapecopy.0    (oncerightcode)  //     and no movement required

        //  we have now navigated to correct tape column in state block, check state validity
11      cond statetapevalid   (statecode)  //
        jump 12     // illegal state, halt in failure
        jump 13 // proceed to application of TM delta function


12      wrt0 success   (oncerightcode) //  indicate failure
        wrt0 busy          //halt

13      cond newsymbol.1  (statecode) //copy new tape symbol second bit into main tape
        jump 15
        wrt1 tape.1    (tapecode)
14      cond newsymbol.0  (statecode)  //copy new tape symbol first bit into main tape
        jump 16
        wrt1 tape.0  (tapecode)
        jump 17             // now jump to shifting cursor of code dealing with tape contents
15      wrt0 tape.1   (tapecode)
        jump 14
16      wrt0 tape.0  (tapecode)

17      cond tapedir.1    (statecode) //  copy delta shift direction
        jump 18                     // L or R
        jump 20             // tapecode shift = C, so jump to state change
18      cond tapedir.0      (statecode) //  move tape cursor left or right?
        jump 19             // left
        mvrt tapecode       // right
        jump 20             //jump to state change
19      mvlt tapecode
20      cond statedir.1    (statecode) //state change, check direction
        jump 21                     // L or R
        jump 31         //  no movement, jump to realign statecode and shiftstatecode back to zero column
21      wrt0 dummy  (oncerightcode)    // harmless dummy instruction, surplus to requirements
        cond statedir.0     (statecode)
        jump 22
        wrt1 statedircopy   (oncerightcode)
        jump 23             // jump to make copy
```

```
22      wrt0 statedircopy   (oncerightcode)


23      mvrt shiftstatecode // make copy of delta block state movement,  mvrt code for recording of delta block shift
        cond statemove    (shiftstatecode)
        jump 24                               // now rewind shiftstatecode
        mvrt statemovecopycode // move once to write the first '1' of 01111110
        wrt1 statemovecopy (statemovecopycode) // we wont rewind as we can use rewind for delta block shift, no
                                                          // need to write final '0'
        jump 23
24      mvlt shiftstatecode
        cond statemove  (shiftstatecode)  //  rewind shiftstatecode to beginning '0' of "0111..10" of delta block col.
        jump 25
        jump 24

25      cond statemovecopy      (statemovecopycode) //  right, so move statecode and shiftstatecode delta blocks to
                                      // the right, 3q+3 times for every unary bit in statemovecopy
        jump 31                       // exit loop, move statemovecopycode one to the right, then realign statecode
                                      // and shiftstatecode back to zero column, if necessary
        mvlt statemovecopycode // move once left to read the rightmost '1' of 01111110

26      cond threeQ    (threeQcode) // now tracking 3q+3 in row 3, must be rewound at the end
        jump 29                       // jump to rewind threeQ
        cond statedircopy   (oncerightcode) //  now test direction
        jump 27             // move statecode and shiftstatecode leftwards
        mvrt statecode                  // move statecode and shiftstatecode rightwards
        mvrt  shiftstatecode
        jump 28
27      mvlt statecode                  // move statecode and shiftstatecode rightwards
        mvlt  shiftstatecode
28      mvrt threeQcode               // also move right rewind mech
        jump 26
29      mvlt threeQcode                 // rewind threeQ
        cond threeQ          (threeQcode)// rewind mechanism
        jump 30      // rewind complete, now shift forwad once to get first 1, then check validity
        jump 29
30      mvrt threeQcode   // realign to leftmost '1' of 0111..10
        jump 25   // repeat loop . statecode and shiftstatecode have been moved (3q+3)x delta block state change
                                      //  movement

31      mvrt statemovecopycode  // realign statemovecopycode to point to leftmost 1 of 0111..10
32      cond tapecopy.1    (oncerightcode) // realign statecode and shiftstatecode back to zero column, if necessary
        jump 36                                   //  0 or 1
33      cond twoQ (twoQ2code)     // #, so move statecode and shiftstatecode 2q+2 blocks to the left,
                            //must be rewound at the end
        jump 34
        mvlt statecode                  // move statecode and shiftstatecode rightwards
        mvlt  shiftstatecode
        mvrt twoQ2code                // also move right rewind mech
        jump 33
34      mvlt twoQ2code                  // rewind twoQ2
        cond twoQ            (twoQ2code)  // rewind mechanism
        jump 35      // rewind complete, now shift forward once to get first 1, then check validity
        jump 34
35      mvrt twoQ2code
        jump 1              //  delta block shift from tape content and rewind completed, repeat main loop

36      cond tapecopy.0    (oncerightcode) // realign statecode and shiftstatecode back to zero column, if necessary
        jump 1   // tape was 0, no realignment needed, so repeat main loop
37      cond oneQ      (oneQ2code) //tape was 1, so move statecode and shiftstatecode q+1 blocks to the left,
                            // must be rewound at the end
        jump 38
        mvlt statecode                  // move statecode and shiftstatecode rightwards
        mvlt  shiftstatecode
        mvrt oneQ2code                // also move right rewind mech
        jump 37
38      mvlt oneQ2code                  // rewind twoQ2
        cond oneQ            (oneQ2code)   // rewind mechanism
```

```
        jump 39     // rewind complete, now shift forward once to get first 1, then check validity
        jump 38
39      mvrt oneQ2code
        jump 1           //  delta block shift from tape content and rewind completed, repeat main loop
```

### A.6.2 ASSEMBLY LANGUAGE CODE.

There are 6 steps to transform B-Language into assembly code:

i. Eliminate blank lines and remove comments.
ii. Let #Codesegment_name be the number of instructions in Codesegment_name. Insert (#Codesegment_name - 1 ) new lines after each **mvrt Codesegment_name** and **mvlt Codesegment_name** in the program. This establishes the final register address for each B-language instruction, including mvrt/mvlt instructions.
iii. Assemble list of register addresses for each instruction in Codesegment_name, for each code segment. Insert a list of mvrt/mvlt instructions for each code segment register into the space occupied by the **mvrt/lt Codesegment_name,** and the new lines introduced in (ii).
iv. Replace variable names and relative offsets by numerical register addresses and absolute offsets.
v. Replace relative jump numbers by absolute ones.

The resulting assembly code is listed below. The final transformation from assembly into machine code is not included here. It is effected by replacing each assembly instruction with the 3 bit representation of the opcode, followed by the 9 bit representation of the destination cell, followed by the 4 bit representation of the offset if the instruction has one, or ‘0000’ otherwise.

CODE BLOCK

| REGISTER | INSTRUCTION |
|---|---|
| 1 | wrt1 0 0 |
| 2 | mvrt 61 |
| 3 | mvrt 65 |
| 4 | mvrt 66 |
| 5 | mvrt 92 |
| 6 | mvrt 93 |
| 7 | mvrt 117 |
| 8 | mvrt 118 |
| 9 | mvrt 122 |
| 10 | mvrt 155 |
| 11 | mvrt 158 |
| 12 | mvrt 160 |
| 13 | mvrt 182 |
| 14 | mvrt 219 |
| 15 | mvrt 244 |
| 16 | mvrt 63 |
| 17 | mvrt 90 |
| 18 | mvrt 126 |
| 19 | mvrt 129 |
| 20 | mvrt 131 |
| 21 | mvrt 133 |
| 22 | mvrt 59 |
| 23 | mvrt 119 |
| 24 | mvrt 124 |
| 25 | mvrt 127 |
| 26 | mvrt 134 |
| 27 | mvrt 137 |
| 28 | mvrt 152 |
| 29 | mvrt 156 |
| 30 | mvrt 163 |
| 31 | mvrt 172 |
| 32 | mvrt 94 |
| 33 | mvrt 111 |
| 34 | mvrt 94 |
| 35 | mvrt 111 |
| 36 | mvrt 248 |
| 37 | mvrt 265 |
| 38 | mvrt 248 |
| 39 | mvrt 265 |
| 40 | mvrt 67 |
| 41 | mvrt 84 |
| 42 | mvrt 67 |
| 43 | mvrt 84 |
| 44 | mvrt 221 |
| 45 | mvrt 238 |
| 46 | mvrt 221 |
| 47 | mvrt 238 |
| 48 | mvrt 180 |
| 49 | mvrt 210 |
| 50 | mvrt 180 |
| 51 | mvrt 210 |
| 52 | mvrt 55 |
| 53 | mvrt 168 |
| 54 | mvrt 175 |
| 55 | wrt0 4 0 |
| 56 | mvrt 55 |
| 57 | mvrt 168 |
| 58 | mvrt 175 |
| 59 | cond 5 7 |
| 60 | jump 64 |
| 61 | wrt1 0 6 |
| 62 | wrt0 0 0 |
| 63 | cond 0 1 |
| 64 | jump 90 |
| 65 | wrt1 0 3 |
| 66 | wrt0 0 2 |
| 67 | cond 2 0 |
| 68 | jump 82 |
| 69 | mvrt 59 |
| 70 | mvrt 119 |
| 71 | mvrt 124 |
| 72 | mvrt 127 |
| 73 | mvrt 134 |
| 74 | mvrt 137 |
| 75 | mvrt 152 |
| 76 | mvrt 156 |
| 77 | mvrt 163 |
| 78 | mvrt 172 |
| 79 | mvrt 67 |
| 80 | mvrt 84 |
| 81 | jump 67 |
| 82 | mvlt 67 |
| 83 | mvlt 84 |
| 84 | cond 2 0 |
| 85 | jump 87 |
| 86 | jump 82 |
| 87 | mvrt 67 |
| 88 | mvrt 84 |
| 89 | jump 119 |
| 90 | cond 0 0 |
| 91 | jump 117 |
| 92 | wrt0 0 3 |
| 93 | wrt1 0 2 |
| 94 | cond 1 0 |
| 95 | jump 109 |
| 96 | mvrt 59 |
| 97 | mvrt 119 |
| 98 | mvrt 124 |
| 99 | mvrt 127 |
| 100 | mvrt 134 |
| 101 | mvrt 137 |
| 102 | mvrt 152 |
| 103 | mvrt 156 |
| 104 | mvrt 163 |
| 105 | mvrt 172 |
| 106 | mvrt 94 |
| 107 | mvrt 111 |
| 108 | jump 94 |
| 109 | mvlt 94 |
| 110 | mvlt 111 |
| 111 | cond 1 0 |
| 112 | jump 114 |
| 113 | jump 109 |
| 114 | mvrt 94 |
| 115 | mvrt 111 |
| 116 | jump 119 |
| 117 | wrt0 0 3 |
| 118 | wrt0 0 2 |
| 119 | cond 5 0 |
| 120 | jump 122 |
| 121 | jump 124 |
| 122 | wrt0 0 6 |
| 123 | wrt0 0 0 |
| 124 | cond 5 4 |
| 125 | jump 131 |
| 126 | wrt1 0 1 |
| 127 | cond 5 3 |
| 128 | jump 133 |
| 129 | wrt1 0 0 |
| 130 | jump 134 |
| 131 | wrt0 0 1 |
| 132 | jump 127 |
| 133 | wrt0 0 0 |
| 134 | cond 5 2 |
| 135 | jump 137 |
| 136 | jump 152 |
| 137 | cond 5 1 |
| 138 | jump 146 |
| 139 | mvrt 63 |
| 140 | mvrt 90 |
| 141 | mvrt 126 |
| 142 | mvrt 129 |
| 143 | mvrt 131 |
| 144 | mvrt 133 |
| 145 | jump 152 |
| 146 | mvlt 63 |
| 147 | mvlt 90 |
| 148 | mvlt 126 |
| 149 | mvlt 129 |
| 150 | mvlt 131 |
| 151 | mvlt 133 |
| 152 | cond 5 6 |
| 153 | jump 155 |
| 154 | jump 216 |
| 155 | wrt0 0 15 |
| 156 | cond 5 5 |
| 157 | jump 160 |
| 158 | wrt1 0 4 |
| 159 | jump 161 |
| 160 | wrt0 0 4 |
| 161 | mvrt 163 |
| 162 | mvrt 172 |
| 163 | cond 6 0 |
| 164 | jump 170 |
| 165 | mvrt 55 |
| 166 | mvrt 168 |
| 167 | mvrt 175 |
| 168 | wrt1 4 0 |
| 169 | jump 161 |
| 170 | mvlt 163 |
| 171 | mvlt 172 |
| 172 | cond 6 0 |
| 173 | jump 175 |
| 174 | jump 170 |
| 175 | cond 4 0 |

```
176    jump 216
177    mvlt 55
178    mvlt 168
179    mvlt 175
180    cond 3 0
181    jump 208
182    cond 0 4
183    jump 195
184    mvrt 59
185    mvrt 119
186    mvrt 124
187    mvrt 127
188    mvrt 134
189    mvrt 137
190    mvrt 152
191    mvrt 156
192    mvrt 163
193    mvrt 172
194    jump 205
195    mvlt 59
196    mvlt 119
197    mvlt 124
198    mvlt 127
199    mvlt 134
200    mvlt 137
201    mvlt 152
202    mvlt 156
203    mvlt 163
204    mvlt 172
205    mvrt 180
206    mvrt 210
207    jump 180
208    mvlt 180
209    mvlt 210
```

```
210    cond 3 0
211    jump 213
212    jump 208
213    mvrt 180
214    mvrt 210
215    jump 175
216    mvrt 55
217    mvrt 168
218    mvrt 175
219    cond 0 3
220    jump 244
221    cond 2 0
222    jump 236
223    mvlt 59
224    mvlt 119
225    mvlt 124
226    mvlt 127
227    mvlt 134
228    mvlt 137
229    mvlt 152
230    mvlt 156
231    mvlt 163
232    mvlt 172
233    mvrt 221
234    mvrt 238
235    jump 221
236    mvlt 221
237    mvlt 238
238    cond 2 0
239    jump 241
240    jump 236
241    mvrt 221
242    mvrt 238
243    jump 59
```

```
244    cond 0 2
245    jump 59
246    cond 1 0
247    jump 261
248    mvlt 59
249    mvlt 119
250    mvlt 124
251    mvlt 127
252    mvlt 134
253    mvlt 137
254    mvlt 152
255    mvlt 156
256    mvlt 163
257    mvlt 172
258    mvrt 248
259    mvrt 265
260    jump 246
261    mvlt 248
262    mvlt 265
263    cond 1 0
264    jump 266
265    jump 261
266    mvrt 248
267    mvrt 265
268    jump 59
```

CODE ENDS

## APPENDIX B.
## MISCELLANEOUS EARTH AND SPACE PROGRAMS.

All of the report's program examples, excepting B.6, are stored in Spatiale's module library, which may be downloaded via a link on www.isynchronise.com.

### B.1 ADDER32: A SERIAL STYLE 32-BIT ADDER FOR UNSIGNED INTEGERS.

The *adder32* module is not an Earth equivalent of a ripple adder, because it reuses the same code segment representing a full adder for all 32 input bit triplets. The module employs another un-named re-usable code segment, which acts as an incrementer which modifies the offsets of four instructions, that enable the reuse of the fulladder. Two instructions (2,3), copy the individual bits of the inputs into the full adder, and the other two (39, 40) transfers the resulting sum bit into individual bits of the output, and the carry out into carry in. The incrementer treats the offset region of instruction 2 as it's *ioput* (input and output). Upon completing the computation, the module is ready for re-use.

Storage names are self-explanatory, with the exception of `FAfb`, which is a bit which is set when a full adder cycle has completed, and `INCfb` , which is set when the incrementation cycle is completed. The `EQ31` bit is used by the incrementer to check if the last offset number 31 has been reached.

The module has no replicative structures, although one could have been used in the incrementer. The module compiles into 138 lines of code, and is quite space efficient. The tradeoff is in running time, which peaks at 736 cycles.

```
NAME: adder32;
BITS: cin private, a private, b private, s private, co private, FAfb private, INCfb
private, EQ31 private, carryout output, busy private;
REGS: addend input, addendum input, sum output;
TIME: 674-736 cycles;

        wrt1 busy
        jump 1 3
1       jump 2 0 // 4 instructions, 2 check 1st two input bits,  3rd checks carry, 4th is
                 //  carry thread
        jump 3 0
        jump 4 0
        jump 5 0      // carry thread
```

```
2       cond addend.0 // offset is modified, so that ith bit is copied from addend to FA.a
        wrt0 a
        wrt1 a
3       cond addendum.0   //offset modified, so that ith bit copied from addendum to FA.b
        wrt0 b
        wrt1 b
4       cond co   // copy FA.cout to FA.cin
        wrt0 cin
        wrt1 cin
5       jump 6 0              // carry thread
6       jump 7 1              //carry thread
7       jump 14 0             //start FA
8       cond FAfb             //check if FA is finished
        jump 8 0
        jump 9 3              // go to next bit cycle
9       jump 10 0             // check if inc has finished
        jump 37 0             // start inc
        wrt0 FAfb
        cond s                       // sum results of FA sum
39      wrt0 sum.0     // offsets modified. these two lines are numbered
40      wrt1 sum.0     // for bracketed addressing only
10      cond INCfb     // is inc finished?
        jump 10 0
        cond EQ31   //now check if bit 31 is set
        jump 1 3
        cond co               // now examine FA cout
        jump 11 1
        jump 12 1
11      wrt0 carryout
12      jump 13 1
        wrt1 carryout
13      wrt0 co
        wrt0 busy
14      cond cin   //this is FA
        jump 17 0
        cond a
        jump 18 0
        cond b
        jump 15 2
        jump 16 2
19      wrt0 s
20      wrt1 FAfb // why is this here?
        wrt0 co
        wrt1 s
15      wrt0 s
16      wrt1 co
        wrt1 FAfb
        wrt1 s
17      cond a
        jump 38 0
18      cond b
        jump 20 2
```

```
        jump 15 2
38      cond b
        jump 19 2
        jump 20 2    // this is where FA ends
37      cond [2] 0     //  INC begin
        jump 21 5
        jump 22 4
21      wrt1 [2] 0
        wrt1 [3] 0
        wrt1 [39] 0
        wrt1 [40] 0   //...break
        wrt1 INCfb // wrt1 to set incfinish bit
        jump 36 0  // to reset finish bit..
22      wrt0 [2] 0
        wrt0 [3] 0
        wrt0 [39] 0
        wrt0 [40] 0
        cond [2] 1
        jump 23 5
        jump 24 4
23      wrt1 [2] 1
        wrt1 [3] 1
        wrt1 [39] 1
        wrt1 [40] 1
        wrt1 INCfb
        jump 36 0
24      wrt0 [2] 1
        wrt0 [3] 1
        wrt0 [39] 1
        wrt0 [40] 1
        cond [2] 2
        jump 25 5
        jump 26 4
25      wrt1 [2] 2
        wrt1 [3] 2
        wrt1 [39] 2
        wrt1 [40] 2
        wrt1 INCfb
        jump 36 0
26      wrt0 [2] 2
        wrt0 [3] 2
        wrt0 [39] 2
        wrt0 [40] 2
        cond [2] 3
        jump 27 5
        jump 28 4
27      wrt1 [2] 3
        wrt1 [3] 3
        wrt1 [39] 3
        wrt1 [40] 3
        wrt1 INCfb
        jump 36 0
```

```
28      wrt0 [2] 3
        wrt0 [3] 3
        wrt0 [39] 3
        wrt0 [40] 3
        cond [2] 4
        jump 29 5
        jump 30 6  // jump to exit sequence
29      wrt1 [2] 4
        wrt1 [3] 4
        wrt1 [39] 4
        wrt1 [40] 4
        wrt1 INCfb
        jump 36 0  //
30      wrt0 [2] 4       // exit sequence
        wrt0 [3] 4
        wrt0 [39] 4
        wrt0 [40] 4
        wrt1 INCfb
        wrt1 EQ31
        jump 33 0  // final reset
36      jump 31 0  // this only resets inc finish bit
31      jump 32 0
32      wrt0 INCfb  // this only resets inc finish bit 2 cycles after bit is set
33      jump 34 0
34      jump 35 2
35      wrt0 FAfb
        wrt0 INCfb
        wrt0 EQ31
        endc
```

## B.2 BARREL SHIFT FOR 32-BIT REGISTER .

The module's activity can be divided into two phases. The first phase identifies the number of register shifts represented in `input`, and in a sense implements a 5x32 decoder. There is not much benefit to be had by implementing the decoder as a direct translation of a logic gate circuit, because another code segment would be required to examine the decoder output, to select the correct barrel shift. Consequently the first phase simply checks the five bits of `input` sequentially. The outcome of the first phase, is used to guide program control to select which barrel shift to perform, by jumping to one of the 31 code segments created by the final nested replicative structure.

As a numex cannot specify a function exotic enough to represent the relevant jump destination for the outcome of the first phase, the programmer was obliged to partially compile the module, and to subsequently fill in the correct jump destinations by hand. The module compiles into 2,190 lines of code, but has a fast completion time of only 8 cycles.

```
NAME: barrelshift32;
BITS: busy private;
OFSTS: input input;
REGS: shiftinput input, shiftoutput output;
TIME: 8-8 cycles;

       wrt1 busy
       jump 1 1
1      jump 33 0
       jump 2 0

2      cond input.0    //  begin decoding
       jump 3 0
       jump 18 0             // first split

3      cond input.1                 // xxxx0
       jump 4 0
       jump 5 0

4      cond input.2                 // xxx00
       jump 6 0
       jump 7 0

5      cond input.2                 // xxx10
       jump 8 0
       jump 9 0

6      cond input.3                 // xx000
       jump 10 0
       jump 11 0

7      cond input.3                 // xx100
       jump 12 0
       jump 13 0

8      cond input.3                 // xx010
       jump 14 0
       jump 15 0

9      cond input.3                 // xx110
       jump 16 0
       jump 17 0

10     cond input.4                 // x0000
       wrt0 input.7                 //dummy instruction exit 0
```

```
        jump 439 0            // exit 16. first jump destination filled in by hand


11      cond input.4                 // x1000
        jump 259 0            // exit 8
        jump 555 0                                                //exit 24

12      cond input.4                 // x0100
        jump 146 0                      //exit 4
        jump 505 0                                                //exit 20


13      cond input.4                 // x1100
        jump 357 0                                                // exit 12

        jump 589 0                                                // exit 28

14      cond input.4                 // x0010
        jump 82     0                            //exit 2
        jump 474 0                                                //exit 18


15      cond input.4                 // x1010
        jump 310 0                   // exit 10
        jump 574 0                                                //exit 26

16      cond input.4                 // x0110
        jump 204 0                   //exit 6
        jump 532 0                                           // exit 22


17      cond input.4                 // x1110
        jump 400 0                                                //exit 14
        jump 600 0                                                //exit 30

18      cond input.1                 // xxxx1
        jump 19 0
        jump 20 0

19      cond input.2                 // xxx01
        jump 21 0
        jump 22 0

20      cond input.2                 // xxx11
        jump 23 0
        jump 24 0

21      cond input.3                 // xx001
        jump 25 0
        jump 26 0

22      cond input.3                 // xx101
```

```
       jump 27 0
       jump 28 0

23     cond input.3              // xx011
       jump 29 0
       jump 30 0

24     cond input.3              // xx111
       jump 31 0
       jump 32 0

25     cond input.4              // x0001
       jump 49 0                                     // exit 1
       jump 457 0                                    // exit 17


26     cond input.4              // x1001
       jump 285 0                                    // exit 9
       jump 565 0                                    //exit 25

27     cond input.4              // x0101
       jump 175 0                //exit 5
       jump 519 0                                                   //exit 21


28     cond input.4              // x1101
       jump 379 0                                                   //exit 13
       jump 595 0                                                   //exit 29

29     cond input.4              // x0011
       jump 114      0                                              //exit 3
       jump 490 0                                                   //exit 19


30     cond input.4              // x1011
       jump 334 0                                                   //exit 11
       jump 582 0                                                   //exit 27

31     cond input.4              // x0111
       jump 232 0                                           // exit 7
       jump 544 0                                           //exit 23


32     cond input.4              // x1111
       jump 420 0                                           //exit 15
       jump 604 0                //exit 31, final jump destination filled in by hand

// end of code for first phase

33 jump 34 0   // carry stream which waits for first phase completion
34 jump 35 0   // jump sequence should have written as an replicative structure,
35 jump 36 0  // it’s a long story.
```

```
36 jump 37 0
37 jump 38 0
38 jump 39 0
39 jump 40 0
40 jump 41 0
41 jump 42 0
42 jump 43 0
43 jump 44 0
44 jump 45 0
45 jump 46 0
46 jump 47 0
47 wrt0 busy

       <0;i;31>{ // code that resets bits that have been shifted left
48     wrt0 shiftoutput.i
}


<0;i;30>{   // second phase

(49+i)  jump (50+i) (31-i)


(50+i) jump 48 i // note that offset is variable to reset appropriate bit range


              <0;k;(30-i)>{
                     jump (51+i+k) 0
              }


              <0;j;(30-i)>{

(51+i+j)      cond shiftinput.j
              wrt0 shiftoutput.(1+i+j)
              wrt1 shiftoutput.(1+i+j)

              }
}

endc
```

## B.3 Programmable Register Copy .

The module `progcopyreg` is a maximally parallel implementation of a programmable copy for all 32 bits in a register, and compiles into the largest Aramaic code in this report at 5,249 lines of code. The module is a natural extension of `progcopybit` that was described in

4.5.

The module accepts two destination addresses as 25 bit values. It is the instructions with linenames `(6+3*i)`,`(7+3*i)`, and `(8+3*i)` which perform the second phase's copying, and whose destination cells are written over by the first phase of the module.

```
NAME: progcopyreg;
META: 2;
BITS: busy private, mbsy private;
REGS: source input, target input;
TIME: 7-9 cycles;

1       wrt1 busy                    // first phase, modifies code
        jump 3 1

2       wrt1 mbsy                    // second phase, executes modified code
        jump 4 1


3       jump 26 0
        jump 9 1

4       jump 22 0
        jump 5 31

<0;i;31>{

5       jump (6+3*i) 0

}

<0;i;31>{

(6+3*i)        cond 0 i       // the destination cell of these 3 instructions
(7+3*i)        wrt0 0 i       //  are written over by the first phase
(8+3*i)        wrt1 0 i

}

9       jump 10 24
        jump 14 24

<0;j;24>{      // jump to begin processing source of programmable copy

10      jump (11+j) 0

}

<0;j;24>{      // begin processing source of programmable copy

(11+j)  cond source.j
        jump (12+2*j) 31
        jump (13+2*j) 31

}       // test each bit of source, to copy into the "0" of (6+3*i) cond 0 i


<0;j;24>{
```

```
       <0;i;31>{
(12+2*j)   wrt0 [(6+3*i)] (5+j) // source bit was reset, so reset destination bit
       }
       <0;i;31>{
(13+2*j)   wrt1 [(6+3*i)] (5+j)  // source bit was set, so set destination bit
       }
}

<0;j;24>{  // jump to begin processing target of programmable copy
14     jump (15+3*j) 0
}
<0;j;24>{ //  begin processing target of programmable copy
(15+3*j)       cond target.j
               jump (16+3*j) 1
               jump (17+3*j) 1

(16+3*j)       jump (18+4*j) 31
               jump (19+4*j) 31

(17+3*j)       jump (20+4*j) 31
               jump (21+4*j) 31
}

<0;j;24>{
               <0;i;31>{
(18+4*j)            wrt0 [(7+3*i)] (5+j)
               }
               <0;i;31>{
(19+4*j)           wrt0 [(8+3*i)] (5+j)
               }
               <0;i;31>{
(20+4*j)           wrt1 [(7+3*i)] (5+j)
               }
               <0;i;31>{
(21+4*j)          wrt1 [(8+3*i)] (5+j)
               }
}
22     jump 23 0             // exit carry stream for second phase
```

```
23      jump 24 0
24      jump 25 0
25      wrt0 mbsy

26      jump 27 0       // exit carry stream for first phase
27      jump 28 0
28      jump 29 0
29      jump 30 0
30      jump 31 0
31      jump 32 0
32      wrt0 busy
        endc
```

Figure B.3 Programmable Register Copy .

## B.4 32-BIT SUBTRACTOR IN SPACE.

The level one SUBTRACT32 module in Figure B.4, is essentially a subtractor version of the serial adder ADDER32 module described iin 6.13.5. It again uses the address operator, meta submodules, cell identifiers, co-active parallelism, activate column parallelism, and the skip instruction. The module employs an array of 3 programmable copy bit modules described in 4.5, and a 5-bit incrementer described in 4.2.2. The incrementer is used to modify the offsets of the first phase inputs of the PCOPYBIT's submodules, whose second phase loads bit triplets into a full subtractor.

```
module SUBTRACT32{

  storage{
       unsigned subtrahend input;
       unsigned minuend input;
       unsigned result output;
       BIT borrow output;
    };

    submodules{
       FULLSUBTRACTOR fullsub;
       inceq5bit inceq;
       PCOPYBIT pcopy[3];
    };

    time: 2714-2714 cycles;

    code{

        1: #0 -> pcopy[0].input0.offst          :: _pcopy[0]    :: jump (2,1) :;
           #0 -> pcopy[1].input0.offst             _pcopy[1]
           #0 -> pcopy[2].input1.offst             _pcopy[2]

           &addend -> pcopy[0].input0.destn
           &fullsub.x -> pcopy[0].input1
           &addendum -> pcopy[1].input0.destn
           &fullsub.y -> pcopy[1].input1
           &fullsub.D -> pcopy[2].input0
           &sum -> pcopy[2].input1.destn


        2:  -pcopy[0]  :: _fullsub  :: fullsub.Bout -> fullsub.Bin :: -pcopy[2] :;
            -pcopy[1]

        3:  _inceq :: inceq.ioput -> pcopy[0].input0.offst   :: skip(2) :: jump (4,0) :;
                      inceq.ioput -> pcopy[1].input0.offst
                      inceq.ioput -> pcopy[2].input1.offst

        4: _pcopy[0]  :: cond_inceq.Eq31 (2,1) (5,0) :;
           _pcopy[1]
           _pcopy[2]

        5:  fullsub.Bout -> borrow :: HALT :;

       };

};
```

Figure B.4 Serial 32-bit Subtractor in Space.

## B.5 8-BIT CARRY ADDER IN SPACE.

A carry adder implemented as an ASIC design, employs a large number of logic gates in a wide and deep dataflow, where the dataflow depth is constrained by the maximum propagation delay allowed by the length of the chip's clock cycle. The processing of the dataflow implemented as a parallel equivalent Synchronic A-Ram program, requires many clock cycles, because Synchronic A-Ram operations are finer-grained even than logic gates. A Space carry adder that had a separate piece of code for each logic gate in the ASIC design, would unnecessarily waste machine resources, because the model is a programmable platform where code may be reused.

For implementational reasons, the storage and submodule declarations in the program in Figure B.5 are somewhat wasteful, in that the input types are *unsigned* rather than BYTE, and the submodules *and* and and *xor* operate on 32 bit values. The inputs must be entered as integers below 255, and their summation is delivered in around 50 cycles to the 8 rightmost bits in *sum*. There is a separate *carrybit* output. The module employs an array of four reusable submodules *pif[4]*, implementing a boolean function $pif:\{0,1\}^4 \rightarrow \{0,1\}^2$. The *pif* function is involved in calculating the carry propagate and the carry generate bits for each pair of binary coefficients, and implements the following boolean specification: $pif(a_0,a_1,b_0,b_1) = \left( (a_0 \wedge b_0), \left( (a_1 \wedge b_0) \vee b_1 \right) \right)$.

## B.6 SOLVER FOR UPPER-TRIANGULAR SYSTEM OF LINEAR EQUATIONS IN SPACE.

The program in figure B.6 is draft code, because the submodules for floating point arithmetic operations have not been written yet. The module *uppertriangle* resolves the $x_i$ values in the upper triangular system of linear equations represented below, in order to exemplify how a reasonably complex piece of parallel programming might be handled in Space.

$$a_{n-1,0}x_0 + a_{n-1,1}x_1 + a_{n-1,2}x_2 + \quad \ldots\ldots \quad a_{n-1,n-1}x_{n-1} = b_{n-1}$$

.

$$a_{2,0}x_0 + a_{2,1}x_1 + a_{2,2}x_2 = b_2$$

$$a_{1,0}x_0 + a_{1,1}x_1 = b_1$$

$$a_{0,0}x_0 = b_0$$

The storage declaration "float S[7] private;" is an array of partial sums that are maintained for each $x_i$ , $1 \le i \le 7$, until the main loop no longer requires them. The other storage declarations are self-explanatory, as are the floating point submodule declarations.

The declaration "TwoDimfloatarrayreturn Areturn[7];" involves a submodule that can return the value of an element of a two dimensional array of floating point numbers, in this case the array $\left[a_{ij}\right]$. The submodule *Onedimfloatarray* is hopefully self-explanatory. The declaration "Onedimfloatarraywrite Xwrite;" provides a submodule for writing a value into an array, in this case $\left[x_i\right]$. The *compare* submodule returns a 1 value if the two input integers are identical. The contractions statement "mult[0-7] ~ sub[0-7];", instructs the compiler that all identical sub-submodules may be pairwise shared between the mult and sub submodule arrays.

The program begins by activating lines 2 and 3. Line 2 is the carry line and primes the PJUMP, inc and dec submodules for the first main loop iteration, and calculates the value of $x_0$. Line 3 primes the array return submodules for $\left[a_{ij}\right]$, and loads the $\left[b_i\right]$ values into the subtractor array.The main loop is in lines 4,5, and 6. Lines 4 and 5 are co-active (notwithstanding the point made in 7.7.3). The activation of the second phase of PJUMP in line 4, activates the grow construct in line 5, which performs the operations needed for the final calculation of an element of $\left[x_i\right]$ in line 6.

```
module CARRYADDER8{  // 8 bit carry adder


  storage{

        unsigned addend input;
        unsigned addendum input;
        unsigned sum output;
        BIT carrybit output; //
        REG temp0 private;
        REG temp1 private;
    };

    submodules{
        pairwiseand32 and;
        pairwisexor32 xor;
        pifunction pif[4];
        boolfun bool;   // ((in0 and in1) or in2 )
    };

    replications{ i/inc, 2*, 2*+1};
    time: 2714-2714 cycles;

    code{

        1: addend -> and.input1    ::  __xor  :: jump (2,0) :;
           addendum -> and.input2      _and
           addend -> xor.input1
           addendum -> xor.input2


        2.1: and.output.i/2* -> pif[i].a0     :: __pif[i] :>  2: deep<i=0;i<=3;inc> (3,0) :;
             xor.output.i/2* -> pif[i].a1
             and.output.i/2*+1 -> pif[i].b0
             xor.output.i/2*+1 -> pif[i].b1

        3:  and.output.0 -> xor.input2.0    :: __pif[0]  :: pif[0].out1 -> xor.input2.2     ::   __pif[0]   ::  pif[0].out1 -> xor.input2.4   :: _xor  :: xor.output -> sum     :: HALT :;
                                                _pif[1]     pif[1].out0 -> pif[0].a0             _pif[1]
            pif[0].out1 -> xor.input2.1         _pif[2]     pif[1].out1 -> pif[0].a1             _pif[2]       xor.output -> xor.input1
                                                                                                                                             _bool    bool.out -> carrybit
                                                _pif[3]     and.output.4 -> pif[0].b0            _pif[3]       xor.output.7 -> bool.in0

                                                            xor.output.4 -> pif[0].b1                          pif[3].out1 -> bool.in1
                                                                                                               and.output.7 -> bool.in2

            pif[0].out0 -> pif[0].a0                        pif[1].out1 -> xor.input2.3                         pif[1].out1 -> xor.input2.5
            pif[0].out1 -> pif[0].a1                        pif[1].out0 -> pif[1].a0
            and.output.2 -> pif[0].b0                       pif[1].out1 -> pif[1].a1
            xor.output.2 -> pif[0].b1                       temp0.2 -> pif[1].b0
                                                            temp1.2 -> pif[1].b1
            pif[0].out0 -> pif[1].a0
            pif[0].out1 -> pif[1].a1                        pif[1].out0 -> pif[2].a0                             pif[2].out1 -> xor.input2.6
                                                            pif[1].out1 -> pif[2].a1
            pif[1].out0 -> pif[1].b0                        pif[2].out0 -> pif[2].b0
            pif[1].out1 -> pif[1].b1                        pif[2].out1 -> pif[2].b1


            pif[2].out0 -> pif[2].a0                        pif[1].out0 -> pif[3].a0                             pif[3].out1 -> xor.input2.7
            pif[2].out1 -> pif[2].a1                        pif[1].out1 -> pif[3].a1
            and.output.6 -> pif[2].b0                       pif[3].out0 -> pif[3].b0
            xor.output.6 -> pif[2].b1                       pif[3].out1 -> pif[3].b1

            pif[2].out0 -> temp0.2
            pif[2].out1 -> temp1.2

            pif[2].out0 -> pif[3].a0
            pif[2].out1 -> pif[3].a1
            pif[3].out0 -> pif[3].b0
            pif[3].out1 -> pif[3].b1


    };
};
```

Figure B.5 8 bit Carry Adder.

```
module uppertriangle{

   storage{

     float A[8][8] input;
     float B[8] input;
     float X[8] output;
     float S[7] private;

   };

   submodules{

     floatsubtract32 sub[7];
     floatmult32 mult[7];
     floatdivide32 divide;
     TwoDimfloatarrayreturn Areturn[7];
     Onedimfloatarrayreturn Sreturn;
     Onedimfloatarraywrite Xwrite;
     inc inc;
     dec dec;
     compare compare;
     PJUMP PJUMP;

   };

  contractions{

    mult[0-7] ~ sub[0-7];

  };

  replications{i / inc};
  time: 0-0 cycles;

  code{

     1: jump (2,1) :;


     2: B[0] -> divide.in0            ::  __divide   :: divide.out -> X[0]          :: jump (4,0) :;  // find x0
        A[0][0] -> divide.in1             _PJUMP(5)     divide.out -> Xwrite.value                      // Xwrite.value also used as temp storage
        #7 -> PJUMP.offset
        #7 -> dec.ioput
        #0 -> inc.ioput

        X[[0]] -> Xwrite.address
        S[[0]] -> Sreturn.address
        #7 -> compare.a
        #1 -> compare.b

     3.1:  #i/inc -> Areturn[i].index0       :>    3: deep<i=0;i<=6; inc > () :;
           A[[0]] -> Areturn[i].address
           B[i] -> sub[i].in0

     4: _inc       :: dec.ioput -> PJUMP.offset :: _PJUMP(5) :;
        _dec
        -PJUMP(5)

  5.1: inc.ioput -> Areturn[i].index1 :: _Areturn[i] :: Areturn[i].value -> mult[i].in0 :: _mult[i] :: mult[i].out -> sub[i].in1 :: _sub[i] :: sub[i].out -> S[i]       :: subhalt(5) :> 5: grow<i=7;i<=1;dec> (6,0) :;
                                                        Xwrite.value -> mult[i].in1                                                           sub[i].out -> sub[i].in0

    6: inc.ioput -> Areturn[0].index0 :: _Areturn[0] :: Areturn[0].value -> divide.in1 :: _divide  :: divide.out -> Xwrite.value :: _Xwrite :: cond_compare.aEQb (4,0) (7,0) :;
       inc.ioput -> Areturn[0].index1    _Sreturn       Sreturn.value -> divide.in0       _compare
       inc.ioput -> Sreturn.index
       inc.ioput -> Xwrite.index
       inc.ioput -> compare.b

     7: HALT:;

      };
};
```

Figure B.7 Space Solver for Upper Triangular System of Equations.

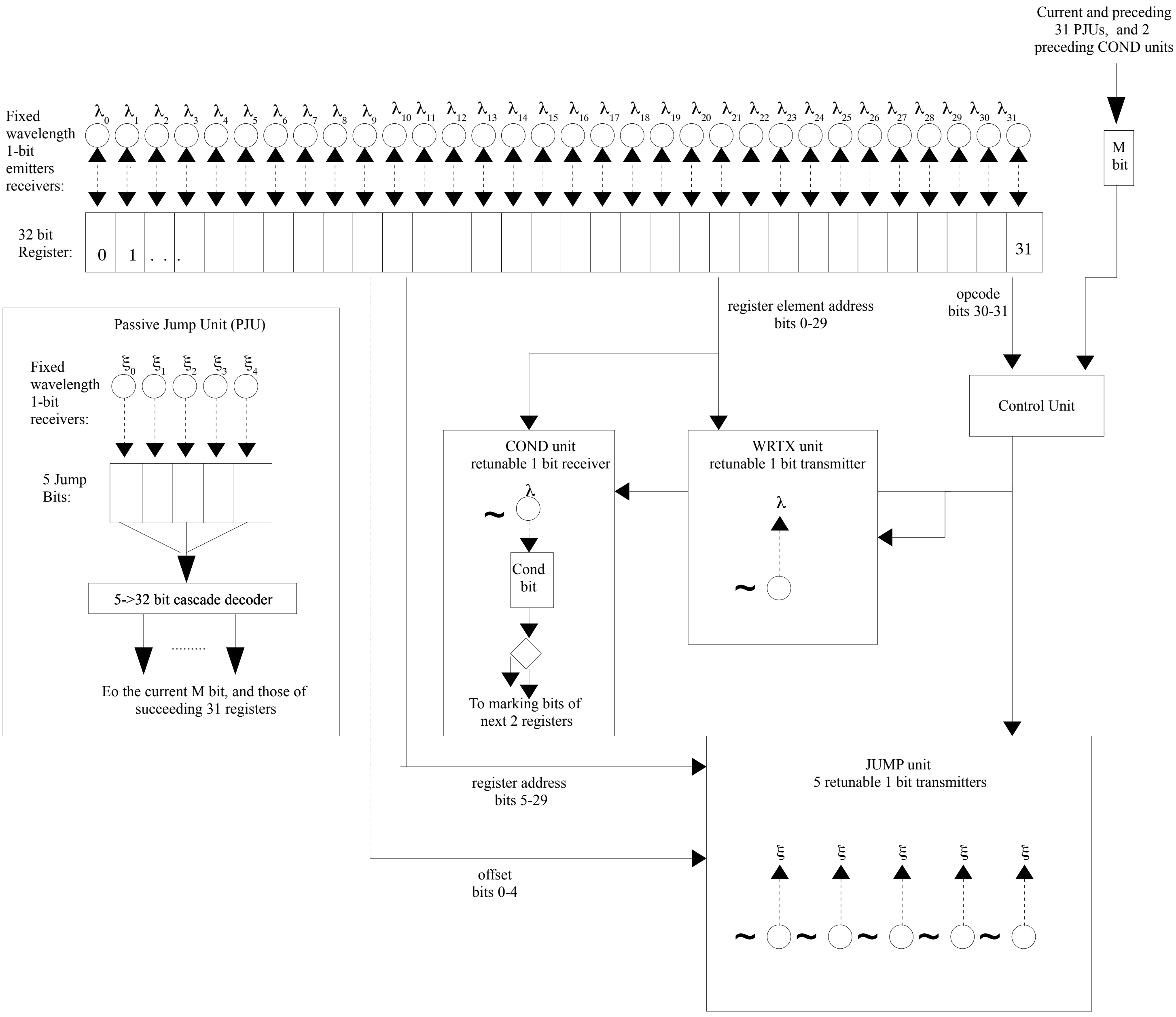

Figure 8.1 Block diagram for a < 5,σ,{1,2},η > register.

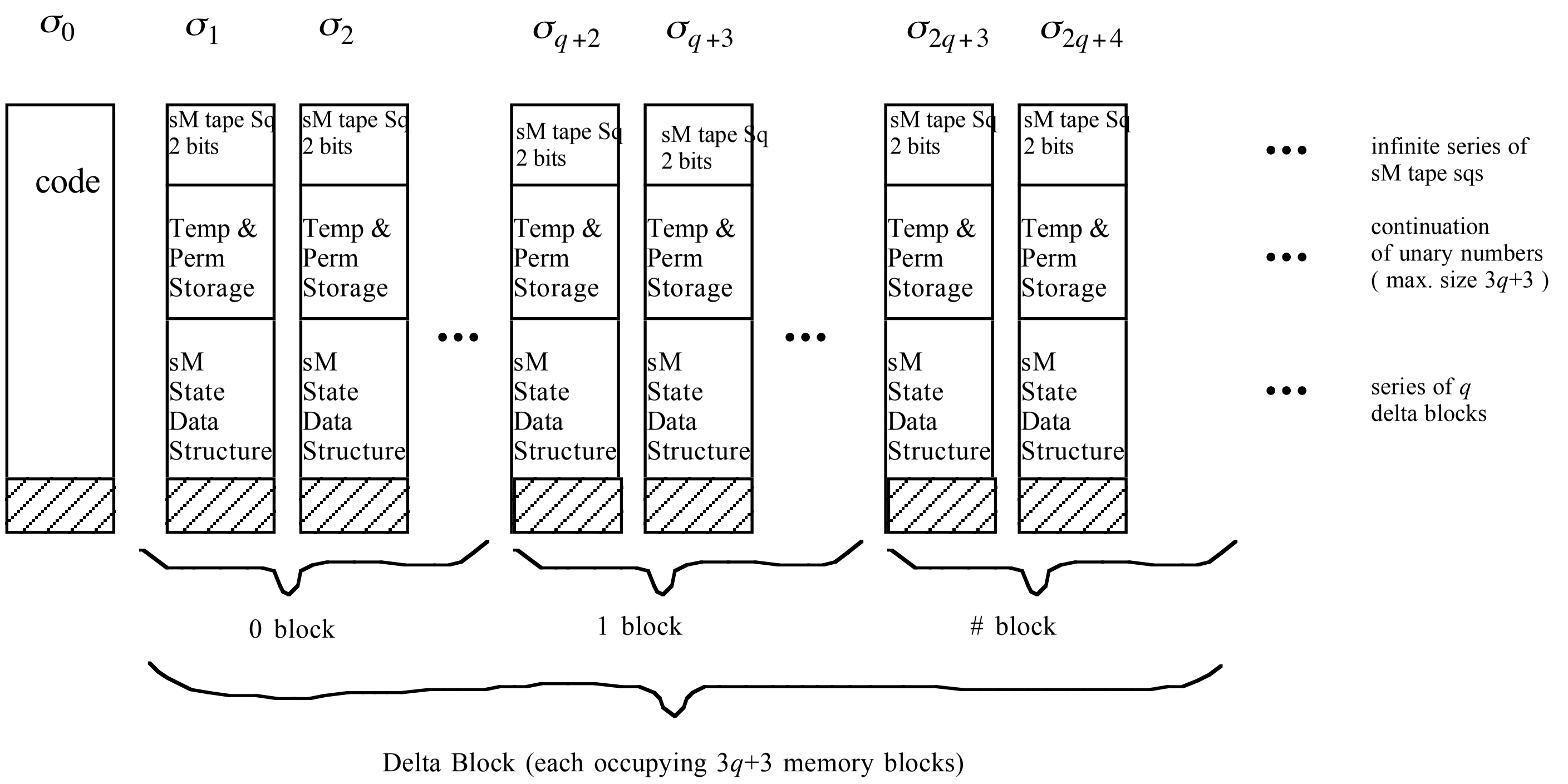

Fig A.1 Memory Map for sM simulation program